\documentclass[aps,prd,10pt,floatfix,twocolumn,superscriptaddress,nofootinbib]{revtex4-2}
\usepackage{amsmath,graphicx,hyperref}
\providecommand{\apj}{ApJ}\providecommand{\apjl}{ApJL}\providecommand{\apjs}{ApJS}
\providecommand{\nat}{Nature}
\providecommand{\prd}{Phys.~Rev.~D}\providecommand{\prl}{Phys.~Rev.~Lett.}
\providecommand{\prc}{Phys.~Rev.~C}

\providecommand{\jcap}{JCAP}

\providecommand{\cqg}{Class.~Quantum Grav.}
\newcommand{\msun}{M_\odot}
\newcommand{\nsat}{n_{\rm sat}}
\newcommand{\cs}{c_s}

\usepackage{xcolor}
\newcommand{\llmfig}{ Generated with LLM assistance per Appendix~\ref{app:llm}.}

\providecommand\result[1]{\ifcsname #1\endcsname\csname #1\endcsname\else\textbf{??#1??}\fi}

\expandafter\providecommand\csname AdCsSqTail\endcsname{}\expandafter\renewcommand\csname AdCsSqTail\endcsname{\ensuremath{0.4}}
\expandafter\providecommand\csname AdEnergyGradRatio\endcsname{}\expandafter\renewcommand\csname AdEnergyGradRatio\endcsname{\ensuremath{1.5}}
\expandafter\providecommand\csname AdFdStep\endcsname{}\expandafter\renewcommand\csname AdFdStep\endcsname{\ensuremath{0.03}}
\expandafter\providecommand\csname AdKsatGradRatio\endcsname{}\expandafter\renewcommand\csname AdKsatGradRatio\endcsname{\ensuremath{3.5}}
\expandafter\providecommand\csname AdMassHi\endcsname{}\expandafter\renewcommand\csname AdMassHi\endcsname{\ensuremath{1.4}}
\expandafter\providecommand\csname AdMassLo\endcsname{}\expandafter\renewcommand\csname AdMassLo\endcsname{\ensuremath{1.2}}
\expandafter\providecommand\csname ArchiveScaleDraws\endcsname{}\expandafter\renewcommand\csname ArchiveScaleDraws\endcsname{\ensuremath{10^{6}}}
\expandafter\providecommand\csname ArxivNestedFullLam\endcsname{}\expandafter\renewcommand\csname ArxivNestedFullLam\endcsname{\ensuremath{309}}
\expandafter\providecommand\csname ArxivNestedFullLamSd\endcsname{}\expandafter\renewcommand\csname ArxivNestedFullLamSd\endcsname{\ensuremath{90}}
\expandafter\providecommand\csname ArxivNestedFullRadius\endcsname{}\expandafter\renewcommand\csname ArxivNestedFullRadius\endcsname{\ensuremath{11.7}}
\expandafter\providecommand\csname ArxivNestedFullRadiusSd\endcsname{}\expandafter\renewcommand\csname ArxivNestedFullRadiusSd\endcsname{\ensuremath{0.4}}
\expandafter\providecommand\csname ArxivSymmetryFig\endcsname{}\expandafter\renewcommand\csname ArxivSymmetryFig\endcsname{\ensuremath{12}}
\expandafter\providecommand\csname BaseStructuralRejectPct\endcsname{}\expandafter\renewcommand\csname BaseStructuralRejectPct\endcsname{\ensuremath{0}}
\expandafter\providecommand\csname BenchAdaptiveEssFour\endcsname{}\expandafter\renewcommand\csname BenchAdaptiveEssFour\endcsname{\ensuremath{705,\ 737}}
\expandafter\providecommand\csname BenchAdaptiveEssOne\endcsname{}\expandafter\renewcommand\csname BenchAdaptiveEssOne\endcsname{\ensuremath{1031}}
\expandafter\providecommand\csname BenchAdaptiveEssThree\endcsname{}\expandafter\renewcommand\csname BenchAdaptiveEssThree\endcsname{\ensuremath{1093}}
\expandafter\providecommand\csname BenchAdaptiveEssTwo\endcsname{}\expandafter\renewcommand\csname BenchAdaptiveEssTwo\endcsname{\ensuremath{851}}
\expandafter\providecommand\csname BenchArchiveDraws\endcsname{}\expandafter\renewcommand\csname BenchArchiveDraws\endcsname{\ensuremath{10^{6}}}
\expandafter\providecommand\csname BenchArchiveEssFour\endcsname{}\expandafter\renewcommand\csname BenchArchiveEssFour\endcsname{\ensuremath{1.29}}
\expandafter\providecommand\csname BenchArchiveEssOne\endcsname{}\expandafter\renewcommand\csname BenchArchiveEssOne\endcsname{\ensuremath{2349}}
\expandafter\providecommand\csname BenchArchiveEssThree\endcsname{}\expandafter\renewcommand\csname BenchArchiveEssThree\endcsname{\ensuremath{11.84}}
\expandafter\providecommand\csname BenchArchiveEssTwo\endcsname{}\expandafter\renewcommand\csname BenchArchiveEssTwo\endcsname{\ensuremath{216}}
\expandafter\providecommand\csname BenchArchiveMaxWtFour\endcsname{}\expandafter\renewcommand\csname BenchArchiveMaxWtFour\endcsname{\ensuremath{87.83}}
\expandafter\providecommand\csname BenchArchiveMaxWtFourShort\endcsname{}\expandafter\renewcommand\csname BenchArchiveMaxWtFourShort\endcsname{\ensuremath{87.8}}
\expandafter\providecommand\csname BenchArchiveMaxWtOne\endcsname{}\expandafter\renewcommand\csname BenchArchiveMaxWtOne\endcsname{\ensuremath{0.14}}
\expandafter\providecommand\csname BenchArchiveMaxWtThree\endcsname{}\expandafter\renewcommand\csname BenchArchiveMaxWtThree\endcsname{\ensuremath{20.77}}
\expandafter\providecommand\csname BenchArchiveMaxWtTwo\endcsname{}\expandafter\renewcommand\csname BenchArchiveMaxWtTwo\endcsname{\ensuremath{1.40}}
\expandafter\providecommand\csname BenchCrustEps\endcsname{}\expandafter\renewcommand\csname BenchCrustEps\endcsname{\ensuremath{150}}
\expandafter\providecommand\csname BenchCrustK\endcsname{}\expandafter\renewcommand\csname BenchCrustK\endcsname{\ensuremath{147}}
\expandafter\providecommand\csname BenchCrustP\endcsname{}\expandafter\renewcommand\csname BenchCrustP\endcsname{\ensuremath{3}}
\expandafter\providecommand\csname BenchDefensiveFrac\endcsname{}\expandafter\renewcommand\csname BenchDefensiveFrac\endcsname{\ensuremath{0.1}}
\expandafter\providecommand\csname BenchDeltaRate\endcsname{}\expandafter\renewcommand\csname BenchDeltaRate\endcsname{\ensuremath{0.5}}
\expandafter\providecommand\csname BenchDeltaScale\endcsname{}\expandafter\renewcommand\csname BenchDeltaScale\endcsname{\ensuremath{150}}
\expandafter\providecommand\csname BenchDim\endcsname{}\expandafter\renewcommand\csname BenchDim\endcsname{\ensuremath{11}}
\expandafter\providecommand\csname BenchDimFour\endcsname{}\expandafter\renewcommand\csname BenchDimFour\endcsname{\ensuremath{4.76}}
\expandafter\providecommand\csname BenchDimFourErr\endcsname{}\expandafter\renewcommand\csname BenchDimFourErr\endcsname{\ensuremath{0.10}}
\expandafter\providecommand\csname BenchDimOne\endcsname{}\expandafter\renewcommand\csname BenchDimOne\endcsname{\ensuremath{3.12}}
\expandafter\providecommand\csname BenchDimOneErr\endcsname{}\expandafter\renewcommand\csname BenchDimOneErr\endcsname{\ensuremath{0.07}}
\expandafter\providecommand\csname BenchDimThree\endcsname{}\expandafter\renewcommand\csname BenchDimThree\endcsname{\ensuremath{4.16}}
\expandafter\providecommand\csname BenchDimThreeErr\endcsname{}\expandafter\renewcommand\csname BenchDimThreeErr\endcsname{\ensuremath{0.08}}
\expandafter\providecommand\csname BenchDimTwo\endcsname{}\expandafter\renewcommand\csname BenchDimTwo\endcsname{\ensuremath{3.42}}
\expandafter\providecommand\csname BenchDimTwoErr\endcsname{}\expandafter\renewcommand\csname BenchDimTwoErr\endcsname{\ensuremath{0.07}}
\expandafter\providecommand\csname BenchEtaDrop\endcsname{}\expandafter\renewcommand\csname BenchEtaDrop\endcsname{\ensuremath{10}}
\expandafter\providecommand\csname BenchGaussFrac\endcsname{}\expandafter\renewcommand\csname BenchGaussFrac\endcsname{\ensuremath{0.9}}
\expandafter\providecommand\csname BenchHpdArchiveDraws\endcsname{}\expandafter\renewcommand\csname BenchHpdArchiveDraws\endcsname{\ensuremath{1}}
\expandafter\providecommand\csname BenchHpdLevelPct\endcsname{}\expandafter\renewcommand\csname BenchHpdLevelPct\endcsname{\ensuremath{95}}
\expandafter\providecommand\csname BenchHpdVolume\endcsname{}\expandafter\renewcommand\csname BenchHpdVolume\endcsname{\ensuremath{1.2\times10^{-6}}}
\expandafter\providecommand\csname BenchIntervals\endcsname{}\expandafter\renewcommand\csname BenchIntervals\endcsname{\ensuremath{8}}
\expandafter\providecommand\csname BenchLogitShift\endcsname{}\expandafter\renewcommand\csname BenchLogitShift\endcsname{\ensuremath{0.5}}
\expandafter\providecommand\csname BenchMassHi\endcsname{}\expandafter\renewcommand\csname BenchMassHi\endcsname{\ensuremath{2}}
\expandafter\providecommand\csname BenchMassLast\endcsname{}\expandafter\renewcommand\csname BenchMassLast\endcsname{\ensuremath{20}}
\expandafter\providecommand\csname BenchMassLo\endcsname{}\expandafter\renewcommand\csname BenchMassLo\endcsname{\ensuremath{1}}
\expandafter\providecommand\csname BenchMassStep\endcsname{}\expandafter\renewcommand\csname BenchMassStep\endcsname{\ensuremath{0.05}}
\expandafter\providecommand\csname BenchMcmcChains\endcsname{}\expandafter\renewcommand\csname BenchMcmcChains\endcsname{\ensuremath{32}}
\expandafter\providecommand\csname BenchMcmcDim\endcsname{}\expandafter\renewcommand\csname BenchMcmcDim\endcsname{\ensuremath{4.73}}
\expandafter\providecommand\csname BenchMcmcDimErr\endcsname{}\expandafter\renewcommand\csname BenchMcmcDimErr\endcsname{\ensuremath{0.04}}
\expandafter\providecommand\csname BenchMcmcDraws\endcsname{}\expandafter\renewcommand\csname BenchMcmcDraws\endcsname{\ensuremath{60\,000}}
\expandafter\providecommand\csname BenchMcmcMaxRhat\endcsname{}\expandafter\renewcommand\csname BenchMcmcMaxRhat\endcsname{\ensuremath{1.010}}
\expandafter\providecommand\csname BenchMcmcMinEss\endcsname{}\expandafter\renewcommand\csname BenchMcmcMinEss\endcsname{\ensuremath{3511}}
\expandafter\providecommand\csname BenchNormRatio\endcsname{}\expandafter\renewcommand\csname BenchNormRatio\endcsname{\ensuremath{3.5}}
\expandafter\providecommand\csname BenchNumMasses\endcsname{}\expandafter\renewcommand\csname BenchNumMasses\endcsname{\ensuremath{21}}
\expandafter\providecommand\csname BenchPressureEdges\endcsname{}\expandafter\renewcommand\csname BenchPressureEdges\endcsname{\ensuremath{(3,8,20,40,80,160,320,640,1800)}}
\expandafter\providecommand\csname BenchProdDraws\endcsname{}\expandafter\renewcommand\csname BenchProdDraws\endcsname{\ensuremath{32\,768}}
\expandafter\providecommand\csname BenchPtRate\endcsname{}\expandafter\renewcommand\csname BenchPtRate\endcsname{\ensuremath{0.25}}
\expandafter\providecommand\csname BenchPtScale\endcsname{}\expandafter\renewcommand\csname BenchPtScale\endcsname{\ensuremath{30}}
\expandafter\providecommand\csname BenchQuarkBase\endcsname{}\expandafter\renewcommand\csname BenchQuarkBase\endcsname{\ensuremath{0.8}}
\expandafter\providecommand\csname BenchQuarkRate\endcsname{}\expandafter\renewcommand\csname BenchQuarkRate\endcsname{\ensuremath{0.7}}
\expandafter\providecommand\csname BenchReliableEss\endcsname{}\expandafter\renewcommand\csname BenchReliableEss\endcsname{\ensuremath{100}}
\expandafter\providecommand\csname BenchRunDraws\endcsname{}\expandafter\renewcommand\csname BenchRunDraws\endcsname{\ensuremath{86\,016}}
\expandafter\providecommand\csname BenchRuns\endcsname{}\expandafter\renewcommand\csname BenchRuns\endcsname{\ensuremath{5}}
\expandafter\providecommand\csname BenchRunsSmallestSigma\endcsname{}\expandafter\renewcommand\csname BenchRunsSmallestSigma\endcsname{\ensuremath{2}}
\expandafter\providecommand\csname BenchScipyMaxDiffM\endcsname{}\expandafter\renewcommand\csname BenchScipyMaxDiffM\endcsname{\ensuremath{2.3\times10^{-5}}}
\expandafter\providecommand\csname BenchSigmaFour\endcsname{}\expandafter\renewcommand\csname BenchSigmaFour\endcsname{\ensuremath{10}}
\expandafter\providecommand\csname BenchSigmaList\endcsname{}\expandafter\renewcommand\csname BenchSigmaList\endcsname{\ensuremath{100,\ 50,\ 20,\ \mbox{or}\ 10}}
\expandafter\providecommand\csname BenchSigmaOne\endcsname{}\expandafter\renewcommand\csname BenchSigmaOne\endcsname{\ensuremath{100}}
\expandafter\providecommand\csname BenchSigmaThree\endcsname{}\expandafter\renewcommand\csname BenchSigmaThree\endcsname{\ensuremath{20}}
\expandafter\providecommand\csname BenchSigmaTwo\endcsname{}\expandafter\renewcommand\csname BenchSigmaTwo\endcsname{\ensuremath{50}}
\expandafter\providecommand\csname BenchSoundSpeeds\endcsname{}\expandafter\renewcommand\csname BenchSoundSpeeds\endcsname{\ensuremath{(0.18,0.28,0.40,0.60,0.70,0.80,0.80,0.80)}}
\expandafter\providecommand\csname BenchTrainDraws\endcsname{}\expandafter\renewcommand\csname BenchTrainDraws\endcsname{\ensuremath{53\,248}}
\expandafter\providecommand\csname BruteRefDraws\endcsname{}\expandafter\renewcommand\csname BruteRefDraws\endcsname{\ensuremath{10^{5}}}
\expandafter\providecommand\csname BruteRefEss\endcsname{}\expandafter\renewcommand\csname BruteRefEss\endcsname{\ensuremath{681}}
\expandafter\providecommand\csname BruteRefLam\endcsname{}\expandafter\renewcommand\csname BruteRefLam\endcsname{\ensuremath{396}}
\expandafter\providecommand\csname BruteRefSd\endcsname{}\expandafter\renewcommand\csname BruteRefSd\endcsname{\ensuremath{12.0}}
\expandafter\providecommand\csname BruteRefWall\endcsname{}\expandafter\renewcommand\csname BruteRefWall\endcsname{\ensuremath{17}}
\expandafter\providecommand\csname CeftCrustEdgeNsat\endcsname{}\expandafter\renewcommand\csname CeftCrustEdgeNsat\endcsname{\ensuremath{0.34}}
\expandafter\providecommand\csname ClipPlainEss\endcsname{}\expandafter\renewcommand\csname ClipPlainEss\endcsname{\ensuremath{401}}
\expandafter\providecommand\csname ClipPlainMaxOffset\endcsname{}\expandafter\renewcommand\csname ClipPlainMaxOffset\endcsname{\ensuremath{0.86}}
\expandafter\providecommand\csname ClipPlainMean\endcsname{}\expandafter\renewcommand\csname ClipPlainMean\endcsname{\ensuremath{400}}
\expandafter\providecommand\csname ClipPlainRefitsHi\endcsname{}\expandafter\renewcommand\csname ClipPlainRefitsHi\endcsname{\ensuremath{3}}
\expandafter\providecommand\csname ClipPlainRefitsLo\endcsname{}\expandafter\renewcommand\csname ClipPlainRefitsLo\endcsname{\ensuremath{2}}
\expandafter\providecommand\csname ClipPlainScatter\endcsname{}\expandafter\renewcommand\csname ClipPlainScatter\endcsname{\ensuremath{2.8}}
\expandafter\providecommand\csname ClipPlainWidthHi\endcsname{}\expandafter\renewcommand\csname ClipPlainWidthHi\endcsname{\ensuremath{1.21}}
\expandafter\providecommand\csname ClipPlainWidthLo\endcsname{}\expandafter\renewcommand\csname ClipPlainWidthLo\endcsname{\ensuremath{0.91}}
\expandafter\providecommand\csname ClipShrunkEss\endcsname{}\expandafter\renewcommand\csname ClipShrunkEss\endcsname{\ensuremath{1136}}
\expandafter\providecommand\csname ClipShrunkMaxOffset\endcsname{}\expandafter\renewcommand\csname ClipShrunkMaxOffset\endcsname{\ensuremath{0.64}}
\expandafter\providecommand\csname ClipShrunkMean\endcsname{}\expandafter\renewcommand\csname ClipShrunkMean\endcsname{\ensuremath{399}}
\expandafter\providecommand\csname ClipShrunkRefitsHi\endcsname{}\expandafter\renewcommand\csname ClipShrunkRefitsHi\endcsname{\ensuremath{5}}
\expandafter\providecommand\csname ClipShrunkRefitsLo\endcsname{}\expandafter\renewcommand\csname ClipShrunkRefitsLo\endcsname{\ensuremath{3}}
\expandafter\providecommand\csname ClipShrunkScatter\endcsname{}\expandafter\renewcommand\csname ClipShrunkScatter\endcsname{\ensuremath{3.6}}
\expandafter\providecommand\csname ClipShrunkWidthHi\endcsname{}\expandafter\renewcommand\csname ClipShrunkWidthHi\endcsname{\ensuremath{1.20}}
\expandafter\providecommand\csname ClipShrunkWidthLo\endcsname{}\expandafter\renewcommand\csname ClipShrunkWidthLo\endcsname{\ensuremath{1.04}}
\expandafter\providecommand\csname CoeffDim\endcsname{}\expandafter\renewcommand\csname CoeffDim\endcsname{\ensuremath{48}}
\expandafter\providecommand\csname CombineAcceptHi\endcsname{}\expandafter\renewcommand\csname CombineAcceptHi\endcsname{\ensuremath{1.000}}
\expandafter\providecommand\csname CombineAcceptLo\endcsname{}\expandafter\renewcommand\csname CombineAcceptLo\endcsname{\ensuremath{1.000}}
\expandafter\providecommand\csname CombineGenZeroDraws\endcsname{}\expandafter\renewcommand\csname CombineGenZeroDraws\endcsname{\ensuremath{884}}
\expandafter\providecommand\csname CombineGenZeroEss\endcsname{}\expandafter\renewcommand\csname CombineGenZeroEss\endcsname{\ensuremath{473}}
\expandafter\providecommand\csname CombineGenZeroLam\endcsname{}\expandafter\renewcommand\csname CombineGenZeroLam\endcsname{\ensuremath{320}}
\expandafter\providecommand\csname CombineGenZeroLamSd\endcsname{}\expandafter\renewcommand\csname CombineGenZeroLamSd\endcsname{\ensuremath{179}}
\expandafter\providecommand\csname CombineTotalEss\endcsname{}\expandafter\renewcommand\csname CombineTotalEss\endcsname{\ensuremath{156}}
\expandafter\providecommand\csname CombineTotalLam\endcsname{}\expandafter\renewcommand\csname CombineTotalLam\endcsname{\ensuremath{237}}
\expandafter\providecommand\csname CombineTotalLamSd\endcsname{}\expandafter\renewcommand\csname CombineTotalLamSd\endcsname{\ensuremath{210}}
\expandafter\providecommand\csname CrustEdgeNsat\endcsname{}\expandafter\renewcommand\csname CrustEdgeNsat\endcsname{\ensuremath{0.34}}
\expandafter\providecommand\csname CrustNt\endcsname{}\expandafter\renewcommand\csname CrustNt\endcsname{\ensuremath{0.055}}
\expandafter\providecommand\csname CsSqDraws\endcsname{}\expandafter\renewcommand\csname CsSqDraws\endcsname{\ensuremath{1000}}
\expandafter\providecommand\csname CsSqEnsMax\endcsname{}\expandafter\renewcommand\csname CsSqEnsMax\endcsname{\ensuremath{0.99974}}
\expandafter\providecommand\csname CsSqEnsMin\endcsname{}\expandafter\renewcommand\csname CsSqEnsMin\endcsname{\ensuremath{2\times10^{-6}}}
\expandafter\providecommand\csname CsSqFracHi\endcsname{}\expandafter\renewcommand\csname CsSqFracHi\endcsname{\ensuremath{8.2}}
\expandafter\providecommand\csname CsSqMedMax\endcsname{}\expandafter\renewcommand\csname CsSqMedMax\endcsname{\ensuremath{0.74}}
\expandafter\providecommand\csname CsSqNSuper\endcsname{}\expandafter\renewcommand\csname CsSqNSuper\endcsname{\ensuremath{0}}
\expandafter\providecommand\csname DimBootCv\endcsname{}\expandafter\renewcommand\csname DimBootCv\endcsname{\ensuremath{14.2}}
\expandafter\providecommand\csname DimBootCvSd\endcsname{}\expandafter\renewcommand\csname DimBootCvSd\endcsname{\ensuremath{0.4}}
\expandafter\providecommand\csname DimComponentOneRecoveryPct\endcsname{}\expandafter\renewcommand\csname DimComponentOneRecoveryPct\endcsname{\ensuremath{94}}
\expandafter\providecommand\csname DimComponentOneVarPct\endcsname{}\expandafter\renewcommand\csname DimComponentOneVarPct\endcsname{\ensuremath{51}}
\expandafter\providecommand\csname DimComponentTwoRecoveryPct\endcsname{}\expandafter\renewcommand\csname DimComponentTwoRecoveryPct\endcsname{\ensuremath{95}}
\expandafter\providecommand\csname DimComponentTwoVarPct\endcsname{}\expandafter\renewcommand\csname DimComponentTwoVarPct\endcsname{\ensuremath{28}}
\expandafter\providecommand\csname DimCvFolds\endcsname{}\expandafter\renewcommand\csname DimCvFolds\endcsname{\ensuremath{10}}
\expandafter\providecommand\csname DimCvPruned\endcsname{}\expandafter\renewcommand\csname DimCvPruned\endcsname{\ensuremath{13.9}}
\expandafter\providecommand\csname DimEnsCv\endcsname{}\expandafter\renewcommand\csname DimEnsCv\endcsname{\ensuremath{22.1}}
\expandafter\providecommand\csname DimEnsCvHi\endcsname{}\expandafter\renewcommand\csname DimEnsCvHi\endcsname{\ensuremath{23.0}}
\expandafter\providecommand\csname DimEnsCvLo\endcsname{}\expandafter\renewcommand\csname DimEnsCvLo\endcsname{\ensuremath{21.7}}
\expandafter\providecommand\csname DimEnsCvSd\endcsname{}\expandafter\renewcommand\csname DimEnsCvSd\endcsname{\ensuremath{0.9}}
\expandafter\providecommand\csname DimEnsCvTight\endcsname{}\expandafter\renewcommand\csname DimEnsCvTight\endcsname{\ensuremath{21.7}}
\expandafter\providecommand\csname DimEnsCvTightSd\endcsname{}\expandafter\renewcommand\csname DimEnsCvTightSd\endcsname{\ensuremath{0.8}}
\expandafter\providecommand\csname DimEnsDropHi\endcsname{}\expandafter\renewcommand\csname DimEnsDropHi\endcsname{\ensuremath{14}}
\expandafter\providecommand\csname DimEnsDropLo\endcsname{}\expandafter\renewcommand\csname DimEnsDropLo\endcsname{\ensuremath{3}}
\expandafter\providecommand\csname DimEnsScatterLoose\endcsname{}\expandafter\renewcommand\csname DimEnsScatterLoose\endcsname{\ensuremath{2.3}}
\expandafter\providecommand\csname DimEnsScatterTight\endcsname{}\expandafter\renewcommand\csname DimEnsScatterTight\endcsname{\ensuremath{0.1}}
\expandafter\providecommand\csname DimHalfCv\endcsname{}\expandafter\renewcommand\csname DimHalfCv\endcsname{\ensuremath{14.9}}
\expandafter\providecommand\csname DimHalfCvSd\endcsname{}\expandafter\renewcommand\csname DimHalfCvSd\endcsname{\ensuremath{0.5}}
\expandafter\providecommand\csname DimLevDropped\endcsname{}\expandafter\renewcommand\csname DimLevDropped\endcsname{\ensuremath{2}}
\expandafter\providecommand\csname DimLeverageMax\endcsname{}\expandafter\renewcommand\csname DimLeverageMax\endcsname{\ensuremath{0.5}}
\expandafter\providecommand\csname DimMaxLeverage\endcsname{}\expandafter\renewcommand\csname DimMaxLeverage\endcsname{\ensuremath{0.7155}}
\expandafter\providecommand\csname DimModeEightRsq\endcsname{}\expandafter\renewcommand\csname DimModeEightRsq\endcsname{\ensuremath{0.44}}
\expandafter\providecommand\csname DimModeOneRsq\endcsname{}\expandafter\renewcommand\csname DimModeOneRsq\endcsname{\ensuremath{0.94}}
\expandafter\providecommand\csname DimModeTwoRsq\endcsname{}\expandafter\renewcommand\csname DimModeTwoRsq\endcsname{\ensuremath{0.95}}
\expandafter\providecommand\csname DimNBoot\endcsname{}\expandafter\renewcommand\csname DimNBoot\endcsname{\ensuremath{20}}
\expandafter\providecommand\csname DimNEnsembles\endcsname{}\expandafter\renewcommand\csname DimNEnsembles\endcsname{\ensuremath{4}}
\expandafter\providecommand\csname DimNHalf\endcsname{}\expandafter\renewcommand\csname DimNHalf\endcsname{\ensuremath{20}}
\expandafter\providecommand\csname DimNullFracPrunedPct\endcsname{}\expandafter\renewcommand\csname DimNullFracPrunedPct\endcsname{\ensuremath{12.1}}
\expandafter\providecommand\csname DimNumRegressors\endcsname{}\expandafter\renewcommand\csname DimNumRegressors\endcsname{\ensuremath{48}}
\expandafter\providecommand\csname DimThrCvHi\endcsname{}\expandafter\renewcommand\csname DimThrCvHi\endcsname{\ensuremath{14.4}}
\expandafter\providecommand\csname DimThrCvLo\endcsname{}\expandafter\renewcommand\csname DimThrCvLo\endcsname{\ensuremath{13.6}}
\expandafter\providecommand\csname DimThrDropHi\endcsname{}\expandafter\renewcommand\csname DimThrDropHi\endcsname{\ensuremath{2.5}}
\expandafter\providecommand\csname DimTopTwoVarPct\endcsname{}\expandafter\renewcommand\csname DimTopTwoVarPct\endcsname{\ensuremath{79}}
\expandafter\providecommand\csname ExactPlainEss\endcsname{}\expandafter\renewcommand\csname ExactPlainEss\endcsname{\ensuremath{3}}
\expandafter\providecommand\csname ExactPlainMaxOffset\endcsname{}\expandafter\renewcommand\csname ExactPlainMaxOffset\endcsname{\ensuremath{1.47}}
\expandafter\providecommand\csname ExactPlainMean\endcsname{}\expandafter\renewcommand\csname ExactPlainMean\endcsname{\ensuremath{404}}
\expandafter\providecommand\csname ExactPlainRefitsHi\endcsname{}\expandafter\renewcommand\csname ExactPlainRefitsHi\endcsname{\ensuremath{3}}
\expandafter\providecommand\csname ExactPlainRefitsLo\endcsname{}\expandafter\renewcommand\csname ExactPlainRefitsLo\endcsname{\ensuremath{2}}
\expandafter\providecommand\csname ExactPlainScatter\endcsname{}\expandafter\renewcommand\csname ExactPlainScatter\endcsname{\ensuremath{7.0}}
\expandafter\providecommand\csname ExactPlainWidthHi\endcsname{}\expandafter\renewcommand\csname ExactPlainWidthHi\endcsname{\ensuremath{1.03}}
\expandafter\providecommand\csname ExactPlainWidthLo\endcsname{}\expandafter\renewcommand\csname ExactPlainWidthLo\endcsname{\ensuremath{0.17}}
\expandafter\providecommand\csname ExactShrunkEss\endcsname{}\expandafter\renewcommand\csname ExactShrunkEss\endcsname{\ensuremath{6}}
\expandafter\providecommand\csname ExactShrunkMaxOffset\endcsname{}\expandafter\renewcommand\csname ExactShrunkMaxOffset\endcsname{\ensuremath{0.88}}
\expandafter\providecommand\csname ExactShrunkMean\endcsname{}\expandafter\renewcommand\csname ExactShrunkMean\endcsname{\ensuremath{392}}
\expandafter\providecommand\csname ExactShrunkRefitsHi\endcsname{}\expandafter\renewcommand\csname ExactShrunkRefitsHi\endcsname{\ensuremath{5}}
\expandafter\providecommand\csname ExactShrunkRefitsLo\endcsname{}\expandafter\renewcommand\csname ExactShrunkRefitsLo\endcsname{\ensuremath{3}}
\expandafter\providecommand\csname ExactShrunkScatter\endcsname{}\expandafter\renewcommand\csname ExactShrunkScatter\endcsname{\ensuremath{4.8}}
\expandafter\providecommand\csname ExactShrunkWidthHi\endcsname{}\expandafter\renewcommand\csname ExactShrunkWidthHi\endcsname{\ensuremath{1.10}}
\expandafter\providecommand\csname ExactShrunkWidthLo\endcsname{}\expandafter\renewcommand\csname ExactShrunkWidthLo\endcsname{\ensuremath{0.68}}
\expandafter\providecommand\csname ExtGpCsDensityRank\endcsname{}\expandafter\renewcommand\csname ExtGpCsDensityRank\endcsname{\ensuremath{68}}
\expandafter\providecommand\csname ExtGpCsEnthalpyRank\endcsname{}\expandafter\renewcommand\csname ExtGpCsEnthalpyRank\endcsname{\ensuremath{34}}
\expandafter\providecommand\csname ExtGpDensityPoints\endcsname{}\expandafter\renewcommand\csname ExtGpDensityPoints\endcsname{\ensuremath{80}}
\expandafter\providecommand\csname ExtGpEnthalpyPoints\endcsname{}\expandafter\renewcommand\csname ExtGpEnthalpyPoints\endcsname{\ensuremath{80}}
\expandafter\providecommand\csname ExtGpLevelPct\endcsname{}\expandafter\renewcommand\csname ExtGpLevelPct\endcsname{\ensuremath{1}}
\expandafter\providecommand\csname ExtGpMassHi\endcsname{}\expandafter\renewcommand\csname ExtGpMassHi\endcsname{\ensuremath{1.8}}
\expandafter\providecommand\csname ExtGpMassLo\endcsname{}\expandafter\renewcommand\csname ExtGpMassLo\endcsname{\ensuremath{1}}
\expandafter\providecommand\csname ExtGpMassPoints\endcsname{}\expandafter\renewcommand\csname ExtGpMassPoints\endcsname{\ensuremath{41}}
\expandafter\providecommand\csname ExtGpRhoHiRatio\endcsname{}\expandafter\renewcommand\csname ExtGpRhoHiRatio\endcsname{\ensuremath{5}}
\expandafter\providecommand\csname ExtGpRhoLo\endcsname{}\expandafter\renewcommand\csname ExtGpRhoLo\endcsname{\ensuremath{2.7\times10^{14}}}
\expandafter\providecommand\csname ExtGpRows\endcsname{}\expandafter\renewcommand\csname ExtGpRows\endcsname{\ensuremath{9959}}
\expandafter\providecommand\csname ExtGpStellarRank\endcsname{}\expandafter\renewcommand\csname ExtGpStellarRank\endcsname{\ensuremath{7}}
\expandafter\providecommand\csname ExtGpTrainPct\endcsname{}\expandafter\renewcommand\csname ExtGpTrainPct\endcsname{\ensuremath{80}}
\expandafter\providecommand\csname FisherArchiveDraws\endcsname{}\expandafter\renewcommand\csname FisherArchiveDraws\endcsname{\ensuremath{1000}}
\expandafter\providecommand\csname FisherBruteFinalEss\endcsname{}\expandafter\renewcommand\csname FisherBruteFinalEss\endcsname{\ensuremath{66}}
\expandafter\providecommand\csname FisherBruteLam\endcsname{}\expandafter\renewcommand\csname FisherBruteLam\endcsname{\ensuremath{381}}
\expandafter\providecommand\csname FisherBruteLamSd\endcsname{}\expandafter\renewcommand\csname FisherBruteLamSd\endcsname{\ensuremath{40}}
\expandafter\providecommand\csname FisherDrawsPerGen\endcsname{}\expandafter\renewcommand\csname FisherDrawsPerGen\endcsname{\ensuremath{200}}
\expandafter\providecommand\csname FisherEvents\endcsname{}\expandafter\renewcommand\csname FisherEvents\endcsname{\ensuremath{40}}
\expandafter\providecommand\csname FisherFinalLam\endcsname{}\expandafter\renewcommand\csname FisherFinalLam\endcsname{\ensuremath{369}}
\expandafter\providecommand\csname FisherFinalLamSd\endcsname{}\expandafter\renewcommand\csname FisherFinalLamSd\endcsname{\ensuremath{38}}
\expandafter\providecommand\csname FisherFmax\endcsname{}\expandafter\renewcommand\csname FisherFmax\endcsname{\ensuremath{1024}}
\expandafter\providecommand\csname FisherFmin\endcsname{}\expandafter\renewcommand\csname FisherFmin\endcsname{\ensuremath{8}}
\expandafter\providecommand\csname FisherMaxDelta\endcsname{}\expandafter\renewcommand\csname FisherMaxDelta\endcsname{\ensuremath{52}}
\expandafter\providecommand\csname FisherRefinements\endcsname{}\expandafter\renewcommand\csname FisherRefinements\endcsname{\ensuremath{4}}
\expandafter\providecommand\csname FisherRefitList\endcsname{}\expandafter\renewcommand\csname FisherRefitList\endcsname{\ensuremath{1,\ 6,\ 7,\ \mbox{and}\ 9}}
\expandafter\providecommand\csname FisherRefitThreshold\endcsname{}\expandafter\renewcommand\csname FisherRefitThreshold\endcsname{\ensuremath{100}}
\expandafter\providecommand\csname FisherSubsampleScatterPct\endcsname{}\expandafter\renewcommand\csname FisherSubsampleScatterPct\endcsname{\ensuremath{10}}
\expandafter\providecommand\csname FribFourCurvAdSd\endcsname{}\expandafter\renewcommand\csname FribFourCurvAdSd\endcsname{\ensuremath{64.2}}
\expandafter\providecommand\csname FribFourCurvSd\endcsname{}\expandafter\renewcommand\csname FribFourCurvSd\endcsname{\ensuremath{75.9}}
\expandafter\providecommand\csname FribFourEnergyAdSd\endcsname{}\expandafter\renewcommand\csname FribFourEnergyAdSd\endcsname{\ensuremath{1.5}}
\expandafter\providecommand\csname FribFourEnergyLab\endcsname{}\expandafter\renewcommand\csname FribFourEnergyLab\endcsname{\ensuremath{1.5}}
\expandafter\providecommand\csname FribFourEnergySd\endcsname{}\expandafter\renewcommand\csname FribFourEnergySd\endcsname{\ensuremath{1.5}}
\expandafter\providecommand\csname FribFourKCurvAdSd\endcsname{}\expandafter\renewcommand\csname FribFourKCurvAdSd\endcsname{\ensuremath{55.8}}
\expandafter\providecommand\csname FribFourKCurvLab\endcsname{}\expandafter\renewcommand\csname FribFourKCurvLab\endcsname{\ensuremath{80}}
\expandafter\providecommand\csname FribFourKCurvSd\endcsname{}\expandafter\renewcommand\csname FribFourKCurvSd\endcsname{\ensuremath{64.1}}
\expandafter\providecommand\csname FribFourKEnergyAdSd\endcsname{}\expandafter\renewcommand\csname FribFourKEnergyAdSd\endcsname{\ensuremath{1.5}}
\expandafter\providecommand\csname FribFourKEnergySd\endcsname{}\expandafter\renewcommand\csname FribFourKEnergySd\endcsname{\ensuremath{1.5}}
\expandafter\providecommand\csname FribFourKKsatAdSd\endcsname{}\expandafter\renewcommand\csname FribFourKKsatAdSd\endcsname{\ensuremath{22.0}}
\expandafter\providecommand\csname FribFourKKsatSd\endcsname{}\expandafter\renewcommand\csname FribFourKKsatSd\endcsname{\ensuremath{23.1}}
\expandafter\providecommand\csname FribFourKSlopeAdSd\endcsname{}\expandafter\renewcommand\csname FribFourKSlopeAdSd\endcsname{\ensuremath{10.4}}
\expandafter\providecommand\csname FribFourKSlopeSd\endcsname{}\expandafter\renewcommand\csname FribFourKSlopeSd\endcsname{\ensuremath{11.0}}
\expandafter\providecommand\csname FribFourKsatAdSd\endcsname{}\expandafter\renewcommand\csname FribFourKsatAdSd\endcsname{\ensuremath{22.4}}
\expandafter\providecommand\csname FribFourKsatSd\endcsname{}\expandafter\renewcommand\csname FribFourKsatSd\endcsname{\ensuremath{23.4}}
\expandafter\providecommand\csname FribFourSlopeAdSd\endcsname{}\expandafter\renewcommand\csname FribFourSlopeAdSd\endcsname{\ensuremath{11.5}}
\expandafter\providecommand\csname FribFourSlopeLab\endcsname{}\expandafter\renewcommand\csname FribFourSlopeLab\endcsname{\ensuremath{15}}
\expandafter\providecommand\csname FribFourSlopeSd\endcsname{}\expandafter\renewcommand\csname FribFourSlopeSd\endcsname{\ensuremath{12.0}}
\expandafter\providecommand\csname FribGwCurvAdSd\endcsname{}\expandafter\renewcommand\csname FribGwCurvAdSd\endcsname{\ensuremath{79.7}}
\expandafter\providecommand\csname FribGwCurvSd\endcsname{}\expandafter\renewcommand\csname FribGwCurvSd\endcsname{\ensuremath{90.5}}
\expandafter\providecommand\csname FribGwEnergyAdSd\endcsname{}\expandafter\renewcommand\csname FribGwEnergyAdSd\endcsname{\ensuremath{2.9}}
\expandafter\providecommand\csname FribGwEnergySd\endcsname{}\expandafter\renewcommand\csname FribGwEnergySd\endcsname{\ensuremath{2.9}}
\expandafter\providecommand\csname FribGwKsatAdSd\endcsname{}\expandafter\renewcommand\csname FribGwKsatAdSd\endcsname{\ensuremath{22.5}}
\expandafter\providecommand\csname FribGwKsatSd\endcsname{}\expandafter\renewcommand\csname FribGwKsatSd\endcsname{\ensuremath{23.5}}
\expandafter\providecommand\csname FribGwSlopeAdSd\endcsname{}\expandafter\renewcommand\csname FribGwSlopeAdSd\endcsname{\ensuremath{15.1}}
\expandafter\providecommand\csname FribGwSlopeSd\endcsname{}\expandafter\renewcommand\csname FribGwSlopeSd\endcsname{\ensuremath{16.1}}
\expandafter\providecommand\csname FribNearCurvAdSd\endcsname{}\expandafter\renewcommand\csname FribNearCurvAdSd\endcsname{\ensuremath{74.2}}
\expandafter\providecommand\csname FribNearCurvSd\endcsname{}\expandafter\renewcommand\csname FribNearCurvSd\endcsname{\ensuremath{85.8}}
\expandafter\providecommand\csname FribNearEnergyAdSd\endcsname{}\expandafter\renewcommand\csname FribNearEnergyAdSd\endcsname{\ensuremath{2.0}}
\expandafter\providecommand\csname FribNearEnergyLab\endcsname{}\expandafter\renewcommand\csname FribNearEnergyLab\endcsname{\ensuremath{2}}
\expandafter\providecommand\csname FribNearEnergySd\endcsname{}\expandafter\renewcommand\csname FribNearEnergySd\endcsname{\ensuremath{2.0}}
\expandafter\providecommand\csname FribNearKsatAdSd\endcsname{}\expandafter\renewcommand\csname FribNearKsatAdSd\endcsname{\ensuremath{22.5}}
\expandafter\providecommand\csname FribNearKsatSd\endcsname{}\expandafter\renewcommand\csname FribNearKsatSd\endcsname{\ensuremath{23.5}}
\expandafter\providecommand\csname FribNearSlopeAdSd\endcsname{}\expandafter\renewcommand\csname FribNearSlopeAdSd\endcsname{\ensuremath{13.8}}
\expandafter\providecommand\csname FribNearSlopeLab\endcsname{}\expandafter\renewcommand\csname FribNearSlopeLab\endcsname{\ensuremath{22}}
\expandafter\providecommand\csname FribNearSlopeSd\endcsname{}\expandafter\renewcommand\csname FribNearSlopeSd\endcsname{\ensuremath{14.8}}
\expandafter\providecommand\csname FribPriorCurvSd\endcsname{}\expandafter\renewcommand\csname FribPriorCurvSd\endcsname{\ensuremath{107.8}}
\expandafter\providecommand\csname FribPriorEnergySd\endcsname{}\expandafter\renewcommand\csname FribPriorEnergySd\endcsname{\ensuremath{2.9}}
\expandafter\providecommand\csname FribPriorKsatSd\endcsname{}\expandafter\renewcommand\csname FribPriorKsatSd\endcsname{\ensuremath{23.8}}
\expandafter\providecommand\csname FribPriorSlopeSd\endcsname{}\expandafter\renewcommand\csname FribPriorSlopeSd\endcsname{\ensuremath{27.0}}
\expandafter\providecommand\csname GenZeroDraws\endcsname{}\expandafter\renewcommand\csname GenZeroDraws\endcsname{\ensuremath{1000}}
\expandafter\providecommand\csname GenZeroUsable\endcsname{}\expandafter\renewcommand\csname GenZeroUsable\endcsname{\ensuremath{884}}
\expandafter\providecommand\csname GpBaseOutlierPct\endcsname{}\expandafter\renewcommand\csname GpBaseOutlierPct\endcsname{\ensuremath{3.9}}
\expandafter\providecommand\csname GpRefinedOutlierPct\endcsname{}\expandafter\renewcommand\csname GpRefinedOutlierPct\endcsname{\ensuremath{0.4}}
\expandafter\providecommand\csname GpTrainOutlierPct\endcsname{}\expandafter\renewcommand\csname GpTrainOutlierPct\endcsname{\ensuremath{1.3}}
\expandafter\providecommand\csname HeroAdaptiveMeanDeltaPct\endcsname{}\expandafter\renewcommand\csname HeroAdaptiveMeanDeltaPct\endcsname{\ensuremath{0.6}}
\expandafter\providecommand\csname HeroAdaptiveWidthRatio\endcsname{}\expandafter\renewcommand\csname HeroAdaptiveWidthRatio\endcsname{\ensuremath{1.0}}
\expandafter\providecommand\csname HeroArchiveWidthRatio\endcsname{}\expandafter\renewcommand\csname HeroArchiveWidthRatio\endcsname{\ensuremath{0.74}}
\expandafter\providecommand\csname HeroBruteCollapse\endcsname{}\expandafter\renewcommand\csname HeroBruteCollapse\endcsname{\ensuremath{56}}
\expandafter\providecommand\csname HeroBruteFinalEss\endcsname{}\expandafter\renewcommand\csname HeroBruteFinalEss\endcsname{\ensuremath{8}}
\expandafter\providecommand\csname HeroBruteStartEss\endcsname{}\expandafter\renewcommand\csname HeroBruteStartEss\endcsname{\ensuremath{470}}
\expandafter\providecommand\csname HeroDrawsPerGen\endcsname{}\expandafter\renewcommand\csname HeroDrawsPerGen\endcsname{\ensuremath{10^{4}}}
\expandafter\providecommand\csname HeroEvents\endcsname{}\expandafter\renewcommand\csname HeroEvents\endcsname{\ensuremath{1000}}
\expandafter\providecommand\csname HeroFinalLamPct\endcsname{}\expandafter\renewcommand\csname HeroFinalLamPct\endcsname{\ensuremath{3}}
\expandafter\providecommand\csname HeroFinalLamSd\endcsname{}\expandafter\renewcommand\csname HeroFinalLamSd\endcsname{\ensuremath{12}}
\expandafter\providecommand\csname HeroFisherAgreePct\endcsname{}\expandafter\renewcommand\csname HeroFisherAgreePct\endcsname{\ensuremath{0}}
\expandafter\providecommand\csname HeroFisherRatio\endcsname{}\expandafter\renewcommand\csname HeroFisherRatio\endcsname{\ensuremath{1.00}}
\expandafter\providecommand\csname HeroFisherSd\endcsname{}\expandafter\renewcommand\csname HeroFisherSd\endcsname{\ensuremath{12.0}}
\expandafter\providecommand\csname HeroMaxDistance\endcsname{}\expandafter\renewcommand\csname HeroMaxDistance\endcsname{\ensuremath{300}}
\expandafter\providecommand\csname HeroRefinements\endcsname{}\expandafter\renewcommand\csname HeroRefinements\endcsname{\ensuremath{2}}
\expandafter\providecommand\csname HeroRefitPoints\endcsname{}\expandafter\renewcommand\csname HeroRefitPoints\endcsname{\ensuremath{1~\mathrm{and}~233}}
\expandafter\providecommand\csname HeroRefitThreshold\endcsname{}\expandafter\renewcommand\csname HeroRefitThreshold\endcsname{\ensuremath{10^{3}}}
\expandafter\providecommand\csname HeroSnrMin\endcsname{}\expandafter\renewcommand\csname HeroSnrMin\endcsname{\ensuremath{8}}
\expandafter\providecommand\csname JOneSixOneFourMass\endcsname{}\expandafter\renewcommand\csname JOneSixOneFourMass\endcsname{\ensuremath{1.908}}
\expandafter\providecommand\csname JOneSixOneFourMassSd\endcsname{}\expandafter\renewcommand\csname JOneSixOneFourMassSd\endcsname{\ensuremath{0.016}}
\expandafter\providecommand\csname JZeroSevenFourZeroMass\endcsname{}\expandafter\renewcommand\csname JZeroSevenFourZeroMass\endcsname{\ensuremath{2.08}}
\expandafter\providecommand\csname JZeroSevenFourZeroMassSd\endcsname{}\expandafter\renewcommand\csname JZeroSevenFourZeroMassSd\endcsname{\ensuremath{0.07}}
\expandafter\providecommand\csname JZeroThreeFourEightMass\endcsname{}\expandafter\renewcommand\csname JZeroThreeFourEightMass\endcsname{\ensuremath{1.806}}
\expandafter\providecommand\csname JZeroThreeFourEightMassSd\endcsname{}\expandafter\renewcommand\csname JZeroThreeFourEightMassSd\endcsname{\ensuremath{0.037}}
\expandafter\providecommand\csname LabSymBaseEss\endcsname{}\expandafter\renewcommand\csname LabSymBaseEss\endcsname{\ensuremath{898}}
\expandafter\providecommand\csname LabSymBaseEsym\endcsname{}\expandafter\renewcommand\csname LabSymBaseEsym\endcsname{\ensuremath{33.2}}
\expandafter\providecommand\csname LabSymBaseEsymSd\endcsname{}\expandafter\renewcommand\csname LabSymBaseEsymSd\endcsname{\ensuremath{2.9}}
\expandafter\providecommand\csname LabSymEss\endcsname{}\expandafter\renewcommand\csname LabSymEss\endcsname{\ensuremath{646}}
\expandafter\providecommand\csname LabSymEsym\endcsname{}\expandafter\renewcommand\csname LabSymEsym\endcsname{\ensuremath{32.2}}
\expandafter\providecommand\csname LabSymEsymSd\endcsname{}\expandafter\renewcommand\csname LabSymEsymSd\endcsname{\ensuremath{1.9}}
\expandafter\providecommand\csname LabSymLam\endcsname{}\expandafter\renewcommand\csname LabSymLam\endcsname{\ensuremath{396}}
\expandafter\providecommand\csname LabSymLamSd\endcsname{}\expandafter\renewcommand\csname LabSymLamSd\endcsname{\ensuremath{78}}
\expandafter\providecommand\csname LabSymRadius\endcsname{}\expandafter\renewcommand\csname LabSymRadius\endcsname{\ensuremath{11.9}}
\expandafter\providecommand\csname LabSymRadiusSd\endcsname{}\expandafter\renewcommand\csname LabSymRadiusSd\endcsname{\ensuremath{0.4}}
\expandafter\providecommand\csname LabSymSZero\endcsname{}\expandafter\renewcommand\csname LabSymSZero\endcsname{\ensuremath{32}}
\expandafter\providecommand\csname LabSymSZeroSd\endcsname{}\expandafter\renewcommand\csname LabSymSZeroSd\endcsname{\ensuremath{2}}
\expandafter\providecommand\csname LdCeftEnvelopeMaxRadius\endcsname{}\expandafter\renewcommand\csname LdCeftEnvelopeMaxRadius\endcsname{\ensuremath{12.76}}
\expandafter\providecommand\csname LdCeftFracAboveThirteen\endcsname{}\expandafter\renewcommand\csname LdCeftFracAboveThirteen\endcsname{\ensuremath{0}}
\expandafter\providecommand\csname LdCeftMaxRadius\endcsname{}\expandafter\renewcommand\csname LdCeftMaxRadius\endcsname{\ensuremath{12.8}}
\expandafter\providecommand\csname LdCeftMmaxMed\endcsname{}\expandafter\renewcommand\csname LdCeftMmaxMed\endcsname{\ensuremath{2.03}}
\expandafter\providecommand\csname LdCeftRadiusHi\endcsname{}\expandafter\renewcommand\csname LdCeftRadiusHi\endcsname{\ensuremath{12.7}}
\expandafter\providecommand\csname LdCeftRadiusLo\endcsname{}\expandafter\renewcommand\csname LdCeftRadiusLo\endcsname{\ensuremath{8.5}}
\expandafter\providecommand\csname LdCeftWidthAnchor\endcsname{}\expandafter\renewcommand\csname LdCeftWidthAnchor\endcsname{\ensuremath{0.16}}
\expandafter\providecommand\csname LdCeftWidthHalfSat\endcsname{}\expandafter\renewcommand\csname LdCeftWidthHalfSat\endcsname{\ensuremath{0.10}}
\expandafter\providecommand\csname LdCeilingBandShift\endcsname{}\expandafter\renewcommand\csname LdCeilingBandShift\endcsname{\ensuremath{0.01}}
\expandafter\providecommand\csname LdMetaFracAboveThirteen\endcsname{}\expandafter\renewcommand\csname LdMetaFracAboveThirteen\endcsname{\ensuremath{33}}
\expandafter\providecommand\csname LdMetaMaxRadius\endcsname{}\expandafter\renewcommand\csname LdMetaMaxRadius\endcsname{\ensuremath{14.5}}
\expandafter\providecommand\csname LdMetaMmaxMed\endcsname{}\expandafter\renewcommand\csname LdMetaMmaxMed\endcsname{\ensuremath{2.02}}
\expandafter\providecommand\csname LdMetaRadiusHi\endcsname{}\expandafter\renewcommand\csname LdMetaRadiusHi\endcsname{\ensuremath{14.0}}
\expandafter\providecommand\csname LdMetaRadiusLo\endcsname{}\expandafter\renewcommand\csname LdMetaRadiusLo\endcsname{\ensuremath{8.7}}
\expandafter\providecommand\csname LdMetaRsqAnchor\endcsname{}\expandafter\renewcommand\csname LdMetaRsqAnchor\endcsname{\ensuremath{13}}
\expandafter\providecommand\csname LdMetaRsqBridge\endcsname{}\expandafter\renewcommand\csname LdMetaRsqBridge\endcsname{\ensuremath{73}}
\expandafter\providecommand\csname LdMetaWidthAnchor\endcsname{}\expandafter\renewcommand\csname LdMetaWidthAnchor\endcsname{\ensuremath{0.50}}
\expandafter\providecommand\csname LdMetaWidthCrust\endcsname{}\expandafter\renewcommand\csname LdMetaWidthCrust\endcsname{\ensuremath{0.01}}
\expandafter\providecommand\csname LdMetaWidthHalfSat\endcsname{}\expandafter\renewcommand\csname LdMetaWidthHalfSat\endcsname{\ensuremath{0.74}}
\expandafter\providecommand\csname LdRadiusShift\endcsname{}\expandafter\renewcommand\csname LdRadiusShift\endcsname{\ensuremath{-0.2}}
\expandafter\providecommand\csname LdRadiusWidthRatio\endcsname{}\expandafter\renewcommand\csname LdRadiusWidthRatio\endcsname{\ensuremath{0.78}}
\expandafter\providecommand\csname LdWidthRatioAnchor\endcsname{}\expandafter\renewcommand\csname LdWidthRatioAnchor\endcsname{\ensuremath{3.2}}
\expandafter\providecommand\csname LdWidthRatioHalfSat\endcsname{}\expandafter\renewcommand\csname LdWidthRatioHalfSat\endcsname{\ensuremath{7.2}}
\expandafter\providecommand\csname LdWidthRatioSat\endcsname{}\expandafter\renewcommand\csname LdWidthRatioSat\endcsname{\ensuremath{10.6}}
\expandafter\providecommand\csname LitAntoniadisMass\endcsname{}\expandafter\renewcommand\csname LitAntoniadisMass\endcsname{\ensuremath{2.01}}
\expandafter\providecommand\csname LitAntoniadisMassSd\endcsname{}\expandafter\renewcommand\csname LitAntoniadisMassSd\endcsname{\ensuremath{0.04}}
\expandafter\providecommand\csname LitJNineFiveTwoMass\endcsname{}\expandafter\renewcommand\csname LitJNineFiveTwoMass\endcsname{\ensuremath{2.35}}
\expandafter\providecommand\csname LitJNineFiveTwoMassSd\endcsname{}\expandafter\renewcommand\csname LitJNineFiveTwoMassSd\endcsname{\ensuremath{0.11}}
\expandafter\providecommand\csname LitJZeroThirtyR\endcsname{}\expandafter\renewcommand\csname LitJZeroThirtyR\endcsname{\ensuremath{12.7}}
\expandafter\providecommand\csname LitJZeroThirtyRMinus\endcsname{}\expandafter\renewcommand\csname LitJZeroThirtyRMinus\endcsname{\ensuremath{1.0}}
\expandafter\providecommand\csname LitJZeroThirtyRPlus\endcsname{}\expandafter\renewcommand\csname LitJZeroThirtyRPlus\endcsname{\ensuremath{1.3}}
\expandafter\providecommand\csname LitWaveformFmax\endcsname{}\expandafter\renewcommand\csname LitWaveformFmax\endcsname{\ensuremath{1700}}
\expandafter\providecommand\csname LitWaveformFmin\endcsname{}\expandafter\renewcommand\csname LitWaveformFmin\endcsname{\ensuremath{23}}
\expandafter\providecommand\csname LitWysockiFrac\endcsname{}\expandafter\renewcommand\csname LitWysockiFrac\endcsname{\ensuremath{10^{-4}}}
\expandafter\providecommand\csname LitZhuEvents\endcsname{}\expandafter\renewcommand\csname LitZhuEvents\endcsname{\ensuremath{2.5\times10^{5}}}
\expandafter\providecommand\csname LitZhuParams\endcsname{}\expandafter\renewcommand\csname LitZhuParams\endcsname{\ensuremath{5}}
\expandafter\providecommand\csname MatchNsat\endcsname{}\expandafter\renewcommand\csname MatchNsat\endcsname{\ensuremath{2}}
\expandafter\providecommand\csname MetaEsat\endcsname{}\expandafter\renewcommand\csname MetaEsat\endcsname{\ensuremath{-16}}
\expandafter\providecommand\csname MetaGammaCrust\endcsname{}\expandafter\renewcommand\csname MetaGammaCrust\endcsname{\ensuremath{1.3}}
\expandafter\providecommand\csname MetaNmin\endcsname{}\expandafter\renewcommand\csname MetaNmin\endcsname{\ensuremath{3\times10^{-4}}}
\expandafter\providecommand\csname MetaNsat\endcsname{}\expandafter\renewcommand\csname MetaNsat\endcsname{\ensuremath{0.16}}
\expandafter\providecommand\csname MetaQsat\endcsname{}\expandafter\renewcommand\csname MetaQsat\endcsname{\ensuremath{-300}}
\expandafter\providecommand\csname MetaQsym\endcsname{}\expandafter\renewcommand\csname MetaQsym\endcsname{\ensuremath{0}}
\expandafter\providecommand\csname MillionArchiveEss\endcsname{}\expandafter\renewcommand\csname MillionArchiveEss\endcsname{\ensuremath{8.3\times10^{3}}}
\expandafter\providecommand\csname MixtureCombineEss\endcsname{}\expandafter\renewcommand\csname MixtureCombineEss\endcsname{\ensuremath{1112}}
\expandafter\providecommand\csname MixtureCombineLam\endcsname{}\expandafter\renewcommand\csname MixtureCombineLam\endcsname{\ensuremath{310}}
\expandafter\providecommand\csname MixtureCombineLamSd\endcsname{}\expandafter\renewcommand\csname MixtureCombineLamSd\endcsname{\ensuremath{162}}
\expandafter\providecommand\csname MixtureGenOneEss\endcsname{}\expandafter\renewcommand\csname MixtureGenOneEss\endcsname{\ensuremath{439}}
\expandafter\providecommand\csname MixtureGenOneLam\endcsname{}\expandafter\renewcommand\csname MixtureGenOneLam\endcsname{\ensuremath{284}}
\expandafter\providecommand\csname MixtureGenOneLamSd\endcsname{}\expandafter\renewcommand\csname MixtureGenOneLamSd\endcsname{\ensuremath{149}}
\expandafter\providecommand\csname MixtureGenOneMultiPct\endcsname{}\expandafter\renewcommand\csname MixtureGenOneMultiPct\endcsname{\ensuremath{11}}
\expandafter\providecommand\csname MixtureGenZeroMultiPct\endcsname{}\expandafter\renewcommand\csname MixtureGenZeroMultiPct\endcsname{\ensuremath{20}}
\expandafter\providecommand\csname MmaxPriorHi\endcsname{}\expandafter\renewcommand\csname MmaxPriorHi\endcsname{\ensuremath{2.8}}
\expandafter\providecommand\csname MmaxPriorLo\endcsname{}\expandafter\renewcommand\csname MmaxPriorLo\endcsname{\ensuremath{1.3}}
\expandafter\providecommand\csname MmaxPriorMedian\endcsname{}\expandafter\renewcommand\csname MmaxPriorMedian\endcsname{\ensuremath{2.02}}
\expandafter\providecommand\csname MultiAnalyticRedraws\endcsname{}\expandafter\renewcommand\csname MultiAnalyticRedraws\endcsname{\ensuremath{4}}
\expandafter\providecommand\csname MultiAnalyticSdEnd\endcsname{}\expandafter\renewcommand\csname MultiAnalyticSdEnd\endcsname{\ensuremath{34}}
\expandafter\providecommand\csname MultiAnalyticSdStart\endcsname{}\expandafter\renewcommand\csname MultiAnalyticSdStart\endcsname{\ensuremath{146}}
\expandafter\providecommand\csname MultiAnalyticShrink\endcsname{}\expandafter\renewcommand\csname MultiAnalyticShrink\endcsname{\ensuremath{4}}
\expandafter\providecommand\csname NestedFullBaseEss\endcsname{}\expandafter\renewcommand\csname NestedFullBaseEss\endcsname{\ensuremath{84}}
\expandafter\providecommand\csname NestedFullEss\endcsname{}\expandafter\renewcommand\csname NestedFullEss\endcsname{\ensuremath{898}}
\expandafter\providecommand\csname NestedFullLam\endcsname{}\expandafter\renewcommand\csname NestedFullLam\endcsname{\ensuremath{398}}
\expandafter\providecommand\csname NestedFullLamSd\endcsname{}\expandafter\renewcommand\csname NestedFullLamSd\endcsname{\ensuremath{77}}
\expandafter\providecommand\csname NestedFullMmax\endcsname{}\expandafter\renewcommand\csname NestedFullMmax\endcsname{\ensuremath{2.4}}
\expandafter\providecommand\csname NestedFullMmaxSd\endcsname{}\expandafter\renewcommand\csname NestedFullMmaxSd\endcsname{\ensuremath{0.2}}
\expandafter\providecommand\csname NestedFullRadius\endcsname{}\expandafter\renewcommand\csname NestedFullRadius\endcsname{\ensuremath{11.9}}
\expandafter\providecommand\csname NestedFullRadiusSd\endcsname{}\expandafter\renewcommand\csname NestedFullRadiusSd\endcsname{\ensuremath{0.4}}
\expandafter\providecommand\csname NestedGwEss\endcsname{}\expandafter\renewcommand\csname NestedGwEss\endcsname{\ensuremath{1427}}
\expandafter\providecommand\csname NestedGwLam\endcsname{}\expandafter\renewcommand\csname NestedGwLam\endcsname{\ensuremath{320}}
\expandafter\providecommand\csname NestedGwLamSd\endcsname{}\expandafter\renewcommand\csname NestedGwLamSd\endcsname{\ensuremath{181}}
\expandafter\providecommand\csname NestedPriorEss\endcsname{}\expandafter\renewcommand\csname NestedPriorEss\endcsname{\ensuremath{8953}}
\expandafter\providecommand\csname NestedPriorLam\endcsname{}\expandafter\renewcommand\csname NestedPriorLam\endcsname{\ensuremath{449}}
\expandafter\providecommand\csname NestedPriorLamSd\endcsname{}\expandafter\renewcommand\csname NestedPriorLamSd\endcsname{\ensuremath{244}}
\expandafter\providecommand\csname NestedPriorRadius\endcsname{}\expandafter\renewcommand\csname NestedPriorRadius\endcsname{\ensuremath{12.2}}
\expandafter\providecommand\csname NestedPriorRadiusSd\endcsname{}\expandafter\renewcommand\csname NestedPriorRadiusSd\endcsname{\ensuremath{1.6}}
\expandafter\providecommand\csname NestedPsrGwLam\endcsname{}\expandafter\renewcommand\csname NestedPsrGwLam\endcsname{\ensuremath{466}}
\expandafter\providecommand\csname NestedPsrGwLamSd\endcsname{}\expandafter\renewcommand\csname NestedPsrGwLamSd\endcsname{\ensuremath{108}}
\expandafter\providecommand\csname NestedRadiusShrink\endcsname{}\expandafter\renewcommand\csname NestedRadiusShrink\endcsname{\ensuremath{4}}
\expandafter\providecommand\csname NicerMassHi\endcsname{}\expandafter\renewcommand\csname NicerMassHi\endcsname{\ensuremath{2.1}}
\expandafter\providecommand\csname NicerMassLo\endcsname{}\expandafter\renewcommand\csname NicerMassLo\endcsname{\ensuremath{0.4}}
\expandafter\providecommand\csname PcaLamMaxErrPct\endcsname{}\expandafter\renewcommand\csname PcaLamMaxErrPct\endcsname{\ensuremath{0.8}}
\expandafter\providecommand\csname PcaLamRmsErrPct\endcsname{}\expandafter\renewcommand\csname PcaLamRmsErrPct\endcsname{\ensuremath{0.4}}
\expandafter\providecommand\csname PooledBaseDraws\endcsname{}\expandafter\renewcommand\csname PooledBaseDraws\endcsname{\ensuremath{1000}}
\expandafter\providecommand\csname PooledDraws\endcsname{}\expandafter\renewcommand\csname PooledDraws\endcsname{\ensuremath{10000}}
\expandafter\providecommand\csname PooledEvidenceRuns\endcsname{}\expandafter\renewcommand\csname PooledEvidenceRuns\endcsname{\ensuremath{2897}}
\expandafter\providecommand\csname PooledExtDraws\endcsname{}\expandafter\renewcommand\csname PooledExtDraws\endcsname{\ensuremath{9000}}
\expandafter\providecommand\csname PooledExtSelected\endcsname{}\expandafter\renewcommand\csname PooledExtSelected\endcsname{\ensuremath{2009}}
\expandafter\providecommand\csname PooledMultiBranchPct\endcsname{}\expandafter\renewcommand\csname PooledMultiBranchPct\endcsname{\ensuremath{18}}
\expandafter\providecommand\csname PqcdMuGeV\endcsname{}\expandafter\renewcommand\csname PqcdMuGeV\endcsname{\ensuremath{2.6}}
\expandafter\providecommand\csname PriorDrawsShown\endcsname{}\expandafter\renewcommand\csname PriorDrawsShown\endcsname{\ensuremath{200}}
\expandafter\providecommand\csname PsrBaseEss\endcsname{}\expandafter\renewcommand\csname PsrBaseEss\endcsname{\ensuremath{485}}
\expandafter\providecommand\csname PsrBaseEssPct\endcsname{}\expandafter\renewcommand\csname PsrBaseEssPct\endcsname{\ensuremath{49}}
\expandafter\providecommand\csname PsrPostMmaxHi\endcsname{}\expandafter\renewcommand\csname PsrPostMmaxHi\endcsname{\ensuremath{2.8}}
\expandafter\providecommand\csname PsrPostMmaxLo\endcsname{}\expandafter\renewcommand\csname PsrPostMmaxLo\endcsname{\ensuremath{2.1}}
\expandafter\providecommand\csname PsrPostMmaxMed\endcsname{}\expandafter\renewcommand\csname PsrPostMmaxMed\endcsname{\ensuremath{2.4}}
\expandafter\providecommand\csname PsrRevisedEss\endcsname{}\expandafter\renewcommand\csname PsrRevisedEss\endcsname{\ensuremath{4776}}
\expandafter\providecommand\csname PsrRevisedEssPct\endcsname{}\expandafter\renewcommand\csname PsrRevisedEssPct\endcsname{\ensuremath{96}}
\expandafter\providecommand\csname PsrRevisedMmaxHi\endcsname{}\expandafter\renewcommand\csname PsrRevisedMmaxHi\endcsname{\ensuremath{2.8}}
\expandafter\providecommand\csname PsrRevisedMmaxLo\endcsname{}\expandafter\renewcommand\csname PsrRevisedMmaxLo\endcsname{\ensuremath{2.1}}
\expandafter\providecommand\csname PsrRevisedMmaxMed\endcsname{}\expandafter\renewcommand\csname PsrRevisedMmaxMed\endcsname{\ensuremath{2.5}}
\expandafter\providecommand\csname PsrRevisedN\endcsname{}\expandafter\renewcommand\csname PsrRevisedN\endcsname{\ensuremath{5000}}
\expandafter\providecommand\csname SeqAllMaxOffset\endcsname{}\expandafter\renewcommand\csname SeqAllMaxOffset\endcsname{\ensuremath{1.71}}
\expandafter\providecommand\csname SeqAllWidthHi\endcsname{}\expandafter\renewcommand\csname SeqAllWidthHi\endcsname{\ensuremath{2.23}}
\expandafter\providecommand\csname SeqAllWidthLo\endcsname{}\expandafter\renewcommand\csname SeqAllWidthLo\endcsname{\ensuremath{0.88}}
\expandafter\providecommand\csname SeqClipNinetyNineEss\endcsname{}\expandafter\renewcommand\csname SeqClipNinetyNineEss\endcsname{\ensuremath{191}}
\expandafter\providecommand\csname SeqClipSqrtEss\endcsname{}\expandafter\renewcommand\csname SeqClipSqrtEss\endcsname{\ensuremath{29}}
\expandafter\providecommand\csname SeqComponents\endcsname{}\expandafter\renewcommand\csname SeqComponents\endcsname{\ensuremath{16}}
\expandafter\providecommand\csname SeqExactEss\endcsname{}\expandafter\renewcommand\csname SeqExactEss\endcsname{\ensuremath{3}}
\expandafter\providecommand\csname SeqFastEss\endcsname{}\expandafter\renewcommand\csname SeqFastEss\endcsname{\ensuremath{7736}}
\expandafter\providecommand\csname SeqFastMaxOffset\endcsname{}\expandafter\renewcommand\csname SeqFastMaxOffset\endcsname{\ensuremath{1.46}}
\expandafter\providecommand\csname SeqFastMean\endcsname{}\expandafter\renewcommand\csname SeqFastMean\endcsname{\ensuremath{406}}
\expandafter\providecommand\csname SeqFastRefitsHi\endcsname{}\expandafter\renewcommand\csname SeqFastRefitsHi\endcsname{\ensuremath{13}}
\expandafter\providecommand\csname SeqFastRefitsLo\endcsname{}\expandafter\renewcommand\csname SeqFastRefitsLo\endcsname{\ensuremath{9}}
\expandafter\providecommand\csname SeqFastScatter\endcsname{}\expandafter\renewcommand\csname SeqFastScatter\endcsname{\ensuremath{4.8}}
\expandafter\providecommand\csname SeqFastThreshold\endcsname{}\expandafter\renewcommand\csname SeqFastThreshold\endcsname{\ensuremath{6\times10^{3}}}
\expandafter\providecommand\csname SeqFastWidthHi\endcsname{}\expandafter\renewcommand\csname SeqFastWidthHi\endcsname{\ensuremath{1.34}}
\expandafter\providecommand\csname SeqFastWidthLo\endcsname{}\expandafter\renewcommand\csname SeqFastWidthLo\endcsname{\ensuremath{0.88}}
\expandafter\providecommand\csname SeqFracHi\endcsname{}\expandafter\renewcommand\csname SeqFracHi\endcsname{\ensuremath{0.95}}
\expandafter\providecommand\csname SeqFracLo\endcsname{}\expandafter\renewcommand\csname SeqFracLo\endcsname{\ensuremath{0.04}}
\expandafter\providecommand\csname SeqMidEss\endcsname{}\expandafter\renewcommand\csname SeqMidEss\endcsname{\ensuremath{6273}}
\expandafter\providecommand\csname SeqMidMaxOffset\endcsname{}\expandafter\renewcommand\csname SeqMidMaxOffset\endcsname{\ensuremath{1.53}}
\expandafter\providecommand\csname SeqMidMean\endcsname{}\expandafter\renewcommand\csname SeqMidMean\endcsname{\ensuremath{406}}
\expandafter\providecommand\csname SeqMidRefitsHi\endcsname{}\expandafter\renewcommand\csname SeqMidRefitsHi\endcsname{\ensuremath{7}}
\expandafter\providecommand\csname SeqMidRefitsLo\endcsname{}\expandafter\renewcommand\csname SeqMidRefitsLo\endcsname{\ensuremath{4}}
\expandafter\providecommand\csname SeqMidScatter\endcsname{}\expandafter\renewcommand\csname SeqMidScatter\endcsname{\ensuremath{6.7}}
\expandafter\providecommand\csname SeqMidThreshold\endcsname{}\expandafter\renewcommand\csname SeqMidThreshold\endcsname{\ensuremath{3\times10^{3}}}
\expandafter\providecommand\csname SeqMidWidthHi\endcsname{}\expandafter\renewcommand\csname SeqMidWidthHi\endcsname{\ensuremath{1.38}}
\expandafter\providecommand\csname SeqMidWidthLo\endcsname{}\expandafter\renewcommand\csname SeqMidWidthLo\endcsname{\ensuremath{0.97}}
\expandafter\providecommand\csname SeqOrderings\endcsname{}\expandafter\renewcommand\csname SeqOrderings\endcsname{\ensuremath{8}}
\expandafter\providecommand\csname SeqPlainEss\endcsname{}\expandafter\renewcommand\csname SeqPlainEss\endcsname{\ensuremath{1436}}
\expandafter\providecommand\csname SeqPlainMaxOffset\endcsname{}\expandafter\renewcommand\csname SeqPlainMaxOffset\endcsname{\ensuremath{1.71}}
\expandafter\providecommand\csname SeqPlainMean\endcsname{}\expandafter\renewcommand\csname SeqPlainMean\endcsname{\ensuremath{399}}
\expandafter\providecommand\csname SeqPlainRefitsHi\endcsname{}\expandafter\renewcommand\csname SeqPlainRefitsHi\endcsname{\ensuremath{3}}
\expandafter\providecommand\csname SeqPlainRefitsLo\endcsname{}\expandafter\renewcommand\csname SeqPlainRefitsLo\endcsname{\ensuremath{2}}
\expandafter\providecommand\csname SeqPlainScatter\endcsname{}\expandafter\renewcommand\csname SeqPlainScatter\endcsname{\ensuremath{9.0}}
\expandafter\providecommand\csname SeqPlainWidthHi\endcsname{}\expandafter\renewcommand\csname SeqPlainWidthHi\endcsname{\ensuremath{2.23}}
\expandafter\providecommand\csname SeqPlainWidthLo\endcsname{}\expandafter\renewcommand\csname SeqPlainWidthLo\endcsname{\ensuremath{0.90}}
\expandafter\providecommand\csname SeqPoints\endcsname{}\expandafter\renewcommand\csname SeqPoints\endcsname{\ensuremath{80}}
\expandafter\providecommand\csname SeqRefitJumpPlain\endcsname{}\expandafter\renewcommand\csname SeqRefitJumpPlain\endcsname{\ensuremath{1.27}}
\expandafter\providecommand\csname SeqRefitJumpShrunk\endcsname{}\expandafter\renewcommand\csname SeqRefitJumpShrunk\endcsname{\ensuremath{2.5}}
\expandafter\providecommand\csname SeqScanSteps\endcsname{}\expandafter\renewcommand\csname SeqScanSteps\endcsname{\ensuremath{40}}
\expandafter\providecommand\csname SeqSeedRuns\endcsname{}\expandafter\renewcommand\csname SeqSeedRuns\endcsname{\ensuremath{5}}
\expandafter\providecommand\csname SeqSeedScatter\endcsname{}\expandafter\renewcommand\csname SeqSeedScatter\endcsname{\ensuremath{1.3}}
\expandafter\providecommand\csname SeqShrinkGamma\endcsname{}\expandafter\renewcommand\csname SeqShrinkGamma\endcsname{\ensuremath{0.1}}
\expandafter\providecommand\csname SeqShrinkPostSd\endcsname{}\expandafter\renewcommand\csname SeqShrinkPostSd\endcsname{\ensuremath{37}}
\expandafter\providecommand\csname SeqShrinkPreSd\endcsname{}\expandafter\renewcommand\csname SeqShrinkPreSd\endcsname{\ensuremath{14}}
\expandafter\providecommand\csname SeqShrinkRefitN\endcsname{}\expandafter\renewcommand\csname SeqShrinkRefitN\endcsname{\ensuremath{690}}
\expandafter\providecommand\csname SeqShrunkEss\endcsname{}\expandafter\renewcommand\csname SeqShrunkEss\endcsname{\ensuremath{4018}}
\expandafter\providecommand\csname SeqShrunkMaxOffset\endcsname{}\expandafter\renewcommand\csname SeqShrunkMaxOffset\endcsname{\ensuremath{2.85}}
\expandafter\providecommand\csname SeqShrunkMean\endcsname{}\expandafter\renewcommand\csname SeqShrunkMean\endcsname{\ensuremath{403}}
\expandafter\providecommand\csname SeqShrunkRefitsHi\endcsname{}\expandafter\renewcommand\csname SeqShrunkRefitsHi\endcsname{\ensuremath{5}}
\expandafter\providecommand\csname SeqShrunkRefitsLo\endcsname{}\expandafter\renewcommand\csname SeqShrunkRefitsLo\endcsname{\ensuremath{3}}
\expandafter\providecommand\csname SeqShrunkScatter\endcsname{}\expandafter\renewcommand\csname SeqShrunkScatter\endcsname{\ensuremath{14.5}}
\expandafter\providecommand\csname SeqShrunkWidthHi\endcsname{}\expandafter\renewcommand\csname SeqShrunkWidthHi\endcsname{\ensuremath{2.37}}
\expandafter\providecommand\csname SeqShrunkWidthLo\endcsname{}\expandafter\renewcommand\csname SeqShrunkWidthLo\endcsname{\ensuremath{1.13}}
\expandafter\providecommand\csname SeqTailFewMedian\endcsname{}\expandafter\renewcommand\csname SeqTailFewMedian\endcsname{\ensuremath{1.25}}
\expandafter\providecommand\csname SeqTailManyHi\endcsname{}\expandafter\renewcommand\csname SeqTailManyHi\endcsname{\ensuremath{1.03}}
\expandafter\providecommand\csname SeqTailManyMedian\endcsname{}\expandafter\renewcommand\csname SeqTailManyMedian\endcsname{\ensuremath{0.94}}
\expandafter\providecommand\csname SubCatCount\endcsname{}\expandafter\renewcommand\csname SubCatCount\endcsname{\ensuremath{4}}
\expandafter\providecommand\csname SubCatEvents\endcsname{}\expandafter\renewcommand\csname SubCatEvents\endcsname{\ensuremath{250}}
\expandafter\providecommand\csname SubClipPlainMaxOffset\endcsname{}\expandafter\renewcommand\csname SubClipPlainMaxOffset\endcsname{\ensuremath{0.85}}
\expandafter\providecommand\csname SubClipPlainWidthHi\endcsname{}\expandafter\renewcommand\csname SubClipPlainWidthHi\endcsname{\ensuremath{1.05}}
\expandafter\providecommand\csname SubClipPlainWidthLo\endcsname{}\expandafter\renewcommand\csname SubClipPlainWidthLo\endcsname{\ensuremath{0.94}}
\expandafter\providecommand\csname SubClipShrunkMaxOffset\endcsname{}\expandafter\renewcommand\csname SubClipShrunkMaxOffset\endcsname{\ensuremath{0.31}}
\expandafter\providecommand\csname SubClipShrunkWidthHi\endcsname{}\expandafter\renewcommand\csname SubClipShrunkWidthHi\endcsname{\ensuremath{1.16}}
\expandafter\providecommand\csname SubClipShrunkWidthLo\endcsname{}\expandafter\renewcommand\csname SubClipShrunkWidthLo\endcsname{\ensuremath{0.97}}
\expandafter\providecommand\csname SubSeqPlainMaxOffset\endcsname{}\expandafter\renewcommand\csname SubSeqPlainMaxOffset\endcsname{\ensuremath{1.04}}
\expandafter\providecommand\csname SubSeqPlainWidthHi\endcsname{}\expandafter\renewcommand\csname SubSeqPlainWidthHi\endcsname{\ensuremath{1.19}}
\expandafter\providecommand\csname SubSeqPlainWidthLo\endcsname{}\expandafter\renewcommand\csname SubSeqPlainWidthLo\endcsname{\ensuremath{0.94}}
\expandafter\providecommand\csname SubSeqShrunkMaxOffset\endcsname{}\expandafter\renewcommand\csname SubSeqShrunkMaxOffset\endcsname{\ensuremath{0.43}}
\expandafter\providecommand\csname SubSeqShrunkWidthHi\endcsname{}\expandafter\renewcommand\csname SubSeqShrunkWidthHi\endcsname{\ensuremath{1.44}}
\expandafter\providecommand\csname SubSeqShrunkWidthLo\endcsname{}\expandafter\renewcommand\csname SubSeqShrunkWidthLo\endcsname{\ensuremath{1.01}}
\expandafter\providecommand\csname SymCurvPostSd\endcsname{}\expandafter\renewcommand\csname SymCurvPostSd\endcsname{\ensuremath{91}}
\expandafter\providecommand\csname SymCurvPriorSd\endcsname{}\expandafter\renewcommand\csname SymCurvPriorSd\endcsname{\ensuremath{108}}
\expandafter\providecommand\csname SymFisherCorr\endcsname{}\expandafter\renewcommand\csname SymFisherCorr\endcsname{\ensuremath{-0.91}}
\expandafter\providecommand\csname SymFisherCurvSd\endcsname{}\expandafter\renewcommand\csname SymFisherCurvSd\endcsname{\ensuremath{80}}
\expandafter\providecommand\csname SymFisherSlopeSd\endcsname{}\expandafter\renewcommand\csname SymFisherSlopeSd\endcsname{\ensuremath{15.1}}
\expandafter\providecommand\csname SymGlobalLooBias\endcsname{}\expandafter\renewcommand\csname SymGlobalLooBias\endcsname{\ensuremath{10.6}}
\expandafter\providecommand\csname SymGlobalLooRms\endcsname{}\expandafter\renewcommand\csname SymGlobalLooRms\endcsname{\ensuremath{20.5}}
\expandafter\providecommand\csname SymGlobalSd\endcsname{}\expandafter\renewcommand\csname SymGlobalSd\endcsname{\ensuremath{21.1}}
\expandafter\providecommand\csname SymHMax\endcsname{}\expandafter\renewcommand\csname SymHMax\endcsname{\ensuremath{0.3}}
\expandafter\providecommand\csname SymHMin\endcsname{}\expandafter\renewcommand\csname SymHMin\endcsname{\ensuremath{0.1}}
\expandafter\providecommand\csname SymKernelEss\endcsname{}\expandafter\renewcommand\csname SymKernelEss\endcsname{\ensuremath{280}}
\expandafter\providecommand\csname SymKernelH\endcsname{}\expandafter\renewcommand\csname SymKernelH\endcsname{\ensuremath{0.2}}
\expandafter\providecommand\csname SymLamCorr\endcsname{}\expandafter\renewcommand\csname SymLamCorr\endcsname{\ensuremath{0.03}}
\expandafter\providecommand\csname SymLamFourteenPct\endcsname{}\expandafter\renewcommand\csname SymLamFourteenPct\endcsname{\ensuremath{2.9}}
\expandafter\providecommand\csname SymLamTwelvePct\endcsname{}\expandafter\renewcommand\csname SymLamTwelvePct\endcsname{\ensuremath{2.8}}
\expandafter\providecommand\csname SymLooBias\endcsname{}\expandafter\renewcommand\csname SymLooBias\endcsname{\ensuremath{0.5}}
\expandafter\providecommand\csname SymLooN\endcsname{}\expandafter\renewcommand\csname SymLooN\endcsname{\ensuremath{281}}
\expandafter\providecommand\csname SymLooPred\endcsname{}\expandafter\renewcommand\csname SymLooPred\endcsname{\ensuremath{16.2}}
\expandafter\providecommand\csname SymLooRms\endcsname{}\expandafter\renewcommand\csname SymLooRms\endcsname{\ensuremath{16.0}}
\expandafter\providecommand\csname SymLooWindow\endcsname{}\expandafter\renewcommand\csname SymLooWindow\endcsname{\ensuremath{0.3}}
\expandafter\providecommand\csname SymLooZ\endcsname{}\expandafter\renewcommand\csname SymLooZ\endcsname{\ensuremath{1.0}}
\expandafter\providecommand\csname SymMassHi\endcsname{}\expandafter\renewcommand\csname SymMassHi\endcsname{\ensuremath{1.4}}
\expandafter\providecommand\csname SymMassLo\endcsname{}\expandafter\renewcommand\csname SymMassLo\endcsname{\ensuremath{1.2}}
\expandafter\providecommand\csname SymPostCorr\endcsname{}\expandafter\renewcommand\csname SymPostCorr\endcsname{\ensuremath{-0.82}}
\expandafter\providecommand\csname SymScanPct\endcsname{}\expandafter\renewcommand\csname SymScanPct\endcsname{\ensuremath{3}}
\expandafter\providecommand\csname SymSlopeAdOneMass\endcsname{}\expandafter\renewcommand\csname SymSlopeAdOneMass\endcsname{\ensuremath{18.7}}
\expandafter\providecommand\csname SymSlopeAdSd\endcsname{}\expandafter\renewcommand\csname SymSlopeAdSd\endcsname{\ensuremath{19}}
\expandafter\providecommand\csname SymSlopeAdSlySd\endcsname{}\expandafter\renewcommand\csname SymSlopeAdSlySd\endcsname{\ensuremath{19}}
\expandafter\providecommand\csname SymSlopeHMax\endcsname{}\expandafter\renewcommand\csname SymSlopeHMax\endcsname{\ensuremath{16.4}}
\expandafter\providecommand\csname SymSlopeHMin\endcsname{}\expandafter\renewcommand\csname SymSlopeHMin\endcsname{\ensuremath{16.0}}
\expandafter\providecommand\csname SymSlopeLowMass\endcsname{}\expandafter\renewcommand\csname SymSlopeLowMass\endcsname{\ensuremath{14.7}}
\expandafter\providecommand\csname SymSlopeOneMassSd\endcsname{}\expandafter\renewcommand\csname SymSlopeOneMassSd\endcsname{\ensuremath{22.8}}
\expandafter\providecommand\csname SymSlopePairMass\endcsname{}\expandafter\renewcommand\csname SymSlopePairMass\endcsname{\ensuremath{16.3}}
\expandafter\providecommand\csname SymSlopePerfectSd\endcsname{}\expandafter\renewcommand\csname SymSlopePerfectSd\endcsname{\ensuremath{15.3}}
\expandafter\providecommand\csname SymSlopePostMean\endcsname{}\expandafter\renewcommand\csname SymSlopePostMean\endcsname{\ensuremath{59.3}}
\expandafter\providecommand\csname SymSlopePostSd\endcsname{}\expandafter\renewcommand\csname SymSlopePostSd\endcsname{\ensuremath{16.1}}
\expandafter\providecommand\csname SymSlopePriorSd\endcsname{}\expandafter\renewcommand\csname SymSlopePriorSd\endcsname{\ensuremath{27.0}}
\expandafter\providecommand\csname SymSlopeSingleMass\endcsname{}\expandafter\renewcommand\csname SymSlopeSingleMass\endcsname{\ensuremath{22.7}}
\expandafter\providecommand\csname SymSlopeSpanMass\endcsname{}\expandafter\renewcommand\csname SymSlopeSpanMass\endcsname{\ensuremath{39.7}}
\expandafter\providecommand\csname SymSlopeTruth\endcsname{}\expandafter\renewcommand\csname SymSlopeTruth\endcsname{\ensuremath{41.9}}
\expandafter\providecommand\csname SymSlopeTruthZ\endcsname{}\expandafter\renewcommand\csname SymSlopeTruthZ\endcsname{\ensuremath{1.1}}
\expandafter\providecommand\csname SymSpanLooPred\endcsname{}\expandafter\renewcommand\csname SymSpanLooPred\endcsname{\ensuremath{38.9}}
\expandafter\providecommand\csname SymSpanLooRms\endcsname{}\expandafter\renewcommand\csname SymSpanLooRms\endcsname{\ensuremath{35.7}}
\expandafter\providecommand\csname SymSpanMasses\endcsname{}\expandafter\renewcommand\csname SymSpanMasses\endcsname{\ensuremath{1.0,\ 1.2,\ 1.4,\ \mbox{and}\ 1.6}}
\expandafter\providecommand\csname SynthTruthLam\endcsname{}\expandafter\renewcommand\csname SynthTruthLam\endcsname{\ensuremath{401}}
\expandafter\providecommand\csname TwoTransDeltaOne\endcsname{}\expandafter\renewcommand\csname TwoTransDeltaOne\endcsname{\ensuremath{65}}
\expandafter\providecommand\csname TwoTransDeltaOneRate\endcsname{}\expandafter\renewcommand\csname TwoTransDeltaOneRate\endcsname{\ensuremath{0.65}}
\expandafter\providecommand\csname TwoTransDeltaTwo\endcsname{}\expandafter\renewcommand\csname TwoTransDeltaTwo\endcsname{\ensuremath{110}}
\expandafter\providecommand\csname TwoTransDeltaTwoRate\endcsname{}\expandafter\renewcommand\csname TwoTransDeltaTwoRate\endcsname{\ensuremath{0.65}}
\expandafter\providecommand\csname TwoTransDim\endcsname{}\expandafter\renewcommand\csname TwoTransDim\endcsname{\ensuremath{14}}
\expandafter\providecommand\csname TwoTransDimMin\endcsname{}\expandafter\renewcommand\csname TwoTransDimMin\endcsname{\ensuremath{6.03}}
\expandafter\providecommand\csname TwoTransDimMinErr\endcsname{}\expandafter\renewcommand\csname TwoTransDimMinErr\endcsname{\ensuremath{0.29}}
\expandafter\providecommand\csname TwoTransDimOne\endcsname{}\expandafter\renewcommand\csname TwoTransDimOne\endcsname{\ensuremath{4.94}}
\expandafter\providecommand\csname TwoTransDimOneErr\endcsname{}\expandafter\renewcommand\csname TwoTransDimOneErr\endcsname{\ensuremath{0.12}}
\expandafter\providecommand\csname TwoTransDimTwo\endcsname{}\expandafter\renewcommand\csname TwoTransDimTwo\endcsname{\ensuremath{5.42}}
\expandafter\providecommand\csname TwoTransDimTwoErr\endcsname{}\expandafter\renewcommand\csname TwoTransDimTwoErr\endcsname{\ensuremath{0.18}}
\expandafter\providecommand\csname TwoTransEtaDrop\endcsname{}\expandafter\renewcommand\csname TwoTransEtaDrop\endcsname{\ensuremath{42}}
\expandafter\providecommand\csname TwoTransEtaMin\endcsname{}\expandafter\renewcommand\csname TwoTransEtaMin\endcsname{\ensuremath{3.9\times10^{-6}}}
\expandafter\providecommand\csname TwoTransOnsetOne\endcsname{}\expandafter\renewcommand\csname TwoTransOnsetOne\endcsname{\ensuremath{1.17}}
\expandafter\providecommand\csname TwoTransOnsetTwo\endcsname{}\expandafter\renewcommand\csname TwoTransOnsetTwo\endcsname{\ensuremath{1.80}}
\expandafter\providecommand\csname TwoTransPOne\endcsname{}\expandafter\renewcommand\csname TwoTransPOne\endcsname{\ensuremath{20}}
\expandafter\providecommand\csname TwoTransPOneRate\endcsname{}\expandafter\renewcommand\csname TwoTransPOneRate\endcsname{\ensuremath{0.4}}
\expandafter\providecommand\csname TwoTransPTwo\endcsname{}\expandafter\renewcommand\csname TwoTransPTwo\endcsname{\ensuremath{65}}
\expandafter\providecommand\csname TwoTransPTwoRate\endcsname{}\expandafter\renewcommand\csname TwoTransPTwoRate\endcsname{\ensuremath{0.35}}
\expandafter\providecommand\csname TwoTransQuarkTwo\endcsname{}\expandafter\renewcommand\csname TwoTransQuarkTwo\endcsname{\ensuremath{0.85}}
\expandafter\providecommand\csname TwoTransSigmaMin\endcsname{}\expandafter\renewcommand\csname TwoTransSigmaMin\endcsname{\ensuremath{35}}
\expandafter\providecommand\csname TwoTransSigmaOne\endcsname{}\expandafter\renewcommand\csname TwoTransSigmaOne\endcsname{\ensuremath{100}}
\expandafter\providecommand\csname TwoTransSigmaTwo\endcsname{}\expandafter\renewcommand\csname TwoTransSigmaTwo\endcsname{\ensuremath{50}}

\begin{document}
\title{Adaptive proposals for nonparametric equation-of-state inference with RIFT:
iterative Gaussian-process refinement in a compressed sequence representation}
\author{R.~O'Shaughnessy}\affiliation{Rochester Institute of Technology}
\author{Z.~Zhu}\affiliation{Rochester Institute of Technology}
\begin{abstract}
Nonparametric equation-of-state (EOS) models have two practical problems for Bayesian inference: too many parameters, introducing
observationally inaccessible degrees of freedom; and no painless route to adaptive refinement, necessary to forecast
future observing scenarios.
In this work we address both problems, providing a framework for adaptive nonparmetric EOS sampling which simultaneously provides insight
into the observationally-accessible effective dimensions.
Applied to a published Gaussian-process ensemble, this analysis
reveals substantial sound-speed freedom beyond what observations demand.
Conversely, by concrete example, we demonstrate that often-used schemes to use many EOS realizations will invariably break down when
challenged by new physics, as the observationally-accessible dimension increases substantially.
These functional adaptive sampling tools provide a path forward to efficient computation, interpretation, and discovery
for future analyses of  cold, beta-equilibrium nuclear matter.
\end{abstract}
\maketitle

\section{Introduction}\label{sec:intro}
The equation of state (EOS) of cold matter above nuclear saturation density is a
central open problem connecting nuclear physics, gravitational-wave (GW)
astronomy, and X-ray timing.  The binary-neutron-star inspirals
GW170817~\cite{2017PhRvL.119p1101A,LIGO-GW170817-SourceProperties,
LIGO-GW170817-EOS} and
GW190425~\cite{2020ApJ...892L...3A,2026arXiv260527225T} provide GW information about the tidal
deformability $\Lambda$.  NICER X-ray pulse-profile modeling constrains the
mass--radius relation for individual
pulsars~\cite{J0030_2026,Salmi2024,Dittmann2024,Choudhury2024,Mauviard2025}, while
radio timing of massive pulsars~\cite{Fonseca2021,Arzoumanian2018} provides lower
bounds on the maximum mass.
Early theoretical investigations  established how stacking binary-neutron-star
observations can reconstruct the EOS~\cite{2015PhRvD..91d3002L}.  More complete
catalog treatments show, however, that the loudest events do not dominate EOS
inference: numerous weaker detections contribute critically, and the mass
distribution controls which density ranges are resolved~\cite{gwastro-ns-eos-Kedia2024,
gwastro-nsnuc-Fisher3g-Zhu2025}.
Third-generation gravitational-wave observatories --- Cosmic
Explorer~\cite{CosmicExplorer2021} and the Einstein
Telescope~\cite{EinsteinTelescope2010,ETscience2020} --- are expected to detect
binary neutron stars in the hundreds of thousands.  Forecasts using a range of
Bayesian samplers and Fisher-matrix approximations project correspondingly sharp
constraints on the equation of state~\cite{WalkerEtAl2024,BandopadhyayEtAl2024,WilliamsEtAl2026,
gwastro-nsnuc-Fisher3g-Zhu2025}.  As they probe complementary density ranges and have distinct systematics, these high-precision astronomical probes will complement
laboratory and theoretical programs on the same matter: parity-violating
electron scattering measurements of the neutron
skin~\cite{PREX2021,CREX2022} and their implications for the symmetry
energy~\cite{ReedEtAl2021}, heavy-ion collisions and rare-isotope
measurements~\cite{HuthEtAl2022,TsangEtAl2024}, and \emph{ab initio}
chiral-effective-field-theory calculations of neutron-rich
matter~\cite{Drischler2019,Drischler2021review}.
Combining these extremely heterogeneous constraints  requires both a
physical EOS model and a statistical representation of uncertainty over EOS
functions.
Joint GW--NICER EOS analyses have already established the importance of a
common EOS, explicit prior choices, and consistent source-level
likelihoods~\cite{gwastro-nsnuc-SteinerJoint-2020,RaaijmakersEtAl2020,
RaaijmakersEtAl2021,PangEtAl2021,RutherfordEtAl2024}.
As these constraints tighten, the range of viable solutions narrows, posing a
practical computational challenge.

Contemporary astronomical constraints are sufficiently uninformative that uniform prior draws have been a highly effective and
widely adopted sampling strategy.
However,
a third-generation network is expected to detect
binary neutron stars at rates orders of magnitude above the current
one~\cite{WalkerEtAl2024,BandopadhyayEtAl2024,WilliamsEtAl2026}.  Particularly when hunting after the increased
complexity and discovery that most motivates studies of dense nuclear matter (e.g., phase transitions; quark matter; et
cetera), however, the complexity of the inference challenge expands substantially -- as we show in this work,
prohibitively so, such that uniform prior draws will no longer remain effective.
[We demonstrate this failure with a constructive example in
Appendix~\ref{app:transition-stress}.]
Even with current observations, however, poor
coordinates can further reduce sampling efficiency: for a simple parametric EOS
family, Wysocki et al.~\cite{gwastro-PopulationReconstruct-EOSStack-Wysocki2019}
found that fewer than $\result{LitWysockiFrac}$ of uniform proposals passed their physical
filters.  A useful approach must therefore establish sampling adequacy while
exposing the physical combinations that the data can distinguish.

Two broad modeling strategies are widely used.  Parametric families, including
piecewise polytropes~\cite{Read2009} and spectral
representations~\cite{Lindblom2010}, are low-dimensional and tractable, but their
restricted functional forms can produce representation-dependent inference
systematics~\cite{Carney2018,GambaReadWade2020,LegredEtAl2022,
Biswas2025extrap}.  Nonparametric constructions
instead place a Gaussian-process (GP) prior over an auxiliary function, commonly
related to the sound speed $\cs^2(n)$~\cite{LandryEssick2019,Essick2020,
LandryEtAl2020,Legred2021};
recent work extends such priors from the crust to perturbative-QCD
densities~\cite{Finch2025}.  Complementary learned representations include
normalizing-flow EOS emulators~\cite{McGinnEtAl2024}, physics-informed flow
priors for compact-binary parameter estimation~\cite{WoutersEtAl2026}, direct mass--radius
priors~\cite{MRprior2026}, and neural EOS inference and stellar-structure
emulators~\cite{FujimotoEtAl2018,ReedEtAl2024,TiwariPai2024,KaushikkEtAl2026},
and neural posterior models~\cite{Carvalho2025NPE,AmortizedRMF2026}.
Microscopic nuclear calculations are also beginning to use differentiable,
correlated GP models~\cite{LeeEtAl2026GPDiff}; unlike the sequence-space
compression developed here, these regress microscopic matter predictions.
These approaches differ in their physical coordinates and inductive
assumptions, but all must resolve a posterior that can occupy a small fraction
of the original functional ensemble.

Existing EOS analyses generally specify the functional prior before conditioning
on the data, as required for a well-defined Bayesian target.  Richer choices ---
including microscopic model mixing~\cite{Semposki2025}, crust-to-pQCD
constructions~\cite{Finch2025}, and semiparametric model-trained
GPs~\cite{Mihaylov2025} --- alter that target model, but do not by themselves
solve a separate computational problem: a finite draw set from the chosen prior
may poorly resolve a concentrated posterior.
Iterative proposal and surrogate
refinement are well developed for GW parameter inference, including the RIFT
pipeline~\cite{gwastro-PENR-RIFT,gwastro-RIFT-Update}.
This motivates an
analogous adaptive proposal in EOS function space, with an explicit correction
back to the fixed base-prior posterior.
While a previous study with RIFT performed iterative refinement for the EOS within the context of a fixed set of EOS prior
draws, for single-event inference ~\cite{gwastro-ns-eos-Vilkha2024}, this proposal did not re-sample EOS realizations
from a common prior.

In this paper, we adapt RIFT's inference-driven refinement strategy to EOS
inference.  First, we formulate
adaptive proposal refinement using multiple importance sampling: draws from
successive data-informed proposals are combined with balance-heuristic weights
to target the posterior defined by a fixed base prior, provided the proposals
retain support over that target.  Second, we implement the loop in two
representations --- a compressed principal-component parametrization of the
neutron-star sequence $\{M,R,\Lambda\}(h_c)$ and a conventional sound-speed
parametrization with a differentiable Tolman--Oppenheimer--Volkoff (TOV) forward
map.  Third, we construct the base prior with the constrained bridge of
Ref.~\cite{GKKS2025bridge}, but supply our own low-density anchor: where that
work fixes a crust below $\result{CrustEdgeNsat}\nsat$ and draws the anchor from a
chiral-effective-field-theory ensemble, we resample a nuclear
metamodel~\cite{Margueron2018}.   We deliberately adopt this broad prior so that the prior is weakly informative and
subsequent measurements make the posterior change appreciably relative to the prior; a prior that conversely
imposed every available constraint provides little room for future measurements to improve upon, particularly GW
measurements as we discuss in our work.
Fourth, for growing catalogues we introduce multiple methods to efficiently perform the reweighting, including a sequential
Gaussian projection; and test them against  a large
fixed-prior reference for up to \result{HeroEvents} events.
We use this large-catalog study both to validate our method and to reassess forecasts for nuclear-parameter constraints
with third-generation GW surveys. 
Fifth, we compare effective dimensions in EOS and stellar-observable
space to identify which matter properties the observations can distinguish.

This paper is organized as follows.  Section~\ref{sec:methods} develops the
EOS representations, evidence evaluation, and refinement procedure, and describes
the synthetic observations and a cross-model validation by differentiation.  Section~\ref{sec:results} tests adaptive refinement on synthetic catalogues of future
GW observations and forecasts the resulting nuclear-parameter constraints.   Section~\ref{sec:information}
examines the observable information retained by the EOS representation, the
computational cost of sampling it, and its implications.  Section~\ref{sec:conclusions} summarizes our
results.  Appendix~\ref{app:observations} demonstrates our tools using current observations,
Appendix~\ref{sec:lowdensity} discusses why we adopt a broad low-density EOS prior and what implications that choice has,
Appendix~\ref{app:transition-stress} gives an explicit demonstration that a $\result{BenchArchiveDraws}$-draw prior archive can
fail to resolve a mass--radius likelihood that adaptive refinement successfully
samples, and the Supplemental Material provides technical derivations and implementation checks.

Our investigation is important because current techniques overparameterize (GP) or under-resolve (large fixed-prior
sampling); our method provides both interpretability and efficiency.  In other words,
flexible GP priors admit sound-speed features that present stellar measurements
need not constrain, and EOS posteriors can be difficult to resolve even with
large prior archives.  Our demonstration in Appendix~\ref{app:transition-stress} proves  that the EOS model space may well have
extremely many observationally accessible dimensions, far more than can be sampled by brute force a priori.
In the main text, we demonstrate how to sample efficiently, show why the EOS model space has low effective dimension,
and provide a concrete algorithm to compute that dimension. 
our proof-of-concept study provides an avenue to more accurate and more interpretable EOS constraints.

\section{Methods}\label{sec:methods}
We use the publicly available NuclearMatter-Backend package
(\texttt{nmbackend}) for EOS generation, stellar-structure calculations,
sequence compression, and adaptive proposal refinement~\cite{NuclearMatterBackend}.

\subsection{Nuclear matter and neutron stars}
\subsubsection{Neutron-star structure and tides}
\label{sec:tov}
For a given cold, $\beta$-equilibrated EOS we obtain the neutron-star sequence by
integrating the Tolman--Oppenheimer--Volkoff (TOV) equations in the
pseudo-enthalpy variable $h$~\cite{Lindblom1992}.
Specifically, with $(r,m)$ the areal radius and the enclosed
gravitational mass, we solve
\begin{equation}
\frac{dr}{dh} = -\frac{r(r-2m)}{m + 4\pi r^3 p},
\qquad
\frac{dm}{dh} = 4\pi r^2 \epsilon \,\frac{dr}{dh},
\label{eq:tov}
\end{equation}
integrated alongside the standard $\ell=2$ tidal equation, which returns the
Love number $k_2$ and hence the dimensionless tidal deformability
$\Lambda=\tfrac{2}{3}k_2C^{-5}$ at compactness
$C=M/R$~\cite{Hinderer2008,Postnikov2010,DamourNagar2009}.
While our internal representation characterizes NS by their central enthalpy $h_c$, enabling multiple stable branches,
in this work we emphasize a single stable branch given the limitations of downstream tools described later.

The solver is implemented in JAX~\cite{BradburyEtAl2021JAX}, integrating
Eq.~\eqref{eq:tov} and the tidal system with an adaptive Dormand--Prince scheme
and PID step control from \texttt{diffrax}~\cite{Kidger2022diffrax}.
Forward-mode automatic differentiation propagates through the discretized
integration and yields $\partial\{M,R,\Lambda\}/\partial\theta$ without finite
differencing, which we use for fixed-mass derivatives via the
implicit-function theorem.  We calibrate the implementation against an
independent radius-domain integrator, the LALSuite~\cite{lalsuite-soft}
routines used by the downstream RIFT consumer, and
RePrimAnd~\cite{Kastaun2021reprimand}.  Supplemental Material, Sec.~{SI} gives the
tidal system, the treatment of first-order transitions and of the stellar
surface, the EOS interpolation, and the validation itself.

\subsubsection{Nonparametric EOS modeling}\label{sec:rep}
The EOS at $\beta$-equilibrium $p(\epsilon)$ is poorly known above nuclear saturation density $\nsat$.  Several
nonparametric strategies have been proposed to flexibly extend the EOS above its low-density and high-density targets
(henceforth abbreviated as anchors). Different approaches achieve different levels of success at satisfying
thermodynamic consistency, causality, and capability of matching  the two  anchors. 

The foundational construction of Landry \&
Essick~\cite{LandryEssick2019} builds a Gaussian process (GP) model starting  with the primitive variable
\begin{equation}
\varphi(x) = \ln\!\left(c^2\,\frac{d\epsilon}{dp} - 1\right),
\qquad x = \log p,
\end{equation}
so that the sound speed is recovered through the sigmoid map
\begin{equation}
\cs^2 = \frac{c^2}{1 + e^{\varphi}},
\label{eq:LEsigmoid}
\end{equation}
which enforces $0<\cs^2<c^2$ for any real $\varphi$.  A GP prior is placed on
$\varphi(x)$ with a squared-exponential kernel
\begin{equation}
K(x,x') = \sigma^2 \exp\!\left[-\frac{(x-x')^2}{2\ell^2}\right]
          + \sigma_n^2\,\delta(x-x'),
\label{eq:SEkernel}
\end{equation}
($\sigma$ the marginal scale, $\ell$ the correlation length, $\sigma_n$ a
white-noise term).  A draw $\varphi(\log p)$ is integrated through
$d\epsilon = (1+e^{\varphi})\,dp$ to give $\epsilon(p)$, with a fixed low-density
crust attached.  Legred et al.~\cite{Legred2021} use the same construction
(hereafter the LCEHL prior; data release Zenodo~6502467), which we adopt as the
fixed-prior baseline; related work enriches the kernel or
training~\cite{Essick2020,LandryEtAl2020,Mroczek2023,Semposki2025,
Finch2025,Mihaylov2025}.

A complementary, purely diffusive construction was given by Gorda, Komoltsev,
Kurkela \& Sunde (GKKS)~\cite{GKKS2025bridge}, working with the chemical
potential $\mu$ and baryon density $n$ so that
\begin{equation}
\cs^2 = \frac{d\ln\mu}{d\ln n}.
\label{eq:cs2mu}
\end{equation}
Smoothness is imposed by evolving $\mu$ under a heat equation in a fictitious
time $\tau$, run from $\tau=0$ to $\tau=1$,
\begin{equation}
\partial_\tau\mu = \partial_n\!\left[D(n)\,\partial_n\mu\right],
\qquad D(n) = A\,n^2 ,
\label{eq:GKKSheat}
\end{equation}
where the single free coefficient $A=\ell^2/4$ fixes the smoothing scale and
$\ell\equiv\sigma/n$ is the resulting sound-speed correlation length, constant
in $\ln n$.  The choice $D\propto n^2$ saturates the causality-preserving
condition $D''-D'/n\le0$.

Our implementation evolves a comparable operator defined using  logarithmic variables:
with $x=\ln n$ and $\upsilon = \ln\mu$ we diffuse at constant coefficient,
\begin{equation}
\partial_\tau \upsilon = \partial_x\!\left[A\,\partial_x \upsilon\right]
 = A\,\partial_x^2 \upsilon ,
\label{eq:ourheat}
\end{equation}
with $\upsilon$ held fixed (Dirichlet) at both anchors.  Because the sound
speed is itself the $x$-derivative of the evolved field,
$\cs^2=\partial_x\upsilon$ [Eq.~\eqref{eq:cs2mu}], differentiating
Eq.~\eqref{eq:ourheat} shows that $\cs^2$ satisfies the \emph{same} equation,
$\partial_\tau\cs^2 = A\,\partial_x^2\cs^2$.  The maximum principle for the
heat equation then applies to the sound speed directly: $\cs^2$ can never leave
the range it started in, so a draw that is causal and stable before smoothing is
causal and stable after it, exactly and not merely to discretization accuracy.
Dirichlet ends also conserve $\int\cs^2\,dx=\ln(\mu_H/\mu_L)$ identically.
Equations~\eqref{eq:GKKSheat} and~\eqref{eq:ourheat} share the correlation
length $\ell$ but are not the same operator; the companion
paper~\cite{OShaughnessyZhu2026prior} compares the two schemes and shows where
the difference shows up.  The anchors are our nuclear metamodel band at
$\result{MatchNsat}\nsat$ and the pQCD endpoint at $\mu=\result{PqcdMuGeV}$\,GeV.

The high-density anchor is the perturbative-QCD (pQCD) constraint of Komoltsev \&
Kurkela~\cite{KomoltsevKurkela2022}.  Given a pQCD endpoint at
$\mu_H\simeq2.6$\,GeV~\cite{Gorda2021}, thermodynamic stability ($dn/d\mu\ge0$),
causality ($(\mu/n)\,dn/d\mu \ge 1$), and the integral constraint
\begin{equation}
\int_{\mu_L}^{\mu_H} n\,d\mu = p_H - p_L
\label{eq:KKintegral}
\end{equation}
confine the EOS to a calculable allowed region in the $(\mu,n,p)$ plane; the
existence of an interpolant connecting low- and high-density endpoints reduces to
an inequality among the boundary data~\cite{KomoltsevKurkela2022,Gorda2023pQCD}.
We use the resulting density and pressure envelopes as a hard prior cut.

The low-density prior differs between our implementation and previous work.  GKKS
supply a tightly constrained low-density sector --- a fixed BPS crust~\cite{BPS1971} up to
$\result{CrustEdgeNsat}\nsat$, and above it the chiral-effective-field-theory ($\chi$EFT)
calculations of Drischler et al.~\cite{Drischler2019} sampled
within their theory band out to $n_L=\result{MatchNsat}\nsat$, where the anchor triplet
$(\mu_L,n_L,p_L)$ is drawn.  We adopt a different framework: we change both the low-density model matching and the
precise smoothing procedure adopted.  The low-density model is
a nuclear metamodel band~\cite{Margueron2018}, evaluated at $\result{MatchNsat}\nsat$
with the nuclear empirical parameters resampled per draw and continued downward
through the crust, making the prior self-contained in the backend's nuclear
physics rather than conditioned on one many-body calculation.  The diffusive
smoothing is carried out in $(\ln\mu,\ln n)$ as above, with the high-density
endpoint the partial-N3LO pQCD result at $\mu=\result{PqcdMuGeV}$\,GeV.   Our model choice is intentionally very conservative at crust
densities and below, consistent with the wide uncertainties adopted by metamodels; as a result, our highly uninformed
prior implies every wide priors on neutron star radius. [For context, Appendix~\ref{sec:lowdensity} rebuilds the ensemble with the
GKKS prescription in place of ours, with everything above $n_L$ held fixed, to illustrate the differences beteween our
approach and others.]  The bridge stage places every draw
inside the Komoltsev--Kurkela volume by construction, and the ensemble is causal.
The diffusive smoothing that follows preserves $\int\cs^2$ but not the pressure
integral of Eq.~\eqref{eq:KKintegral}, so the smoothed curve no longer lands
exactly on the pQCD endpoint; see 
\cite{OShaughnessyZhu2026prior} for further discussion, and the Supplemental Material, Sec.~{SI}
for additional numerical details. 

\begin{figure}[t]\centering
  \includegraphics[width=\columnwidth]{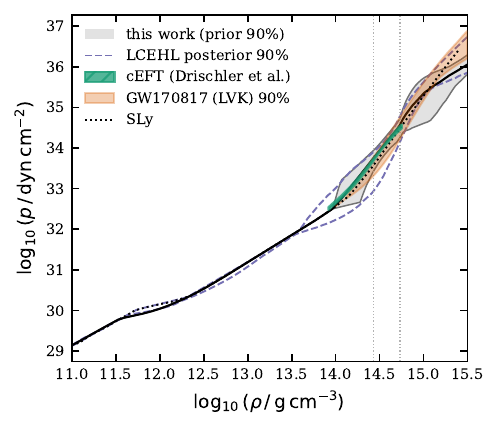}
  \caption{\label{fig:val_prior} The prior we infer with, against the equations
  of state it has to be able to contain.  Grey band and black line: the $90\%$
  interval and the median of our bridge ensemble, from crust to core; vertical
  dashed lines mark $\nsat$ and the matching density $\result{MatchNsat}\nsat$.  Overlaid for
  scale: SLy (dotted), the Drischler et al.\ $\chi$EFT pure-neutron-matter band
  (green), and two published \emph{posteriors} --- the LVK GW170817 $p(\rho)$
  result (orange) and the LCEHL nonparametric result (dashed).  The posteriors
  are drawn to check that the prior covers them; they are not inputs to it, and
  nothing in this paper conditions on them.\llmfig}
\end{figure}

Figure~\ref{fig:val_prior} shows the bridge
ensemble's
pressure--density band from crust to core, overlaid with reference EOS: the SLy
crust/EOS~\cite{lalsuite-soft}, the chiral-effective-field-theory ($\chi$EFT)
pure-neutron-matter band of Drischler et al.~\cite{Drischler2019}, the LVK GW170817
$p(\rho)$ posterior~\cite{LIGO-GW170817-EOS}, and the LCEHL nonparametric
\emph{posterior} (Legred et al.~\cite{Legred2021}).  The construction enforces
causality and the Komoltsev--Kurkela endpoint constraints; the numerical checks
and sampled sound-speed extrema are reported in Supplemental Material, Sec.~{SI}.
The constrained bridge only operates \emph{above} the matching density
$n_L=\result{MatchNsat}\nsat$; below it each draw is the nuclear metamodel with its empirical
parameters (symmetry energy and its slope, incompressibility) resampled per
draw, continued through a polytropic crust.  The metamodel band plays the role
that a $\chi$EFT band plays in GKKS or a fixed SLy crust plays in the LVK
analyses~\cite{LIGO-GW170817-EOS}, and the near-saturation width of that band
strongly controls the attainable radius.  We selected a broad symmetry-sector
calibration whose envelope covers the published LCEHL and GW170817 posterior
bands shown in Fig.~\ref{fig:val_prior}.  This is a coverage design choice, not
an update by those data, so every inference below is conditional on it; that is
why we carry out the prior-sensitivity study of Sec.~\ref{sec:lowdensity} rather
than defer it.

The bridge is a differentiable function of its arguments: the anchor triplet, the
nuclear parameters that set it, and the random variates that place the nodes.  The
nodes come from inverse-CDF maps of the Komoltsev--Kurkela volume, and the smoothing
is a linear solve.  Together with the \textsc{jax} metamodel and TOV solver, a
differentiable implementation lets a Fisher forecast be composed through the same
prior used for sampling.  We use this differentiability to validate the
sample-based results; Sec.~\ref{sec:differentiation} begins with a fixed parametric
continuation in place of the bridge.

\subsection{EOS evidence from observations}\label{sec:evidence}
Each EOS $\theta$ maps to a one-parameter sequence
$\{M(h_c),R(h_c),\Lambda(h_c)\}$, and in particular to the functions
$\Lambda_\theta(m)$ and $R_\theta(m)$ along the stable branch.  Independent
observations contribute multiplicatively, so the EOS evidence is the
product~\cite{gwastro-ns-eos-Kedia2024}
\begin{equation}
\mathcal{L}(d\,|\,\theta) = \prod_e \mathcal{L}_e(d_e\,|\,\theta),
\qquad
p(\theta\,|\,d) \propto \mathcal{L}(d\,|\,\theta)\,\pi(\theta),
\label{eq:eosevidence}
\end{equation}
with $\pi(\theta)$ the EOS prior of Sec.~\ref{sec:rep}.

\begin{widetext}
For a gravitational-wave (GW) event the per-EOS likelihood marginalizes the
event likelihood over the binary parameters with the tidal deformabilities fixed
to the EOS,
\begin{equation}
\mathcal{L}_{\rm GW}(\theta) = \!\int\! dm_1\,dm_2\,d\chi_1\,d\chi_2\;
  \mathcal{L}_e\big(d_e\,|\,m_1,m_2,\chi_1,\chi_2,
  \Lambda_\theta(m_1),\Lambda_\theta(m_2)\big)\,
  \pi_{\rm pop}(m_1,m_2,\chi_1,\chi_2).
\label{eq:LGW}
\end{equation}
The population prior $\pi_{\rm pop}$ over masses and spins enters \emph{every}
EOS evidence and is not optional: it weights the mass range over which the EOS
$\Lambda$--$m$ relation is sampled.
\end{widetext}

The observation-specific electromagnetic factors and optional laboratory
constraints used in the current-data demonstrations are given in
Appendix~\ref{app:observation-factors}.

\subsection{Representations of the equation of state}\label{sec:representation}
The refinement method needs a parameter vector $c$, an evaluable sampling
density $q(c)$, and the observables $\Lambda(m)$, $R(m)$, and $M_{\rm max}$
needed to compute the likelihood.  We compare two choices for $c$: coefficients
of the stellar sequence, which avoid a new TOV solve for each draw, and EOS
coefficients, which give the microscopic EOS directly.  Both use the same
sampling and reweighting procedure in Sec.~\ref{sec:adapt}.

\subsubsection{How EOS changes affect stellar observables}
\label{sec:linear}
An EOS determines a stellar sequence through the TOV and tidal equations.
We can therefore sample either EOS parameters and solve those equations, or
coefficients describing the resulting $M$, $R$, and $\Lambda$ curves.
A complete stable mass--radius relation determines the EOS under the
assumptions of the inverse stellar-structure problem~\cite{Lindblom1992};
a finite set of compressed sequence coefficients need not do so.

Near a reference EOS, a small parameter change produces
\begin{equation}
\delta f_\alpha = \sum_i
\frac{\partial f_\alpha}{\partial c_{{\rm eos},i}}
\delta c_{{\rm eos},i},
\label{eq:linresp}
\end{equation}
where $f_\alpha$ is a stellar observable at a specified point on the sequence.
The derivatives are evaluated at the reference EOS.  Our differentiable TOV
solver computes them directly; Supplemental Material, Sec.~{SI}
gives the numerical checks.  Projection onto the retained sequence basis gives
the corresponding derivatives of $c_{{\rm obs},a}$.

For a sufficiently narrow distribution of EOS parameters, the observable
means and covariances follow the usual first-order propagation of errors:
\begin{align}
\langle c_{{\rm obs},a}\rangle &\simeq c_{{\rm obs},a,\star}
 + \sum_i \frac{\partial c_{{\rm obs},a}}{\partial c_{{\rm eos},i}}
 (\langle c_{{\rm eos},i}\rangle-c_{{\rm eos},i,\star}),\nonumber\\
\Sigma^{\rm obs}_{ab} &\simeq \sum_{ij}
\frac{\partial c_{{\rm obs},a}}{\partial c_{{\rm eos},i}}
\Sigma^{\rm eos}_{ij}
\frac{\partial c_{{\rm obs},b}}{\partial c_{{\rm eos},j}}.
\label{eq:pushforward}
\end{align}
Here $\star$ denotes the reference model and $\Sigma$ is the covariance of the
indicated coefficients.  These expressions require small enough excursions
that curvature and discarded basis components are negligible.  The
importance-sampling method below does not require this linear approximation.
We use the same derivatives in the symmetry-energy forecast of
Sec.~\ref{sec:differentiation}.

The stellar-structure equations also explain why sound-speed priors can look more
complex than observable-space ones --- posterior draws in the
$\cs^2(n)$-GP program~\cite{LandryEssick2019,Legred2021} carry rich small-scale
structure, while \result{SeqComponents} sequence components per channel suffice here.  Two effects are at
work.  The first is prior choice: those constructions mix kernels down to short
correlation lengths, populating many $\cs^2$ modes by fiat, whereas our diffusive
bridge sets a long correlation length.  The second is the stellar-structure map itself: it
integrates thermodynamic information on the way to $\{M,R,\Lambda\}$ and
therefore smooths some fine-scale $\cs^2$ structure.  Sequence compression can
therefore be accurate without resolving every EOS-space mode; we measure how much
$\cs^2$ structure the map actually discards in Sec.~\ref{sec:validation}.

The derivatives also identify which small changes in stellar observables
can be produced by a nearby EOS.  A component outside that set signals a
local inconsistency.  We describe this optional test, including its uncertainty
scale and limitations, in Supplemental Material, Sec.~{SII\,C}.
It is not an acceptance rule in the calculations reported here.

\subsubsection{Observable-space representation (compressed sequences)}
\label{sec:obs-space}
The first choice parametrizes the neutron-star \emph{sequence} directly: the curve
$\{M,R,\log\Lambda\}(h_c)$ traced as central enthalpy $h_c$ varies.  We map $h_c$
to a normalized coordinate $u\in[0,1]$ affinely in each EOS's own enthalpy range,
\begin{equation}
h_c(u) = h_{\rm lo} + (h_{\rm hi}-h_{\rm lo})\,u .
\label{eq:ucoord}
\end{equation}
The lower end $h_{\rm lo}$ is the central pseudo-enthalpy of the minimum-mass star,
or $f_{\rm lo}h_c^{\rm max}$ if that is larger.  The upper end $h_{\rm hi}$ lies two
steps of a \result{SeqScanSteps}-point scan past the last maximum of $M(h_c)$, so every
stable branch is inside the range, or at $f_{\rm hi}h_c^{\rm max}$ if the mass still
rises there.  Here $(f_{\rm lo},f_{\rm hi})=(\result{SeqFracLo},\result{SeqFracHi})$ and
$h_c^{\rm max}$ is the maximum tabulated pseudo-enthalpy of that EOS.  Each sequence
is sampled at $\result{SeqPoints}$ equally spaced $u$ values (the map is monotone, so it is single-valued
and branch-explicit; per-EOS normalization aligns corresponding structural
features --- low-mass stars, the turning point --- at similar $u$ across the
ensemble, which is what makes a shared PCA basis efficient).  Collecting an
ensemble of sequences, we compress each channel with principal-component
analysis (PCA):
\begin{equation}
\{M,R,\log\Lambda\}_e(u) \approx \bar{f}(u) + \sum_{k} c_{e,k}\,\phi_k(u),
\label{eq:pca}
\end{equation}
with a shared mean $\bar f$, basis $\{\phi_k\}$, and per-EOS coefficients $c_e$.  A
Gaussian over the coefficient vector,
\begin{equation}
c \sim \mathcal{N}(\boldsymbol{\mu}_c, \boldsymbol{\Sigma}_c),
\label{eq:gaussc}
\end{equation}
defines a finite-rank functional GP over sequence space, with mean
$\bar f+\sum_k(\mu_c)_k\phi_k$ and covariance set by $\boldsymbol\Sigma_c$.
The bridge ensemble calibrates this Gaussian; conditioning it on structural
validity gives the operational prior $\pi_0^A$.  Reusing the calibration archive
as generation zero introduces a finite-ensemble approximation to this density
(Supplemental Material, Sec.~{SIII\,B}).  Figure~\ref{fig:val_pca}
quantifies the reconstruction accuracy.  Refinement updates the coefficient
mean and covariance in a fixed basis.

Because the parameters describe the \emph{observables} themselves, evaluating a
proposal draw needs no new stellar-structure solve: $\Lambda(m)$, $R(m)$ and
$M_{\rm max}$ are reconstructed from Eq.~\eqref{eq:pca}.  This is why the
multi-event and large-catalogue studies below use this representation.
Observable-space inference returns stellar sequences directly.  Microscopic
$p(\rho)$ bands, where shown (Fig.~\ref{fig:val_prior}), come from the forward
EOS tables of the generating draws; recovering an EOS from a sampled sequence
would require inverse stellar structure~\cite{Lindblom1992}.

\paragraph*{Physical consistency and stability.}
We evaluate observables on the stable rising-mass branch of each reconstructed
sequence, located from the sign of $dM/dh_c$.  Draws with non-finite or negative
curves, or without a nontrivial rising branch, are rejected.  The base Gaussian
and all subsequent proposals use the same validity set, with its normalization
included in Eq.~\eqref{eq:balance}.  These structural checks leave possible
cross-channel departures from the physical EOS manifold.

We retain $\Lambda(m)$ as an independent channel measured by the GW likelihood.
An empirical compactness--Love relation would impose an additional prior;
Supplemental Material, Sec.~{SIII\,D} discusses its use as
a diagnostic.

\subsubsection{Equation-of-state-space representation (sound speed)}
\label{sec:eos-space}
If instead one uses a generative model in EOS space --- for instance a
Gaussian process in the sound speed $\cs^2(n)$ under a logit-type transformation,
as in the nonparametric program of Landry, Essick \& Legred~\cite{LandryEssick2019,
Legred2021} --- the same fixed-target importance construction applies: $c$ now parametrizes
$\cs^2(n)$, and the forward map is a TOV solve per draw.  This pays a
stellar-structure integration that the observable-space scheme avoids, and it
returns the microscopic EOS \emph{directly}, with no inverse to solve.  The
JESTER framework similarly uses differentiable programming through the TOV
equations for EOS inference and inversion~\cite{WoutersEtAl2025JESTER}.
Our TOV solver is written in \textsc{jax}, so the entire map from
EOS (or nuclear) parameters to observables is algorithmically differentiable
end-to-end: sensitivities such as $\partial\Lambda(m)/\partial L_{\rm sym}$ are
available for the discretized calculation by automatic differentiation.  We
exploit this in Sec.~\ref{sec:differentiation} to forecast the symmetry-energy reach.  Because
it retains a physical EOS representation, this is the natural choice whenever
EOS-space or nuclear-parameter posteriors ---
$p(\rho)$, $L_{\rm sym}$, $K_{\rm sym}$ --- are themselves the target, rather than
the stellar observables.

\paragraph*{Resolution and data-driven basis growth.}
The basis sets the finest EOS structure available to inference.  In the
diffusive GKKS construction, the correlation length $\ell$ in $(\ln\mu,\ln n)$
[Eq.~\eqref{eq:GKKSheat}] limits the sound-speed features supported by the prior.
Resolving finer structure requires additional basis functions and observations
spanning the corresponding central densities.  Concentrating observations at
one mass mainly improves precision along existing directions
(Sec.~\ref{sec:hero}).  Extending the basis using new EOS curves is a natural
further refinement.

\subsubsection{Effective dimension and reconstruction accuracy}
\label{sec:effective-dimension}
An observation resolves a direction in EOS space when the likelihood varies
appreciably across the range allowed by the prior.  This gives a volume-based
interpretation of effective dimension~\cite{OShaughnessy2013ModelDimension}:
for an allowed loss $t$ in log likelihood, define
\begin{align}
 V(t)&=\int_{\ln\mathcal L_{\max}-\ln\mathcal L(c)\leq t}
             \pi_0(c)\,dc,\nonumber\\
 d_{\rm vol}(t)&=2\frac{d\ln V}{d\ln t}.
 \label{eq:volume_dimension}
\end{align}
A quadratic likelihood resolving $d$ directions within the prior gives
$V\propto t^{d/2}$.  Directions over which the likelihood scarcely varies
are instead limited by the prior boundaries.  Once a contour encloses the
whole prior, $V$ saturates and no further dimensions contribute to its growth.
When likelihood precision scales as squared signal-to-noise ratio
$\varrho^2$, the equivalent contraction is
$d_{\rm vol}=-d\ln V/d\ln\varrho$ at fixed $t$.

To apply this idea to stellar curves, choose a mass interval $[M_-,M_+]$
and a weight $w(M)$ with $\int_{M_-}^{M_+}w(M)\,dM=1$.
The weight has units of inverse mass.  Let $s_R(M)$ and
$s_{\ln\Lambda}(M)$ be reference accuracies, with units of length and no
units, respectively.  For two sequences $f$ and $g$, define
\begin{align}
 D_\varrho^2(f,g)&=\varrho^2\int_{M_-}^{M_+}dM\,w(M)\bigg[
 \nonumber\\
 &\quad\frac{[R_f(M)-R_g(M)]^2}{s_R^2(M)}\nonumber\\
 &\quad+\frac{[\ln\Lambda_f(M)-\ln\Lambda_g(M)]^2}
                  {s_{\ln\Lambda}^2(M)}\bigg].
 \label{eq:physical_curve_distance}
\end{align}
The \emph{resolution likelihood}
$\mathcal L_\varrho^{\rm res}(g|f)\propto\exp[-D_\varrho^2(f,g)/2]$
has the form of an approximate Gaussian likelihood for simultaneous radius
and logarithmic tidal measurements, taking their errors to be uncorrelated.
The mass weight allocates measurement sensitivity across the reference masses.  It measures
our reconstruction accuracy at a chosen measurement resolution.  The
SNR-like factor $\varrho$ rescales the reference accuracies together:
only $s_R/\varrho$ and $s_{\ln\Lambda}/\varrho$ determine the resolution.
Each difference is divided by a scale with matching units before being
combined.  In practice, finite reference grids and quadrature weights
realize the integral.

The relative importance of radius and tidal measurements, and the masses at
which they matter, are physical choices.  They should reflect observational
capabilities and scientific priorities, and evolve as those circumstances
change.  Appendix~\ref{app:resolution-weighting} describes the choices made
in our calculations and their implications.

PCA uses this same quadratic form for the joint representation to find
the natural directions of variation: it expresses each channel in units of its reference accuracy
and finds patterns orthogonal under the mass-weighted integral.  It
orders these patterns by their contributions to the mean log-likelihood
loss across the family.  Keeping the first $d$ therefore gives the
$d$-pattern linear representation with the smallest average loss in this
resolution likelihood.  Increasing $\varrho$ leaves the patterns unchanged
but requires more of them to meet a fixed log-likelihood error budget.

We can therefore measure reconstruction error directly as a loss in this
resolution likelihood.  For curves excluded from the basis fit, define
\begin{equation}
 \epsilon_d^2=
 \frac{\sum_e D_\varrho^2(f_e,\hat f_e^{(d)})}
      {\sum_e D_\varrho^2(f_e,\bar f)},
 \label{eq:dimension_error}
\end{equation}
where $\hat f_e^{(d)}$ is the reconstruction and $\bar f$ the training mean.
This is the mean log-likelihood loss from truncation divided by the loss
from replacing every curve by the mean; the common resolution factor
$\varrho^2$ cancels.  Figure~\ref{fig:external_gp_dimension} plots
$100\epsilon_d$, the percentage of family variation left unreconstructed.
The number of patterns needed at a chosen accuracy is the effective
representation dimension.

For Fig.~\ref{fig:dimension}(b), stellar coefficients instead predict the
amplitude $a_{e,k}$ of each sound-speed pattern.  Define its recovered fraction
by
\begin{equation}
 R_k^2=1-
 \frac{\sum_e(a_{e,k}-\hat a_{e,k})^2}
      {\sum_e(a_{e,k}-\bar a_k)^2},
 \label{eq:mode_recovery}
\end{equation}
where $\bar a_k$ is the family's mean amplitude.  Both sums have the same
squared amplitude units.  The plotted $100R_k^2$ is $100\%$ for perfect
recovery, zero for no improvement over the mean, and negative when the
predicted amplitudes do worse than the mean.

The reconstruction criterion connects to the prior-volume picture through
the allowed PCA amplitudes.  For illustration, write
$f=\bar f+\sum_k a_k e_k$, with patterns orthonormal in the
accuracy-scaled integral of Eq.~\eqref{eq:physical_curve_distance} at
$\varrho=1$, and impose uniform priors $|a_k|\le A_k$.
The resolution likelihood then has log-likelihood loss
$\varrho^2\sum_k a_k^2/2$ about $\bar f$.
In direction $k$, a contour of allowed loss $t$ reaches the prior boundary
when $\sqrt{2t}/\varrho\simeq A_k$: beyond that scale, the allowed range
in this direction is set by the prior rather than by measurement accuracy.
PCA orders these directions by their prior mean squared amplitudes,
$\lambda_k=\langle a_k^2\rangle=A_k^2/3$ in this example.
Keeping $d$ patterns leaves the mean loss
\begin{equation}
 \left\langle\Delta\ln\mathcal L^{\rm res}_{d}\right\rangle
 =\frac{\varrho^2}{2}\sum_{k>d}\lambda_k
 \simeq\epsilon_d^2\,\frac{\varrho^2}{2}\sum_k\lambda_k.
 \label{eq:dimension_loss_budget}
\end{equation}
Thus the required coefficient count retains the directions whose allowed
variation matters at the chosen resolution, until the combined discarded
variation fits within the loss budget.  The volume dimension describes the
same competition between likelihood resolution and prior extent through
contour growth; the coefficient count controls the mean reconstruction loss.

The independent bounded PCA-amplitude prior used in this interpretation is
generally different from the prior induced by the EOS model family.  The
associated volume dimension therefore refers to this auxiliary prior and the
chosen resolution likelihood; the reconstruction errors themselves are still
evaluated on held-out family draws.  In practice, the family prior is usually
available through EOS draws and a GP construction rather than a tractable
density in stellar PCA coordinates.  The amplitude-prior ansatz provides a
practical connection between reconstruction accuracy and prior-limited
dimension; the inference itself retains the physical EOS prior.

\subsection{Adaptive refinement}\label{sec:adapt}
The usual inference cycle evaluates a marginal evidence $Z_e$ for each EOS
and uses these evidences to weight the ensemble.  Here the gravitational-wave
evidences are computed with RIFT, with the EOS held fixed while the binary
parameters are marginalized~\cite{gwastro-PE-AlternativeArchitectures,
gwastro-PENR-RIFT,gwastro-RIFT-Update}.  Independent observations contribute
additional factors as described in Sec.~\ref{sec:evidence}.  For an ensemble
drawn from the base prior, the resulting weight is simply proportional to the
product of these factors.  Reweighting the existing EOS draws is therefore the
ordinary calculation; it requires no change to their representation or to the
prior.

Sometimes, however, too few draws retain appreciable weight, and we trigger a
refinement stage.  The usual refinement updates the mean and covariance of
the distribution from which new EOS realizations are drawn.  In our compressed
representation, these are the hyperparameters of the Gaussian over sequence
coefficients $c$.  We keep the basis fixed, re-fit its coefficient mean and
covariance using the evidence-weighted draws, and generate fresh sequences from
the resulting distribution, evaluating their evidences as before.

A mean and covariance provide an efficient description when the EOS
parametrization is well aligned with the stellar observables: the region
supported by the data should be compact and only weakly curved in the chosen
coordinates.  The local response of Sec.~\ref{sec:linear} explains why a
suitable sequence basis can make this possible.  A strongly curved model
manifold or several separated regions of support would be poorly described by
a single Gaussian, even if its moments were accurately measured.  The representation therefore determines how effectively a Gaussian refit
concentrates the draws.

The default moment refit acts on the accepted draws $c_e$ of generation $g$,
with evidences $Z_e$.  Each draw receives the normalized batch weight
\begin{equation}
w_e^{\rm fit}=\frac{Z_e^{\beta}}{\sum_{e'}Z_{e'}^{\beta}}.
\label{eq:fitweight}
\end{equation}
Tempering with $\beta<1$ broadens the evidence-weighted distribution used for
the fit.  The weighted moments of the draws are
\begin{align}
\hat{\boldsymbol{\mu}}_c&=\sum_e w_e^{\rm fit}\,c_e,\nonumber\\
\hat{\boldsymbol{\Sigma}}_c&=\sum_e w_e^{\rm fit}\,
(c_e-\hat{\boldsymbol{\mu}}_c)(c_e-\hat{\boldsymbol{\mu}}_c)^{T}.
\label{eq:fitmoments}
\end{align}
These weights contain no ratio of prior to proposal.  With $\beta=1$, the
moments therefore estimate those of the current sampling density multiplied by
$Z$.  For stability, shrinkage with fraction $\gamma$ blends them with the
moments of generation $g$:
\begin{align}
\boldsymbol{\mu}_c^{(g+1)}&=(1-\gamma)\,\hat{\boldsymbol{\mu}}_c
+\gamma\,\boldsymbol{\mu}_c^{(g)},\nonumber\\
\boldsymbol{\Sigma}_c^{(g+1)}&=(1-\gamma)\,\hat{\boldsymbol{\Sigma}}_c
+\gamma\,\boldsymbol{\Sigma}_c^{(g)}.
\label{eq:shrink}
\end{align}
Generation $g+1$ draws from
$\mathcal{N}(\boldsymbol{\mu}_c^{(g+1)},\boldsymbol{\Sigma}_c^{(g+1)})$ and
keeps the draws that pass the validity checks of Supplemental Material,
Sec.~{SIII\,B}.

An optional second form of refinement re-fits the PCA basis itself from the
weighted sequences, shifting the mean curve and reorienting the components.
Because those sequences were generated in the existing span, this update
redistributes the retained directions rather than adding new physical ones.
Expanding the span would require fresh external training curves and is left
to future work.

The new distribution is an EOS \emph{proposal}; the scientific prior remains
the fixed physical base density $\pi_0^A$.  When we combine several generations
for the same data, we must account for their different sampling densities.
For fixed data, let $q_g^A(c)$ be the density that generated the $N_g$
accepted draws in batch $g$.  The density of the pooled sample is their
mixture, weighted by the batch sizes.  Each draw then receives weight
\begin{align}
 w(c)&\propto Z(c)\,\frac{\pi_0^A(c)}{q_{\rm mix}^A(c)},\nonumber\\
 q_{\rm mix}^A(c)&=\sum_g\frac{N_g}{\sum_{g'}N_{g'}}q_g^A(c).
\label{eq:balance}
\end{align}
This corrects for the increased sampling of regions that fit the data well.
For a single batch drawn from the prior, it reduces to $w\propto Z$.
The mixture must cover the target distribution, and its densities must be
evaluated in the same coefficient coordinates as the prior.

The quality of any such weighted ensemble is summarized by its effective sample
size (ESS),
\begin{equation}
N_{\rm eff} = \frac{\big(\sum_e w_e\big)^2}{\sum_e w_e^2},
\label{eq:ess}
\end{equation}
which runs from $1$ (a single draw dominates) to the number of draws (uniform
weights), summarizing weight concentration.  In the catalogue demonstrations, a refinement is triggered when
$N_{\rm eff}$ of the current set falls below a threshold.  The refinement
operation itself can also be invoked on a chosen schedule.  For the fixed-data
combination, Eq.~\eqref{eq:balance} recomputes the weights of all accumulated
draws.  Supplemental Material, Sec.~{SIII} relates threshold and batch
choices to Monte Carlo precision.

\subsubsection{Sequential Gaussian projection for accumulating observations}
\label{sec:sequential}
Equation~\eqref{eq:balance} targets a fixed data set, and the GW170817 example
of Appendix~\ref{app:observations} uses it; implementation details are given in
Supplemental Material, Sec.~{SIII\,B}.  For growing catalogues
we use a second, approximate method.  After each refit, the fitted Gaussian
replaces the accumulated posterior, and later draws carry only the likelihoods
of later events.  The Gaussian therefore holds the information from earlier
events: it acts as a working prior as well as a proposal.

Let events $1,\dots,r$ be the data used in the latest refit and redraw,
$p_r=\pi_0\mathcal L_{1:r}$ the accumulated posterior, and $q_r$ the fitted
Gaussian.  For newly arriving events $r+1,\dots,n$, keeping the physical prior
$\pi_0$ requires the weight
\begin{equation}
w \;\propto\; \frac{\pi_0\,\mathcal{L}_{1:n}}{q_r}
\;=\;\rho_r\,\mathcal{L}_{r+1:n},
\qquad \rho_r\equiv\frac{p_r}{q_r}.
\label{eq:refitweight}
\end{equation}
The fixed-data analyses have no new events and apply this correction through
Eq.~\eqref{eq:balance}.  Sequential projection instead replaces $p_r$ by $q_r$,
so that $\rho_r=1$ and
\begin{equation}
w \;\propto\; \mathcal{L}_{r+1:n}.
\label{eq:seqweight}
\end{equation}
At the next refit, after event $r'$, we fit a Gaussian to these weighted draws:
\begin{equation}
q_{r'} = \mathcal{G}_\gamma\!\left[q_r\,\mathcal{L}_{r+1:r'}\right],
\label{eq:seqrecursion}
\end{equation}
where $\mathcal{G}_\gamma$ denotes the moment fit of Eqs.~\eqref{eq:fitweight}
to \eqref{eq:shrink} with $\beta=1$ and $Z_e=\mathcal{L}_{r+1:r'}(c_e)$.  Each
event enters the ensemble once.  After the first refit, $q_r$ approximates
$p_r$ only through this chain of fits.  The recursion is assumed-density
filtering~\cite{Minka2001EP}, applied to blocks of events with a Gaussian
assumed density in coefficient space.

Equation~\eqref{eq:seqweight} is exact only if $q_r=p_r$.  Its large-sample limit
is the chain of Eq.~\eqref{eq:seqrecursion}, so additional draws reduce Monte
Carlo noise and leave the projection error unchanged.  That error has two
sources: non-Gaussian structure in the accumulated posterior, including the
validity truncation of Supplemental Material, Sec.~{SIII\,B},
and shrinkage.
For the fixed-data estimator, shrinkage changes only the sampling variance.
Here it changes the target, because blending in the wider previous covariance
discards information from earlier events.

We drop $\rho_r$ because it is heavy tailed.  For the
$\result{HeroEvents}$-event catalogue of Sec.~\ref{sec:seqtest},
Eq.~\eqref{eq:refitweight} leaves a median
$N_{\rm eff}=\result{SeqExactEss}$ among $\result{HeroDrawsPerGen}$ draws in the \result{CoeffDim}-dimensional
coefficient space.  A standard treatment of heavy-tailed importance weights is
to clip them~\cite{Ionides2008TruncatedIS,Vehtari2024PSIS}:
\begin{equation}
w_\tau \;\propto\; \min(\rho_r,\tau)\,\mathcal{L}_{r+1:n}.
\label{eq:clipweight}
\end{equation}
Large $\tau$ recovers Eq.~\eqref{eq:refitweight}, and $\tau$ below the smallest
$\rho_r$ recovers Eq.~\eqref{eq:seqweight}, so clipping trades variance against
bias between the two estimators.  Clipping needs the likelihoods of all earlier
events for every new draw.  Sequential projection needs only the new events, so
its cost per draw does not grow with the catalogue.  Section~\ref{sec:seqtest}
tests both against a converged fixed-prior posterior.

\subsection{Synthetic observations}\label{sec:obs-synth}
For the multi-event demonstration we generate synthetic catalogues with a
deliberately simple model, with two components. First, the \emph{error model}:
at the simplest level each synthetic event carries an analytic Gaussian
likelihood in the tidal observable; the more detailed demonstrations upgrade this to
a \textsc{gwfish} 
Fisher-matrix covariance in $(\mathcal{M},q,\tilde\Lambda)$
--- chirp mass, mass ratio, and binary tidal deformability --- computed from the
frequency-domain tidal waveform
\textsc{IMRPhenomD\_NRTidalv2}~\cite{DietrichEtAl2019NRTidal}.  The injected neutron stars
are non-spinning, while aligned
spins are included in the Fisher matrix with a low-spin prior and marginalized
from the reported $(\mathcal{M},q,\tilde\Lambda)$ covariance.  The Fisher
calculation uses the LIGO Hanford--LIGO Livingston--Virgo (HLV) geometry and the
same bilby \texttt{aLIGO\_O4\_high} noise curve at all three sites over
$\result{FisherFmin}$--$\result{FisherFmax}$ Hz~\cite{2015CQGra..32g4001T,2015CQGra..32b4001A,
gwastro-pe-bilby-2018}.  Thus Virgo contributes its site geometry and arm
length but is not assigned a Virgo-specific O4 sensitivity; KAGRA is
intentionally omitted~\cite{2021PTEP.2021eA103A}.  These catalogues are controlled method tests, not
projections for a specific LVK observing scenario.  We also hold the waveform
model fixed, so the known propagation of waveform systematics into joint
population--EOS inference is outside this test~\cite{gwastro-ns-eos-pop_systematics-Yelikar2024}.
Second, the \emph{population model}, which enters every per-EOS evidence through
$\pi_{\rm pop}$ in Eq.~\eqref{eq:LGW}.  We adopt the skew-normal binary
neutron-star mass distribution of Anik, Steiner \&
O'Shaughnessy~\cite{gwastro-nsnuc-AnikSteiner-EOSPopMaxMass-2025},
\begin{equation}
p(M) = \frac{2}{\sigma}\,
  \phi\!\left(\frac{M-\mu}{\sigma}\right)
  \Phi\!\left(\alpha\,\frac{M-\mu}{\sigma}\right),
\label{eq:skewnormal}
\end{equation}
with $\phi$ and $\Phi$ the standard normal density and CDF, location
$\mu\simeq1.35\,\msun$, scale $\sigma\simeq0.14\,\msun$, skewness $\alpha$, and
mass support $[1\,\msun,\,M_{\rm max}]$.

\subsection{Cross-model validation by differentiation}\label{sec:differentiation}
The representation of Sec.~\ref{sec:eos-space} provides a cross-model check on
sample-based statements about nuclear parameters.  The comparison is between two EOS
models, so agreement makes a conclusion plausible without validating either
calculation.  Let $\boldsymbol{\nu}=(E_{\rm sym},L_{\rm sym},K_{\rm sym},K_{\rm sat})$
denote the nuclear empirical parameters, with prior widths $\sigma_a$, and let
$\boldsymbol{y}$ collect $\ln\Lambda(m)$ at $m=\result{SymMassLo}$ and $\result{SymMassHi}\,\msun$.

The sample-based calculation conditions $\boldsymbol{\nu}$ on a measurement
$\hat{\boldsymbol{y}}$ with covariance $\boldsymbol{N}$, using the prior draws
$(\boldsymbol{\nu}_i,\boldsymbol{y}_i)$ near $\hat{\boldsymbol{y}}$.  Each draw has
the weight
\begin{equation}
w_i=\exp\!\left[-\tfrac12(\boldsymbol{y}_i-\hat{\boldsymbol{y}})^{T}
\left(\boldsymbol{N}+h^2\boldsymbol{I}\right)^{-1}
(\boldsymbol{y}_i-\hat{\boldsymbol{y}})\right],
\label{eq:nepkernel}
\end{equation}
and weighted least squares fits
$\boldsymbol{\nu}_i=\boldsymbol{a}+\boldsymbol{B}(\boldsymbol{y}_i-\hat{\boldsymbol{y}})+\boldsymbol{r}_i$.
The conditional mean and covariance are
\begin{equation}
\boldsymbol{\mu}_{\nu|y}=\boldsymbol{a},\qquad
\boldsymbol{\Sigma}_{\nu|y}=\boldsymbol{\Sigma}_{r}+\boldsymbol{B}\boldsymbol{N}\boldsymbol{B}^{T},
\label{eq:nepcondition}
\end{equation}
where $\boldsymbol{\Sigma}_r$ is the weighted covariance of the residuals
$\boldsymbol{r}_i$, scaled by $n_{\rm eff}/(n_{\rm eff}-1-d)$ for $d$ measured masses
and the kernel's effective number of draws $n_{\rm eff}$.  The kernel width $h$ sets the range of
$\ln\Lambda$ over which the linear map must hold.  With uniform weights and
$\boldsymbol{N}$ small, Eq.~\eqref{eq:nepcondition} reduces to the global
linear-Gaussian result
$\boldsymbol{\Sigma}_{\nu\nu}-\boldsymbol{\Sigma}_{\nu y}\boldsymbol{\Sigma}_{yy}^{-1}\boldsymbol{\Sigma}_{\nu y}^{T}$
over all draws.  The map from $\boldsymbol{y}$ to $\boldsymbol{\nu}$ is not linear
across the prior, so the global result mixes in scatter from draws far from the
measurement.  We test calibration by leave-one-out.  Each archive draw within
$\result{SymLooWindow}$ of $\hat{\boldsymbol{y}}$ in every $\ln\Lambda$ is held out, given a simulated
measurement with covariance $\boldsymbol{N}$, and predicted from the remaining draws.

The differentiated calculation uses a parametric \textsc{jax} map: the metamodel up to $n_L=\result{MatchNsat}\nsat$, a
constant-sound-speed continuation with $\cs^2=\result{AdCsSqTail}$, and the TOV solve (Supplemental
Material, Sec.~{SI\,B}).  The parameters
$\boldsymbol{\nu}$ affect that continuation only through its starting point.  At the
prior mean $\bar{\boldsymbol{\nu}}$, we evaluate the fixed-mass derivative
\begin{equation}
J_{ka}=\left.\frac{\partial\ln\Lambda(m_k)}{\partial\nu_a}\right|_{\bar{\boldsymbol{\nu}}}
\label{eq:nepjacobian}
\end{equation}
at $m_k=\result{AdMassLo}$ and $\result{AdMassHi}\,\msun$
by central differences, which agree with forward-mode derivatives for the SLy crust
(Supplemental Material, Sec.~{SI\,B}).
For a measurement of $\boldsymbol{y}$ with covariance $\boldsymbol{N}$, the linearized
forecast is
\begin{equation}
\boldsymbol{\Sigma}_{\rm AD}=\boldsymbol{F}^{-1},\qquad
\boldsymbol{F}=\boldsymbol{J}^{T}\boldsymbol{N}^{-1}\boldsymbol{J}
+{\rm diag}\!\left(\sigma_a^{-2}\right).
\label{eq:nepfisher}
\end{equation}
For $\Lambda(1.4)$ alone, $\boldsymbol{J}$ has one row and $\boldsymbol{N}$ is the
squared fractional width $\sigma_y^2$.
The two calculations share the prior widths and the low-density model, and differ
above $n_L$.
Equation~\eqref{eq:nepcondition} averages over the bridge draws, whose random
structure enters the covariance blocks.  Equation~\eqref{eq:nepfisher} fixes the
high-density EOS to one parametric form.

Both calculations can also include laboratory information.  Rare-isotope and
heavy-ion experiments probe the same symmetry sector directly and at densities not
covered by the observed neutron-star mass range~\cite{SorensenEtAl2024,BrownEtAl2025,
LeFevre2026}.  We represent such a scenario by replacement widths
$\sigma_a^{\rm lab}$ for selected parameters.  In Eq.~\eqref{eq:nepfisher},
$\sigma_a^{\rm lab}$ replaces $\sigma_a$.  In the sample-based calculation, we add the
corresponding information to the conditional result:
\begin{align}
\boldsymbol{\Sigma}_{\rm lab}^{-1}&=\boldsymbol{\Sigma}_{\nu|y}^{-1}+\boldsymbol{\Delta},
\nonumber\\
\Delta_{ab}&=\delta_{ab}\max\!\left[0,\,
\frac{1}{(\sigma_a^{\rm lab})^{2}}-\frac{1}{\sigma_a^{2}}\right],
\label{eq:neplab}
\end{align}
with $\Delta_{aa}=0$ for parameters without a scenario width.
This treatment is an ad hoc, proof-of-concept argument that assumes direct
constraints on symmetry parameters.  A real joint analysis requires a suitable
likelihood, as in Refs.~\cite{HuthEtAl2022,TsangEtAl2024}, to express the ability of
these measurements to constrain the EOS directly.

\section{Results and forecast demonstrations}\label{sec:results}
We use synthetic catalogues to test the sequential refinement technique of
Sec.~\ref{sec:sequential} against fixed-prior reweighting.  Section~\ref{sec:assess}
illustrates our assessment with two small catalogues, Sec.~\ref{sec:hero} applies
the method to a large catalogue, and Sec.~\ref{sec:seqtest} quantifies its accuracy.
The implementation
details are given in Supplemental Material, Sec.~{SIII\,B}.
Appendix~\ref{app:observations} gives the separate GW170817 and multimessenger
checks, including a comparison of ways to spend evidence runs on a fixed target.

\subsection{Assessing adaptive refinement: illustrations with small catalogues}\label{sec:assess}
In this section, we perform EOS inference on two small collections of synthetic observations, comparing the conventional
benchmark (fixed EOS prior draws) against our adaptive refinement technique.  These examples show how we assess adaptive
refinement in Secs.~\ref{sec:hero} and \ref{sec:seqtest}, where we test it quantitatively.
To assess performance, we examine each methods' estimated posterior for a single quantity -- the tidal deformability of a $1.4 M_\odot$
neutron star $\Lambda(1.4)$; for each estimate, we assess its mean, standard deviation, and effective sample size.

\begin{figure}[t]\centering
  \includegraphics[width=\columnwidth]{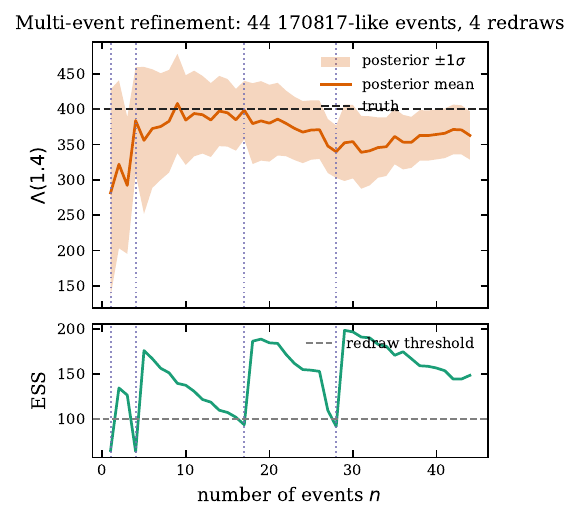}
  \caption{\label{fig:lambda_vs_n} Evidence evolution with analytic likelihoods.
  Top: $\Lambda(1.4)$ posterior (mean $\pm1\sigma$) vs number of GW170817-like events
  $n$, converging to the injected truth (dashed); dotted lines mark GP redraws.
  Bottom: effective sample size, which starves the fixed draw set and is restored
  at each redraw --- why refinement is required as measurements accumulate.\llmfig}
\end{figure}

Figure \ref{fig:lambda_vs_n} illustrates the kinds of calculations performed in this study.  To be concrete,  this
study adopts a simplified physical scenario: repeated observations of a GW170817-like event, with uncertainty
effectively only in $\Lambda$. Specifically, this calculation's synthetic measurements  are one-dimensional likelihoods
for each synthetic event versus $\Lambda$, comparable to several previous proof-of-concept studies, which reduced GW170817
to a single limit on a tidal parameter~\cite{gwastro-ns-eos-170817-Annala2018,
gwastro-mergers-ns-MostNewConstraintsRadii2018,
gwastro-TewsCriticalExaminationConstraints2018,
gwastro-mergers-ns-TewsConfrontingGravitationalWave2019}.  In the top figure, we see the mean and standard deviation of the posterior converge as the number of
observations increases.  In the bottom panel, the effective sample size of the posterior also changes as the number of
events increases. The two dashed vertical lines indicate our two redraws, where we reassess our sampling prior.  At
these two events, we rejuvinate sampling, without appreciably changing the inferred properties.

\begin{figure*}[t]\centering
  \includegraphics[width=\textwidth]{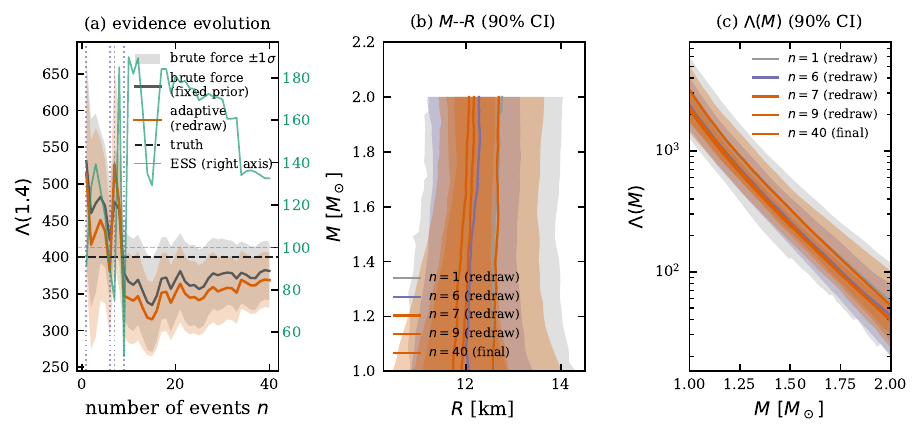}
  \caption{\label{fig:synth_fisher} Multi-event refinement with the
  mass-marginalized Fisher likelihood, used in all three panels (Anik et al.\
  BNS population; \textsc{gwfish} per-event $(\mathcal{M},q,\tilde\Lambda)$
  covariances; $\result{FisherEvents}$ events).  (a)~$\Lambda(1.4)$ vs event count $n$: the
  sequential projection (orange) and reweighting of the full
  \result{GenZeroDraws}-draw archive (grey).  The twin axis shows the effective
  sample size of the adaptive set.  The adaptive set holds \result{FisherDrawsPerGen} draws and is redrawn when
  $N_{\rm eff}$ falls below \result{FisherRefitThreshold}; this run redraws after events
  \result{FisherRefitList} (dotted).  (b)~$M$--$R$ and (c)~$\Lambda(M)$ $90\%$ credible bands of the
  posterior sequence just after the first redraw and at $n=\result{FisherEvents}$.\llmfig}
\end{figure*}

Figure~\ref{fig:synth_fisher} performs a comparable calculation, but now directly comparing (in the left panel) the
conclusions derived using a fixed prior versus our method.  This calculation also differs modestly in its underlying
synthetic observations: to increase realism, we use random observations
from  the population and detector assumptions of
Sec.~\ref{sec:obs-synth}, and use more realistic mass-marginalized
Fisher likelihoods, so  each of $\result{FisherEvents}$ events has a \textsc{gwfish} covariance in
$(\mathcal{M},q,\tilde\Lambda)$.  We integrate each event likelihood over the
binary masses using Eq.~\eqref{eq:LGW}, with the tidal relation supplied by each
EOS.  Sequential projection and reweighting of the
$\result{GenZeroDraws}$-draw archive are in excellent agreement, both in mean and credible interval;
at $n=\result{FisherEvents}$ they give
$\result{FisherFinalLam}\pm\result{FisherFinalLamSd}$ and
$\result{FisherBruteLam}\pm\result{FisherBruteLamSd}$, respectively.
Both this illustration and the previous figure adopt extremely small prior draw realizations, to facilitate
visualization of the key features of our argument with relatively few observations.
An archive of this size cannot resolve a narrow posterior: for the
$\result{HeroEvents}$-event catalogue of Sec.~\ref{sec:hero}, its effective
sample size falls to $N_{\rm eff}=\result{HeroBruteFinalEss}$.

\subsection{Large-catalogs and projected science capabilities}\label{sec:hero}
\begin{figure*}[t]\centering
  \includegraphics[width=\textwidth]{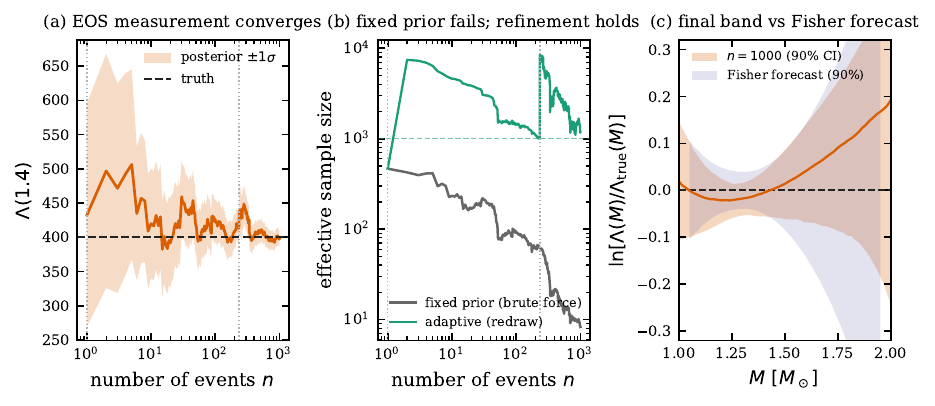}
  \caption{\label{fig:hero} Large-catalogue method test (\result{HeroEvents} second-generation HLV BNS,
  lazy cosmology). (a)~$\Lambda(1.4)$ vs event count $n$: the adaptive posterior
  ($\pm1\sigma$) contracts to the injected truth (dashed; dotted verticals mark
  refinements). (b)~effective sample size: the \emph{fixed}-prior brute force (grey)
  collapses by $\sim\!\result{HeroBruteCollapse}\times$ while the adaptive draw set (teal) is restored at
  each redraw. (c)~the final ($n=\result{HeroEvents}$) $\Lambda(M)$ $90\%$ band
  relative to the truth, against the Fisher forecast (shaded;
  derivatives with respect to sequence coefficients, accumulated over all events, cf.\
  Ref.~\cite{gwastro-nsnuc-Fisher3g-Zhu2025}), whose mass dependence tracks the
  posterior's and gives a width consistent with the converged fixed-prior
  calculation for this catalogue.  The adaptive standard deviation is
  \result{HeroAdaptiveWidthRatio} times that reference (text).\llmfig}
\end{figure*}

Figure \ref{fig:hero} shows our refinement technique applied to a larger catalog of  $\result{HeroEvents}$ binary
neutron stars, again intentionally starting with an
extremely small number of prior EOS realizations.  
Sources are uniform in comoving volume to
$\result{HeroMaxDistance}$\,Mpc, with redshift neglected and an SNR threshold of \result{HeroSnrMin}.   This run uses
Eq.~\eqref{eq:seqweight} with $\gamma=0$, a refit threshold
$N_{\rm eff}=\result{HeroRefitThreshold}$, and $\result{HeroDrawsPerGen}$ draws per generation.  It is the original-order
run of Sec.~\ref{sec:seqtest}.
In the left panel, we demonstrate the posterior converges towards truth, with only $\result{HeroRefinements}$  refinement stages.
The center panel shows how the effective sample size of our (intentionally small) set of fixed prior draws versus
adaptive refinement: we retain high, tunable efficiency in excess of the intitial set of prior draws as observations
accumulate.
In the right panel, we compare our prediction for the distribution of $\Lambda(M)$ at each $M$ against a direct
linear Fisher-matrix calculation, accumulated over all \result{HeroEvents} events, following the approach of Ref.~\cite{gwastro-nsnuc-Fisher3g-Zhu2025}.

The final width should also be compared against the Fisher
expectation.  Following the approach of
Ref.~\cite{gwastro-nsnuc-Fisher3g-Zhu2025}, we work in the sequence-coefficient
space.  Each catalogue event $k$ contributes its fractional $\tilde\Lambda$ width
$\sigma_k$ and the derivatives $g_{k,i}=\partial\ln\tilde\Lambda_k/\partial c_i$ about
the truth.  With $h_i(M)=\partial\ln\Lambda(M)/\partial c_i$ and the prior coefficient
covariance $\boldsymbol{\Sigma}_c$, the forecast is
\begin{align}
F_{ij}&=\sum_k\frac{g_{k,i}\,g_{k,j}}{\sigma_k^2},\nonumber\\
\sigma^2_{\ln\Lambda(M)}&=\boldsymbol{h}^{T}
\left(\boldsymbol{\Sigma}_c^{-1}+\boldsymbol{F}\right)^{-1}\boldsymbol{h}.
\label{eq:coefffisher}
\end{align}
These derivatives act on the reconstructed sequences, which are explicit functions of
$c$, and we evaluate them by finite differences.  This forecast therefore needs no
derivative through the bridge or the TOV solve.  The bridge enters only through the
basis and $\boldsymbol{\Sigma}_c$.  This forecast predicts
$\sigma_{\Lambda(1.4)}\approx\result{HeroFisherSd}$, and its mass-resolved band
tracks the posterior's, widening off the sampled masses in the same way
(Fig.~\ref{fig:hero}c).  Compared against the converged reference of
Sec.~\ref{sec:seqtest} --- $\pm\result{BruteRefSd}$ from
$\result{BruteRefDraws}$ fixed-prior draws, the appropriate yardstick because
the archive-sized reweighting is starved --- their standard deviations differ
by \result{HeroFisherAgreePct}\%.  For this catalogue, the linearized
calculation and the fixed-prior nonparametric calculation give consistent
widths.

\begin{figure*}[t]\centering
  \includegraphics[width=\textwidth]{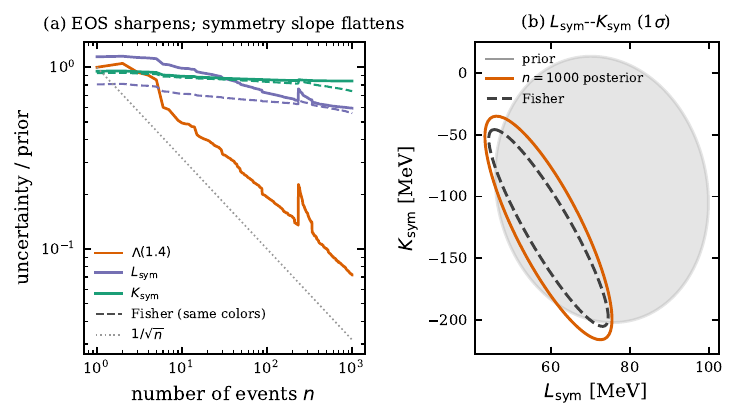}
  \caption{\label{fig:symmetry} The EOS sharpens; the symmetry slope narrows to a
  floor and the curvature narrows less.
  (a)~Fractional uncertainty vs $n$: $\Lambda(1.4)$ shrinks $\sim n^{-1/2}$, while
  $\sigma_{L_{\rm sym}}$ (NEPs conditioned on the measured $\Lambda(1.2)$ and
  $\Lambda(1.4)$ with Eq.~\eqref{eq:nepcondition}) flattens toward
  \result{SymSlopePerfectSd}\,MeV and $\sigma_{K_{\rm sym}}$ narrows slowly.
  For $n<\result{HeroEvents}$ the measurement covariance is scaled by the
  $\Lambda(1.4)$ width of the loop. (b)~$L_{\rm sym}$--$K_{\rm sym}$
  $1\sigma$ region, prior vs the \result{HeroEvents}-event posterior: both parameters
  narrow, and the posterior is anticorrelated.  Dashed: the
  Fisher forecast of Eq.~\eqref{eq:nepfisher} for the same measurement, with the
  parametric metamodel and constant-sound-speed tail; in (b) it is centred on the
  sample-based posterior mean.  The jaggedness of the $\Lambda(1.4)$
  trace in (a) is Monte Carlo noise in the weighted-variance estimate at finite
  effective sample size --- largest just before each redraw, where the ESS is
  smallest, and reset when a fresh ensemble is drawn.  An upward step at a redraw is
  the refit broadening of Sec.~\ref{sec:seqtest}.\llmfig}
\end{figure*}

Figure~\ref{fig:symmetry} shows what this EOS measurement implies for the nuclear
symmetry parameters.  Symmetric matter is comparatively well known, so a neutron-star
measurement is nominally a measurement of the symmetry sector.  In our model, the
catalogue narrows the slope $L_{\rm sym}$ to a floor that more events barely lower.
It narrows the curvature $K_{\rm sym}$ less, along a direction anticorrelated with
$L_{\rm sym}$, and leaves $E_{\rm sym}$ at its prior width.

The limiting uncertainty reflects the range of nuclear parameters compatible
with the measured tidal deformabilities.  We condition on the joint
$\Lambda(1.2)$--$\Lambda(1.4)$ measurement, retaining its covariance
[Eq.~\eqref{eq:nepcondition}].  Even exact measurements of these two quantities
leave scatter in $L_{\rm sym}$ within the adopted ensemble; the catalogue is
already approaching that limit (Fig.~\ref{fig:symmetry}a).  More events can
therefore sharpen the tidal measurement without a comparable gain in the
symmetry slope.  This limit depends on the prior and the measured masses.

Local leave-one-out tests support the predicted $L_{\rm sym}$ uncertainty
near the measured point, and varying the conditioning kernel has little effect
on it.  These checks assess local predictive scatter; they do not establish
coverage across the EOS prior.  We have tested one injection.  Supplemental
Material, Sec.~{SI\,C} gives the calibration, the offset
from the injected slope, and the mass-coverage checks.

The differentiated parametric model of Sec.~\ref{sec:differentiation} gives a
similar constraint despite replacing the high-density bridge with a
constant-sound-speed continuation.  In both calculations, the tidal
measurements constrain mainly one combination of $L_{\rm sym}$ and
$K_{\rm sym}$; the forecasts agree in the orientation and approximate width
of that constraint (Fig.~\ref{fig:symmetry}b).  The improvement in
$L_{\rm sym}$ and the weak response of $E_{\rm sym}$ also agree qualitatively
with Ref.~\cite{gwastro-nsnuc-Fisher3g-Zhu2025}.  Its relativistic mean-field
model derives $K_{\rm sym}$ from the other parameters, however, so its
parameter correlations cannot be compared directly with ours.

Laboratory information constrains complementary directions
(Table~\ref{tab:frib}).  The illustrative scenarios tighten $E_{\rm sym}$,
which the GW catalogue leaves nearly unchanged.  Tightening $L_{\rm sym}$
also narrows $K_{\rm sym}$ through their correlation, even when no direct
curvature constraint is imposed; $K_{\rm sat}$ remains close to its prior width.
Both EOS calculations give this pattern.  These scenarios demonstrate
complementarity under assumed parameter constraints, rather than forecast
the reach of a particular experiment.

\begin{table*}[t]
\caption{\label{tab:frib}
Symmetry-sector widths (MeV, $1\sigma$) for the \result{HeroEvents}-event
catalogue combined with laboratory scenario priors: sample-based local-linear
conditioning of the nuclear empirical parameters on the measured $\Lambda(1.2)$ and
$\Lambda(1.4)$ [Eq.~\eqref{eq:nepcondition}], with the linearized Jacobian forecast
of Eq.~\eqref{eq:nepfisher} for the same measurement in parentheses.  Each
scenario \emph{replaces} the marginal prior width of the listed parameters ---
$\sigma_{L_{\rm sym}}\!=\!\result{FribNearSlopeLab}$, $\sigma_{E_{\rm sym}}\!=\!\result{FribNearEnergyLab}$ for FRIB-2028;
$\result{FribFourSlopeLab}$ and $\result{FribFourEnergyLab}$ for FRIB400; $+K_{\rm sym}$ adds
$\sigma_{K_{\rm sym}}\!=\!\result{FribFourKCurvLab}$ --- and leaves the others at their prior values.
These widths are assumed scenario values chosen to span the reach of the
rare-isotope and heavy-ion programs~\cite{SorensenEtAl2024,BrownEtAl2025,
LeFevre2026,TsangEtAl2024}; they are not published commitments, and no
projection of that form exists.  The GW-only column reproduces the widths of
Sec.~\ref{sec:hero}.}
\begin{ruledtabular}
\begin{tabular}{lccccc}
 & Prior & GW-only & $+$FRIB-2028 & $+$FRIB400 & $+$FRIB400$+K_{\rm sym}$ \\
\hline
$\sigma_{L_{\rm sym}}$ & \result{FribPriorSlopeSd} &
  \result{FribGwSlopeSd} (\result{FribGwSlopeAdSd}) &
  \result{FribNearSlopeSd} (\result{FribNearSlopeAdSd}) &
  \result{FribFourSlopeSd} (\result{FribFourSlopeAdSd}) &
  \result{FribFourKSlopeSd} (\result{FribFourKSlopeAdSd}) \\
$\sigma_{K_{\rm sym}}$ & \result{FribPriorCurvSd} &
  \result{FribGwCurvSd} (\result{FribGwCurvAdSd}) &
  \result{FribNearCurvSd} (\result{FribNearCurvAdSd}) &
  \result{FribFourCurvSd} (\result{FribFourCurvAdSd}) &
  \result{FribFourKCurvSd} (\result{FribFourKCurvAdSd}) \\
$\sigma_{E_{\rm sym}}$ & \result{FribPriorEnergySd} &
  \result{FribGwEnergySd} (\result{FribGwEnergyAdSd}) &
  \result{FribNearEnergySd} (\result{FribNearEnergyAdSd}) &
  \result{FribFourEnergySd} (\result{FribFourEnergyAdSd}) &
  \result{FribFourKEnergySd} (\result{FribFourKEnergyAdSd}) \\
$\sigma_{K_{\rm sat}}$ & \result{FribPriorKsatSd} &
  \result{FribGwKsatSd} (\result{FribGwKsatAdSd}) &
  \result{FribNearKsatSd} (\result{FribNearKsatAdSd}) &
  \result{FribFourKsatSd} (\result{FribFourKsatAdSd}) &
  \result{FribFourKKsatSd} (\result{FribFourKKsatAdSd}) \\
\end{tabular}
\end{ruledtabular}
\end{table*}

\subsection{Accuracy of exact, clipped, and sequential estimators}\label{sec:seqtest}
Our approach introduced several parameters and optional features; we therefore also
demonstrate how performance changes as some of those choices are varied.
Table~\ref{tab:seqtest} describes more systematic, quantitative tests of our approach
using the $\result{HeroEvents}$-event catalogue of Sec.~\ref{sec:hero} and a
significantly larger initial prior archive: we start with $\result{BruteRefDraws}$ draws from $\pi_0$, sufficient for the fair-draw posterior
to converge.
This gives the converged posterior
$\Lambda(1.4)=\result{BruteRefLam}\pm\result{BruteRefSd}$ with
$N_{\rm eff}=\result{BruteRefEss}$.  We measure each estimator against this
reference.  The exact posterior does not depend on event order, so we repeat the
calculation for $\result{SeqOrderings}$ orderings of the catalogue, and the spread
over orderings measures reliability.  Table~\ref{tab:seqtest} gives the results.

The three estimators differ only in their treatment of $\rho_r$.  The exact weight
of Eq.~\eqref{eq:refitweight} keeps it in full, so its error is variance from the
tail of $\rho_r$.  Sequential projection, Eq.~\eqref{eq:seqweight}, sets $\rho_r=1$,
so its error is bias from the Gaussian fit.  Clipping, Eq.~\eqref{eq:clipweight}, lies
between them.

\begin{table*}[t]
\caption{\label{tab:seqtest}
Recovery of the converged fixed-prior posterior for $\Lambda(1.4)$.  The upper
block uses $\result{SeqOrderings}$ orderings of the
$\result{HeroEvents}$-event catalogue and $\result{HeroDrawsPerGen}$ draws per generation, with the
stated $N_{\rm eff}$ refit threshold.  The scatter is the standard deviation of
the means over orderings.  Offsets are in units of the reference standard
deviation, and width ratios compare with that standard deviation.  The lower block uses \result{SubCatCount} disjoint
$\result{SubCatEvents}$-event sub-catalogues, each compared with its own reference.  Importance
corrections are evaluated at the last event on the final generation.  Clipping
uses $\tau$ at the 90th percentile of $\rho_r$.}
\begin{ruledtabular}
\begin{tabular}{lccccc}
Estimator & Refits & Mean $\pm$ scatter & Largest offset & Width ratio & Median $N_{\rm eff}$ \\
\hline
Exact $\rho_r$, $\gamma=0$ &
$\result{ExactPlainRefitsLo}$ &
$\result{ExactPlainMean}\pm\result{ExactPlainScatter}$ &
$\result{ExactPlainMaxOffset}$ &
$\result{ExactPlainWidthLo}$--$\result{ExactPlainWidthHi}$ &
$\result{ExactPlainEss}$ \\
Exact $\rho_r$, $\gamma=\result{SeqShrinkGamma}$ &
$\result{ExactShrunkRefitsLo}$--$\result{ExactShrunkRefitsHi}$ &
$\result{ExactShrunkMean}\pm\result{ExactShrunkScatter}$ &
$\result{ExactShrunkMaxOffset}$ &
$\result{ExactShrunkWidthLo}$--$\result{ExactShrunkWidthHi}$ &
$\result{ExactShrunkEss}$ \\
Clipped $\rho_r$, $\gamma=0$ &
$\result{ClipPlainRefitsLo}$ &
$\result{ClipPlainMean}\pm\result{ClipPlainScatter}$ &
$\result{ClipPlainMaxOffset}$ &
$\result{ClipPlainWidthLo}$--$\result{ClipPlainWidthHi}$ &
$\result{ClipPlainEss}$ \\
Clipped $\rho_r$, $\gamma=\result{SeqShrinkGamma}$ &
$\result{ClipShrunkRefitsLo}$--$\result{ClipShrunkRefitsHi}$ &
$\result{ClipShrunkMean}\pm\result{ClipShrunkScatter}$ &
$\result{ClipShrunkMaxOffset}$ &
$\result{ClipShrunkWidthLo}$--$\result{ClipShrunkWidthHi}$ &
$\result{ClipShrunkEss}$ \\
Sequential, $\gamma=0$, threshold $\result{HeroRefitThreshold}$ &
$\result{SeqPlainRefitsLo}$ &
$\result{SeqPlainMean}\pm\result{SeqPlainScatter}$ &
$\result{SeqPlainMaxOffset}$ &
$\result{SeqPlainWidthLo}$--$\result{SeqPlainWidthHi}$ &
$\result{SeqPlainEss}$ \\
Sequential, $\gamma=0$, threshold $\result{SeqMidThreshold}$ &
$\result{SeqMidRefitsLo}$--$\result{SeqMidRefitsHi}$ &
$\result{SeqMidMean}\pm\result{SeqMidScatter}$ &
$\result{SeqMidMaxOffset}$ &
$\result{SeqMidWidthLo}$--$\result{SeqMidWidthHi}$ &
$\result{SeqMidEss}$ \\
Sequential, $\gamma=0$, threshold $\result{SeqFastThreshold}$ &
$\result{SeqFastRefitsLo}$--$\result{SeqFastRefitsHi}$ &
$\result{SeqFastMean}\pm\result{SeqFastScatter}$ &
$\result{SeqFastMaxOffset}$ &
$\result{SeqFastWidthLo}$--$\result{SeqFastWidthHi}$ &
$\result{SeqFastEss}$ \\
Sequential, $\gamma=\result{SeqShrinkGamma}$, threshold $\result{HeroRefitThreshold}$ &
$\result{SeqShrunkRefitsLo}$--$\result{SeqShrunkRefitsHi}$ &
$\result{SeqShrunkMean}\pm\result{SeqShrunkScatter}$ &
$\result{SeqShrunkMaxOffset}$ &
$\result{SeqShrunkWidthLo}$--$\result{SeqShrunkWidthHi}$ &
$\result{SeqShrunkEss}$ \\
\hline
Sub-catalogues: clipped, $\gamma=0$ & & &
$\result{SubClipPlainMaxOffset}$ &
$\result{SubClipPlainWidthLo}$--$\result{SubClipPlainWidthHi}$ & \\
Sub-catalogues: clipped, $\gamma=\result{SeqShrinkGamma}$ & & &
$\result{SubClipShrunkMaxOffset}$ &
$\result{SubClipShrunkWidthLo}$--$\result{SubClipShrunkWidthHi}$ & \\
Sub-catalogues: sequential, $\gamma=0$ & & &
$\result{SubSeqPlainMaxOffset}$ &
$\result{SubSeqPlainWidthLo}$--$\result{SubSeqPlainWidthHi}$ & \\
Sub-catalogues: sequential, $\gamma=\result{SeqShrinkGamma}$ & & &
$\result{SubSeqShrunkMaxOffset}$ &
$\result{SubSeqShrunkWidthLo}$--$\result{SubSeqShrunkWidthHi}$ & \\
\end{tabular}
\end{ruledtabular}
\end{table*}

\emph{Exact weights.}  An attempt to use the raw weights introduces tail resampling
error.  A few draws in the tail of $\rho_r$ carry most of the weight, leaving
$N_{\rm eff}$ of order ten.  The mean remains near the reference, but the width
depends on which tail draws appear, and its ratio to the reference falls as low as
$\result{ExactPlainWidthLo}$.  We need a scheme that stabilizes this result, either
clipping or sequential projection.

\emph{Clipping.}  Clipping is a well-established protection against this tail
risk~\cite{Ionides2008TruncatedIS,Vehtari2024PSIS}.  Clipping $\rho_r$ at its 90th
percentile gives the most stable results in Table~\ref{tab:seqtest}: the smallest
scatter over orderings, with widths close to the reference and
$N_{\rm eff}$ near $\result{ClipPlainEss}$.  The choice of threshold is not critical.
A higher threshold or the rule $\tau=\sqrt{N}\,\langle\rho_r\rangle$ gives similar
means with fewer effective draws, and a lower threshold moves toward the sequential
result.  Clipping keeps most of the correction for proposal mismatch, so it also
protects the width after a late refit.  Its cost is that every new draw needs the
likelihoods of all earlier events.

\emph{Sequential projection.}  Setting $\rho_r=1$ removes the tail altogether and
needs only the new events.  With $\gamma=0$, it recovers the reference for every
ordering, threshold, and sub-catalogue: the mean stays within one reference width,
and the width ratio lies between $\result{SeqAllWidthLo}$ and
$\result{SeqAllWidthHi}$.  Repeating one ordering with different random seeds moves
the mean much less than changing the ordering does, so the scatter is projection
error and not sampling noise.  The result is stable under changes of the refit
threshold.  Raising it takes a run from $\result{SeqPlainRefitsLo}$ refits to as many
as $\result{SeqFastRefitsHi}$, and the mean and its scatter do not change appreciably.
Even with many refits we obtain the same answer, so repeated projection does not
accumulate a visible error.  We expect this for a compact, single-peaked posterior,
whose first two moments carry most of its information.  Section~\ref{sec:adapt}
describes when a Gaussian would fail.

\emph{Shrinkage.}  Shrinkage can help when a refit rests on few effective draws,
because it limits how far the proposal moves.  In this test it is associated with the
largest outliers in width and mean in Table~\ref{tab:seqtest}.  It also discards
information from earlier events under sequential projection
(Sec.~\ref{sec:sequential}).  Here $N_{\rm eff}$ at each refit far exceeds the number
of fitted coefficients, so we see no compelling reason to use it and good reasons not
to.  Section~\ref{sec:hero} uses $\gamma=0$.

\emph{Width after a refit.}  The width error is most likely just after a refit.  Each
refit broadens the distribution of $\Lambda(1.4)$ among the new draws, relative to the
weighted draws it fits, and later events narrow it again.  We have not isolated the
cause of this broadening at $\gamma=0$.  Runs with several hundred events after the
last refit match the reference, and runs with a recent refit are conservative.  This
need not limit an analysis, because several overlapping estimators can run at once
and the analyst can report the one judged most reliable.  A new refit can begin early
while the existing ensemble supplies the science result.  The analysis then switches
to the new ensemble once enough later events have narrowed it.  We have not
demonstrated this switch here.  Clipping is preferable when the likelihoods of earlier
events are cheap to evaluate on new draws, as they are here.  Sequential projection
with $\gamma=0$ applies when each per-EOS evidence is expensive.

\clearpage
\section{Observable degrees of freedom and EOS model complexity}
\label{sec:validation}\label{sec:information}
EOS inference constrains the combinations of matter properties that affect
stellar observables over the measured mass range and at the available precision.
The complexity of an EOS prior can substantially exceed that of its image in
mass, radius, and tidal deformability.  We examine this relation using sequence
compression and the recovery of sound-speed structure, then discuss its
implications for sampling adequacy and the interpretation of prior and model
changes.

\subsection{Structure retained by the stellar observables}
\begin{figure}[t]\centering
  \includegraphics[width=\columnwidth]{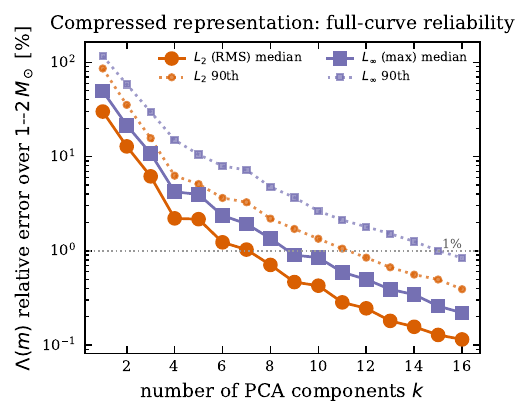}
  \caption{\label{fig:val_pca} Compressed-representation reliability across the
  full $\Lambda(m)$ curve ($1$--$2\,\msun$): relative-error $L_2$ (RMS) and
  $L_\infty$ (max) norms versus the number of PCA components (median and 90th
  percentile over the bridge ensemble). The vertical marker identifies the
  \result{SeqComponents}-component production setting.\llmfig}
\end{figure}

\emph{Compressed representation.} Figure~\ref{fig:val_pca} tests the PCA
representation across the full tidal sequence $\Lambda(m)$ over
$1$--$2\,\msun$.  It summarizes each EOS by the $L_2$ (RMS) and $L_\infty$
(maximum) relative errors over the bridge ensemble.  The median and upper-tail
errors decrease monotonically through the \result{SeqComponents}-component production setting.
This test measures truncation error within the calibration ensemble, not the
accuracy of proposal draws far outside it.

\begin{figure*}[t]\centering
  \includegraphics[width=\textwidth]{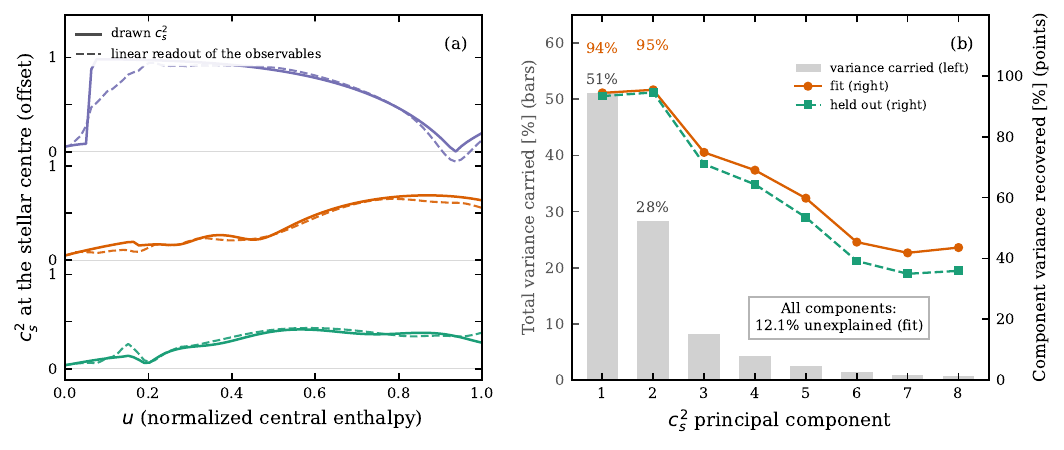}
  \caption{\label{fig:dimension} Linear reconstruction of central sound speed
  $\cs^2$ from observable-space coefficients.  (a)~Three draws versus normalized
  central enthalpy $u$ [Eq.~\eqref{eq:ucoord}], offset vertically: solid curves
  show the draws and dashed curves their reconstructions.  The reconstruction
  retains the overall shape and the step at the matching density $n_L$, but
  smooths finer structure.  (b)~Sound-speed principal components, ordered by
  decreasing variance.  Grey bars use the \emph{left axis}: each gives the
  percentage of total sound-speed variance carried by that component; the first
  two carry approximately $\result{DimComponentOneVarPct}\%$ and
  $\result{DimComponentTwoVarPct}\%$.  Circles (fit) and squares (held-out
  predictions) use the \emph{right axis}: $100R^2$ is the percentage of each
  component's amplitude variance explained by the stellar coefficients
  [Eq.~\eqref{eq:mode_recovery}].  Labels on the first two circles identify the
  approximately $\result{DimComponentOneRecoveryPct}\%$ and
  $\result{DimComponentTwoRecoveryPct}\%$ fitted recovery.  Zero means no
  improvement over predicting the mean amplitude.  The annotation reports the
  variance-weighted unrecovered fraction of sound-speed variance, $\sum_k f_k(1-R_k^2)$, summed over
  all components, including those not plotted.  All fits and
  \result{DimCvFolds}-fold held-out predictions use the same leverage-screened
  ensemble (Supplemental Material, Sec.~{SII}).\llmfig}
\end{figure*}

\emph{Sound-speed information retained by the observable map.}
Figure~\ref{fig:dimension} tests how accurately the central sound-speed curve
$\cs^2(u)$ can be reconstructed linearly from the coefficients describing the
corresponding mass, radius, and tidal-deformability curves.  Panel~(a) compares
three sound-speed curves with their reconstructions: the overall shape is
reproduced more accurately than the finer structure.

Panel~(b) summarizes this comparison across the ensemble.  Here variance
measures the differences among the sound-speed curves, rather than measurement
uncertainty.  Principal components describe orthogonal patterns of variation
about the mean curve and are ordered by their contribution to the total
variance.  The grey bars, read against the \emph{left} axis, give these
contributions: the first two components carry approximately
$\result{DimComponentOneVarPct}\%$ and $\result{DimComponentTwoVarPct}\%$,
respectively, or $\result{DimTopTwoVarPct}\%$ together.
The circles and squares, read against the \emph{right} axis, instead show how
accurately each component's amplitude can be predicted from the stellar
coefficients.  The first two circles lie at approximately
$\result{DimComponentOneRecoveryPct}\%$ and
$\result{DimComponentTwoRecoveryPct}\%$: these are the fractions of each
component's amplitude variance explained by the fit ($100R^2$), not shares
of the total sound-speed variance.  The squares show the corresponding
predictions for held-out draws.  Recovery decreases for subsequent components.
The separately annotated $\result{DimNullFracPrunedPct}\%$ unexplained variance
is an aggregate over \emph{all} components, including those beyond the eight
shown: each component's variance share is multiplied by its unrecovered
fraction, and the products are summed.  Each component therefore contributes
$f_k(1-R_k^2)$ to the total normalized mean-squared reconstruction error.
This comparison is consistent with the linear
argument of Sec.~\ref{sec:linear}.

This diagnostic applies only to the adopted prior ensemble, forward map, and
linear reconstruction.  It is not a posterior constraint on sound-speed components;
identifiability under a particular likelihood is a separate
question~\cite{gwastro-nsnuc-Fisher3g-Zhu2025}.  Supplemental Material, Sec.~{SII}
describes the construction, held-out validation, and a near-degenerate draw.

\subsection{Prior complexity can obscure the measured problem}
The original GP inference construction samples a functional EOS prior and
compares its realizations with the observations by direct Monte Carlo
integration~\cite{LandryEssick2019}; subsequent applications use broad GP
priors to combine stellar observations~\cite{Legred2021}.  This freedom helps
avoid committing to a restrictive microscopic model.  It also creates many
explicit degrees of freedom that the observations may not resolve.  Merely
generating more of them does not reveal which combinations are measured.

For cold stellar observations, the relevant changes are those in $\Lambda(m)$,
$R(m)$, and $M_{\rm max}$ over the observed masses.  Derivatives such as
$\partial\Lambda(m)/\partial c_i$ and $\partial R(m)/\partial c_i$, compared
with the measurement errors, identify the parameter combinations to which the
experiment is sensitive.  A sound-speed excursion that leaves these predictions
indistinguishable at that precision contributes prior freedom, not an additional
observationally accessible degree of freedom.

Figure~\ref{fig:dimension} illustrates the distinction between the complexity
of the sound-speed curves and that of their stellar predictions.  As discussed
above, most of the differences between sound-speed curves are described by two
principal components whose amplitudes can be predicted accurately from the
stellar coefficients.  Subsequent components contribute less to the total
variance and are recovered less accurately.  Thus, an accurate representation
of the mass, radius, and tidal-deformability curves need not reproduce all the
sound-speed structure present in the underlying EOS ensemble.
Conversely, the systematic error (relative to sound-speed differences) associated with our projection  can be estimated as a balance between mode strength and
accuracy, with expected value roughly $f_k(1-R_k^2)$ where $f_k$ (the
fraction of the total variance) is indicated in gray and $R_k^2$ (the recovery accuracy) is indicated with colored
points.  Because we predict the strongest modes reliably, only about $12\%$ of the overall variance isn't captured in
our projections; in part, that variance is driven by our dimensional reduction, which loses some information about weak phase
transitions and other strong sound-speed variations as illustrated in the left panel, without significantly changing our
ability to predict observables. 
Figure~\ref{fig:val_pca} tests the accuracy of the compressed stellar predictions,
whereas Fig.~\ref{fig:dimension} tests \emph{linear} recovery of sound-speed structure
from that representation.
Refinement complements this compression by concentrating new draws in
the region supported by the data, using coordinates aligned with the retained
stellar predictions.

\subsection{Consequences for sampling and interpretation}
\begin{figure}[t]\centering
  \includegraphics[width=\columnwidth]{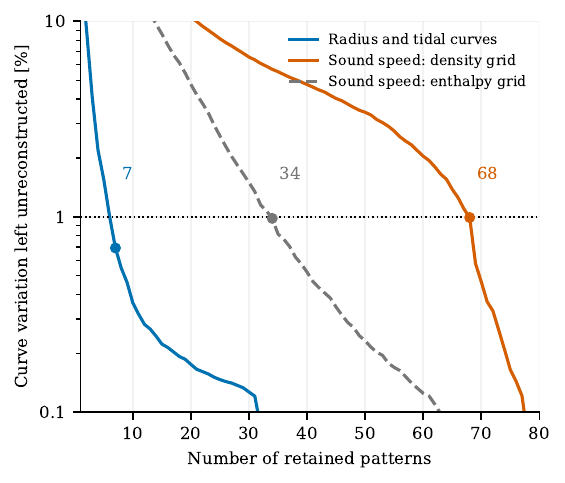}
  \caption{\label{fig:external_gp_dimension}
  Effective dimension of the pulsar-conditioned GP family of
  Ref.~\cite{Legred2021}, reused as a starting prior.  The vertical axis is
  the percentage of family variation left unreconstructed,
  $100\epsilon_d$ [Eq.~\eqref{eq:dimension_error}]; the horizontal axis counts
  retained principal-component patterns.  At the dotted $\result{ExtGpLevelPct}\%$ line the stellar
  curves need \result{ExtGpStellarRank} patterns, while the sound-speed curves need \result{ExtGpCsDensityRank} on a fixed
  density grid and \result{ExtGpCsEnthalpyRank} in the enthalpy representation.
  Stellar curves contain $R(M)$ and $\ln\Lambda(M)$ at \result{ExtGpMassPoints} masses between
  $\result{ExtGpMassLo}$ and $\result{ExtGpMassHi}\,\msun$, each channel normalized by its training-family spread.
  Sound speed is sampled at \result{ExtGpDensityPoints} rest-mass densities between
  $\rho_{\rm sat}$ and $\result{ExtGpRhoHiRatio}\rho_{\rm sat}$
  ($\rho_{\rm sat}=\result{ExtGpRhoLo}\,\mathrm{g\,cm^{-3}}$), or at \result{ExtGpEnthalpyPoints} points
  along each complete archived central-enthalpy sequence.
  Bases are fitted on \result{ExtGpTrainPct}\% of unique EOS IDs; all errors are evaluated on the
  remaining IDs.  All three representations use the same \result{ExtGpRows} sample rows
  with an unambiguous increasing-mass branch across the mass interval;
  repeated EOS retain their multiplicity and stay in the same partition.\llmfig}
\end{figure}

By characterizing the effective dimensions of EOS families in observation
space and EOS space, our framework provides a powerful way to understand what
inferences constrain and what prior choices introduce.  As a concrete example,
Fig.~\ref{fig:external_gp_dimension} applies the framework to the published
pulsar-conditioned GP ensemble of Ref.~\cite{Legred2021}, used here as an
empirical starting prior.  At every accuracy threshold shown, its
sound-speed curves require many more patterns than its radius and tidal
curves.  At a relative reconstruction error of $1\%$
[Eq.~\eqref{eq:dimension_error}], \result{ExtGpStellarRank} patterns describe the stellar curves,
whereas \result{ExtGpCsDensityRank} are needed for the sound speed on a fixed density grid.  The
sound-speed representation along the archived central-enthalpy sequences also
requires substantially more patterns.  This published GP family therefore
provides a direct example of a rich EOS model with a much simpler image in
stellar observables.

Multiple phase transitions make the sampling consequence particularly clear. As
the number of transitions grows, so does the number of independently adjustable
transition densities, strengths, and intervening stiffnesses.  GP priors can
admit many analogous localized softenings and stiffenings, including
near-transition features~\cite{Legred2021,Mroczek2023Transitions}.  When
these features affect radii or tidal deformabilities independently over the
measured mass range, they increase the observationally accessible dimension $d$,
the number of directions the likelihood resolves within the prior
[Eq.~\eqref{eq:volume_dimension}].  If $f<1$ denotes the fraction of prior
probability the data retain along each resolved direction, roughly the ratio of
posterior to prior width, then   the jointly supported fraction of the prior 
scales as $f^d$, and an archive of $N$ independent prior draws supplies only
$N_{\rm eff}\approx Nf^d$ relevant realizations on average.  Increasing model
complexity can raise $d$, so direct prior sampling eventually becomes
intractable: for fixed $N$, the probability of finding even one supported EOS,
$1-(1-f^d)^N$, tends to zero with increasing $d$.
Millions of draws provide no general safeguard.  Adaptive sampling or MCMC
can instead explore this region directly, generating the combinations of
features favored by the data rather than waiting for them to occur in a
fixed prior archive.  Figure~\ref{fig:external_gp_dimension}
shows the reservoir of microscopic freedom available in a GP family and
measures how much of it reaches the stellar curves at a chosen resolution.
Our approach identifies those accessible combinations and refines the sampling
accordingly. 

\section{Conclusions}\label{sec:conclusions}
We have introduced an adaptive-proposal procedure for nonparametric
neutron-star equation-of-state inference that leaves the scientific prior fixed
while concentrating new draws in the region supported by the data.  We
implemented the procedure both in a compressed principal-component
representation of the neutron-star sequence, whose draws require no
stellar-structure solve, and in a sound-speed representation with a
\textsc{jax}-differentiable TOV solver.
We assess the informativeness of our procedure with an effective local dimension.
We test our porcedures against synthetic catalogs in the main text.
In an Appendix, we construct two concrete, plausible EOS model families with very large effective dimension, for which
our adaptive refinement approach will be necessary to avoid prohibitive numbers of prior EOS realizations.

Our calculations reveal inspiral-dominated GW observables provide access to only a handful of local dimensions,
These few dimensions enable both enhanced computational efficiency and insight: many fine details in principle
accessible to the EOS family are effectively observationally-inaccessible epicycles.
Our concrete demonstrations show that nature however need not be kind: phase transitions in particular will dramatically
increase the observationally-accessible effective dimension, to the point that existing strategies using large numbers
of fixed prior draws will inevitably fail.  Unless nature is kind, the framework introduced in this work could be
critical to address EOS inference in the presence of phase transitions. 
In brief, while a fixed ensemble
remains adequate for the binary neutron stars observed so far
~\cite{2017PhRvL.119p1101A,2020ApJ...892L...3A},  adaptive proposals become
useful when a large catalogue or additional electromagnetic and nuclear
information concentrates the posterior relative to the prior.

Our work also has direct implications for the degree of complexity any physically-parameterized EOS model family can
extract from observations.  At fixed experimental
precision, the relation between EOS and observations sets the number of EOS or nuclear-parameter
combinations that the data can resolve.  Many EOS model families have significantly more free parameters than we
anticipate are accessible via EM and inspiral-dominated GW observations of NS.   Too, at present and using inspiral-dominated GW observations, this relationship
does not seem to enable meaningful improvements on some EOS parameters like the symmetry energy from GW observations
of cold, beta-equilibrated nuclear matter.  Our proof-of-concept investigations however focused on
low-effective-dimension scenarios without phase transitions, for which observations have relatively little leverage; the
discovery and characterization of phase transitions could enable sharper constraints.

\begin{acknowledgments}
  The authors thank our anonymous referee for constructive feedback.
The authors are grateful for computational resources provided by the LIGO
Laboratory and supported by National Science Foundation Grants PHY-0757058 and
PHY-0823459.  This material is based upon work supported by NSF's LIGO
Laboratory which is a major facility fully funded by the National Science
Foundation.
The authors acknowledge Research Computing at the Rochester Institute of Technology for providing computational resources and support that have contributed to the research results reported in this publication.
ROS and ZZ also acknowledge assistance from Claude/Codex during the implementation, operation, and drafting of the analyses in this manuscript; Appendix~\ref{app:llm} describes that use.
\end{acknowledgments}

\appendix
\section{Current-data demonstrations}\label{app:observations}
These examples check the sampling method with current observations; their
radius summaries remain conditional on the adopted prior and NICER products.

\subsection{Observation-specific evidence factors}\label{app:observation-factors}
For the current-data demonstrations, we supplement the GW likelihood of
Eq.~\eqref{eq:LGW} with the following electromagnetic and laboratory factors.
Products containing the NICER posterior-overlap factor defined below are
operational multimessenger scores rather than prior-independent posteriors.

A NICER mass--radius data product contributes the posterior-overlap factor
\begin{equation}
\mathcal{S}_{\rm NICER}(\theta) = \!\int\! dm\;
  p_{\rm pub}\!\big(R_\theta(m),\,m\big)\,\omega(m),
\label{eq:LNICER}
\end{equation}
the published two-dimensional $(R,m)$ posterior density integrated along the EOS
mass--radius curve.  Our line measure $\omega(m)$ is uniform over the
EOS-supported range,
\begin{equation}
\omega(m) \propto
\begin{cases}
1, & m_{\rm lo} \le m \le \min\!\big(m_{\rm hi},\,M_{\rm max}(\theta)\big),\\
0, & \text{otherwise},
\end{cases}
\label{eq:massprior}
\end{equation}
with $(m_{\rm lo},m_{\rm hi})=(\result{NicerMassLo},\result{NicerMassHi})\,\msun$.  This line measure differs
from the population model used in GW evidences.  Integrating the released
posterior densities retains their sampling priors, so Eq.~\eqref{eq:LNICER}
is a posterior-overlap score~\cite{gwastro-ns-eos-Kedia2024}.

Radio timing of massive pulsars enters as support constraints on the maximum
mass.  Each measurement $\bar m_j \pm \sigma_j$ requires $M_{\rm max}(\theta) \ge
\bar m_j$, contributing a cumulative-distribution factor
$\Phi\!\big((M_{\rm max}-\bar m_j)/\sigma_j\big)$, with $\Phi$ the standard
normal CDF.  The analyses reported here do not apply a remnant-derived upper
bound on $M_{\rm max}$.

The same product admits optional laboratory constraints on the symmetry energy,
which act on the nuclear empirical parameters carried by each draw rather than
on its stellar observables.  Three encodings are available, covering the forms
in which nuclear experiments report: a one-dimensional Gaussian on the symmetry
energy $E_{\rm sym}$ at saturation; a correlated two-dimensional Gaussian on
$(E_{\rm sym},L_{\rm sym})$; and a Gaussian on the metamodel symmetry pressure
\[
P_{\rm sym}(n) = \frac{n^2}{3\nsat}
  \left[L_{\rm sym} + K_{\rm sym}\,\frac{n-\nsat}{3\nsat}\right],
\]
tagged at a stated density and reducing to $P_{\rm sym}(\nsat)=\nsat L_{\rm
sym}/3$ at saturation.  The last is admitted only for $n\le n_L=\result{MatchNsat}\nsat$: above
the matching density the nonparametric bridge, not the empirical expansion,
determines the draw, so a metamodel symmetry pressure evaluated there would not
describe the equation of state actually used.  All three terms are disabled
throughout this paper except in the robustness variant of
Sec.~\ref{sec:nested}.

\subsection{Astrophysical observations}\label{sec:obs-real}
The real data that inform the EOS are heterogeneous.  Gravitational-wave tidal
measurements come from the binary-neutron-star inspirals
GW170817~\cite{2017PhRvL.119p1101A,LIGO-GW170817-SourceProperties,
LIGO-GW170817-EOS}
and GW190425~\cite{2020ApJ...892L...3A}, which constrain $\Lambda$ through
Eq.~\eqref{eq:LGW}; no additional confirmed BNS detections have been reported
through the O4b catalog release~\cite{2026arXiv260527225T}.  For GW170817 this
work reuses the archived RIFT intrinsic-likelihood grid constructed for the
joint GW--electromagnetic analysis of Risti\'{c} et al.~\cite{2026PhRvR...8b3117R},
rather than reprocessing detector strain.  That grid used open GWOSC O2
data~\cite{ligo-O1O2-opendata}, the GWTC-1 power spectral densities, the
\textsc{IMRPhenomPv2\_NRTidalv2} waveform from $\result{LitWaveformFmin}$ to $\result{LitWaveformFmax}$ Hz, and the
electromagnetically determined sky position and luminosity distance.
NICER X-ray timing yields
two-dimensional mass--radius posteriors, which enter through
Eq.~\eqref{eq:LNICER}.  We use the current headline analysis for each source:
for PSR J0030+0451 the six-year reanalysis~\cite{J0030_2026}, whose preferred
\textsc{PDT-U} mode ($R=\result{LitJZeroThirtyR}^{+\result{LitJZeroThirtyRPlus}}_{-\result{LitJZeroThirtyRMinus}}$\,km) supersedes the original
measurement~\cite{Miller2019} and largely resolves the mode ambiguity of the
intermediate reanalysis~\cite{Vinciguerra2024}; for the massive pulsar
J0740+6620 the joint NICER--XMM result of Salmi et al.~\cite{Salmi2024},
concordant with the independent pipeline of Dittmann et al.~\cite{Dittmann2024};
for the nearby J0437--4715 the baseline analysis of Choudhury et
al.~\cite{Choudhury2024}, noting that a recent independent treatment including a
modulated nonthermal component finds a radius roughly $2$\,km
larger~\cite{MillerJ0437_2025} --- an unresolved cross-group systematic we carry
as a caveat rather than an input; and for the compact source J0614--3329 the
discovery analysis~\cite{Mauviard2025}.  We omit J1231--1411, whose inference
converges only under a restrictive, EOS-informed radius
prior~\cite{Salmi2024J1231} and would be circular here.  J0030 enters through a
kernel-density estimate of the published posterior samples~\cite{J0030_2026},
which carry the strong positive $(R,M)$ correlation ($\rho\simeq0.8$) of a
compactness-pinned source with no radio mass prior; the remaining three sources,
whose masses are radio-pinned and whose $(R,M)$ correlations are correspondingly
small, enter as two-dimensional Gaussians with symmetrized $68\%$ intervals and
zero correlation.  Radio timing of massive pulsars sets lower bounds on
$M_{\rm max}$ through the support factors of Appendix~\ref{app:observation-factors}: J0740+6620
($\result{JZeroSevenFourZeroMass}\pm\result{JZeroSevenFourZeroMassSd}\,\msun$~\cite{Fonseca2021}) and J1614--2230
($\result{JOneSixOneFourMass}\pm\result{JOneSixOneFourMassSd}\,\msun$~\cite{Arzoumanian2018}).  We retain J0348+0432 with the
revised CHIME mass $\result{JZeroThreeFourEightMass}\pm\result{JZeroThreeFourEightMassSd}\,\msun$~\cite{Saffer2025}, which supersedes the
earlier $\result{LitAntoniadisMass}\pm\result{LitAntoniadisMassSd}\,\msun$~\cite{Antoniadis2013}; at the revised value it no
longer functions as an independent two-solar-mass anchor.  Black-widow
candidates such as J0952--0607 ($\result{LitJNineFiveTwoMass}\pm\result{LitJNineFiveTwoMassSd}\,\msun$~\cite{J0952_2025}) are
excluded, following standard practice, because their masses rest on optical
light-curve modeling rather than clean orbital dynamics.

\subsection{GW170817: evidence runs on a fixed target}\label{app:gw170817}
We first analyse GW170817 alone.  The target is the archive prior times the
GW170817 evidence.  For each archive EOS we compute a per-EOS evidence $Z(\theta)$
from the tidal likelihood $\mathcal{L}_{\rm GW}$ via Eq.~\eqref{eq:eosevidence} with
the RIFT tabular pipeline, summing over every stable branch of the sequence.  The
uninformed prior gives
$\Lambda(1.4)=\result{NestedPriorLam}\pm\result{NestedPriorLamSd}$.  Reweighting
the \result{PooledBaseDraws} base-archive draws (gen-0) by $w\propto Z$ gives
$\Lambda(1.4)=\result{CombineGenZeroLam}\pm\result{CombineGenZeroLamSd}$ at
effective sample size $N_{\rm eff}=\result{CombineGenZeroEss}$ of
\result{GenZeroUsable} usable draws.

\begin{figure}[t]\centering
  \includegraphics[width=\columnwidth]{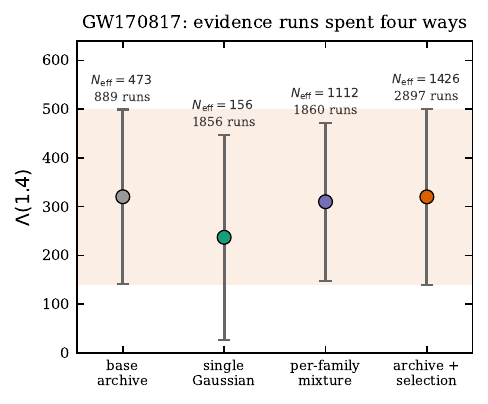}
  \caption{\label{fig:res_loop} GW170817 $\Lambda(1.4)$ from four ways of spending
  evidence runs on the archive posterior.  From left: the base archive
  [$w\propto Z$, Eq.~\eqref{eq:eosevidence}]; a single-Gaussian refit combined by
  Eq.~\eqref{eq:balance}; a refit with one Gaussian per branch family, combined the
  same way; and the enlarged archive with selected evidence runs
  [$w\propto Z/\pi$].  Annotations give $N_{\rm eff}$ and the number of finite
  evidence runs.  The shaded band is the enlarged-archive result.  The two refits
  target their fitted densities.\llmfig}
\end{figure}

Figure~\ref{fig:res_loop} compares three ways to spend further evidence runs on the
same target.  First, we refit the coefficient Gaussian to the gen-0 posterior, draw a
generation-1 ensemble from it, and combine both generations with the
multiple-importance weights of Eq.~\eqref{eq:balance}, $w\propto Z\,\pi_0/q_{\rm
mix}$.  This combination needs the archive's density, which we represent by the
fitted Gaussian (Sec.~\ref{sec:obs-space}).  With \result{PooledMultiBranchPct}\% of
the draws on sequences with more than one stable branch, that representation fails.
The multi-branch draws lie in a separate region of coefficient space, the refit
drops most of them, and the combined weights concentrate on a few draws:
$\Lambda(1.4)=\result{CombineTotalLam}\pm\result{CombineTotalLamSd}$ at
$N_{\rm eff}=\result{CombineTotalEss}$.  Second, a refit with one Gaussian for each
family (one stable branch, or more than one) avoids this collapse and gives
$\Lambda(1.4)=\result{MixtureCombineLam}\pm\result{MixtureCombineLamSd}$ at
$N_{\rm eff}=\result{MixtureCombineEss}$.  It targets the fitted mixture rather than
the archive.  Third, we enlarge the archive to \result{PooledDraws} draws and compute
evidences only for a subset $S$.  Each draw enters $S$ independently with a known
probability $\pi_i$, set from a cheap estimate of $Z_i$ and the exact
electromagnetic factors of Sec.~\ref{sec:nested}.  The weights $Z_i/\pi_i$ on $S$
estimate the full-archive sums without bias, so the estimate
$\sum_{i\in S}(Z_i/\pi_i)f_i/\sum_{i\in S}Z_i/\pi_i$ of a quantity $f$ is consistent
for the archive posterior.  With \result{PooledEvidenceRuns} evidence runs this gives
$\Lambda(1.4)=\result{NestedGwLam}\pm\result{NestedGwLamSd}$ at
$N_{\rm eff}=\result{NestedGwEss}$.  It agrees with the gen-0 estimate.
On this likelihood GW170817 alone is weakly informative: its effective tidal
width is comparable to the prior.  This motivates the multimessenger combination
below.

\subsection{Nested PSR / PSR+GW / PSR+GW+NICER weights}\label{sec:nested}
NICER mass--radius posterior-overlap scores and radio maximum-mass factors enter
as per-EOS weights
on the enlarged archive of Appendix~\ref{app:gw170817}, using the factors defined in
Appendix~\ref{app:observation-factors}; each is cheap post-processing on the
precomputed sequence.  The GW170817 factor is $Z_i/\pi_i$ on the selected draws.
No redraw is triggered: this nested analysis \emph{is} the
finite-archive (``brute-force'') reference used for the fixed-data comparisons.
With the selected evidence runs, the PSR+GW+NICER weights reach
$N_{\rm eff}=\result{NestedFullEss}$, against \result{NestedFullBaseEss} for the
base archive alone.
The NICER-weighted distribution is not a prior-independent posterior because the
published source posteriors have not been divided by their sampling priors
(Appendix~\ref{app:observation-factors}).
weighted bands contracting through the nested data sets. The radius at $1.4\,\msun$
tightens from
$R(1.4)=\result{NestedPriorRadius}\pm\result{NestedPriorRadiusSd}$\,km (prior)
to $\result{NestedFullRadius}\pm\result{NestedFullRadiusSd}$\,km (PSR+GW+NICER),
with $\Lambda(1.4)=\result{NestedFullLam}\pm\result{NestedFullLamSd}$ and
$M_{\rm max}=\result{NestedFullMmax}\pm\result{NestedFullMmaxSd}\,\msun$; the
contraction is driven by the NICER overlap scores and radio factors rather than
the comparatively weak GW170817 tidal likelihood.  The quoted summaries are
conditional on the bridge calibration, finite archive, Gaussian
approximations for three NICER sources, and inherited NICER sampling priors.

The low-density anchor controls the prior radius support.  The adopted
calibration spans the J0030 credible region; Sec.~\ref{sec:lowdensity}
compares it with an explicit $\chi$EFT pressure anchor to assess the resulting
radius sensitivity.

The macroscopic posterior is, by contrast, insensitive to the laboratory
constraint on the symmetry energy tested here.
Reweighting the same PSR+GW+NICER draws by the one-dimensional term of
Appendix~\ref{app:observation-factors}, with the saturation symmetry energy
$E_{\rm sym}=\result{LabSymSZero}\pm\result{LabSymSZeroSd}$\,MeV~\cite{Tsang2012} adopted in
Ref.~\cite{Mroczek2023}, changes
$E_{\rm sym}=\result{LabSymBaseEsym}\pm\result{LabSymBaseEsymSd}$ to
$\result{LabSymEsym}\pm\result{LabSymEsymSd}$\,MeV, but leaves the stellar
observables nearly unchanged:
$\Lambda(1.4)=\result{NestedFullLam}\pm\result{NestedFullLamSd}$ becomes
$\result{LabSymLam}\pm\result{LabSymLamSd}$, and
$R(1.4)=\result{NestedFullRadius}\pm\result{NestedFullRadiusSd}$\,km becomes
$\result{LabSymRadius}\pm\result{LabSymRadiusSd}$\,km, with
$N_{\rm eff}=\result{LabSymBaseEss}\!\to\!\result{LabSymEss}$.  This weak
coupling is consistent with the forecast of Sec.~\ref{sec:hero}:
$E_{\rm sym}$ is a weak lever on $\Lambda(m)$, whereas $L_{\rm sym}$ carries
most of the symmetry-sector sensitivity.  Table~\ref{tab:frib} tests the latter
with illustrative laboratory prior widths; incorporating measured heavy-ion
constraints through the correlated $(E_{\rm sym},L_{\rm sym})$ or
density-tagged $P_{\rm sym}(n)$ likelihoods is left to future work.

\begin{figure}[t]\centering
  \includegraphics[width=\columnwidth]{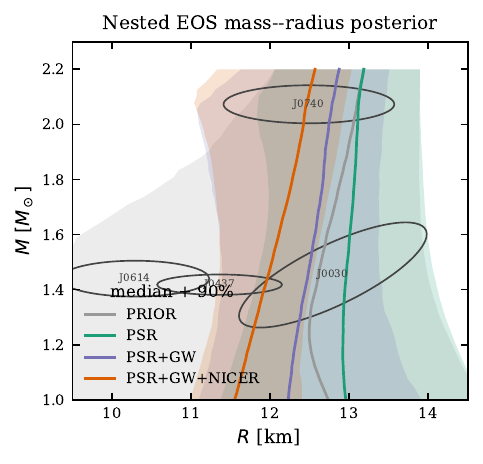}
  \caption{\label{fig:res_mr} Mass--radius weighted bands (median + 90\%) under
  nested data sets (prior / PSR / PSR+GW / PSR+GW+NICER), with NICER credible
  ellipses (J0030, J0740, J0437) overlaid. Radius at $1.4\,\msun$ tightens
  $\sim\result{NestedRadiusShrink}\times$ from prior to the full data set.\llmfig}
\end{figure}

\begin{figure}[t]\centering
  \includegraphics[width=\columnwidth]{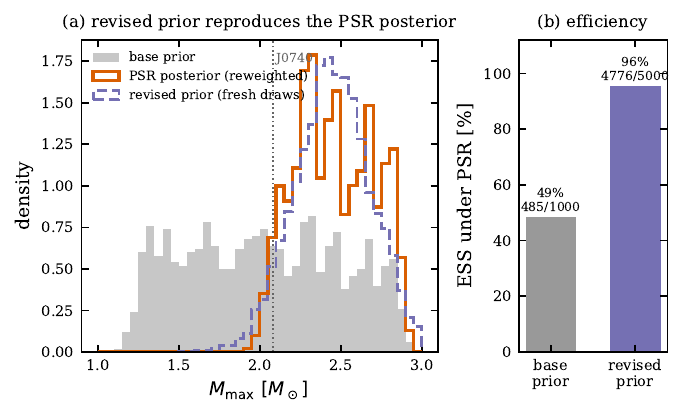}
  \caption{\label{fig:psr_refine} Refinement on the real maximum-mass constraint.
  (a)~The base prior (grey) is broad in $M_{\rm max}$; the radio likelihood
  (J0740/J0348/J1614) concentrates it (PSR posterior, orange). A single refit of
  the coefficient-GP gives a refined proposal whose fresh draws (dashed)
  reproduce that posterior. (b)~Effective sample size under the PSR likelihood:
  fresh draws from the refined proposal are $\result{PsrRevisedEssPct}\%$ efficient,
  versus $\result{PsrBaseEssPct}\%$ for the
  base prior --- the soft tail that the data discards is precisely what the
  refinement recovers.\llmfig}
\end{figure}

The broad maximum-mass prior also illustrates the benefit of refinement
(Fig.~\ref{fig:psr_refine}).  The radio likelihood downweights soft EOS,
leaving a prior-draw efficiency of $\result{PsrBaseEssPct}\%$.  A Gaussian
refit concentrates fresh draws in the supported region and raises the
efficiency to $\result{PsrRevisedEssPct}\%$ under the same likelihood.

\section{Sensitivity to the low-density anchor}
\label{sec:lowdensity}
To isolate this dependence, we replace the sector below $n_L=\result{MatchNsat}\nsat$ and
rebuild the ensemble while leaving the bridge unchanged (Sec.~\ref{sec:rep}).
Following the released GKKS construction, we use a fixed crust to
$\result{CeftCrustEdgeNsat}\nsat$ and the chiral-effective-field-theory calculation of Drischler et
al.~\cite{Drischler2019} above it, continued to $\beta$ equilibrium and sampled
across its theory band to $n_L$.  The fractal refinement, hierarchical
correlation length, pQCD endpoint, truncation density, and random seed remain at
their production settings.  Two implementation differences are required:
\textsc{lalsuite} provides no BPS table, so we use the SLy
crust~\cite{lalsuite-soft}; and because GKKS do not specify sampling within the
chiral-EFT band, we interpolate with one scalar per draw and use the standard
order-by-order truncation estimate, $|E_{\rm N3LO}-E_{\rm N2LO}|$ per cutoff.
We also test the min--max envelope of the four (order, cutoff) curves, which is
half as wide in $E/A$ at the anchor.

\begin{figure*}[t]\centering
  \includegraphics[width=\textwidth]{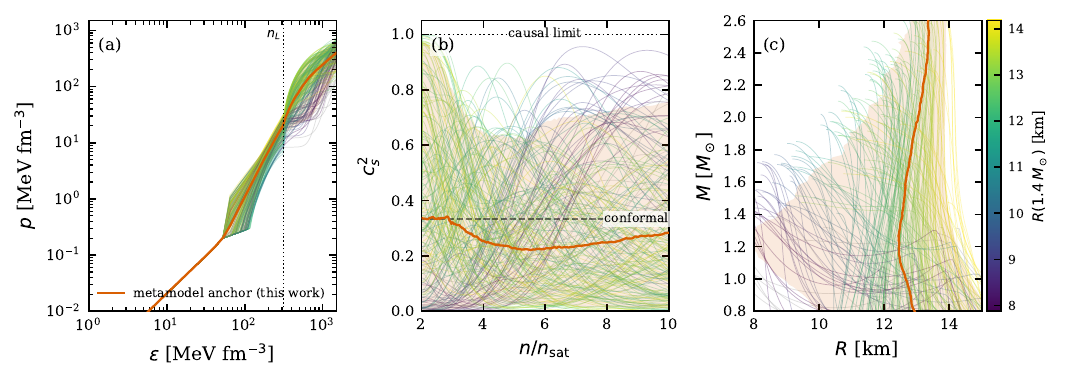}\\[2ex]
  \includegraphics[width=\textwidth]{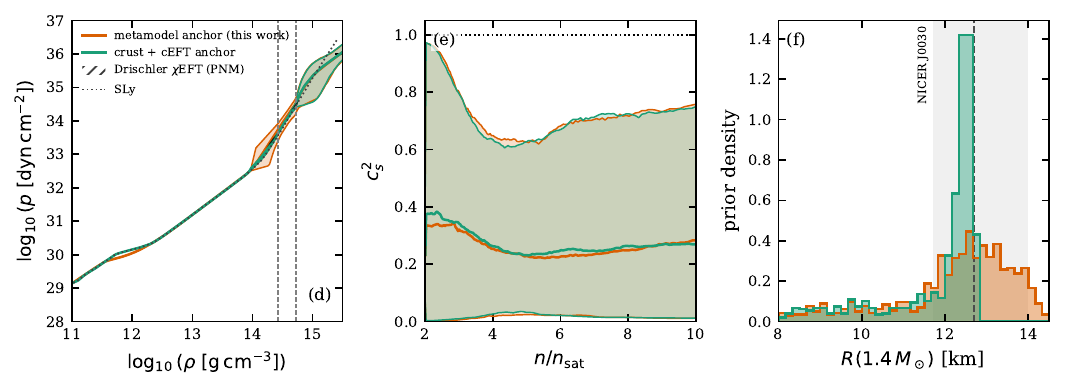}
  \caption{\label{fig:prior_anatomy} Bridge-prior realizations and sensitivity
  to the low-density anchor.  \textbf{Top row}: (a)~pressure versus energy
  density, (b)~sound speed versus baryon density above $n_L=\result{MatchNsat}\nsat$, and
  (c)~mass--radius relations.  Thin curves are $\result{PriorDrawsShown}$ production-ensemble draws,
  coloured by $R(1.4\,\msun)$ (grey curves do not reach $1.4\,\msun$); heavy lines
  and bands show the median and $90\%$ interval.  \textbf{Bottom row}
  (Sec.~\ref{sec:lowdensity}): ensembles with the same bridge but either the
  resampled nuclear metamodel (orange) or a GKKS-style fixed crust and
  chiral-EFT anchor (green).  (d)~Pressure--density relations, with the Drischler
  et al.\ $\chi$EFT pure-neutron-matter band (hatched), SLy crust (dotted), and
  $\nsat$ and $\result{MatchNsat}\nsat$ (vertical lines).  (e)~Sound-speed distributions above
  the anchor, which are statistically consistent.  (f)~Prior-predictive radii:
  the $\chi$EFT anchor ends at $\result{LdCeftMaxRadius}$\,km, near the NICER
  J0030 central value (dashed; shaded $68\%$ interval), whereas the metamodel
  prior extends to $\result{LdMetaMaxRadius}$\,km.\llmfig}
\end{figure*}

The metamodel has a substantially broader pressure distribution below
saturation than the crust-plus-$\chi$EFT anchor
[Fig.~\ref{fig:prior_anatomy}(d)].  Above $3\nsat$, the sound-speed priors
and maximum-mass distributions remain consistent: these are controlled by
the unchanged bridge.

The radius support is sensitive to the anchor
[Fig.~\ref{fig:prior_anatomy}(f)].  The crust-plus-$\chi$EFT ensemble ends at
$\result{LdCeftMaxRadius}$\,km, compared with
$\result{LdMetaMaxRadius}$\,km for the metamodel.  The narrower chiral-EFT
envelope gives a similar ceiling.  Although the bridge controls much of the
radius variation within either ensemble, the anchor shifts the upper support
boundary by more than a kilometre.

The radius result in Sec.~\ref{sec:nested} is therefore sensitive to the
low-density prior.  The NICER J0030 constraint is
$\result{LitJZeroThirtyR}^{+\result{LitJZeroThirtyRPlus}}_{-\result{LitJZeroThirtyRMinus}}$\,km (Sec.~\ref{sec:obs-real}), whereas the
crust-plus-$\chi$EFT prior ends near its central value and assigns no support to
much of the upper credible region.  The broader metamodel avoids this upper
truncation, but the reported
$R(1.4\,\msun)=\result{NestedFullRadius}\pm\result{NestedFullRadiusSd}$\,km
(Sec.~\ref{sec:nested}) is conditional on a low-density prior much wider than
a realistic crust model.  Neither anchor is therefore suitable for a precision
radius measurement.  Such an analysis requires a low-density sector calibrated
to nuclear theory and crust physics, with its prior width reported alongside
the inferred radius.

Figure~\ref{fig:prior_anatomy}(a--c) shows individual realizations in
$p(\epsilon)$, $\cs^2(n)$, and mass--radius space, coloured by
$R(1.4\,\msun)$.  The sound-speed curves span most of the causal interval
and contain substantially more structure than the observable map recovers.

\section{Resolution weighting and implementation}
\label{app:resolution-weighting}
Resolution weights encode observational priorities and physical judgments
about which aspects of an EOS matter.  Their choice should be guided by
measurement capabilities, the mass range of available sources, and domain
knowledge of the dense-matter question being asked.  These priorities can
change: improved radius measurements, more informative tidal data, or sources
in a different mass range can favor a different comparison.  A weight suited
to characterizing an entire model family need not be the weight best suited
to a particular observing program.

In the external GP comparison of Fig.~\ref{fig:external_gp_dimension}, we
use uniform mass weight over $1$--$1.8\,\msun$ and constant channel scales.
Each scale is the square root of the training ensemble's mean squared
departure from its mean curve, averaged over the mass interval.  Thus radius
and $\ln\Lambda$ receive equal weight relative to their available variation;
this characterizes the family rather than specifying NICER or GW error bars.
We approximate the mass average by \result{ExtGpMassPoints} equally weighted reference masses.
For sound speed, the averages use \result{ExtGpDensityPoints} points uniform in logarithmic density
or normalized central enthalpy, over the domains stated in the figure.
These choices specify which features and which parts of the curves enter the
reported dimension comparison.

More generally, quadrature replaces the mass integral by weights $q_j$
with $\sum_jq_j=1$.  The joint PCA implements this quadratic form by
multiplying each channel by $\sqrt{q_j}/s_a(M_j)$ before decomposing the
ensemble; a common factor $\varrho$ changes eigenvalues but not the patterns.  Its Gaussian interpretation assigns joint reference
constraints on $R(M_j)$ and $\ln\Lambda(M_j)$ with precisions
$\varrho^2q_j/s_R^2(M_j)$ and
$\varrho^2q_j/s_{\ln\Lambda}^2(M_j)$, respectively.
Equivalently, their standard deviations are
$s_a(M_j)/(\varrho\sqrt{q_j})$ for either channel $a$.
The quadrature therefore refines a fixed resolution test; each additional
point is not an additional measurement at unchanged accuracy.  Correlated
radius--tidal errors can instead be included through their joint Gaussian
precision.  The resolution weights specify this comparison; the EOS prior
$\pi_0$ remains the prior used in the inference.

The production compressed-sequence representation uses a different weighting:
$M(u)$, $R(u)$ and $\ln\Lambda(u)$ are each fitted separately with uniform
weight in the normalized enthalpy coordinate of Eq.~\eqref{eq:ucoord}.
We realize these integrals on \result{SeqPoints} equally spaced points and retain \result{SeqComponents}
patterns per channel.  Multiplying a channel by a constant reference scale
does not change its separate PCA basis.  Fixing its pattern count separately,
however, differs from sharing one pattern budget among jointly weighted
radius and tidal curves.  Uniform enthalpy weighting follows sequence geometry
and implies an EOS-dependent mass weighting; it does not represent the
observed mass distribution.  The subsequent inference uses the source
likelihoods to impose the actual observational priorities, while the
reconstruction tests assess whether the retained basis is adequate.

\clearpage
\section{Effective dimension from likelihood-weighted draws}
\label{app:ess-dimension}
A Gaussian resolution likelihood permits a direct dimension estimate using
the physical prior, without fitting a prior density in PCA coordinates.
Let $\mathcal L_\sigma(\theta)=\exp[-Q(\theta)/(2\sigma^2)]$, with
$Q$ the sum of squared residuals and $\sigma$ their common measurement
accuracy.  The residuals and $\sigma$ have matching units.  Define
$Z(\sigma)=\langle\mathcal L_\sigma\rangle_{\pi_0}$, taking the likelihood
maximum to be unity.  For $N$ independent prior draws,
\begin{equation}
 \eta(\sigma)\equiv\lim_{N\to\infty}\frac{N_{\rm eff}}{N}
 =\frac{\langle\mathcal L_\sigma\rangle_{\pi_0}^2}
        {\langle\mathcal L_\sigma^2\rangle_{\pi_0}}
 =\frac{Z(\sigma)^2}{Z(\sigma/\sqrt{2})}.
 \label{eq:ess-volume}
\end{equation}
This effective prior-volume fraction scales as $\sigma^d$ when the
likelihood resolves $d$ approximately quadratic directions within the prior,
just as the contour volume in Eq.~\eqref{eq:volume_dimension}.
Its logarithmic slope therefore supplies a sampling-based dimension:
\begin{align}
 d_{\rm ESS}(\sigma)&=\frac{d\ln\eta}{d\ln\sigma}\nonumber\\
 &=\frac{2}{\sigma^2}\left[
 \langle Q\rangle_\sigma-\langle Q\rangle_{\sigma/\sqrt{2}}\right].
 \label{eq:ess-dimension}
\end{align}
Here each mean uses $\pi_0\mathcal L$ at the indicated accuracy.
The second expression follows by differentiating $Z$; it evaluates the
slope without finite differences or an absolute evidence normalization.
Gaussianity is required in the residuals, not in the EOS parameters.

Empirically, draw $N$ EOS from the prior, weight them by the resolution
likelihood at several measurement accuracies, and compute $N_{\rm eff}/N$.
The logarithmic slope of this fraction with respect to the resolution size
estimates the effective dimension.  At coarse resolution, only the largest
physical variations are distinguishable.  Improving the resolution reveals
further independent variations, so the effective dimension rises toward the
underlying dimension of the identifiable model family.
A fixed sample eventually runs out of draws in the shrinking supported region.
Its ESS then approaches the one-draw floor: the estimated dimension
artificially turns over and decreases, instead of following the physical rise.
This turnover signals exhausted sampling, not a reduction in physical
complexity.  Increasing the prior sample or using adaptive sampling or MCMC
is then necessary to continue the dimension estimate at finer resolution.

\section{A nonlinear transition-EOS sampling benchmark}
\label{app:transition-stress}
\begin{figure*}[t]\centering
 \includegraphics[width=\textwidth]{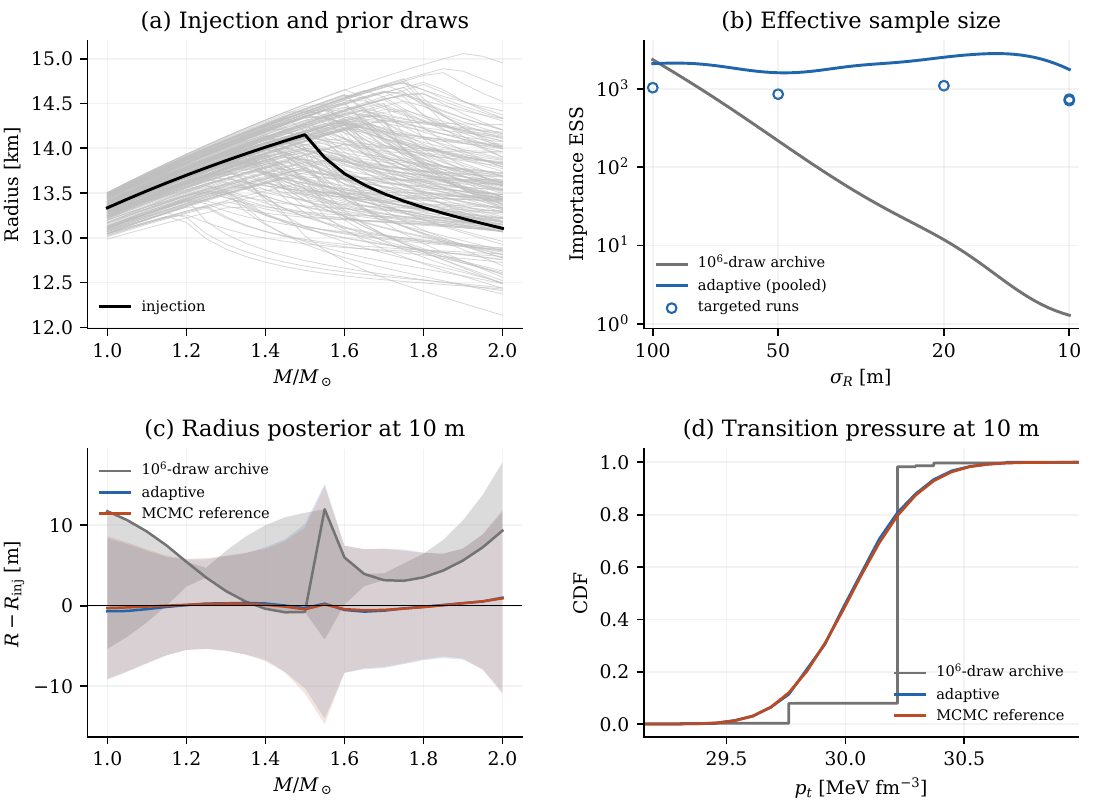}
 \caption{\label{fig:mr_archive_vs_adaptive}
 A $\result{BenchArchiveDraws}$-draw prior archive versus refinement for the same nonlinear
 transition-EOS likelihood. (a)~Injected mass--radius curve (black) and prior
 draws (grey). (b)~Importance ESS across the radius-precision controls; adaptive
 points come from separately targeted runs, with \result{BenchRunsSmallestSigma} seeds at $\result{BenchSigmaFour}$ m.
 Each uses $\result{BenchTrainDraws}$ training and $\result{BenchProdDraws}$ independent production proposals.
 (c)~Posterior medians and pointwise $90\%$ radius intervals at $\result{BenchSigmaFour}$ m,
 relative to the injection. (d)~Transition-pressure posterior CDFs at $\result{BenchSigmaFour}$ m.
 The archive has ESS $\result{BenchArchiveEssFour}$, \result{BenchHpdArchiveDraws} draw in the reference $\result{BenchHpdLevelPct}\%$ HPD region,
 and $\result{BenchArchiveMaxWtFourShort}\%$ of its weight on one EOS; refinement agrees with independent
 MCMC. Radii are obtained from TOV solutions at the $\result{BenchNumMasses}$ reference masses.\llmfig}
\end{figure*}

Figure~\ref{fig:mr_archive_vs_adaptive} compares a $\result{BenchArchiveDraws}$-draw prior archive
with adaptive refinement under the same physical EOS model and explicit
mass--radius likelihood.  Both calculations include the transition and use the
same connected stellar branch.  The radius precision is varied
from a well-sampled control to an intentionally optimistic stress test.

\subsection{EOS family, prior, and likelihood}
In units $c=1$, we anchor the EOS at
$(p_L,\epsilon_L)=(\result{BenchCrustP},\result{BenchCrustEps})$ MeV\,fm$^{-3}$ and write
\begin{equation}
 \epsilon(p)=\epsilon_L+\int_{p_L}^{p}\frac{dp'}{s(p')}
                 +\Delta\epsilon H(p-p_t),\quad 0<s=c_s^2<1.
\label{eq:benchmarkEOS}
\end{equation}
Below the anchor, $p=K\rho^2$ and $\epsilon=\rho+p$, with
$K=\result{BenchCrustP}/\result{BenchCrustK}^2$ in these units.  This illustrative crust joins continuously to the
anchor.
The fixed pressure edges are $\result{BenchPressureEdges}$
MeV\,fm$^{-3}$.  Let $\ell(s)=\ln[s/(1-s)]$ and
$\ell^{-1}(x)=(1+e^{-x})^{-1}$.  In interval $j$, the \result{BenchIntervals} fiducial
hadronic sound speeds squared are
\begin{equation}
 s_j^{H}=\result{BenchSoundSpeeds}_j.
\end{equation}
\result{BenchIntervals} coordinates $\theta_j$ set
$s(p)=\ell^{-1}[\ell(s_j^H)+\result{BenchLogitShift}\theta_j]$ below $p_t$ and
$s(p)=\ell^{-1}[\ell(s_Q)+\result{BenchLogitShift}\theta_j]$ above it, splitting an interval at
$p_t$ when necessary.  The remaining coordinates set
\begin{align}
 p_t&=\result{BenchPtScale}e^{\result{BenchPtRate}\theta_t}\ {\rm MeV\,fm^{-3}},\\
 \Delta\epsilon&=\result{BenchDeltaScale}e^{\result{BenchDeltaRate}\theta_\Delta}\ {\rm MeV\,fm^{-3}},\\
 s_Q&=\ell^{-1}[\ell(\result{BenchQuarkBase})+\result{BenchQuarkRate}\theta_Q].
\end{align}
The last sound-speed segment continues beyond the final pressure edge.
The cold first law determines baryon density, with chemical-potential
continuity requiring $n_+/n_-=(\epsilon_++p_t)/(\epsilon_-+p_t)$ at the jump.
These transition controls can generate connected or disconnected hybrid
branches~\cite{AlfordHanPrakash2013}.

For this benchmark the scientific prior is uniform on
$\theta\in[-1,1]^{\result{BenchDim}}$, conditioned on a subcritical onset
$\Delta\epsilon<(\epsilon_-+3p_t)/2$ and a connected stable first branch
supporting $1$--$2\,M_\odot$.  Write this fixed domain as $A_B$ and its
uniform-cube probability as $C_B$, so
$\pi_B(\theta)=2^{-\result{BenchDim}}{\bf1}_{[-1,1]^{\result{BenchDim}}}{\bf1}_{A_B}/C_B$.
This explicit branch and mass restriction differs from the production bridge
prior, which has no hard maximum-mass cut (Supplemental Material, Sec.~{SIII\,C}).
It is identical for the archive, refinement, and independent reference.

Synthetic radii are generated at $\theta=0$ for exactly known masses
$M_i=(\result{BenchMassLo}+\result{BenchMassStep}i)M_\odot$, $i=0,\ldots,\result{BenchMassLast}$, with zero noise.  The likelihood is
\begin{equation}
 \ln\mathcal L_{\sigma_R}(\theta)
 =-\frac12\sum_{i=0}^{\result{BenchMassLast}}
 \left[\frac{R_{\rm TOV}(M_i;\theta)-R_{\rm TOV}(M_i;0)}{\sigma_R}\right]^2,
\label{eq:benchmarkLikelihood}
\end{equation}
where $\sigma_R=\result{BenchSigmaList}$ m.
Varying $\sigma_R$
controls the resolution at which the same model family is measured.

\subsection{Archive, refinement, and independent checks}
The archive contains $\result{BenchArchiveDraws}$ independent draws from the conditioned prior
$\pi_B$.  Adaptive runs start from uniform prior candidates and update Gaussian
means and covariances using likelihood-weighted moments.  After training, the
proposal
\begin{equation}
 q(\theta)=\result{BenchDefensiveFrac}\,2^{-\result{BenchDim}}{\bf1}_{[-1,1]^{\result{BenchDim}}}(\theta)
                  +\result{BenchGaussFrac}\,\mathcal N(\theta;\mu,\Sigma)
\label{eq:benchmarkProposal}
\end{equation}
is frozen.  Final estimates use independent production draws with weights
$\pi_B\mathcal L/q$; candidates outside the cube or the accepted EOS family
have zero weight.  Each radius precision has a separately targeted run.

\result{BenchMcmcChains} independent Metropolis chains of \result{BenchMcmcDraws} draws provide a posterior
reference, with split $\hat R\le\result{BenchMcmcMaxRhat}$ for all coordinates and radii and
effective sample counts of at least \result{BenchMcmcMinEss}.  Independent adaptive runs agree with this reference.  Fixed-mass
radius solutions agree with high-accuracy SciPy roots to within $\result{BenchScipyMaxDiffM}$ m.

\begin{table}[t]
 \caption{\label{tab:benchmarkESS}Sampling at fixed prior and radius likelihood.
 Adaptive ESS uses independent production draws; each run includes
 $\result{BenchRunDraws}$ total candidates.}
 \begin{ruledtabular}
 \begin{tabular}{rrrr}
 $\sigma_R$ [m] & Archive ESS & Largest weight & Adaptive ESS\\
 \hline
 $\result{BenchSigmaOne}$ & $\result{BenchArchiveEssOne}$ & $\result{BenchArchiveMaxWtOne}\%$ & $\result{BenchAdaptiveEssOne}$\\
 $\result{BenchSigmaTwo}$ & $\result{BenchArchiveEssTwo}$ & $\result{BenchArchiveMaxWtTwo}\%$ & $\result{BenchAdaptiveEssTwo}$\\
 $\result{BenchSigmaThree}$ & $\result{BenchArchiveEssThree}$ & $\result{BenchArchiveMaxWtThree}\%$ & $\result{BenchAdaptiveEssThree}$\\
 $\result{BenchSigmaFour}$ & $\result{BenchArchiveEssFour}$ & $\result{BenchArchiveMaxWtFour}\%$ & $\result{BenchAdaptiveEssFour}$\\
 \end{tabular}
 \end{ruledtabular}
\end{table}

At $\result{BenchSigmaOne}$ m the archive and adaptive estimates agree
(Table~\ref{tab:benchmarkESS}).  At $\result{BenchSigmaFour}$ m the archive supplies only one draw
in the reference $\result{BenchHpdLevelPct}\%$ highest-posterior-density region, which occupies
approximately $\result{BenchHpdVolume}$ of the prior volume.  Its radius posterior and
transition-pressure distribution differ visibly from the independent reference
(Fig.~\ref{fig:mr_archive_vs_adaptive}), and its likelihood normalization is underestimated
by a factor of about $\result{BenchNormRatio}$.  Adaptive refinement resolves both distributions.

\subsection{Effective dimension at the measurement resolution}
\label{app:benchmark-dimension}
The Gaussian radius likelihood of Eq.~\eqref{eq:benchmarkLikelihood} lets us
measure the effective dimension that controls prior-sampling cost, using the
physical prior.
We apply the logarithmic slope of the prior-sampling ESS fraction derived in
Appendix~\ref{app:ess-dimension}, with
$Q(\theta)=\sum_i[R_{\rm TOV}(M_i;\theta)-R_{\rm TOV}(M_i;0)]^2$
and $\sigma=\sigma_R$.

\begin{figure}[t]\centering
 \includegraphics[width=\columnwidth]{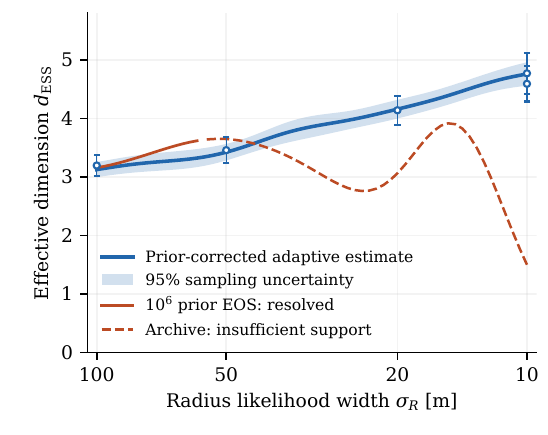}
 \caption{\label{fig:benchmark-dimension}
 Effective dimension versus radius likelihood width
 [Eq.~\eqref{eq:ess-dimension}], retaining the physical
 transition-family prior.  The adaptive curve combines independent frozen-proposal production samples;
 its band shows $95\%$ Monte Carlo intervals.  Circles show separately targeted runs with the same intervals.
 The $\result{BenchArchiveDraws}$-draw archive slope is dashed wherever either likelihood-weighted
 ESS entering the derivative falls below $\result{BenchReliableEss}$.  Its small-$\sigma_R$
 downturn reflects sample exhaustion.  All estimates reuse the nonlinear
 TOV radii of the benchmark.\llmfig}
\end{figure}

Figure~\ref{fig:benchmark-dimension} evaluates this slope by pooling the \result{BenchRuns} independent
frozen-proposal production batches, with the physical prior retained through
balance-mixture importance weights.  The dimension increases from $\result{BenchDimOne}\pm\result{BenchDimOneErr}$ at $\result{BenchSigmaOne}$ m to
$\result{BenchDimFour}\pm\result{BenchDimFourErr}$ at $\result{BenchSigmaFour}$ m; the \result{BenchRunsSmallestSigma} independent $\result{BenchSigmaFour}$ m runs agree with
this pooled estimate.  Independent MCMC at $\result{BenchSigmaFour}$ m gives $\result{BenchMcmcDim}\pm\result{BenchMcmcDimErr}$;
both estimates support the growth in dimension.  The quoted errors describe
Monte Carlo precision.
The direct $\result{BenchArchiveDraws}$-draw archive estimate becomes unreliable as its ESS
approaches unity.  Its eventual flattening is sample exhaustion, not a
physical reduction in dimension.  The prior-corrected adaptive estimates
continue to resolve the likelihood-weighted residuals at both precisions
entering Eq.~\eqref{eq:ess-dimension}.

The model has \result{BenchDim} adjustable coordinates.  Just ten reference values
along each coordinate would already give $10^{\result{BenchDim}}$ combinations, illustrating
the impractical cost of dense coverage of the full family.
As measurement precision improves, additional independent EOS combinations
become distinguishable and the effective dimension grows.  The associated
contraction of the supported prior volume explains why uniform prior draws
become increasingly inadequate; adaptive refinement instead generates draws
in that contracting region.

\subsection{Two independently moving phase transitions}
\label{app:two-transitions}
\begin{figure*}[t]\centering
 \includegraphics[width=\textwidth]{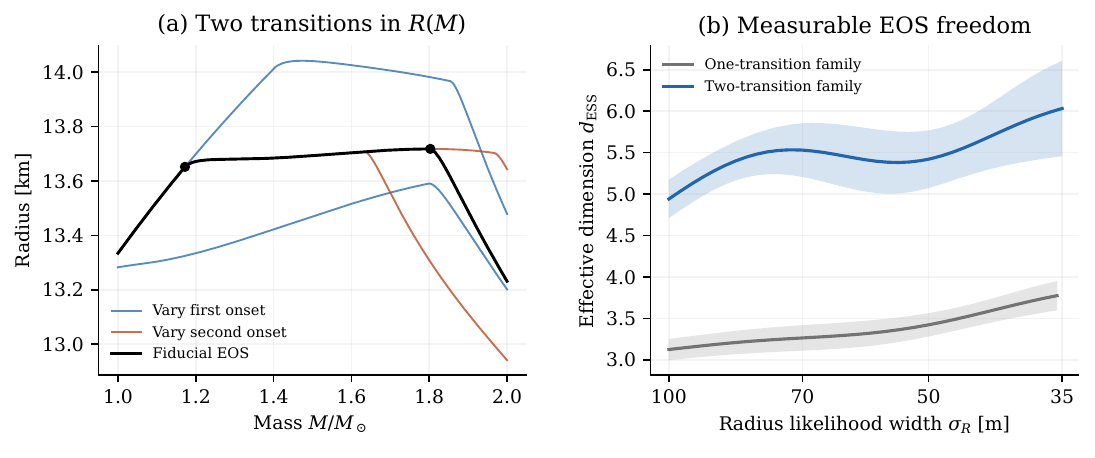}
 \caption{\label{fig:two-transition-dimension}
 Two accessible phase transitions produce more measurable EOS structure.
 (a)~Exact TOV radius curves for the fiducial EOS (black) and independent
 shifts of either transition onset (colors, $\theta_8=\pm0.7$ or
 $\theta_{11}=\pm0.7$); dots mark the fiducial onset masses.
 (b)~The ESS-volume dimension at the same radius precision for the two
 illustrative families; bands show $95\%$ Monte Carlo intervals.
 The likelihood uses the same $\result{BenchNumMasses}$ exactly known masses in both cases.
 The two families have distinct physical priors and fiducial EOS.\llmfig}
\end{figure*}

Additional transition features can increase the accessible dimension already
at less demanding radius precision.  Figure~\ref{fig:two-transition-dimension}
shows a second family with two first-order jumps,
\begin{equation}
 \epsilon(p)=\epsilon_L+\int_{p_L}^{p}\frac{dp'}{s(p')}
       +\sum_{a=1}^{2}\Delta\epsilon_a H(p-p_a).
 \label{eq:two-transition-eos}
\end{equation}
We retain the crust, anchor, pressure intervals and \result{BenchIntervals} interval-wise
sound-speed shifts of Appendix~\ref{app:transition-stress}, labeled
$\theta_0,\ldots,\theta_7$, but set
\begin{align}
 p_1&=\result{TwoTransPOne}e^{\result{TwoTransPOneRate}\theta_8}, &
 \Delta\epsilon_1&=\result{TwoTransDeltaOne}e^{\result{TwoTransDeltaOneRate}\theta_9},\nonumber\\
 p_2&=\result{TwoTransPTwo}e^{\result{TwoTransPTwoRate}\theta_{11}}, &
 \Delta\epsilon_2&=\result{TwoTransDeltaTwo}e^{\result{TwoTransDeltaTwoRate}\theta_{12}},
 \label{eq:two-transition-controls}
\end{align}
with pressures and energy densities in MeV\,fm$^{-3}$.
Between the jumps the base sound speed is
$\ell^{-1}(\theta_{10})$; above the second it is
$\ell^{-1}[\ell(\result{TwoTransQuarkTwo})+\theta_{13}]$.
Each interval still adds $\result{BenchLogitShift}\theta_j$ to the logit of its base sound speed.
The fourteen coordinates have uniform priors on $[-1,1]$, conditioned on
subcritical jumps and a connected stable first branch supporting
$1$--$2\,M_\odot$.  The fiducial transition onsets occur at $\result{TwoTransOnsetOne}$ and
$1.80\,M_\odot$, so moving either onset changes a different part of the
measured mass--radius curve.  This is a separate illustrative family, with a
different fiducial EOS and prior from the one-transition benchmark.

We use the same $\result{BenchNumMasses}$ masses and Gaussian radius likelihood, with independently
trained proposals and production draws.  The effective dimension is
$\result{TwoTransDimOne}\pm\result{TwoTransDimOneErr}$ at $\result{TwoTransSigmaOne}$ m and $\result{TwoTransDimTwo}\pm\result{TwoTransDimTwoErr}$ at $\result{TwoTransSigmaTwo}$ m, compared with
$\result{BenchDimOne}\pm\result{BenchDimOneErr}$ and $\result{BenchDimTwo}\pm\result{BenchDimTwoErr}$ for the first family.  At $\result{TwoTransSigmaMin}$ m it reaches
$\result{TwoTransDimMin}\pm\result{TwoTransDimMinErr}$.  These estimates use balance-mixture weights to retain the
physical prior.

The limiting prior-draw efficiency $\eta$ falls about $\result{TwoTransEtaDrop}$-fold between
$\result{TwoTransSigmaOne}$ and $\result{TwoTransSigmaTwo}$ m in this family, compared with about $\result{BenchEtaDrop}$-fold in the first.
At $\result{TwoTransSigmaMin}$ m, $\eta\simeq\result{TwoTransEtaMin}$: a fixed archive must grow rapidly
even to maintain a modest effective sample count.  This is the consequence
of additional \emph{measurable} EOS freedom, quantified through actual stellar
predictions and the physical prior, rather than inferred from the number
of model coordinates.

\subsection{When a fixed archive is sufficient}\label{app:archive-scaling}
For smooth configurations and modest data size, such as the  analysis in Sec.~\ref{sec:results},
a large catalog would be sufficient: even with only $O(10^3)$ realizations we found several draws matched the data well,
so scaling the final brute-force effective-sample fraction of Sec.~\ref{sec:hero} to
$\result{ArchiveScaleDraws}$ prior draws would provide about $\result{MillionArchiveEss}$
effective draws.

Our constructive example demonstrates that large-draw configurations break down when the effective dimension is large.

\section{Use of large language models}
\label{app:llm}

We used large language model (LLM) agents throughout this work, following the
APS policy on the appropriate use of AI tools. We directed these agents in a
role qualitatively equivalent to that of research staff. They carried out
assigned calculations, returned preliminary reports and findings, and drafted
initial proposals for material to incorporate into the text, including first
drafts of many sections. LLM agents operated the analysis and plotting tools
that produced every figure in this paper, and generated those figures with
those tools. The authors revised all drafted text and decided what entered the
paper. All calculations were carried out under exhaustive human direction.
Both the calculations and the code developments behind them were subjected to
adversarial review and to repeated cross-validation tests. The workflow, its
documentation conventions, and its review procedures are described in
Ref.~\cite{whitepaper-transparent-agentic-science}. The authors take full
responsibility for the content of this paper.


\begin{thebibliography}{110}\makeatletter
\providecommand \@ifxundefined [1]{\@ifx{#1\undefined}
}\providecommand \@ifnum [1]{\ifnum #1\expandafter \@firstoftwo
 \else \expandafter \@secondoftwo
 \fi
}\providecommand \@ifx [1]{\ifx #1\expandafter \@firstoftwo
 \else \expandafter \@secondoftwo
 \fi
}\providecommand \natexlab [1]{#1}\providecommand \enquote  [1]{``#1''}\providecommand \bibnamefont  [1]{#1}\providecommand \bibfnamefont [1]{#1}\providecommand \citenamefont [1]{#1}\providecommand \href@noop [0]{\@secondoftwo}\providecommand \href [0]{\begingroup \@sanitize@url \@href}\providecommand \@href[1]{\@@startlink{#1}\@@href}\providecommand \@@href[1]{\endgroup#1\@@endlink}\providecommand \@sanitize@url [0]{\catcode `\\12\catcode `\$12\catcode
  `\&12\catcode `\#12\catcode `\^12\catcode `\_12\catcode `\%12\relax}\providecommand \@@startlink[1]{}\providecommand \@@endlink[0]{}\providecommand \url  [0]{\begingroup\@sanitize@url \@url }\providecommand \@url [1]{\endgroup\@href {#1}{\urlprefix }}\providecommand \urlprefix  [0]{URL }\providecommand \Eprint [0]{\href }\providecommand \doibase [0]{https://doi.org/}\providecommand \selectlanguage [0]{\@gobble}\providecommand \bibinfo  [0]{\@secondoftwo}\providecommand \bibfield  [0]{\@secondoftwo}\providecommand \translation [1]{[#1]}\providecommand \BibitemOpen [0]{}\providecommand \bibitemStop [0]{}\providecommand \bibitemNoStop [0]{.\EOS\space}\providecommand \EOS [0]{\spacefactor3000\relax}\providecommand \BibitemShut  [1]{\csname bibitem#1\endcsname}\let\auto@bib@innerbib\@empty
\bibitem [{\citenamefont {{Abbott}}\ \emph {et~al.}(2017)\citenamefont
  {{Abbott}} \emph {et~al.}}]{2017PhRvL.119p1101A}\BibitemOpen
  \bibfield  {author} {\bibinfo {author} {\bibfnamefont {B.~P.}\ \bibnamefont
  {{Abbott}}} \emph {et~al.} (\bibinfo {collaboration} {LIGO Scientific
  Collaboration and Virgo Collaboration}),\ }\bibfield  {title} {\bibinfo
  {title} {{GW170817: Observation of Gravitational Waves from a Binary Neutron
  Star Inspiral}},\ }\href {https://doi.org/10.1103/PhysRevLett.119.161101}
  {\bibfield  {journal} {\bibinfo  {journal} {Physical Review Letters}\
  }\textbf {\bibinfo {volume} {119}},\ \bibinfo {eid} {161101} (\bibinfo {year}
  {2017})},\ \Eprint {https://arxiv.org/abs/1710.05832} {arXiv:1710.05832
  [gr-qc]} \BibitemShut {NoStop}\bibitem [{\citenamefont {{Abbott}}\ \emph {et~al.}(2019)\citenamefont
  {{Abbott}} \emph {et~al.}}]{LIGO-GW170817-SourceProperties}\BibitemOpen
  \bibfield  {author} {\bibinfo {author} {\bibfnamefont {B.~P.}\ \bibnamefont
  {{Abbott}}} \emph {et~al.} (\bibinfo {collaboration} {LIGO Scientific
  Collaboration and Virgo Collaboration}),\ }\bibfield  {title} {\bibinfo
  {title} {{Properties of the Binary Neutron Star Merger GW170817}},\ }\href
  {https://doi.org/10.1103/PhysRevX.9.011001} {\bibfield  {journal} {\bibinfo
  {journal} {Physical Review X}\ }\textbf {\bibinfo {volume} {9}},\ \bibinfo
  {eid} {011001} (\bibinfo {year} {2019})},\ \Eprint
  {https://arxiv.org/abs/1805.11579} {arXiv:1805.11579 [gr-qc]} \BibitemShut
  {NoStop}\bibitem [{\citenamefont {{Abbott}}\ \emph {et~al.}(2018)\citenamefont
  {{Abbott}} \emph {et~al.}}]{LIGO-GW170817-EOS}\BibitemOpen
  \bibfield  {author} {\bibinfo {author} {\bibfnamefont {B.~P.}\ \bibnamefont
  {{Abbott}}} \emph {et~al.} (\bibinfo {collaboration} {LIGO Scientific
  Collaboration and Virgo Collaboration}),\ }\bibfield  {title} {\bibinfo
  {title} {{GW170817: Measurements of neutron star radii and equation of
  state}},\ }\href {https://doi.org/10.1103/PhysRevLett.121.161101} {\bibfield
  {journal} {\bibinfo  {journal} {\prl}\ }\textbf {\bibinfo {volume} {121}},\
  \bibinfo {pages} {161101} (\bibinfo {year} {2018})},\ \Eprint
  {https://arxiv.org/abs/1805.11581} {arXiv:1805.11581 [gr-qc]} \BibitemShut
  {NoStop}\bibitem [{\citenamefont {{Abbott}}\ \emph {et~al.}(2020)\citenamefont
  {{Abbott}} \emph {et~al.}}]{2020ApJ...892L...3A}\BibitemOpen
  \bibfield  {author} {\bibinfo {author} {\bibfnamefont {B.~P.}\ \bibnamefont
  {{Abbott}}} \emph {et~al.} (\bibinfo {collaboration} {LIGO Scientific
  Collaboration and Virgo Collaboration}),\ }\bibfield  {title} {\bibinfo
  {title} {{GW190425: Observation of a Compact Binary Coalescence with Total
  Mass {\ensuremath{\sim}} 3.4 M$_{\odot}$}},\ }\href
  {https://doi.org/10.3847/2041-8213/ab75f5} {\bibfield  {journal} {\bibinfo
  {journal} {\apjl}\ }\textbf {\bibinfo {volume} {892}},\ \bibinfo {eid} {L3}
  (\bibinfo {year} {2020})},\ \Eprint {https://arxiv.org/abs/2001.01761}
  {arXiv:2001.01761 [astro-ph.HE]} \BibitemShut {NoStop}\bibitem [{\citenamefont {{Abac}}\ \emph {et~al.}(2026)\citenamefont {{Abac}}
  \emph {et~al.}}]{2026arXiv260527225T}\BibitemOpen
  \bibfield  {author} {\bibinfo {author} {\bibfnamefont {A.~G.}\ \bibnamefont
  {{Abac}}} \emph {et~al.} (\bibinfo {collaboration} {LIGO Scientific
  Collaboration, Virgo Collaboration, KAGRA Collaboration}),\ }\bibfield
  {title} {\bibinfo {title} {{GWTC-5.0: Observations from the Second Part of
  the Fourth LIGO-Virgo-KAGRA Observing Run and Updates to the
  Gravitational-Wave Transient Catalog}},\ }\href
  {https://doi.org/10.48550/arXiv.2605.27225} {\bibfield  {journal} {\bibinfo
  {journal} {arXiv e-prints}\ ,\ \bibinfo {pages} {arXiv:2605.27225}} (\bibinfo
  {year} {2026})},\ \Eprint {https://arxiv.org/abs/2605.27225}
  {arXiv:2605.27225 [gr-qc]} \BibitemShut {NoStop}\bibitem [{\citenamefont {Kini}\ \emph {et~al.}(2026)\citenamefont {Kini},
  \citenamefont {Mauviard}, \citenamefont {Salmi}, \citenamefont {Watts},
  \citenamefont {Guillot}, \citenamefont {Dorsman}, \citenamefont {Choudhury},
  \citenamefont {Gonz{\'a}lez-Caniulef}, \citenamefont {Hoogkamer},
  \citenamefont {Huppenkothen}, \citenamefont {Kazantsev}, \citenamefont
  {Kerr}, \citenamefont {Nissanke}, \citenamefont {Ray}, \citenamefont
  {Stammler},\ and\ \citenamefont {Vinciguerra}}]{J0030_2026}\BibitemOpen
  \bibfield  {author} {\bibinfo {author} {\bibfnamefont {Y.}~\bibnamefont
  {Kini}}, \bibinfo {author} {\bibfnamefont {L.}~\bibnamefont {Mauviard}},
  \bibinfo {author} {\bibfnamefont {T.}~\bibnamefont {Salmi}}, \bibinfo
  {author} {\bibfnamefont {A.~L.}\ \bibnamefont {Watts}}, \bibinfo {author}
  {\bibfnamefont {S.}~\bibnamefont {Guillot}}, \bibinfo {author} {\bibfnamefont
  {B.}~\bibnamefont {Dorsman}}, \bibinfo {author} {\bibfnamefont
  {D.}~\bibnamefont {Choudhury}}, \bibinfo {author} {\bibfnamefont
  {D.}~\bibnamefont {Gonz{\'a}lez-Caniulef}}, \bibinfo {author} {\bibfnamefont
  {M.}~\bibnamefont {Hoogkamer}}, \bibinfo {author} {\bibfnamefont
  {D.}~\bibnamefont {Huppenkothen}}, \bibinfo {author} {\bibfnamefont
  {C.}~\bibnamefont {Kazantsev}}, \bibinfo {author} {\bibfnamefont
  {M.}~\bibnamefont {Kerr}}, \bibinfo {author} {\bibfnamefont {S.}~\bibnamefont
  {Nissanke}}, \bibinfo {author} {\bibfnamefont {P.~S.}\ \bibnamefont {Ray}},
  \bibinfo {author} {\bibfnamefont {P.}~\bibnamefont {Stammler}},\ and\
  \bibinfo {author} {\bibfnamefont {S.}~\bibnamefont {Vinciguerra}},\
  }\bibfield  {title} {\bibinfo {title} {A {NICER} view of {PSR} {J0030+0451}:
  Updated constraints from 6 yr of {NICER} observations},\ }\href
  {https://doi.org/10.3847/1538-4357/ae733e} {\bibfield  {journal} {\bibinfo
  {journal} {\apj}\ }\textbf {\bibinfo {volume} {1005}},\ \bibinfo {pages}
  {201} (\bibinfo {year} {2026})},\ \Eprint {https://arxiv.org/abs/2602.23743}
  {arXiv:2602.23743} \BibitemShut {NoStop}\bibitem [{\citenamefont {Salmi}\ \emph {et~al.}(2024)\citenamefont {Salmi}
  \emph {et~al.}}]{Salmi2024}\BibitemOpen
  \bibfield  {author} {\bibinfo {author} {\bibfnamefont {T.}~\bibnamefont
  {Salmi}} \emph {et~al.},\ }\bibfield  {title} {\bibinfo {title} {The radius
  of the high-mass pulsar {PSR} {J0740+6620} with 3.6 yr of {NICER} data},\
  }\href {https://doi.org/10.3847/1538-4357/ad5f1f} {\bibfield  {journal}
  {\bibinfo  {journal} {\apj}\ }\textbf {\bibinfo {volume} {974}},\ \bibinfo
  {pages} {294} (\bibinfo {year} {2024})},\ \Eprint
  {https://arxiv.org/abs/2406.14466} {arXiv:2406.14466} \BibitemShut {NoStop}\bibitem [{\citenamefont {Dittmann}\ \emph {et~al.}(2024)\citenamefont
  {Dittmann} \emph {et~al.}}]{Dittmann2024}\BibitemOpen
  \bibfield  {author} {\bibinfo {author} {\bibfnamefont {A.~J.}\ \bibnamefont
  {Dittmann}} \emph {et~al.},\ }\bibfield  {title} {\bibinfo {title} {A more
  precise measurement of the radius of {PSR} {J0740+6620} using updated {NICER}
  data},\ }\href {https://doi.org/10.3847/1538-4357/ad5f1e} {\bibfield
  {journal} {\bibinfo  {journal} {\apj}\ }\textbf {\bibinfo {volume} {974}},\
  \bibinfo {pages} {295} (\bibinfo {year} {2024})},\ \Eprint
  {https://arxiv.org/abs/2406.14467} {arXiv:2406.14467} \BibitemShut {NoStop}\bibitem [{\citenamefont {Choudhury}\ \emph {et~al.}(2024)\citenamefont
  {Choudhury} \emph {et~al.}}]{Choudhury2024}\BibitemOpen
  \bibfield  {author} {\bibinfo {author} {\bibfnamefont {D.}~\bibnamefont
  {Choudhury}} \emph {et~al.},\ }\bibfield  {title} {\bibinfo {title} {A
  {NICER} view of the nearest and brightest millisecond pulsar: {PSR}
  {J0437-4715}},\ }\href {https://doi.org/10.3847/2041-8213/ad5a6f} {\bibfield
  {journal} {\bibinfo  {journal} {\apj}\ }\textbf {\bibinfo {volume} {971}},\
  \bibinfo {pages} {L20} (\bibinfo {year} {2024})},\ \Eprint
  {https://arxiv.org/abs/2407.06789} {arXiv:2407.06789} \BibitemShut {NoStop}\bibitem [{\citenamefont {Mauviard}\ \emph {et~al.}(2025)\citenamefont
  {Mauviard} \emph {et~al.}}]{Mauviard2025}\BibitemOpen
  \bibfield  {author} {\bibinfo {author} {\bibfnamefont {L.}~\bibnamefont
  {Mauviard}} \emph {et~al.},\ }\bibfield  {title} {\bibinfo {title} {A {NICER}
  view of the 1.4 {$M_\odot$} edge-on pulsar {PSR} {J0614-3329}},\ }\href
  {https://doi.org/10.3847/1538-4357/ae145d} {\bibfield  {journal} {\bibinfo
  {journal} {\apj}\ }\textbf {\bibinfo {volume} {995}},\ \bibinfo {pages} {60}
  (\bibinfo {year} {2025})},\ \Eprint {https://arxiv.org/abs/2506.14883}
  {arXiv:2506.14883} \BibitemShut {NoStop}\bibitem [{\citenamefont {Fonseca}\ \emph {et~al.}(2021)\citenamefont {Fonseca}
  \emph {et~al.}}]{Fonseca2021}\BibitemOpen
  \bibfield  {author} {\bibinfo {author} {\bibfnamefont {E.}~\bibnamefont
  {Fonseca}} \emph {et~al.},\ }\bibfield  {title} {\bibinfo {title} {Refined
  mass and geometric measurements of the high-mass {PSR} {J0740+6620}},\
  }\href@noop {} {\bibfield  {journal} {\bibinfo  {journal} {Astrophys. J.
  Lett.}\ }\textbf {\bibinfo {volume} {915}},\ \bibinfo {pages} {L12} (\bibinfo
  {year} {2021})}\BibitemShut {NoStop}\bibitem [{\citenamefont {{Arzoumanian}}\ \emph {et~al.}(2018)\citenamefont
  {{Arzoumanian}} \emph {et~al.}}]{Arzoumanian2018}\BibitemOpen
  \bibfield  {author} {\bibinfo {author} {\bibfnamefont {Z.}~\bibnamefont
  {{Arzoumanian}}} \emph {et~al.},\ }\bibfield  {title} {\bibinfo {title} {{The
  NANOGrav 11-year Data Set: High-precision Timing of 45 Millisecond
  Pulsars}},\ }\href {https://doi.org/10.3847/1538-4365/aab5b0} {\bibfield
  {journal} {\bibinfo  {journal} {\apjs}\ }\textbf {\bibinfo {volume} {235}},\
  \bibinfo {pages} {37} (\bibinfo {year} {2018})},\ \Eprint
  {https://arxiv.org/abs/1801.01837} {1801.01837} \BibitemShut {NoStop}\bibitem [{\citenamefont {{Lackey}}\ and\ \citenamefont
  {{Wade}}(2015)}]{2015PhRvD..91d3002L}\BibitemOpen
  \bibfield  {author} {\bibinfo {author} {\bibfnamefont {B.~D.}\ \bibnamefont
  {{Lackey}}}\ and\ \bibinfo {author} {\bibfnamefont {L.}~\bibnamefont
  {{Wade}}},\ }\bibfield  {title} {\bibinfo {title} {{Reconstructing the
  neutron-star equation of state with gravitational-wave detectors from a
  realistic population of inspiralling binary neutron stars}},\ }\href
  {https://doi.org/10.1103/PhysRevD.91.043002} {\bibfield  {journal} {\bibinfo
  {journal} {\prd}\ }\textbf {\bibinfo {volume} {91}},\ \bibinfo {eid} {043002}
  (\bibinfo {year} {2015})},\ \Eprint {https://arxiv.org/abs/1410.8866}
  {arXiv:1410.8866 [gr-qc]} \BibitemShut {NoStop}\bibitem [{\citenamefont {{Kedia}}\ \emph {et~al.}(2024)\citenamefont
  {{Kedia}}, \citenamefont {{O'Shaughnessy}}, \citenamefont {{Wade}},\ and\
  \citenamefont {{Yelikar}}}]{gwastro-ns-eos-Kedia2024}\BibitemOpen
  \bibfield  {author} {\bibinfo {author} {\bibfnamefont {A.}~\bibnamefont
  {{Kedia}}}, \bibinfo {author} {\bibfnamefont {R.}~\bibnamefont
  {{O'Shaughnessy}}}, \bibinfo {author} {\bibfnamefont {L.}~\bibnamefont
  {{Wade}}},\ and\ \bibinfo {author} {\bibfnamefont {A.}~\bibnamefont
  {{Yelikar}}},\ }\bibfield  {title} {\bibinfo {title} {{Astrophysical
  assumptions and equation of state framework have larger impact on equation of
  state inference than individual neutron star observations}},\ }\bibfield
  {journal} {\bibinfo  {journal} {arXiv e-prints}\ }\href
  {https://doi.org/10.48550/arXiv.2405.17326} {10.48550/arXiv.2405.17326}
  (\bibinfo {year} {2024}),\ \Eprint {https://arxiv.org/abs/2405.17326}
  {arXiv:2405.17326} \BibitemShut {NoStop}\bibitem [{\citenamefont {{Zhu}}\ and\ \citenamefont
  {{O'Shaughnessy}}(2026)}]{gwastro-nsnuc-Fisher3g-Zhu2025}\BibitemOpen
  \bibfield  {author} {\bibinfo {author} {\bibfnamefont {Z.}~\bibnamefont
  {{Zhu}}}\ and\ \bibinfo {author} {\bibfnamefont {R.}~\bibnamefont
  {{O'Shaughnessy}}},\ }\bibfield  {title} {\bibinfo {title} {{Limitations in
  constraining neutron star radii and nuclear properties from inspiral
  gravitational wave detections}},\ }\href {https://doi.org/10.1103/f166-khyg}
  {\bibfield  {journal} {\bibinfo  {journal} {\prd}\ }\textbf {\bibinfo
  {volume} {113}},\ \bibinfo {eid} {103037} (\bibinfo {year} {2026})},\ \Eprint
  {https://arxiv.org/abs/2508.11875} {arXiv:2508.11875 [astro-ph.HE]}
  \BibitemShut {NoStop}\bibitem [{\citenamefont {Evans}\ \emph {et~al.}(2021)\citenamefont {Evans}
  \emph {et~al.}}]{CosmicExplorer2021}\BibitemOpen
  \bibfield  {author} {\bibinfo {author} {\bibfnamefont {M.}~\bibnamefont
  {Evans}} \emph {et~al.},\ }\bibfield  {title} {\bibinfo {title} {A horizon
  study for {Cosmic Explorer}: Science, observatories, and community},\
  }\bibfield  {journal} {\bibinfo  {journal} {arXiv e-prints}\ }\href
  {https://doi.org/10.48550/arXiv.2109.09882} {10.48550/arXiv.2109.09882}
  (\bibinfo {year} {2021}),\ \Eprint {https://arxiv.org/abs/2109.09882}
  {arXiv:2109.09882} \BibitemShut {NoStop}\bibitem [{\citenamefont {Punturo}\ \emph {et~al.}(2010)\citenamefont {Punturo}
  \emph {et~al.}}]{EinsteinTelescope2010}\BibitemOpen
  \bibfield  {author} {\bibinfo {author} {\bibfnamefont {M.}~\bibnamefont
  {Punturo}} \emph {et~al.},\ }\bibfield  {title} {\bibinfo {title} {The
  {Einstein Telescope}: a third-generation gravitational wave observatory},\
  }\href {https://doi.org/10.1088/0264-9381/27/19/194002} {\bibfield  {journal}
  {\bibinfo  {journal} {\cqg}\ }\textbf {\bibinfo {volume} {27}},\ \bibinfo
  {pages} {194002} (\bibinfo {year} {2010})}\BibitemShut {NoStop}\bibitem [{\citenamefont {Maggiore}\ \emph {et~al.}(2020)\citenamefont
  {Maggiore} \emph {et~al.}}]{ETscience2020}\BibitemOpen
  \bibfield  {author} {\bibinfo {author} {\bibfnamefont {M.}~\bibnamefont
  {Maggiore}} \emph {et~al.},\ }\bibfield  {title} {\bibinfo {title} {Science
  case for the {Einstein} telescope},\ }\href
  {https://doi.org/10.1088/1475-7516/2020/03/050} {\bibfield  {journal}
  {\bibinfo  {journal} {\jcap}\ }\textbf {\bibinfo {volume} {2020}},\ \bibinfo
  {pages} {050} (\bibinfo {year} {2020})}\BibitemShut {NoStop}\bibitem [{\citenamefont {Walker}\ \emph {et~al.}(2024)\citenamefont {Walker},
  \citenamefont {Smith}, \citenamefont {Thrane},\ and\ \citenamefont
  {Reardon}}]{WalkerEtAl2024}\BibitemOpen
  \bibfield  {author} {\bibinfo {author} {\bibfnamefont {K.}~\bibnamefont
  {Walker}}, \bibinfo {author} {\bibfnamefont {R.}~\bibnamefont {Smith}},
  \bibinfo {author} {\bibfnamefont {E.}~\bibnamefont {Thrane}},\ and\ \bibinfo
  {author} {\bibfnamefont {D.~J.}\ \bibnamefont {Reardon}},\ }\bibfield
  {title} {\bibinfo {title} {Precision constraints on the neutron star equation
  of state with third-generation gravitational-wave observatories},\ }\href
  {https://doi.org/10.1103/PhysRevD.110.043013} {\bibfield  {journal} {\bibinfo
   {journal} {\prd}\ }\textbf {\bibinfo {volume} {110}},\ \bibinfo {pages}
  {043013} (\bibinfo {year} {2024})},\ \Eprint
  {https://arxiv.org/abs/2401.02604} {arXiv:2401.02604} \BibitemShut {NoStop}\bibitem [{\citenamefont {Bandopadhyay}\ \emph {et~al.}(2024)\citenamefont
  {Bandopadhyay}, \citenamefont {Kacanja}, \citenamefont {Somasundaram},
  \citenamefont {Nitz},\ and\ \citenamefont {Brown}}]{BandopadhyayEtAl2024}\BibitemOpen
  \bibfield  {author} {\bibinfo {author} {\bibfnamefont {A.}~\bibnamefont
  {Bandopadhyay}}, \bibinfo {author} {\bibfnamefont {K.}~\bibnamefont
  {Kacanja}}, \bibinfo {author} {\bibfnamefont {R.}~\bibnamefont
  {Somasundaram}}, \bibinfo {author} {\bibfnamefont {A.~H.}\ \bibnamefont
  {Nitz}},\ and\ \bibinfo {author} {\bibfnamefont {D.~A.}\ \bibnamefont
  {Brown}},\ }\bibfield  {title} {\bibinfo {title} {Measuring neutron star
  radius with second and third generation gravitational wave detector
  networks},\ }\href {https://doi.org/10.1088/1361-6382/ad828a} {\bibfield
  {journal} {\bibinfo  {journal} {\cqg}\ }\textbf {\bibinfo {volume} {41}},\
  \bibinfo {pages} {225003} (\bibinfo {year} {2024})},\ \Eprint
  {https://arxiv.org/abs/2402.05056} {arXiv:2402.05056} \BibitemShut {NoStop}\bibitem [{\citenamefont {Williams}\ \emph {et~al.}(2026)\citenamefont
  {Williams}, \citenamefont {Puecher}, \citenamefont {Grams}, \citenamefont
  {Flores},\ and\ \citenamefont {Dietrich}}]{WilliamsEtAl2026}\BibitemOpen
  \bibfield  {author} {\bibinfo {author} {\bibfnamefont {N.}~\bibnamefont
  {Williams}}, \bibinfo {author} {\bibfnamefont {A.}~\bibnamefont {Puecher}},
  \bibinfo {author} {\bibfnamefont {G.}~\bibnamefont {Grams}}, \bibinfo
  {author} {\bibfnamefont {C.~V.}\ \bibnamefont {Flores}},\ and\ \bibinfo
  {author} {\bibfnamefont {T.}~\bibnamefont {Dietrich}},\ }\bibfield  {title}
  {\bibinfo {title} {Investigating the impact of quasiuniversal relations on
  neutron star constraints in third-generation detectors},\ }\href
  {https://doi.org/10.1103/c862-2dvz} {\bibfield  {journal} {\bibinfo
  {journal} {Phys. Rev. D}\ }\textbf {\bibinfo {volume} {114}},\ \bibinfo
  {pages} {024063} (\bibinfo {year} {2026})},\ \Eprint
  {https://arxiv.org/abs/2602.14659} {arXiv:2602.14659} \BibitemShut {NoStop}\bibitem [{\citenamefont {{Adhikari}}\ \emph {et~al.}(2021)\citenamefont
  {{Adhikari}}, \citenamefont {others},\ and\ \citenamefont {{PREX
  Collaboration}}}]{PREX2021}\BibitemOpen
  \bibfield  {author} {\bibinfo {author} {\bibfnamefont {D.}~\bibnamefont
  {{Adhikari}}}, \bibinfo {author} {\bibnamefont {others}},\ and\ \bibinfo
  {author} {\bibnamefont {{PREX Collaboration}}},\ }\bibfield  {title}
  {\bibinfo {title} {Accurate determination of the neutron skin thickness of
  $^{208}$pb through parity-violation in electron scattering},\ }\href
  {https://doi.org/10.1103/PhysRevLett.126.172502} {\bibfield  {journal}
  {\bibinfo  {journal} {\prl}\ }\textbf {\bibinfo {volume} {126}},\ \bibinfo
  {pages} {172502} (\bibinfo {year} {2021})},\ \Eprint
  {https://arxiv.org/abs/2102.10767} {arXiv:2102.10767} \BibitemShut {NoStop}\bibitem [{\citenamefont {{Adhikari}}\ \emph {et~al.}(2022)\citenamefont
  {{Adhikari}}, \citenamefont {others},\ and\ \citenamefont {{CREX
  Collaboration}}}]{CREX2022}\BibitemOpen
  \bibfield  {author} {\bibinfo {author} {\bibfnamefont {D.}~\bibnamefont
  {{Adhikari}}}, \bibinfo {author} {\bibnamefont {others}},\ and\ \bibinfo
  {author} {\bibnamefont {{CREX Collaboration}}},\ }\bibfield  {title}
  {\bibinfo {title} {Precision determination of the neutral weak form factor of
  $^{48}$ca},\ }\href {https://doi.org/10.1103/PhysRevLett.129.042501}
  {\bibfield  {journal} {\bibinfo  {journal} {\prl}\ }\textbf {\bibinfo
  {volume} {129}},\ \bibinfo {pages} {042501} (\bibinfo {year}
  {2022})}\BibitemShut {NoStop}\bibitem [{\citenamefont {Reed}\ \emph {et~al.}(2021)\citenamefont {Reed},
  \citenamefont {Fattoyev}, \citenamefont {Horowitz},\ and\ \citenamefont
  {Piekarewicz}}]{ReedEtAl2021}\BibitemOpen
  \bibfield  {author} {\bibinfo {author} {\bibfnamefont {B.~T.}\ \bibnamefont
  {Reed}}, \bibinfo {author} {\bibfnamefont {F.~J.}\ \bibnamefont {Fattoyev}},
  \bibinfo {author} {\bibfnamefont {C.~J.}\ \bibnamefont {Horowitz}},\ and\
  \bibinfo {author} {\bibfnamefont {J.}~\bibnamefont {Piekarewicz}},\
  }\bibfield  {title} {\bibinfo {title} {Implications of {PREX-2} on the
  equation of state of neutron-rich matter},\ }\href
  {https://doi.org/10.1103/PhysRevLett.126.172503} {\bibfield  {journal}
  {\bibinfo  {journal} {\prl}\ }\textbf {\bibinfo {volume} {126}},\ \bibinfo
  {pages} {172503} (\bibinfo {year} {2021})}\BibitemShut {NoStop}\bibitem [{\citenamefont {Huth}\ \emph {et~al.}(2022)\citenamefont {Huth},
  \citenamefont {Pang}, \citenamefont {Tews}, \citenamefont {Dietrich},
  \citenamefont {Le~F{\`e}vre}, \citenamefont {Schwenk}, \citenamefont
  {Trautmann}, \citenamefont {Agarwal}, \citenamefont {Bulla}, \citenamefont
  {Coughlin},\ and\ \citenamefont {Van Den~Broeck}}]{HuthEtAl2022}\BibitemOpen
  \bibfield  {author} {\bibinfo {author} {\bibfnamefont {S.}~\bibnamefont
  {Huth}}, \bibinfo {author} {\bibfnamefont {P.~T.~H.}\ \bibnamefont {Pang}},
  \bibinfo {author} {\bibfnamefont {I.}~\bibnamefont {Tews}}, \bibinfo {author}
  {\bibfnamefont {T.}~\bibnamefont {Dietrich}}, \bibinfo {author}
  {\bibfnamefont {A.}~\bibnamefont {Le~F{\`e}vre}}, \bibinfo {author}
  {\bibfnamefont {A.}~\bibnamefont {Schwenk}}, \bibinfo {author} {\bibfnamefont
  {W.}~\bibnamefont {Trautmann}}, \bibinfo {author} {\bibfnamefont
  {K.}~\bibnamefont {Agarwal}}, \bibinfo {author} {\bibfnamefont
  {M.}~\bibnamefont {Bulla}}, \bibinfo {author} {\bibfnamefont {M.~W.}\
  \bibnamefont {Coughlin}},\ and\ \bibinfo {author} {\bibfnamefont
  {C.}~\bibnamefont {Van Den~Broeck}},\ }\bibfield  {title} {\bibinfo {title}
  {Constraining neutron-star matter with microscopic and macroscopic
  collisions},\ }\href {https://doi.org/10.1038/s41586-022-04750-w} {\bibfield
  {journal} {\bibinfo  {journal} {\nat}\ }\textbf {\bibinfo {volume} {606}},\
  \bibinfo {pages} {276} (\bibinfo {year} {2022})}\BibitemShut {NoStop}\bibitem [{\citenamefont {Tsang}\ \emph {et~al.}(2024)\citenamefont {Tsang},
  \citenamefont {Tsang}, \citenamefont {Lynch}, \citenamefont {Kumar},\ and\
  \citenamefont {Horowitz}}]{TsangEtAl2024}\BibitemOpen
  \bibfield  {author} {\bibinfo {author} {\bibfnamefont {C.~Y.}\ \bibnamefont
  {Tsang}}, \bibinfo {author} {\bibfnamefont {M.~B.}\ \bibnamefont {Tsang}},
  \bibinfo {author} {\bibfnamefont {W.~G.}\ \bibnamefont {Lynch}}, \bibinfo
  {author} {\bibfnamefont {R.}~\bibnamefont {Kumar}},\ and\ \bibinfo {author}
  {\bibfnamefont {C.~J.}\ \bibnamefont {Horowitz}},\ }\bibfield  {title}
  {\bibinfo {title} {Determination of the equation of state from nuclear
  experiments and neutron star observations},\ }\href
  {https://doi.org/10.1038/s41550-023-02161-z} {\bibfield  {journal} {\bibinfo
  {journal} {Nature Astronomy}\ }\textbf {\bibinfo {volume} {8}},\ \bibinfo
  {pages} {328} (\bibinfo {year} {2024})}\BibitemShut {NoStop}\bibitem [{\citenamefont {Drischler}\ \emph {et~al.}(2019)\citenamefont
  {Drischler}, \citenamefont {Hebeler},\ and\ \citenamefont
  {Schwenk}}]{Drischler2019}\BibitemOpen
  \bibfield  {author} {\bibinfo {author} {\bibfnamefont {C.}~\bibnamefont
  {Drischler}}, \bibinfo {author} {\bibfnamefont {K.}~\bibnamefont {Hebeler}},\
  and\ \bibinfo {author} {\bibfnamefont {A.}~\bibnamefont {Schwenk}},\
  }\bibfield  {title} {\bibinfo {title} {Chiral interactions up to
  next-to-next-to-next-to-leading order and nuclear saturation},\ }\href
  {https://doi.org/10.1103/PhysRevLett.122.042501} {\bibfield  {journal}
  {\bibinfo  {journal} {Phys. Rev. Lett.}\ }\textbf {\bibinfo {volume} {122}},\
  \bibinfo {pages} {042501} (\bibinfo {year} {2019})},\ \bibinfo {note} {cEFT
  pure-neutron-matter band; data via BUQEYE collaboration},\ \Eprint
  {https://arxiv.org/abs/1710.08220} {arXiv:1710.08220} \BibitemShut {NoStop}\bibitem [{\citenamefont {Drischler}\ \emph {et~al.}(2021)\citenamefont
  {Drischler}, \citenamefont {Holt},\ and\ \citenamefont
  {Wellenhofer}}]{Drischler2021review}\BibitemOpen
  \bibfield  {author} {\bibinfo {author} {\bibfnamefont {C.}~\bibnamefont
  {Drischler}}, \bibinfo {author} {\bibfnamefont {J.~W.}\ \bibnamefont
  {Holt}},\ and\ \bibinfo {author} {\bibfnamefont {C.}~\bibnamefont
  {Wellenhofer}},\ }\bibfield  {title} {\bibinfo {title} {Chiral effective
  field theory and the high-density nuclear equation of state},\ }\href
  {https://doi.org/10.1146/annurev-nucl-102419-041903} {\bibfield  {journal}
  {\bibinfo  {journal} {Annual Review of Nuclear and Particle Science}\
  }\textbf {\bibinfo {volume} {71}},\ \bibinfo {pages} {403} (\bibinfo {year}
  {2021})}\BibitemShut {NoStop}\bibitem [{\citenamefont {{Al-Mamun}}\ \emph {et~al.}(2021)\citenamefont
  {{Al-Mamun}}, \citenamefont {{Steiner}}, \citenamefont {{N{\"a}ttil{\"a}}},
  \citenamefont {{Lange}}, \citenamefont {{O'Shaughnessy}}, \citenamefont
  {{Tews}}, \citenamefont {{Gandolfi}}, \citenamefont {{Heinke}},\ and\
  \citenamefont {{Han}}}]{gwastro-nsnuc-SteinerJoint-2020}\BibitemOpen
  \bibfield  {author} {\bibinfo {author} {\bibfnamefont {M.}~\bibnamefont
  {{Al-Mamun}}}, \bibinfo {author} {\bibfnamefont {A.~W.}\ \bibnamefont
  {{Steiner}}}, \bibinfo {author} {\bibfnamefont {J.}~\bibnamefont
  {{N{\"a}ttil{\"a}}}}, \bibinfo {author} {\bibfnamefont {J.}~\bibnamefont
  {{Lange}}}, \bibinfo {author} {\bibfnamefont {R.}~\bibnamefont
  {{O'Shaughnessy}}}, \bibinfo {author} {\bibfnamefont {I.}~\bibnamefont
  {{Tews}}}, \bibinfo {author} {\bibfnamefont {S.}~\bibnamefont {{Gandolfi}}},
  \bibinfo {author} {\bibfnamefont {C.}~\bibnamefont {{Heinke}}},\ and\
  \bibinfo {author} {\bibfnamefont {S.}~\bibnamefont {{Han}}},\ }\bibfield
  {title} {\bibinfo {title} {{Combining Electromagnetic and Gravitational-Wave
  Constraints on Neutron-Star Masses and Radii}},\ }\href
  {https://doi.org/10.1103/PhysRevLett.126.061101} {\bibfield  {journal}
  {\bibinfo  {journal} {\prl}\ }\textbf {\bibinfo {volume} {126}},\ \bibinfo
  {eid} {061101} (\bibinfo {year} {2021})},\ \Eprint
  {https://arxiv.org/abs/2008.12817} {arXiv:2008.12817 [astro-ph.HE]}
  \BibitemShut {NoStop}\bibitem [{\citenamefont {Raaijmakers}\ \emph {et~al.}(2020)\citenamefont
  {Raaijmakers}, \citenamefont {Greif}, \citenamefont {Riley}, \citenamefont
  {Hinderer}, \citenamefont {Hebeler}, \citenamefont {Schwenk}, \citenamefont
  {Watts}, \citenamefont {Nissanke}, \citenamefont {Guillot}, \citenamefont
  {Lattimer},\ and\ \citenamefont {Ludlam}}]{RaaijmakersEtAl2020}\BibitemOpen
  \bibfield  {author} {\bibinfo {author} {\bibfnamefont {G.}~\bibnamefont
  {Raaijmakers}}, \bibinfo {author} {\bibfnamefont {S.~K.}\ \bibnamefont
  {Greif}}, \bibinfo {author} {\bibfnamefont {T.~E.}\ \bibnamefont {Riley}},
  \bibinfo {author} {\bibfnamefont {T.}~\bibnamefont {Hinderer}}, \bibinfo
  {author} {\bibfnamefont {K.}~\bibnamefont {Hebeler}}, \bibinfo {author}
  {\bibfnamefont {A.}~\bibnamefont {Schwenk}}, \bibinfo {author} {\bibfnamefont
  {A.~L.}\ \bibnamefont {Watts}}, \bibinfo {author} {\bibfnamefont
  {S.}~\bibnamefont {Nissanke}}, \bibinfo {author} {\bibfnamefont
  {S.}~\bibnamefont {Guillot}}, \bibinfo {author} {\bibfnamefont {J.~M.}\
  \bibnamefont {Lattimer}},\ and\ \bibinfo {author} {\bibfnamefont {R.~M.}\
  \bibnamefont {Ludlam}},\ }\bibfield  {title} {\bibinfo {title} {Constraining
  the dense matter equation of state with joint analysis of {NICER} and
  {LIGO/Virgo} measurements},\ }\href
  {https://doi.org/10.3847/2041-8213/ab822f} {\bibfield  {journal} {\bibinfo
  {journal} {Astrophys. J. Lett.}\ }\textbf {\bibinfo {volume} {893}},\
  \bibinfo {pages} {L21} (\bibinfo {year} {2020})},\ \Eprint
  {https://arxiv.org/abs/1912.11031} {arXiv:1912.11031} \BibitemShut {NoStop}\bibitem [{\citenamefont {Raaijmakers}\ \emph {et~al.}(2021)\citenamefont
  {Raaijmakers}, \citenamefont {Greif}, \citenamefont {Hebeler}, \citenamefont
  {Hinderer}, \citenamefont {Nissanke}, \citenamefont {Schwenk}, \citenamefont
  {Riley}, \citenamefont {Watts}, \citenamefont {Lattimer},\ and\ \citenamefont
  {Ho}}]{RaaijmakersEtAl2021}\BibitemOpen
  \bibfield  {author} {\bibinfo {author} {\bibfnamefont {G.}~\bibnamefont
  {Raaijmakers}}, \bibinfo {author} {\bibfnamefont {S.~K.}\ \bibnamefont
  {Greif}}, \bibinfo {author} {\bibfnamefont {K.}~\bibnamefont {Hebeler}},
  \bibinfo {author} {\bibfnamefont {T.}~\bibnamefont {Hinderer}}, \bibinfo
  {author} {\bibfnamefont {S.}~\bibnamefont {Nissanke}}, \bibinfo {author}
  {\bibfnamefont {A.}~\bibnamefont {Schwenk}}, \bibinfo {author} {\bibfnamefont
  {T.~E.}\ \bibnamefont {Riley}}, \bibinfo {author} {\bibfnamefont {A.~L.}\
  \bibnamefont {Watts}}, \bibinfo {author} {\bibfnamefont {J.~M.}\ \bibnamefont
  {Lattimer}},\ and\ \bibinfo {author} {\bibfnamefont {W.~C.~G.}\ \bibnamefont
  {Ho}},\ }\bibfield  {title} {\bibinfo {title} {Constraints on the dense
  matter equation of state and neutron star properties from {NICER}'s
  mass--radius estimate of {PSR J0740+6620} and multimessenger observations},\
  }\href {https://doi.org/10.3847/2041-8213/ac089a} {\bibfield  {journal}
  {\bibinfo  {journal} {Astrophys. J. Lett.}\ }\textbf {\bibinfo {volume}
  {918}},\ \bibinfo {pages} {L29} (\bibinfo {year} {2021})},\ \Eprint
  {https://arxiv.org/abs/2105.06981} {arXiv:2105.06981} \BibitemShut {NoStop}\bibitem [{\citenamefont {Pang}\ \emph {et~al.}(2021)\citenamefont {Pang},
  \citenamefont {Tews}, \citenamefont {Coughlin}, \citenamefont {Bulla},
  \citenamefont {Van Den~Broeck},\ and\ \citenamefont
  {Dietrich}}]{PangEtAl2021}\BibitemOpen
  \bibfield  {author} {\bibinfo {author} {\bibfnamefont {P.~T.~H.}\
  \bibnamefont {Pang}}, \bibinfo {author} {\bibfnamefont {I.}~\bibnamefont
  {Tews}}, \bibinfo {author} {\bibfnamefont {M.~W.}\ \bibnamefont {Coughlin}},
  \bibinfo {author} {\bibfnamefont {M.}~\bibnamefont {Bulla}}, \bibinfo
  {author} {\bibfnamefont {C.}~\bibnamefont {Van Den~Broeck}},\ and\ \bibinfo
  {author} {\bibfnamefont {T.}~\bibnamefont {Dietrich}},\ }\bibfield  {title}
  {\bibinfo {title} {Nuclear physics multimessenger astrophysics constraints on
  the neutron star equation of state: Adding {NICER}'s {PSR J0740+6620}
  measurement},\ }\href {https://doi.org/10.3847/1538-4357/ac19ab} {\bibfield
  {journal} {\bibinfo  {journal} {Astrophys. J.}\ }\textbf {\bibinfo {volume}
  {922}},\ \bibinfo {pages} {14} (\bibinfo {year} {2021})},\ \Eprint
  {https://arxiv.org/abs/2105.08688} {arXiv:2105.08688} \BibitemShut {NoStop}\bibitem [{\citenamefont {Rutherford}\ \emph {et~al.}(2024)\citenamefont
  {Rutherford}, \citenamefont {Mendes}, \citenamefont {Svensson}, \citenamefont
  {Schwenk}, \citenamefont {Watts}, \citenamefont {Hebeler}, \citenamefont
  {Keller}, \citenamefont {Prescod-Weinstein}, \citenamefont {Choudhury},
  \citenamefont {Raaijmakers}, \citenamefont {Salmi}, \citenamefont
  {Timmerman}, \citenamefont {Vinciguerra}, \citenamefont {Guillot},\ and\
  \citenamefont {Lattimer}}]{RutherfordEtAl2024}\BibitemOpen
  \bibfield  {author} {\bibinfo {author} {\bibfnamefont {N.}~\bibnamefont
  {Rutherford}}, \bibinfo {author} {\bibfnamefont {M.}~\bibnamefont {Mendes}},
  \bibinfo {author} {\bibfnamefont {I.}~\bibnamefont {Svensson}}, \bibinfo
  {author} {\bibfnamefont {A.}~\bibnamefont {Schwenk}}, \bibinfo {author}
  {\bibfnamefont {A.~L.}\ \bibnamefont {Watts}}, \bibinfo {author}
  {\bibfnamefont {K.}~\bibnamefont {Hebeler}}, \bibinfo {author} {\bibfnamefont
  {J.}~\bibnamefont {Keller}}, \bibinfo {author} {\bibfnamefont
  {C.}~\bibnamefont {Prescod-Weinstein}}, \bibinfo {author} {\bibfnamefont
  {D.}~\bibnamefont {Choudhury}}, \bibinfo {author} {\bibfnamefont
  {G.}~\bibnamefont {Raaijmakers}}, \bibinfo {author} {\bibfnamefont
  {T.}~\bibnamefont {Salmi}}, \bibinfo {author} {\bibfnamefont
  {P.}~\bibnamefont {Timmerman}}, \bibinfo {author} {\bibfnamefont
  {S.}~\bibnamefont {Vinciguerra}}, \bibinfo {author} {\bibfnamefont
  {S.}~\bibnamefont {Guillot}},\ and\ \bibinfo {author} {\bibfnamefont {J.~M.}\
  \bibnamefont {Lattimer}},\ }\bibfield  {title} {\bibinfo {title}
  {Constraining the dense matter equation of state with new {NICER}
  mass--radius measurements and new chiral effective field theory inputs},\
  }\href {https://doi.org/10.3847/2041-8213/ad5f02} {\bibfield  {journal}
  {\bibinfo  {journal} {Astrophys. J. Lett.}\ }\textbf {\bibinfo {volume}
  {971}},\ \bibinfo {pages} {L19} (\bibinfo {year} {2024})},\ \Eprint
  {https://arxiv.org/abs/2407.06790} {arXiv:2407.06790} \BibitemShut {NoStop}\bibitem [{\citenamefont {{Wysocki}}\ \emph {et~al.}(2020)\citenamefont
  {{Wysocki}}, \citenamefont {{O'Shaughnessy}}, \citenamefont {{Wade}},\ and\
  \citenamefont
  {{Lange}}}]{gwastro-PopulationReconstruct-EOSStack-Wysocki2019}\BibitemOpen
  \bibfield  {author} {\bibinfo {author} {\bibfnamefont {D.}~\bibnamefont
  {{Wysocki}}}, \bibinfo {author} {\bibfnamefont {R.}~\bibnamefont
  {{O'Shaughnessy}}}, \bibinfo {author} {\bibfnamefont {L.}~\bibnamefont
  {{Wade}}},\ and\ \bibinfo {author} {\bibfnamefont {J.}~\bibnamefont
  {{Lange}}},\ }\bibfield  {title} {\bibinfo {title} {{Inferring the neutron
  star equation of state simultaneously with the population of merging neutron
  stars}},\ }\href {https://doi.org/10.48550/arXiv.2001.01747} {\bibfield
  {journal} {\bibinfo  {journal} {arXiv e-prints}\ ,\ \bibinfo {eid}
  {arXiv:2001.01747}} (\bibinfo {year} {2020})},\ \Eprint
  {https://arxiv.org/abs/2001.01747} {arXiv:2001.01747 [astro-ph.HE]}
  \BibitemShut {NoStop}\bibitem [{\citenamefont {Read}\ \emph {et~al.}(2009)\citenamefont {Read},
  \citenamefont {Lackey}, \citenamefont {Owen},\ and\ \citenamefont
  {Friedman}}]{Read2009}\BibitemOpen
  \bibfield  {author} {\bibinfo {author} {\bibfnamefont {J.~S.}\ \bibnamefont
  {Read}}, \bibinfo {author} {\bibfnamefont {B.~D.}\ \bibnamefont {Lackey}},
  \bibinfo {author} {\bibfnamefont {B.~J.}\ \bibnamefont {Owen}},\ and\
  \bibinfo {author} {\bibfnamefont {J.~L.}\ \bibnamefont {Friedman}},\
  }\bibfield  {title} {\bibinfo {title} {Constraints on a phenomenologically
  parametrized neutron-star equation of state},\ }\href@noop {} {\bibfield
  {journal} {\bibinfo  {journal} {Phys. Rev. D}\ }\textbf {\bibinfo {volume}
  {79}},\ \bibinfo {pages} {124032} (\bibinfo {year} {2009})}\BibitemShut
  {NoStop}\bibitem [{\citenamefont {Lindblom}(2010)}]{Lindblom2010}\BibitemOpen
  \bibfield  {author} {\bibinfo {author} {\bibfnamefont {L.}~\bibnamefont
  {Lindblom}},\ }\bibfield  {title} {\bibinfo {title} {Spectral representations
  of neutron-star equations of state},\ }\href@noop {} {\bibfield  {journal}
  {\bibinfo  {journal} {Phys. Rev. D}\ }\textbf {\bibinfo {volume} {82}},\
  \bibinfo {pages} {103011} (\bibinfo {year} {2010})}\BibitemShut {NoStop}\bibitem [{\citenamefont {Carney}\ \emph {et~al.}(2018)\citenamefont {Carney},
  \citenamefont {Wade},\ and\ \citenamefont {Irwin}}]{Carney2018}\BibitemOpen
  \bibfield  {author} {\bibinfo {author} {\bibfnamefont {M.~F.}\ \bibnamefont
  {Carney}}, \bibinfo {author} {\bibfnamefont {L.~E.}\ \bibnamefont {Wade}},\
  and\ \bibinfo {author} {\bibfnamefont {B.~S.}\ \bibnamefont {Irwin}},\
  }\bibfield  {title} {\bibinfo {title} {Comparing two models for measuring the
  neutron star equation of state from gravitational-wave signals},\ }\href@noop
  {} {\bibfield  {journal} {\bibinfo  {journal} {Phys. Rev. D}\ }\textbf
  {\bibinfo {volume} {98}},\ \bibinfo {pages} {063004} (\bibinfo {year}
  {2018})},\ \Eprint {https://arxiv.org/abs/1805.11217} {arXiv:1805.11217}
  \BibitemShut {NoStop}\bibitem [{\citenamefont {Gamba}\ \emph {et~al.}(2020)\citenamefont {Gamba},
  \citenamefont {Read},\ and\ \citenamefont {Wade}}]{GambaReadWade2020}\BibitemOpen
  \bibfield  {author} {\bibinfo {author} {\bibfnamefont {R.}~\bibnamefont
  {Gamba}}, \bibinfo {author} {\bibfnamefont {J.~S.}\ \bibnamefont {Read}},\
  and\ \bibinfo {author} {\bibfnamefont {L.~E.}\ \bibnamefont {Wade}},\
  }\bibfield  {title} {\bibinfo {title} {The impact of the crust equation of
  state on the analysis of {GW170817}},\ }\href
  {https://doi.org/10.1088/1361-6382/ab5ba4} {\bibfield  {journal} {\bibinfo
  {journal} {Class. Quantum Grav.}\ }\textbf {\bibinfo {volume} {37}},\
  \bibinfo {pages} {025008} (\bibinfo {year} {2020})},\ \Eprint
  {https://arxiv.org/abs/1902.04616} {arXiv:1902.04616} \BibitemShut {NoStop}\bibitem [{\citenamefont {Legred}\ \emph {et~al.}(2022)\citenamefont {Legred},
  \citenamefont {Chatziioannou}, \citenamefont {Essick},\ and\ \citenamefont
  {Landry}}]{LegredEtAl2022}\BibitemOpen
  \bibfield  {author} {\bibinfo {author} {\bibfnamefont {I.}~\bibnamefont
  {Legred}}, \bibinfo {author} {\bibfnamefont {K.}~\bibnamefont
  {Chatziioannou}}, \bibinfo {author} {\bibfnamefont {R.}~\bibnamefont
  {Essick}},\ and\ \bibinfo {author} {\bibfnamefont {P.}~\bibnamefont
  {Landry}},\ }\bibfield  {title} {\bibinfo {title} {Implicit correlations
  within phenomenological parametric models of the neutron star equation of
  state},\ }\href {https://doi.org/10.1103/PhysRevD.105.043016} {\bibfield
  {journal} {\bibinfo  {journal} {Phys. Rev. D}\ }\textbf {\bibinfo {volume}
  {105}},\ \bibinfo {pages} {043016} (\bibinfo {year} {2022})},\ \Eprint
  {https://arxiv.org/abs/2201.06791} {arXiv:2201.06791} \BibitemShut {NoStop}\bibitem [{\citenamefont {{Biswas}}(2026)}]{Biswas2025extrap}\BibitemOpen
  \bibfield  {author} {\bibinfo {author} {\bibfnamefont {B.}~\bibnamefont
  {{Biswas}}},\ }\bibfield  {title} {\bibinfo {title} {{Equation of State
  Extrapolation Systematics: Parametric vs. Nonparametric Inference of Neutron
  Star Structure}},\ }\href@noop {} {\bibfield  {journal} {\bibinfo  {journal}
  {\prd}\ }\textbf {\bibinfo {volume} {113}},\ \bibinfo {pages} {123032}
  (\bibinfo {year} {2026})},\ \Eprint {https://arxiv.org/abs/2509.06145}
  {2509.06145} \BibitemShut {NoStop}\bibitem [{\citenamefont {Landry}\ and\ \citenamefont
  {Essick}(2019)}]{LandryEssick2019}\BibitemOpen
  \bibfield  {author} {\bibinfo {author} {\bibfnamefont {P.}~\bibnamefont
  {Landry}}\ and\ \bibinfo {author} {\bibfnamefont {R.}~\bibnamefont
  {Essick}},\ }\bibfield  {title} {\bibinfo {title} {Nonparametric inference of
  the neutron star equation of state from gravitational wave observations},\
  }\href@noop {} {\bibfield  {journal} {\bibinfo  {journal} {Phys. Rev. D}\
  }\textbf {\bibinfo {volume} {99}},\ \bibinfo {pages} {084049} (\bibinfo
  {year} {2019})},\ \Eprint {https://arxiv.org/abs/1811.12529}
  {arXiv:1811.12529} \BibitemShut {NoStop}\bibitem [{\citenamefont {Essick}\ \emph {et~al.}(2020)\citenamefont {Essick},
  \citenamefont {Landry},\ and\ \citenamefont {Holz}}]{Essick2020}\BibitemOpen
  \bibfield  {author} {\bibinfo {author} {\bibfnamefont {R.}~\bibnamefont
  {Essick}}, \bibinfo {author} {\bibfnamefont {P.}~\bibnamefont {Landry}},\
  and\ \bibinfo {author} {\bibfnamefont {D.~E.}\ \bibnamefont {Holz}},\
  }\bibfield  {title} {\bibinfo {title} {Nonparametric inference of neutron
  star composition, equation of state, and maximum mass with {GW170817}},\
  }\href@noop {} {\bibfield  {journal} {\bibinfo  {journal} {Phys. Rev. D}\
  }\textbf {\bibinfo {volume} {101}},\ \bibinfo {pages} {063007} (\bibinfo
  {year} {2020})}\BibitemShut {NoStop}\bibitem [{\citenamefont {Landry}\ \emph {et~al.}(2020)\citenamefont {Landry},
  \citenamefont {Essick},\ and\ \citenamefont
  {Chatziioannou}}]{LandryEtAl2020}\BibitemOpen
  \bibfield  {author} {\bibinfo {author} {\bibfnamefont {P.}~\bibnamefont
  {Landry}}, \bibinfo {author} {\bibfnamefont {R.}~\bibnamefont {Essick}},\
  and\ \bibinfo {author} {\bibfnamefont {K.}~\bibnamefont {Chatziioannou}},\
  }\bibfield  {title} {\bibinfo {title} {Nonparametric constraints on neutron
  star matter with existing and upcoming gravitational wave and pulsar
  observations},\ }\href {https://doi.org/10.1103/PhysRevD.101.123007}
  {\bibfield  {journal} {\bibinfo  {journal} {Phys. Rev. D}\ }\textbf {\bibinfo
  {volume} {101}},\ \bibinfo {pages} {123007} (\bibinfo {year} {2020})},\
  \Eprint {https://arxiv.org/abs/2003.04880} {arXiv:2003.04880} \BibitemShut
  {NoStop}\bibitem [{\citenamefont {Legred}\ \emph {et~al.}(2021)\citenamefont {Legred},
  \citenamefont {Chatziioannou}, \citenamefont {Essick}, \citenamefont {Han},\
  and\ \citenamefont {Landry}}]{Legred2021}\BibitemOpen
  \bibfield  {author} {\bibinfo {author} {\bibfnamefont {I.}~\bibnamefont
  {Legred}}, \bibinfo {author} {\bibfnamefont {K.}~\bibnamefont
  {Chatziioannou}}, \bibinfo {author} {\bibfnamefont {R.}~\bibnamefont
  {Essick}}, \bibinfo {author} {\bibfnamefont {S.}~\bibnamefont {Han}},\ and\
  \bibinfo {author} {\bibfnamefont {P.}~\bibnamefont {Landry}},\ }\bibfield
  {title} {\bibinfo {title} {Impact of the {PSR} {J0740+6620} radius constraint
  on the properties of high-density matter},\ }\href@noop {} {\bibfield
  {journal} {\bibinfo  {journal} {Phys. Rev. D}\ }\textbf {\bibinfo {volume}
  {104}},\ \bibinfo {pages} {063003} (\bibinfo {year} {2021})},\ \bibinfo
  {note} {data release: Zenodo 6502467},\ \Eprint
  {https://arxiv.org/abs/2106.05313} {arXiv:2106.05313} \BibitemShut {NoStop}\bibitem [{\citenamefont {Finch}\ \emph {et~al.}(2025)\citenamefont {Finch},
  \citenamefont {Landry}, \citenamefont {Essick}, \citenamefont {Legred},
  \citenamefont {Chatziioannou},\ and\ \citenamefont {Han}}]{Finch2025}\BibitemOpen
  \bibfield  {author} {\bibinfo {author} {\bibfnamefont {E.}~\bibnamefont
  {Finch}}, \bibinfo {author} {\bibfnamefont {P.}~\bibnamefont {Landry}},
  \bibinfo {author} {\bibfnamefont {R.}~\bibnamefont {Essick}}, \bibinfo
  {author} {\bibfnamefont {I.}~\bibnamefont {Legred}}, \bibinfo {author}
  {\bibfnamefont {K.}~\bibnamefont {Chatziioannou}},\ and\ \bibinfo {author}
  {\bibfnamefont {S.}~\bibnamefont {Han}},\ }\bibfield  {title} {\bibinfo
  {title} {Unified nonparametric equation-of-state inference from the
  neutron-star crust to perturbative-{QCD} densities},\ }\href
  {https://doi.org/10.1103/krc7-kz2l} {\bibfield  {journal} {\bibinfo
  {journal} {\prd}\ }\textbf {\bibinfo {volume} {112}},\ \bibinfo {pages}
  {103023} (\bibinfo {year} {2025})},\ \Eprint
  {https://arxiv.org/abs/2505.13691} {arXiv:2505.13691} \BibitemShut {NoStop}\bibitem [{\citenamefont {McGinn}\ \emph {et~al.}(2024)\citenamefont {McGinn},
  \citenamefont {Irwin}, \citenamefont {Mukherjee}, \citenamefont {Williams},
  \citenamefont {Messenger},\ and\ \citenamefont {Heng}}]{McGinnEtAl2024}\BibitemOpen
  \bibfield  {author} {\bibinfo {author} {\bibfnamefont {J.}~\bibnamefont
  {McGinn}}, \bibinfo {author} {\bibfnamefont {B.}~\bibnamefont {Irwin}},
  \bibinfo {author} {\bibfnamefont {S.}~\bibnamefont {Mukherjee}}, \bibinfo
  {author} {\bibfnamefont {D.}~\bibnamefont {Williams}}, \bibinfo {author}
  {\bibfnamefont {C.}~\bibnamefont {Messenger}},\ and\ \bibinfo {author}
  {\bibfnamefont {I.~S.}\ \bibnamefont {Heng}},\ }\bibfield  {title} {\bibinfo
  {title} {Rapid neutron star equation of state inference with normalising
  flows},\ }\href {https://doi.org/10.48550/arXiv.2403.17462} {\bibfield
  {journal} {\bibinfo  {journal} {arXiv e-prints}\ ,\ \bibinfo {pages}
  {arXiv:2403.17462}} (\bibinfo {year} {2024})},\ \Eprint
  {https://arxiv.org/abs/2403.17462} {arXiv:2403.17462} \BibitemShut {NoStop}\bibitem [{\citenamefont {Wouters}\ \emph {et~al.}(2026)\citenamefont
  {Wouters}, \citenamefont {Pang}, \citenamefont {Dietrich},\ and\
  \citenamefont {Van Den~Broeck}}]{WoutersEtAl2026}\BibitemOpen
  \bibfield  {author} {\bibinfo {author} {\bibfnamefont {T.}~\bibnamefont
  {Wouters}}, \bibinfo {author} {\bibfnamefont {P.~T.~H.}\ \bibnamefont
  {Pang}}, \bibinfo {author} {\bibfnamefont {T.}~\bibnamefont {Dietrich}},\
  and\ \bibinfo {author} {\bibfnamefont {C.}~\bibnamefont {Van Den~Broeck}},\
  }\bibfield  {title} {\bibinfo {title} {Incorporating neutron star physics
  into gravitational wave inference with physics-informed priors using
  normalizing flows},\ }\href {https://doi.org/10.1103/8t5h-y4lf} {\bibfield
  {journal} {\bibinfo  {journal} {Phys. Rev. D}\ }\textbf {\bibinfo {volume}
  {114}},\ \bibinfo {pages} {043005} (\bibinfo {year} {2026})},\ \Eprint
  {https://arxiv.org/abs/2511.22987} {arXiv:2511.22987} \BibitemShut {NoStop}\bibitem [{\citenamefont {Sun}\ \emph {et~al.}(2026)\citenamefont {Sun},
  \citenamefont {Zhao},\ and\ \citenamefont {Lattimer}}]{MRprior2026}\BibitemOpen
  \bibfield  {author} {\bibinfo {author} {\bibfnamefont {B.}~\bibnamefont
  {Sun}}, \bibinfo {author} {\bibfnamefont {T.}~\bibnamefont {Zhao}},\ and\
  \bibinfo {author} {\bibfnamefont {J.~M.}\ \bibnamefont {Lattimer}},\
  }\bibfield  {title} {\bibinfo {title} {A new bayesian framework with natural
  priors to constrain the neutron star equation of state},\ }\href
  {https://doi.org/10.48550/arXiv.2601.04294} {\bibfield  {journal} {\bibinfo
  {journal} {arXiv e-prints}\ ,\ \bibinfo {pages} {arXiv:2601.04294}} (\bibinfo
  {year} {2026})},\ \Eprint {https://arxiv.org/abs/2601.04294}
  {arXiv:2601.04294} \BibitemShut {NoStop}\bibitem [{\citenamefont {Fujimoto}\ \emph {et~al.}(2018)\citenamefont
  {Fujimoto}, \citenamefont {Fukushima},\ and\ \citenamefont
  {Murase}}]{FujimotoEtAl2018}\BibitemOpen
  \bibfield  {author} {\bibinfo {author} {\bibfnamefont {Y.}~\bibnamefont
  {Fujimoto}}, \bibinfo {author} {\bibfnamefont {K.}~\bibnamefont
  {Fukushima}},\ and\ \bibinfo {author} {\bibfnamefont {K.}~\bibnamefont
  {Murase}},\ }\bibfield  {title} {\bibinfo {title} {Methodology study of
  machine learning for the neutron star equation of state},\ }\href
  {https://doi.org/10.1103/PhysRevD.98.023019} {\bibfield  {journal} {\bibinfo
  {journal} {Phys. Rev. D}\ }\textbf {\bibinfo {volume} {98}},\ \bibinfo
  {pages} {023019} (\bibinfo {year} {2018})},\ \Eprint
  {https://arxiv.org/abs/1711.06748} {arXiv:1711.06748 [nucl-th]} \BibitemShut
  {NoStop}\bibitem [{\citenamefont {Reed}\ \emph {et~al.}(2024)\citenamefont {Reed},
  \citenamefont {Somasundaram}, \citenamefont {De}, \citenamefont {Armstrong},
  \citenamefont {Giuliani}, \citenamefont {Capano}, \citenamefont {Brown},\
  and\ \citenamefont {Tews}}]{ReedEtAl2024}\BibitemOpen
  \bibfield  {author} {\bibinfo {author} {\bibfnamefont {B.~T.}\ \bibnamefont
  {Reed}}, \bibinfo {author} {\bibfnamefont {R.}~\bibnamefont {Somasundaram}},
  \bibinfo {author} {\bibfnamefont {S.}~\bibnamefont {De}}, \bibinfo {author}
  {\bibfnamefont {C.~L.}\ \bibnamefont {Armstrong}}, \bibinfo {author}
  {\bibfnamefont {P.}~\bibnamefont {Giuliani}}, \bibinfo {author}
  {\bibfnamefont {C.}~\bibnamefont {Capano}}, \bibinfo {author} {\bibfnamefont
  {D.~A.}\ \bibnamefont {Brown}},\ and\ \bibinfo {author} {\bibfnamefont
  {I.}~\bibnamefont {Tews}},\ }\bibfield  {title} {\bibinfo {title} {Toward
  accelerated nuclear-physics parameter estimation from binary neutron star
  mergers: Emulators for the tolman--oppenheimer--volkoff equations},\ }\href
  {https://doi.org/10.3847/1538-4357/ad737c} {\bibfield  {journal} {\bibinfo
  {journal} {Astrophys. J.}\ }\textbf {\bibinfo {volume} {974}},\ \bibinfo
  {pages} {285} (\bibinfo {year} {2024})},\ \Eprint
  {https://arxiv.org/abs/2405.20558} {arXiv:2405.20558 [astro-ph.HE]}
  \BibitemShut {NoStop}\bibitem [{\citenamefont {Tiwari}\ and\ \citenamefont
  {Pai}(2024)}]{TiwariPai2024}\BibitemOpen
  \bibfield  {author} {\bibinfo {author} {\bibfnamefont {P.}~\bibnamefont
  {Tiwari}}\ and\ \bibinfo {author} {\bibfnamefont {A.}~\bibnamefont {Pai}},\
  }\bibfield  {title} {\bibinfo {title} {Deep {TOV} to characterize neutron
  stars},\ }\href {https://doi.org/10.48550/arXiv.2405.08163} {\bibfield
  {journal} {\bibinfo  {journal} {arXiv e-prints}\ ,\ \bibinfo {pages}
  {arXiv:2405.08163}} (\bibinfo {year} {2024})},\ \Eprint
  {https://arxiv.org/abs/2405.08163} {arXiv:2405.08163 [astro-ph.HE]}
  \BibitemShut {NoStop}\bibitem [{\citenamefont {{Kaushikk V. N.}}\ \emph {et~al.}(2026)\citenamefont
  {{Kaushikk V. N.}}, \citenamefont {Biswas},\ and\ \citenamefont
  {Rosswog}}]{KaushikkEtAl2026}\BibitemOpen
  \bibfield  {author} {\bibinfo {author} {\bibnamefont {{Kaushikk V. N.}}},
  \bibinfo {author} {\bibfnamefont {B.}~\bibnamefont {Biswas}},\ and\ \bibinfo
  {author} {\bibfnamefont {S.}~\bibnamefont {Rosswog}},\ }\bibfield  {title}
  {\bibinfo {title} {Fast and accurate prediction of neutron star structure
  with deep neural networks},\ }\href
  {https://doi.org/10.48550/arXiv.2608.07698} {\bibfield  {journal} {\bibinfo
  {journal} {arXiv e-prints}\ ,\ \bibinfo {pages} {arXiv:2608.07698}} (\bibinfo
  {year} {2026})},\ \Eprint {https://arxiv.org/abs/2608.07698}
  {arXiv:2608.07698} \BibitemShut {NoStop}\bibitem [{\citenamefont {{Carvalho}}\ \emph {et~al.}(2025)\citenamefont
  {{Carvalho}}, \citenamefont {{Ferreira}}, \citenamefont {{Bejger}},\ and\
  \citenamefont {{Provid{\^e}ncia}}}]{Carvalho2025NPE}\BibitemOpen
  \bibfield  {author} {\bibinfo {author} {\bibfnamefont {V.}~\bibnamefont
  {{Carvalho}}}, \bibinfo {author} {\bibfnamefont {M.}~\bibnamefont
  {{Ferreira}}}, \bibinfo {author} {\bibfnamefont {M.}~\bibnamefont
  {{Bejger}}},\ and\ \bibinfo {author} {\bibfnamefont {C.}~\bibnamefont
  {{Provid{\^e}ncia}}},\ }\bibfield  {title} {\bibinfo {title} {{Neural
  Posterior Estimation of Neutron Star Equations of State}},\ }\href
  {https://doi.org/10.1103/6396-4dcy} {\bibfield  {journal} {\bibinfo
  {journal} {\prd}\ }\textbf {\bibinfo {volume} {112}},\ \bibinfo {pages}
  {083044} (\bibinfo {year} {2025})},\ \Eprint
  {https://arxiv.org/abs/2507.23506} {2507.23506} \BibitemShut {NoStop}\bibitem [{\citenamefont {{Thakur}}\ and\ \citenamefont
  {{Malik}}(2026)}]{AmortizedRMF2026}\BibitemOpen
  \bibfield  {author} {\bibinfo {author} {\bibfnamefont {P.}~\bibnamefont
  {{Thakur}}}\ and\ \bibinfo {author} {\bibfnamefont {T.}~\bibnamefont
  {{Malik}}},\ }\bibfield  {title} {\bibinfo {title} {{Amortized
  Simulation-Based Inference of Relativistic Mean-Field Couplings for
  Neutron-Star Equations of State}},\ }\href
  {https://doi.org/10.48550/arXiv.2606.25446} {\bibfield  {journal} {\bibinfo
  {journal} {arXiv e-prints}\ ,\ \bibinfo {pages} {arXiv:2606.25446}} (\bibinfo
  {year} {2026})},\ \Eprint {https://arxiv.org/abs/2606.25446} {2606.25446}
  \BibitemShut {NoStop}\bibitem [{\citenamefont {Lee}\ \emph {et~al.}(2026)\citenamefont {Lee},
  \citenamefont {Kim}, \citenamefont {Zhao},\ and\ \citenamefont
  {Drischler}}]{LeeEtAl2026GPDiff}\BibitemOpen
  \bibfield  {author} {\bibinfo {author} {\bibfnamefont {Y.~G.}\ \bibnamefont
  {Lee}}, \bibinfo {author} {\bibfnamefont {J.}~\bibnamefont {Kim}}, \bibinfo
  {author} {\bibfnamefont {T.}~\bibnamefont {Zhao}},\ and\ \bibinfo {author}
  {\bibfnamefont {C.}~\bibnamefont {Drischler}},\ }\bibfield  {title} {\bibinfo
  {title} {A {Gaussian} process framework for constraining the nuclear equation
  of state from microscopic calculations with correlated uncertainties},\
  }\href {https://doi.org/10.48550/arXiv.2608.09678} {\bibfield  {journal}
  {\bibinfo  {journal} {arXiv e-prints}\ ,\ \bibinfo {pages}
  {arXiv:2608.09678}} (\bibinfo {year} {2026})},\ \Eprint
  {https://arxiv.org/abs/2608.09678} {arXiv:2608.09678} \BibitemShut {NoStop}\bibitem [{\citenamefont {Semposki}\ \emph {et~al.}(2026)\citenamefont
  {Semposki}, \citenamefont {Drischler}, \citenamefont {Furnstahl},\ and\
  \citenamefont {Phillips}}]{Semposki2025}\BibitemOpen
  \bibfield  {author} {\bibinfo {author} {\bibfnamefont {A.~C.}\ \bibnamefont
  {Semposki}}, \bibinfo {author} {\bibfnamefont {C.}~\bibnamefont {Drischler}},
  \bibinfo {author} {\bibfnamefont {R.~J.}\ \bibnamefont {Furnstahl}},\ and\
  \bibinfo {author} {\bibfnamefont {D.~R.}\ \bibnamefont {Phillips}},\
  }\bibfield  {title} {\bibinfo {title} {Microscopic constraints for the
  equation of state and structure of neutron stars: A {Bayesian} model mixing
  framework},\ }\href {https://doi.org/10.1103/fxv6-gdnw} {\bibfield  {journal}
  {\bibinfo  {journal} {\prc}\ }\textbf {\bibinfo {volume} {113}},\ \bibinfo
  {pages} {015808} (\bibinfo {year} {2026})},\ \Eprint
  {https://arxiv.org/abs/2505.18921} {arXiv:2505.18921} \BibitemShut {NoStop}\bibitem [{\citenamefont {Ng}\ \emph {et~al.}(2025)\citenamefont {Ng},
  \citenamefont {Legred}, \citenamefont {Suleiman}, \citenamefont {Landry},
  \citenamefont {Traylor},\ and\ \citenamefont {Read}}]{Mihaylov2025}\BibitemOpen
  \bibfield  {author} {\bibinfo {author} {\bibfnamefont {S.}~\bibnamefont
  {Ng}}, \bibinfo {author} {\bibfnamefont {I.}~\bibnamefont {Legred}}, \bibinfo
  {author} {\bibfnamefont {L.}~\bibnamefont {Suleiman}}, \bibinfo {author}
  {\bibfnamefont {P.}~\bibnamefont {Landry}}, \bibinfo {author} {\bibfnamefont
  {L.}~\bibnamefont {Traylor}},\ and\ \bibinfo {author} {\bibfnamefont
  {J.}~\bibnamefont {Read}},\ }\bibfield  {title} {\bibinfo {title} {Inferring
  the neutron star equation of state with nuclear-physics informed
  semiparametric models},\ }\href {https://doi.org/10.1088/1361-6382/ae1094}
  {\bibfield  {journal} {\bibinfo  {journal} {\cqg}\ }\textbf {\bibinfo
  {volume} {42}},\ \bibinfo {pages} {205008} (\bibinfo {year} {2025})},\
  \Eprint {https://arxiv.org/abs/2507.03232} {arXiv:2507.03232} \BibitemShut
  {NoStop}\bibitem [{\citenamefont {{Lange}}\ \emph {et~al.}(2018)\citenamefont
  {{Lange}}, \citenamefont {{O'Shaughnessy}},\ and\ \citenamefont
  {{Rizzo}}}]{gwastro-PENR-RIFT}\BibitemOpen
  \bibfield  {author} {\bibinfo {author} {\bibfnamefont {J.}~\bibnamefont
  {{Lange}}}, \bibinfo {author} {\bibfnamefont {R.}~\bibnamefont
  {{O'Shaughnessy}}},\ and\ \bibinfo {author} {\bibfnamefont {M.}~\bibnamefont
  {{Rizzo}}},\ }\bibfield  {title} {\bibinfo {title} {Rapid and accurate
  parameter inference for coalescing, precessing compact binaries},\ }\href
  {https://doi.org/10.48550/arXiv.1805.10457} {\bibfield  {journal} {\bibinfo
  {journal} {arXiv e-prints}\ ,\ \bibinfo {pages} {arXiv:1805.10457}} (\bibinfo
  {year} {2018})},\ \Eprint {https://arxiv.org/abs/1805.10457}
  {arXiv:1805.10457} \BibitemShut {NoStop}\bibitem [{\citenamefont {{Wofford}}\ \emph {et~al.}(2023)\citenamefont
  {{Wofford}}, \citenamefont {{Yelikar}}, \citenamefont {{Gallagher}},
  \citenamefont {{Champion}}, \citenamefont {{Wysocki}}, \citenamefont
  {{Delfavero}}, \citenamefont {{Lange}}, \citenamefont {{Rose}}, \citenamefont
  {{Valsan}}, \citenamefont {{Morisaki}}, \citenamefont {{Read}}, \citenamefont
  {{Henshaw}},\ and\ \citenamefont {{O'Shaughnessy}}}]{gwastro-RIFT-Update}\BibitemOpen
  \bibfield  {author} {\bibinfo {author} {\bibfnamefont {J.}~\bibnamefont
  {{Wofford}}}, \bibinfo {author} {\bibfnamefont {A.~B.}\ \bibnamefont
  {{Yelikar}}}, \bibinfo {author} {\bibfnamefont {H.}~\bibnamefont
  {{Gallagher}}}, \bibinfo {author} {\bibfnamefont {E.}~\bibnamefont
  {{Champion}}}, \bibinfo {author} {\bibfnamefont {D.}~\bibnamefont
  {{Wysocki}}}, \bibinfo {author} {\bibfnamefont {V.}~\bibnamefont
  {{Delfavero}}}, \bibinfo {author} {\bibfnamefont {J.}~\bibnamefont
  {{Lange}}}, \bibinfo {author} {\bibfnamefont {C.}~\bibnamefont {{Rose}}},
  \bibinfo {author} {\bibfnamefont {V.}~\bibnamefont {{Valsan}}}, \bibinfo
  {author} {\bibfnamefont {S.}~\bibnamefont {{Morisaki}}}, \bibinfo {author}
  {\bibfnamefont {J.}~\bibnamefont {{Read}}}, \bibinfo {author} {\bibfnamefont
  {C.}~\bibnamefont {{Henshaw}}},\ and\ \bibinfo {author} {\bibfnamefont
  {R.}~\bibnamefont {{O'Shaughnessy}}},\ }\bibfield  {title} {\bibinfo {title}
  {{Improving performance for gravitational-wave parameter inference with an
  efficient and highly-parallelized algorithm}},\ }\href
  {https://doi.org/10.1103/PhysRevD.107.024040} {\bibfield  {journal} {\bibinfo
   {journal} {\prd}\ }\textbf {\bibinfo {volume} {107}},\ \bibinfo {eid}
  {024040} (\bibinfo {year} {2023})}\BibitemShut {NoStop}\bibitem [{\citenamefont {{Vilkha}}\ \emph {et~al.}(2024)\citenamefont
  {{Vilkha}}, \citenamefont {{Yelikar}}, \citenamefont {{O'Shaughnessy}},\ and\
  \citenamefont {{Read}}}]{gwastro-ns-eos-Vilkha2024}\BibitemOpen
  \bibfield  {author} {\bibinfo {author} {\bibfnamefont {A.}~\bibnamefont
  {{Vilkha}}}, \bibinfo {author} {\bibfnamefont {A.}~\bibnamefont {{Yelikar}}},
  \bibinfo {author} {\bibfnamefont {R.}~\bibnamefont {{O'Shaughnessy}}},\ and\
  \bibinfo {author} {\bibfnamefont {J.}~\bibnamefont {{Read}}},\ }\bibfield
  {title} {\bibinfo {title} {{Inference on neutron star parameters and the
  nuclear equation of state with RIFT, using prior EOS information}},\ }\href
  {https://doi.org/10.48550/arXiv.2407.15753} {\bibfield  {journal} {\bibinfo
  {journal} {arXiv e-prints}\ ,\ \bibinfo {eid} {arXiv:2407.15753}} (\bibinfo
  {year} {2024})},\ \Eprint {https://arxiv.org/abs/2407.15753}
  {arXiv:2407.15753 [gr-qc]} \BibitemShut {NoStop}\bibitem [{\citenamefont {Gorda}\ \emph {et~al.}(2026)\citenamefont {Gorda},
  \citenamefont {Komoltsev}, \citenamefont {Kurkela},\ and\ \citenamefont
  {Sunde}}]{GKKS2025bridge}\BibitemOpen
  \bibfield  {author} {\bibinfo {author} {\bibfnamefont {T.}~\bibnamefont
  {Gorda}}, \bibinfo {author} {\bibfnamefont {O.}~\bibnamefont {Komoltsev}},
  \bibinfo {author} {\bibfnamefont {A.}~\bibnamefont {Kurkela}},\ and\ \bibinfo
  {author} {\bibfnamefont {E.}~\bibnamefont {Sunde}},\ }\bibfield  {title}
  {\bibinfo {title} {Constrained {Gaussian}-process-bridge prior for
  neutron-star equation-of-state inference},\ }\href
  {https://doi.org/10.3847/1538-4357/ae552a} {\bibfield  {journal} {\bibinfo
  {journal} {\apj}\ }\textbf {\bibinfo {volume} {1002}},\ \bibinfo {pages} {40}
  (\bibinfo {year} {2026})},\ \Eprint {https://arxiv.org/abs/2512.18044}
  {arXiv:2512.18044} \BibitemShut {NoStop}\bibitem [{\citenamefont {Margueron}\ \emph {et~al.}(2018)\citenamefont
  {Margueron}, \citenamefont {Hoffmann~Casali},\ and\ \citenamefont
  {Gulminelli}}]{Margueron2018}\BibitemOpen
  \bibfield  {author} {\bibinfo {author} {\bibfnamefont {J.}~\bibnamefont
  {Margueron}}, \bibinfo {author} {\bibfnamefont {R.}~\bibnamefont
  {Hoffmann~Casali}},\ and\ \bibinfo {author} {\bibfnamefont {F.}~\bibnamefont
  {Gulminelli}},\ }\bibfield  {title} {\bibinfo {title} {Equation of state for
  dense nucleonic matter from metamodeling. i. foundational aspects},\
  }\href@noop {} {\bibfield  {journal} {\bibinfo  {journal} {Phys. Rev. C}\
  }\textbf {\bibinfo {volume} {97}},\ \bibinfo {pages} {025805} (\bibinfo
  {year} {2018})}\BibitemShut {NoStop}\bibitem [{\citenamefont {O'Shaughnessy}\ and\ \citenamefont {{RIT CCRG
  nuclear-matter group}}(2026)}]{NuclearMatterBackend}\BibitemOpen
  \bibfield  {author} {\bibinfo {author} {\bibfnamefont {R.}~\bibnamefont
  {O'Shaughnessy}}\ and\ \bibinfo {author} {\bibnamefont {{RIT CCRG
  nuclear-matter group}}},\ }\href@noop {} {\bibinfo {title}
  {{NuclearMatter-Backend}}},\ \bibinfo {howpublished}
  {\url{https://github.com/oshaughnessy-junior/NuclearMatter-Backend}}
  (\bibinfo {year} {2026}),\ \bibinfo {note} {public software repository,
  accessed 22 October 2026}\BibitemShut {NoStop}\bibitem [{\citenamefont {Lindblom}(1992)}]{Lindblom1992}\BibitemOpen
  \bibfield  {author} {\bibinfo {author} {\bibfnamefont {L.}~\bibnamefont
  {Lindblom}},\ }\bibfield  {title} {\bibinfo {title} {Determining the nuclear
  equation of state from neutron-star masses and radii},\ }\href@noop {}
  {\bibfield  {journal} {\bibinfo  {journal} {Astrophys. J.}\ }\textbf
  {\bibinfo {volume} {398}},\ \bibinfo {pages} {569} (\bibinfo {year}
  {1992})}\BibitemShut {NoStop}\bibitem [{\citenamefont {Hinderer}(2008)}]{Hinderer2008}\BibitemOpen
  \bibfield  {author} {\bibinfo {author} {\bibfnamefont {T.}~\bibnamefont
  {Hinderer}},\ }\bibfield  {title} {\bibinfo {title} {Tidal love numbers of
  neutron stars},\ }\href@noop {} {\bibfield  {journal} {\bibinfo  {journal}
  {Astrophys. J.}\ }\textbf {\bibinfo {volume} {677}},\ \bibinfo {pages} {1216}
  (\bibinfo {year} {2008})}\BibitemShut {NoStop}\bibitem [{\citenamefont {Postnikov}\ \emph {et~al.}(2010)\citenamefont
  {Postnikov}, \citenamefont {Prakash},\ and\ \citenamefont
  {Lattimer}}]{Postnikov2010}\BibitemOpen
  \bibfield  {author} {\bibinfo {author} {\bibfnamefont {S.}~\bibnamefont
  {Postnikov}}, \bibinfo {author} {\bibfnamefont {M.}~\bibnamefont {Prakash}},\
  and\ \bibinfo {author} {\bibfnamefont {J.~M.}\ \bibnamefont {Lattimer}},\
  }\bibfield  {title} {\bibinfo {title} {Tidal love numbers of neutron and
  self-bound quark stars},\ }\href@noop {} {\bibfield  {journal} {\bibinfo
  {journal} {Phys. Rev. D}\ }\textbf {\bibinfo {volume} {82}},\ \bibinfo
  {pages} {024016} (\bibinfo {year} {2010})}\BibitemShut {NoStop}\bibitem [{\citenamefont {Damour}\ and\ \citenamefont
  {Nagar}(2009)}]{DamourNagar2009}\BibitemOpen
  \bibfield  {author} {\bibinfo {author} {\bibfnamefont {T.}~\bibnamefont
  {Damour}}\ and\ \bibinfo {author} {\bibfnamefont {A.}~\bibnamefont {Nagar}},\
  }\bibfield  {title} {\bibinfo {title} {Relativistic tidal properties of
  neutron stars},\ }\href@noop {} {\bibfield  {journal} {\bibinfo  {journal}
  {Phys. Rev. D}\ }\textbf {\bibinfo {volume} {80}},\ \bibinfo {pages} {084035}
  (\bibinfo {year} {2009})}\BibitemShut {NoStop}\bibitem [{\citenamefont {Bradbury}\ \emph {et~al.}(2021)\citenamefont
  {Bradbury}, \citenamefont {Frostig}, \citenamefont {Hawkins}, \citenamefont
  {Johnson}, \citenamefont {Leary}, \citenamefont {Maclaurin}, \citenamefont
  {Necula}, \citenamefont {Paszke}, \citenamefont {VanderPlas}, \citenamefont
  {Wanderman-Milne},\ and\ \citenamefont {Zhang}}]{BradburyEtAl2021JAX}\BibitemOpen
  \bibfield  {author} {\bibinfo {author} {\bibfnamefont {J.}~\bibnamefont
  {Bradbury}}, \bibinfo {author} {\bibfnamefont {R.}~\bibnamefont {Frostig}},
  \bibinfo {author} {\bibfnamefont {P.}~\bibnamefont {Hawkins}}, \bibinfo
  {author} {\bibfnamefont {M.~J.}\ \bibnamefont {Johnson}}, \bibinfo {author}
  {\bibfnamefont {C.}~\bibnamefont {Leary}}, \bibinfo {author} {\bibfnamefont
  {D.}~\bibnamefont {Maclaurin}}, \bibinfo {author} {\bibfnamefont
  {G.}~\bibnamefont {Necula}}, \bibinfo {author} {\bibfnamefont
  {A.}~\bibnamefont {Paszke}}, \bibinfo {author} {\bibfnamefont
  {J.}~\bibnamefont {VanderPlas}}, \bibinfo {author} {\bibfnamefont
  {S.}~\bibnamefont {Wanderman-Milne}},\ and\ \bibinfo {author} {\bibfnamefont
  {Q.}~\bibnamefont {Zhang}},\ }\href@noop {} {\bibinfo {title} {{JAX}:
  Autograd and {XLA}}},\ \bibinfo {howpublished}
  {\url{https://github.com/jax-ml/jax}} (\bibinfo {year} {2021})\BibitemShut
  {NoStop}\bibitem [{\citenamefont {Kidger}(2022)}]{Kidger2022diffrax}\BibitemOpen
  \bibfield  {author} {\bibinfo {author} {\bibfnamefont {P.}~\bibnamefont
  {Kidger}},\ }\bibfield  {title} {\bibinfo {title} {On neural differential
  equations},\ }\href {https://doi.org/10.48550/arXiv.2202.02435} {\bibfield
  {journal} {\bibinfo  {journal} {arXiv e-prints}\ ,\ \bibinfo {pages}
  {arXiv:2202.02435}} (\bibinfo {year} {2022})},\ \bibinfo {note} {dPhil
  thesis, University of Oxford (2021); {\tt diffrax} library},\ \Eprint
  {https://arxiv.org/abs/2202.02435} {arXiv:2202.02435} \BibitemShut {NoStop}\bibitem [{\citenamefont {{LIGO Scientific
  Collaboration}}(2020)}]{lalsuite-soft}\BibitemOpen
  \bibfield  {author} {\bibinfo {author} {\bibnamefont {{LIGO Scientific
  Collaboration}}},\ }\href {https://doi.org/10.7935/GT1W-FZ16} {\bibinfo
  {title} {{LALSuite}: {LIGO} scientific collaboration algorithm library
  suite}} (\bibinfo {year} {2020})\BibitemShut {NoStop}\bibitem [{\citenamefont {Kastaun}\ \emph {et~al.}(2021)\citenamefont
  {Kastaun}, \citenamefont {Kalinani},\ and\ \citenamefont
  {Ciolfi}}]{Kastaun2021reprimand}\BibitemOpen
  \bibfield  {author} {\bibinfo {author} {\bibfnamefont {W.}~\bibnamefont
  {Kastaun}}, \bibinfo {author} {\bibfnamefont {J.~V.}\ \bibnamefont
  {Kalinani}},\ and\ \bibinfo {author} {\bibfnamefont {R.}~\bibnamefont
  {Ciolfi}},\ }\bibfield  {title} {\bibinfo {title} {Robust recovery of
  primitive variables in relativistic ideal magnetohydrodynamics},\ }\href@noop
  {} {\bibfield  {journal} {\bibinfo  {journal} {Phys. Rev. D}\ }\textbf
  {\bibinfo {volume} {103}},\ \bibinfo {pages} {023018} (\bibinfo {year}
  {2021})},\ \bibinfo {note} {rePrimAnd library}\BibitemShut {NoStop}\bibitem [{\citenamefont {Mroczek}\ \emph {et~al.}(2024)\citenamefont
  {Mroczek}, \citenamefont {Miller}, \citenamefont {Noronha-Hostler},\ and\
  \citenamefont {Yunes}}]{Mroczek2023}\BibitemOpen
  \bibfield  {author} {\bibinfo {author} {\bibfnamefont {D.}~\bibnamefont
  {Mroczek}}, \bibinfo {author} {\bibfnamefont {M.~C.}\ \bibnamefont {Miller}},
  \bibinfo {author} {\bibfnamefont {J.}~\bibnamefont {Noronha-Hostler}},\ and\
  \bibinfo {author} {\bibfnamefont {N.}~\bibnamefont {Yunes}},\ }\bibfield
  {title} {\bibinfo {title} {Nontrivial features in the speed of sound inside
  neutron stars},\ }\href {https://doi.org/10.1103/PhysRevD.110.123009}
  {\bibfield  {journal} {\bibinfo  {journal} {Phys. Rev. D}\ }\textbf {\bibinfo
  {volume} {110}},\ \bibinfo {pages} {123009} (\bibinfo {year} {2024})},\
  \Eprint {https://arxiv.org/abs/2309.02345} {arXiv:2309.02345} \BibitemShut
  {NoStop}\bibitem [{\citenamefont {{O'Shaughnessy}}\ and\ \citenamefont
  {{Zhu}}(2026)}]{OShaughnessyZhu2026prior}\BibitemOpen
  \bibfield  {author} {\bibinfo {author} {\bibfnamefont {R.}~\bibnamefont
  {{O'Shaughnessy}}}\ and\ \bibinfo {author} {\bibfnamefont {Z.}~\bibnamefont
  {{Zhu}}},\ }\bibfield  {title} {\bibinfo {title} {{Assessing the impact of
  smoothing priors on nonparametric inference of the nuclear equation of
  state}},\ }\href@noop {} {\bibfield  {journal} {\bibinfo  {journal} {in
  preparation}\ } (\bibinfo {year} {2026})},\ \bibinfo {note} {companion
  paper}\BibitemShut {NoStop}\bibitem [{\citenamefont {Komoltsev}\ and\ \citenamefont
  {Kurkela}(2022)}]{KomoltsevKurkela2022}\BibitemOpen
  \bibfield  {author} {\bibinfo {author} {\bibfnamefont {O.}~\bibnamefont
  {Komoltsev}}\ and\ \bibinfo {author} {\bibfnamefont {A.}~\bibnamefont
  {Kurkela}},\ }\bibfield  {title} {\bibinfo {title} {How perturbative {QCD}
  constrains the equation of state at neutron-star densities},\ }\href@noop {}
  {\bibfield  {journal} {\bibinfo  {journal} {Phys. Rev. Lett.}\ }\textbf
  {\bibinfo {volume} {128}},\ \bibinfo {pages} {202701} (\bibinfo {year}
  {2022})},\ \Eprint {https://arxiv.org/abs/2111.05350} {arXiv:2111.05350}
  \BibitemShut {NoStop}\bibitem [{\citenamefont {Gorda}\ \emph {et~al.}(2021)\citenamefont {Gorda},
  \citenamefont {Kurkela}, \citenamefont {Paatelainen}, \citenamefont
  {S{\"a}ppi},\ and\ \citenamefont {Vuorinen}}]{Gorda2021}\BibitemOpen
  \bibfield  {author} {\bibinfo {author} {\bibfnamefont {T.}~\bibnamefont
  {Gorda}}, \bibinfo {author} {\bibfnamefont {A.}~\bibnamefont {Kurkela}},
  \bibinfo {author} {\bibfnamefont {R.}~\bibnamefont {Paatelainen}}, \bibinfo
  {author} {\bibfnamefont {S.}~\bibnamefont {S{\"a}ppi}},\ and\ \bibinfo
  {author} {\bibfnamefont {A.}~\bibnamefont {Vuorinen}},\ }\bibfield  {title}
  {\bibinfo {title} {Soft interactions in cold quark matter},\ }\href@noop {}
  {\bibfield  {journal} {\bibinfo  {journal} {Phys. Rev. Lett.}\ }\textbf
  {\bibinfo {volume} {127}},\ \bibinfo {pages} {162003} (\bibinfo {year}
  {2021})}\BibitemShut {NoStop}\bibitem [{\citenamefont {Gorda}\ \emph {et~al.}(2023)\citenamefont {Gorda},
  \citenamefont {Komoltsev},\ and\ \citenamefont {Kurkela}}]{Gorda2023pQCD}\BibitemOpen
  \bibfield  {author} {\bibinfo {author} {\bibfnamefont {T.}~\bibnamefont
  {Gorda}}, \bibinfo {author} {\bibfnamefont {O.}~\bibnamefont {Komoltsev}},\
  and\ \bibinfo {author} {\bibfnamefont {A.}~\bibnamefont {Kurkela}},\
  }\bibfield  {title} {\bibinfo {title} {Ab-initio {QCD} calculations impact
  the inference of the neutron-star-matter equation of state},\ }\href@noop {}
  {\bibfield  {journal} {\bibinfo  {journal} {Astrophys. J.}\ }\textbf
  {\bibinfo {volume} {950}},\ \bibinfo {pages} {107} (\bibinfo {year}
  {2023})},\ \Eprint {https://arxiv.org/abs/2204.11877} {arXiv:2204.11877}
  \BibitemShut {NoStop}\bibitem [{\citenamefont {Baym}\ \emph {et~al.}(1971)\citenamefont {Baym},
  \citenamefont {Pethick},\ and\ \citenamefont {Sutherland}}]{BPS1971}\BibitemOpen
  \bibfield  {author} {\bibinfo {author} {\bibfnamefont {G.}~\bibnamefont
  {Baym}}, \bibinfo {author} {\bibfnamefont {C.}~\bibnamefont {Pethick}},\ and\
  \bibinfo {author} {\bibfnamefont {P.}~\bibnamefont {Sutherland}},\ }\bibfield
   {title} {\bibinfo {title} {The ground state of matter at high densities:
  Equation of state and stellar models},\ }\href
  {https://doi.org/10.1086/151216} {\bibfield  {journal} {\bibinfo  {journal}
  {\apj}\ }\textbf {\bibinfo {volume} {170}},\ \bibinfo {pages} {299} (\bibinfo
  {year} {1971})},\ \bibinfo {note} {outer-crust EOS; the fixed low-density
  sector of GKKS2025bridge}\BibitemShut {NoStop}\bibitem [{\citenamefont {Wouters}\ \emph {et~al.}(2025)\citenamefont
  {Wouters}, \citenamefont {Pang}, \citenamefont {Koehn}, \citenamefont {Rose},
  \citenamefont {Somasundaram}, \citenamefont {Tews}, \citenamefont
  {Dietrich},\ and\ \citenamefont {Van Den~Broeck}}]{WoutersEtAl2025JESTER}\BibitemOpen
  \bibfield  {author} {\bibinfo {author} {\bibfnamefont {T.}~\bibnamefont
  {Wouters}}, \bibinfo {author} {\bibfnamefont {P.~T.~H.}\ \bibnamefont
  {Pang}}, \bibinfo {author} {\bibfnamefont {H.}~\bibnamefont {Koehn}},
  \bibinfo {author} {\bibfnamefont {H.}~\bibnamefont {Rose}}, \bibinfo {author}
  {\bibfnamefont {R.}~\bibnamefont {Somasundaram}}, \bibinfo {author}
  {\bibfnamefont {I.}~\bibnamefont {Tews}}, \bibinfo {author} {\bibfnamefont
  {T.}~\bibnamefont {Dietrich}},\ and\ \bibinfo {author} {\bibfnamefont
  {C.}~\bibnamefont {Van Den~Broeck}},\ }\bibfield  {title} {\bibinfo {title}
  {Leveraging differentiable programming in the inverse problem of neutron
  stars},\ }\href {https://doi.org/10.1103/v2y8-kxvx} {\bibfield  {journal}
  {\bibinfo  {journal} {Phys. Rev. D}\ }\textbf {\bibinfo {volume} {112}},\
  \bibinfo {pages} {043037} (\bibinfo {year} {2025})},\ \Eprint
  {https://arxiv.org/abs/2504.15893} {arXiv:2504.15893 [astro-ph.HE]}
  \BibitemShut {NoStop}\bibitem [{\citenamefont
  {O'Shaughnessy}(2013)}]{OShaughnessy2013ModelDimension}\BibitemOpen
  \bibfield  {author} {\bibinfo {author} {\bibfnamefont {R.}~\bibnamefont
  {O'Shaughnessy}},\ }\bibfield  {title} {\bibinfo {title} {Comparing compact
  binary parameter distributions i: Methods},\ }\href
  {https://doi.org/10.1103/PhysRevD.88.084061} {\bibfield  {journal} {\bibinfo
  {journal} {Phys. Rev. D}\ }\textbf {\bibinfo {volume} {88}},\ \bibinfo
  {pages} {084061} (\bibinfo {year} {2013})},\ \Eprint
  {https://arxiv.org/abs/1204.3117} {arXiv:1204.3117 [astro-ph.CO]}
  \BibitemShut {NoStop}\bibitem [{\citenamefont {{Pankow}}\ \emph {et~al.}(2015)\citenamefont
  {{Pankow}}, \citenamefont {{Brady}}, \citenamefont {{Ochsner}},\ and\
  \citenamefont {{O'Shaughnessy}}}]{gwastro-PE-AlternativeArchitectures}\BibitemOpen
  \bibfield  {author} {\bibinfo {author} {\bibfnamefont {C.}~\bibnamefont
  {{Pankow}}}, \bibinfo {author} {\bibfnamefont {P.}~\bibnamefont {{Brady}}},
  \bibinfo {author} {\bibfnamefont {E.}~\bibnamefont {{Ochsner}}},\ and\
  \bibinfo {author} {\bibfnamefont {R.}~\bibnamefont {{O'Shaughnessy}}},\
  }\bibfield  {title} {\bibinfo {title} {{Novel scheme for rapid parallel
  parameter estimation of gravitational waves from compact binary
  coalescences}},\ }\href {https://doi.org/10.1103/PhysRevD.92.023002}
  {\bibfield  {journal} {\bibinfo  {journal} {\prd}\ }\textbf {\bibinfo
  {volume} {92}},\ \bibinfo {eid} {023002} (\bibinfo {year} {2015})},\ \Eprint
  {https://arxiv.org/abs/1502.04370} {arXiv:1502.04370 [gr-qc]} \BibitemShut
  {NoStop}\bibitem [{\citenamefont {Minka}(2001)}]{Minka2001EP}\BibitemOpen
  \bibfield  {author} {\bibinfo {author} {\bibfnamefont {T.~P.}\ \bibnamefont
  {Minka}},\ }\bibfield  {title} {\bibinfo {title} {Expectation propagation for
  approximate {B}ayesian inference},\ }in\ \href@noop {} {\emph {\bibinfo
  {booktitle} {Proceedings of the Seventeenth Conference on Uncertainty in
  Artificial Intelligence (UAI 2001)}}}\ (\bibinfo {year} {2001})\ pp.\
  \bibinfo {pages} {362--369},\ \Eprint {https://arxiv.org/abs/1301.2294}
  {arXiv:1301.2294 [cs.AI]} \BibitemShut {NoStop}\bibitem [{\citenamefont {Ionides}(2008)}]{Ionides2008TruncatedIS}\BibitemOpen
  \bibfield  {author} {\bibinfo {author} {\bibfnamefont {E.~L.}\ \bibnamefont
  {Ionides}},\ }\bibfield  {title} {\bibinfo {title} {Truncated importance
  sampling},\ }\href {https://doi.org/10.1198/106186008X320456} {\bibfield
  {journal} {\bibinfo  {journal} {Journal of Computational and Graphical
  Statistics}\ }\textbf {\bibinfo {volume} {17}},\ \bibinfo {pages} {295}
  (\bibinfo {year} {2008})}\BibitemShut {NoStop}\bibitem [{\citenamefont {Vehtari}\ \emph {et~al.}(2024)\citenamefont
  {Vehtari}, \citenamefont {Simpson}, \citenamefont {Gelman}, \citenamefont
  {Yao},\ and\ \citenamefont {Gabry}}]{Vehtari2024PSIS}\BibitemOpen
  \bibfield  {author} {\bibinfo {author} {\bibfnamefont {A.}~\bibnamefont
  {Vehtari}}, \bibinfo {author} {\bibfnamefont {D.}~\bibnamefont {Simpson}},
  \bibinfo {author} {\bibfnamefont {A.}~\bibnamefont {Gelman}}, \bibinfo
  {author} {\bibfnamefont {Y.}~\bibnamefont {Yao}},\ and\ \bibinfo {author}
  {\bibfnamefont {J.}~\bibnamefont {Gabry}},\ }\bibfield  {title} {\bibinfo
  {title} {Pareto smoothed importance sampling},\ }\href@noop {} {\bibfield
  {journal} {\bibinfo  {journal} {Journal of Machine Learning Research}\
  }\textbf {\bibinfo {volume} {25}},\ \bibinfo {pages} {1} (\bibinfo {year}
  {2024})},\ \Eprint {https://arxiv.org/abs/1507.02646} {arXiv:1507.02646
  [stat.CO]} \BibitemShut {NoStop}\bibitem [{\citenamefont {Dietrich}\ \emph {et~al.}(2019)\citenamefont
  {Dietrich}, \citenamefont {Samajdar}, \citenamefont {Khan}, \citenamefont
  {Johnson-McDaniel}, \citenamefont {Dudi},\ and\ \citenamefont
  {Tichy}}]{DietrichEtAl2019NRTidal}\BibitemOpen
  \bibfield  {author} {\bibinfo {author} {\bibfnamefont {T.}~\bibnamefont
  {Dietrich}}, \bibinfo {author} {\bibfnamefont {A.}~\bibnamefont {Samajdar}},
  \bibinfo {author} {\bibfnamefont {S.}~\bibnamefont {Khan}}, \bibinfo {author}
  {\bibfnamefont {N.~K.}\ \bibnamefont {Johnson-McDaniel}}, \bibinfo {author}
  {\bibfnamefont {R.}~\bibnamefont {Dudi}},\ and\ \bibinfo {author}
  {\bibfnamefont {W.}~\bibnamefont {Tichy}},\ }\bibfield  {title} {\bibinfo
  {title} {Improving the {NRTidal} model for binary neutron star systems},\
  }\href {https://doi.org/10.1103/PhysRevD.100.044003} {\bibfield  {journal}
  {\bibinfo  {journal} {Phys. Rev. D}\ }\textbf {\bibinfo {volume} {100}},\
  \bibinfo {pages} {044003} (\bibinfo {year} {2019})},\ \Eprint
  {https://arxiv.org/abs/1905.06011} {arXiv:1905.06011} \BibitemShut {NoStop}\bibitem [{\citenamefont {{Aasi}}\ \emph {et~al.}(2015)\citenamefont {{Aasi}}
  \emph {et~al.}}]{2015CQGra..32g4001T}\BibitemOpen
  \bibfield  {author} {\bibinfo {author} {\bibfnamefont {J.}~\bibnamefont
  {{Aasi}}} \emph {et~al.} (\bibinfo {collaboration} {LIGO Scientific
  Collaboration}),\ }\bibfield  {title} {\bibinfo {title} {{Advanced LIGO}},\
  }\href {https://doi.org/10.1088/0264-9381/32/7/074001} {\bibfield  {journal}
  {\bibinfo  {journal} {\cqg}\ }\textbf {\bibinfo {volume} {32}},\ \bibinfo
  {eid} {074001} (\bibinfo {year} {2015})},\ \Eprint
  {https://arxiv.org/abs/1411.4547} {arXiv:1411.4547 [gr-qc]} \BibitemShut
  {NoStop}\bibitem [{\citenamefont {{Acernese}}\ \emph {et~al.}(2015)\citenamefont
  {{Acernese}} \emph {et~al.}}]{2015CQGra..32b4001A}\BibitemOpen
  \bibfield  {author} {\bibinfo {author} {\bibfnamefont {F.}~\bibnamefont
  {{Acernese}}} \emph {et~al.} (\bibinfo {collaboration} {Virgo
  Collaboration}),\ }\bibfield  {title} {\bibinfo {title} {{Advanced Virgo: a
  second-generation interferometric gravitational wave detector}},\ }\href
  {https://doi.org/10.1088/0264-9381/32/2/024001} {\bibfield  {journal}
  {\bibinfo  {journal} {Classical and Quantum Gravity}\ }\textbf {\bibinfo
  {volume} {32}},\ \bibinfo {eid} {024001} (\bibinfo {year} {2015})},\ \Eprint
  {https://arxiv.org/abs/1408.3978} {arXiv:1408.3978 [gr-qc]} \BibitemShut
  {NoStop}\bibitem [{\citenamefont {{Ashton}}\ \emph {et~al.}(2019)\citenamefont
  {{Ashton}}, \citenamefont {{H{\"u}bner}}, \citenamefont {{Lasky}},
  \citenamefont {{Talbot}}, \citenamefont {{Ackley}}, \citenamefont
  {{Biscoveanu}}, \citenamefont {{Chu}}, \citenamefont {{Divakarla}},
  \citenamefont {{Easter}}, \citenamefont {{Goncharov}}, \citenamefont
  {{Hernandez Vivanco}}, \citenamefont {{Harms}}, \citenamefont {{Lower}},
  \citenamefont {{Meadors}}, \citenamefont {{Melchor}}, \citenamefont
  {{Payne}}, \citenamefont {{Pitkin}}, \citenamefont {{Powell}}, \citenamefont
  {{Sarin}}, \citenamefont {{Smith}},\ and\ \citenamefont
  {{Thrane}}}]{gwastro-pe-bilby-2018}\BibitemOpen
  \bibfield  {author} {\bibinfo {author} {\bibfnamefont {G.}~\bibnamefont
  {{Ashton}}}, \bibinfo {author} {\bibfnamefont {M.}~\bibnamefont
  {{H{\"u}bner}}}, \bibinfo {author} {\bibfnamefont {P.~D.}\ \bibnamefont
  {{Lasky}}}, \bibinfo {author} {\bibfnamefont {C.}~\bibnamefont {{Talbot}}},
  \bibinfo {author} {\bibfnamefont {K.}~\bibnamefont {{Ackley}}}, \bibinfo
  {author} {\bibfnamefont {S.}~\bibnamefont {{Biscoveanu}}}, \bibinfo {author}
  {\bibfnamefont {Q.}~\bibnamefont {{Chu}}}, \bibinfo {author} {\bibfnamefont
  {A.}~\bibnamefont {{Divakarla}}}, \bibinfo {author} {\bibfnamefont {P.~J.}\
  \bibnamefont {{Easter}}}, \bibinfo {author} {\bibfnamefont {B.}~\bibnamefont
  {{Goncharov}}}, \bibinfo {author} {\bibfnamefont {F.}~\bibnamefont
  {{Hernandez Vivanco}}}, \bibinfo {author} {\bibfnamefont {J.}~\bibnamefont
  {{Harms}}}, \bibinfo {author} {\bibfnamefont {M.~E.}\ \bibnamefont
  {{Lower}}}, \bibinfo {author} {\bibfnamefont {G.~D.}\ \bibnamefont
  {{Meadors}}}, \bibinfo {author} {\bibfnamefont {D.}~\bibnamefont
  {{Melchor}}}, \bibinfo {author} {\bibfnamefont {E.}~\bibnamefont {{Payne}}},
  \bibinfo {author} {\bibfnamefont {M.~D.}\ \bibnamefont {{Pitkin}}}, \bibinfo
  {author} {\bibfnamefont {J.}~\bibnamefont {{Powell}}}, \bibinfo {author}
  {\bibfnamefont {N.}~\bibnamefont {{Sarin}}}, \bibinfo {author} {\bibfnamefont
  {R.~J.~E.}\ \bibnamefont {{Smith}}},\ and\ \bibinfo {author} {\bibfnamefont
  {E.}~\bibnamefont {{Thrane}}},\ }\bibfield  {title} {\bibinfo {title}
  {{BILBY: A User-friendly Bayesian Inference Library for Gravitational-wave
  Astronomy}},\ }\href {https://doi.org/10.3847/1538-4365/ab06fc} {\bibfield
  {journal} {\bibinfo  {journal} {\apjs}\ }\textbf {\bibinfo {volume} {241}},\
  \bibinfo {eid} {27} (\bibinfo {year} {2019})},\ \Eprint
  {https://arxiv.org/abs/1811.02042} {arXiv:1811.02042 [astro-ph.IM]}
  \BibitemShut {NoStop}\bibitem [{\citenamefont {{Akutsu}}\ \emph {et~al.}(2021)\citenamefont
  {{Akutsu}}, \citenamefont {{Ando}}, \citenamefont {{Arai}}, \citenamefont
  {{Arai}}, \citenamefont {{Araki}}, \citenamefont {{Araya}}, \citenamefont
  {{Aritomi}}, \citenamefont {{Asada}}, \citenamefont {{Aso}}, \citenamefont
  {{Bae}}, \citenamefont {{Bae}}, \citenamefont {{Baiotti}}, \citenamefont
  {{Bajpai}}, \citenamefont {{Barton}}, \citenamefont {{Cannon}}, \citenamefont
  {{Cao}}, \citenamefont {{Capocasa}}, \citenamefont {{Chan}}, \citenamefont
  {{Chen}}, \citenamefont {{Chen}}, \citenamefont {{Chen}}, \citenamefont
  {{Chiang}}, \citenamefont {{Chu}}, \citenamefont {{Chu}}, \citenamefont
  {{Eguchi}}, \citenamefont {{Enomoto}}, \citenamefont {{Flaminio}},
  \citenamefont {{Fujii}}, \citenamefont {{Fujikawa}}, \citenamefont
  {{Fukunaga}}, \citenamefont {{Fukushima}}, \citenamefont {{Gao}},
  \citenamefont {{Ge}}, \citenamefont {{Ha}}, \citenamefont {{Hagiwara}},
  \citenamefont {{Haino}}, \citenamefont {{Han}}, \citenamefont {{Hasegawa}},
  \citenamefont {{Hattori}}, \citenamefont {{Hayakawa}}, \citenamefont
  {{Hayama}}, \citenamefont {{Himemoto}}, \citenamefont {{Hiranuma}},
  \citenamefont {{Hirata}}, \citenamefont {{Hirose}}, \citenamefont {{Hong}},
  \citenamefont {{Hsieh}}, \citenamefont {{Huang}}, \citenamefont {{Huang}},
  \citenamefont {{Huang}}, \citenamefont {{Huang}}, \citenamefont {{Huang}},
  \citenamefont {{Hui}}, \citenamefont {{Ide}}, \citenamefont {{Ikenoue}},
  \citenamefont {{Imam}}, \citenamefont {{Inayoshi}}, \citenamefont {{Inoue}},
  \citenamefont {{Ioka}}, \citenamefont {{Ito}}, \citenamefont {{Itoh}},
  \citenamefont {{Izumi}}, \citenamefont {{Jeon}}, \citenamefont {{Jin}},
  \citenamefont {{Jung}}, \citenamefont {{Jung}}, \citenamefont {{Kaihotsu}},
  \citenamefont {{Kajita}}, \citenamefont {{Kakizaki}}, \citenamefont
  {{Kamiizumi}}, \citenamefont {{Kanda}}, \citenamefont {{Kang}}, \citenamefont
  {{Kashiyama}}, \citenamefont {{Kawaguchi}}, \citenamefont {{Kawai}},
  \citenamefont {{Kawasaki}}, \citenamefont {{Kim}}, \citenamefont {{Kim}},
  \citenamefont {{Kim}}, \citenamefont {{Kim}}, \citenamefont {{Kim}},
  \citenamefont {{Kimura}}, \citenamefont {{Kita}}, \citenamefont {{Kitazawa}},
  \citenamefont {{Kojima}}, \citenamefont {{Kokeyama}}, \citenamefont
  {{Komori}}, \citenamefont {{Kong}}, \citenamefont {{Kotake}}, \citenamefont
  {{Kozakai}}, \citenamefont {{Kozu}}, \citenamefont {{Kumar}}, \citenamefont
  {{Kume}}, \citenamefont {{Kuo}}, \citenamefont {{Kuo}}, \citenamefont
  {{Kuromiya}}, \citenamefont {{Kuroyanagi}}, \citenamefont {{Kusayanagi}},
  \citenamefont {{Kwak}}, \citenamefont {{Lee}}, \citenamefont {{Lee}},
  \citenamefont {{Lee}}, \citenamefont {{Leonardi}}, \citenamefont {{Li}},
  \citenamefont {{Li}}, \citenamefont {{Lin}}, \citenamefont {{Lin}},
  \citenamefont {{Lin}}, \citenamefont {{Lin}}, \citenamefont {{Lin}},
  \citenamefont {{Liu}}, \citenamefont {{Luo}}, \citenamefont {{Majorana}},
  \citenamefont {{Marchio}}, \citenamefont {{Michimura}}, \citenamefont
  {{Mio}}, \citenamefont {{Miyakawa}}, \citenamefont {{Miyamoto}},
  \citenamefont {{Miyazaki}}, \citenamefont {{Miyo}}, \citenamefont {{Miyoki}},
  \citenamefont {{Mori}}, \citenamefont {{Morisaki}}, \citenamefont
  {{Moriwaki}}, \citenamefont {{Nagano}}, \citenamefont {{Nagano}},
  \citenamefont {{Nakamura}}, \citenamefont {{Nakano}}, \citenamefont
  {{Nakano}}, \citenamefont {{Nakashima}}, \citenamefont {{Nakayama}},
  \citenamefont {{Narikawa}}, \citenamefont {{Naticchioni}}, \citenamefont
  {{Negishi}}, \citenamefont {{Nguyen Quynh}}, \citenamefont {{Ni}},
  \citenamefont {{Nishizawa}}, \citenamefont {{Nozaki}}, \citenamefont
  {{Obuchi}}, \citenamefont {{Ogaki}}, \citenamefont {{Oh}}, \citenamefont
  {{Oh}}, \citenamefont {{Oh}}, \citenamefont {{Ohashi}}, \citenamefont
  {{Ohishi}}, \citenamefont {{Ohkawa}}, \citenamefont {{Ohta}}, \citenamefont
  {{Okutani}}, \citenamefont {{Okutomi}}, \citenamefont {{Oohara}},
  \citenamefont {{Ooi}}, \citenamefont {{Oshino}}, \citenamefont {{Otabe}},
  \citenamefont {{Pan}}, \citenamefont {{Pang}}, \citenamefont {{Parisi}},
  \citenamefont {{Park}}, \citenamefont {{Pe na Arellano}}, \citenamefont
  {{Pinto}}, \citenamefont {{Sago}}, \citenamefont {{Saito}}, \citenamefont
  {{Saito}}, \citenamefont {{Sakai}}, \citenamefont {{Sakai}}, \citenamefont
  {{Sakuno}}, \citenamefont {{Sato}}, \citenamefont {{Sato}}, \citenamefont
  {{Sawada}}, \citenamefont {{Sekiguchi}}, \citenamefont {{Sekiguchi}},
  \citenamefont {{Shao}}, \citenamefont {{Shibagaki}}, \citenamefont
  {{Shimizu}}, \citenamefont {{Shimoda}}, \citenamefont {{Shimode}},
  \citenamefont {{Shinkai}}, \citenamefont {{Shishido}}, \citenamefont
  {{Shoda}}, \citenamefont {{Somiya}}, \citenamefont {{Son}}, \citenamefont
  {{Sotani}}, \citenamefont {{Sugimoto}}, \citenamefont {{Suresh}},
  \citenamefont {{Suzuki}}, \citenamefont {{Suzuki}}, \citenamefont
  {{Tagoshi}}, \citenamefont {{Takahashi}}, \citenamefont {{Takahashi}},
  \citenamefont {{Takamori}}, \citenamefont {{Takano}}, \citenamefont
  {{Takeda}}, \citenamefont {{Takeda}}, \citenamefont {{Tanaka}}, \citenamefont
  {{Tanaka}}, \citenamefont {{Tanaka}}, \citenamefont {{Tanaka}}, \citenamefont
  {{Tanaka}}, \citenamefont {{Tanioka}}, \citenamefont {{Tapia San Martin}},\
  and\ \citenamefont {{Telada}}}]{2021PTEP.2021eA103A}\BibitemOpen
  \bibfield  {author} {\bibinfo {author} {\bibfnamefont {T.}~\bibnamefont
  {{Akutsu}}}, \bibinfo {author} {\bibfnamefont {M.}~\bibnamefont {{Ando}}},
  \bibinfo {author} {\bibfnamefont {K.}~\bibnamefont {{Arai}}}, \bibinfo
  {author} {\bibfnamefont {Y.}~\bibnamefont {{Arai}}}, \bibinfo {author}
  {\bibfnamefont {S.}~\bibnamefont {{Araki}}}, \bibinfo {author} {\bibfnamefont
  {A.}~\bibnamefont {{Araya}}}, \bibinfo {author} {\bibfnamefont
  {N.}~\bibnamefont {{Aritomi}}}, \bibinfo {author} {\bibfnamefont
  {H.}~\bibnamefont {{Asada}}}, \bibinfo {author} {\bibfnamefont
  {Y.}~\bibnamefont {{Aso}}}, \bibinfo {author} {\bibfnamefont
  {S.}~\bibnamefont {{Bae}}}, \bibinfo {author} {\bibfnamefont
  {Y.}~\bibnamefont {{Bae}}}, \bibinfo {author} {\bibfnamefont
  {L.}~\bibnamefont {{Baiotti}}}, \bibinfo {author} {\bibfnamefont
  {R.}~\bibnamefont {{Bajpai}}}, \bibinfo {author} {\bibfnamefont {M.~A.}\
  \bibnamefont {{Barton}}}, \bibinfo {author} {\bibfnamefont {K.}~\bibnamefont
  {{Cannon}}}, \bibinfo {author} {\bibfnamefont {Z.}~\bibnamefont {{Cao}}},
  \bibinfo {author} {\bibfnamefont {E.}~\bibnamefont {{Capocasa}}}, \bibinfo
  {author} {\bibfnamefont {M.}~\bibnamefont {{Chan}}}, \bibinfo {author}
  {\bibfnamefont {C.}~\bibnamefont {{Chen}}}, \bibinfo {author} {\bibfnamefont
  {K.}~\bibnamefont {{Chen}}}, \bibinfo {author} {\bibfnamefont
  {Y.}~\bibnamefont {{Chen}}}, \bibinfo {author} {\bibfnamefont {C.-Y.}\
  \bibnamefont {{Chiang}}}, \bibinfo {author} {\bibfnamefont {H.}~\bibnamefont
  {{Chu}}}, \bibinfo {author} {\bibfnamefont {Y.-K.}\ \bibnamefont {{Chu}}},
  \bibinfo {author} {\bibfnamefont {S.}~\bibnamefont {{Eguchi}}}, \bibinfo
  {author} {\bibfnamefont {Y.}~\bibnamefont {{Enomoto}}}, \bibinfo {author}
  {\bibfnamefont {R.}~\bibnamefont {{Flaminio}}}, \bibinfo {author}
  {\bibfnamefont {Y.}~\bibnamefont {{Fujii}}}, \bibinfo {author} {\bibfnamefont
  {F.}~\bibnamefont {{Fujikawa}}}, \bibinfo {author} {\bibfnamefont
  {M.}~\bibnamefont {{Fukunaga}}}, \bibinfo {author} {\bibfnamefont
  {M.}~\bibnamefont {{Fukushima}}}, \bibinfo {author} {\bibfnamefont
  {D.}~\bibnamefont {{Gao}}}, \bibinfo {author} {\bibfnamefont
  {G.}~\bibnamefont {{Ge}}}, \bibinfo {author} {\bibfnamefont {S.}~\bibnamefont
  {{Ha}}}, \bibinfo {author} {\bibfnamefont {A.}~\bibnamefont {{Hagiwara}}},
  \bibinfo {author} {\bibfnamefont {S.}~\bibnamefont {{Haino}}}, \bibinfo
  {author} {\bibfnamefont {W.-B.}\ \bibnamefont {{Han}}}, \bibinfo {author}
  {\bibfnamefont {K.}~\bibnamefont {{Hasegawa}}}, \bibinfo {author}
  {\bibfnamefont {K.}~\bibnamefont {{Hattori}}}, \bibinfo {author}
  {\bibfnamefont {H.}~\bibnamefont {{Hayakawa}}}, \bibinfo {author}
  {\bibfnamefont {K.}~\bibnamefont {{Hayama}}}, \bibinfo {author}
  {\bibfnamefont {Y.}~\bibnamefont {{Himemoto}}}, \bibinfo {author}
  {\bibfnamefont {Y.}~\bibnamefont {{Hiranuma}}}, \bibinfo {author}
  {\bibfnamefont {N.}~\bibnamefont {{Hirata}}}, \bibinfo {author}
  {\bibfnamefont {E.}~\bibnamefont {{Hirose}}}, \bibinfo {author}
  {\bibfnamefont {Z.}~\bibnamefont {{Hong}}}, \bibinfo {author} {\bibfnamefont
  {B.~H.}\ \bibnamefont {{Hsieh}}}, \bibinfo {author} {\bibfnamefont {C.-Z.}\
  \bibnamefont {{Huang}}}, \bibinfo {author} {\bibfnamefont {H.-Y.}\
  \bibnamefont {{Huang}}}, \bibinfo {author} {\bibfnamefont {P.}~\bibnamefont
  {{Huang}}}, \bibinfo {author} {\bibfnamefont {Y.}~\bibnamefont {{Huang}}},
  \bibinfo {author} {\bibfnamefont {Y.-C.}\ \bibnamefont {{Huang}}}, \bibinfo
  {author} {\bibfnamefont {D.~C.~Y.}\ \bibnamefont {{Hui}}}, \bibinfo {author}
  {\bibfnamefont {S.}~\bibnamefont {{Ide}}}, \bibinfo {author} {\bibfnamefont
  {B.}~\bibnamefont {{Ikenoue}}}, \bibinfo {author} {\bibfnamefont
  {S.}~\bibnamefont {{Imam}}}, \bibinfo {author} {\bibfnamefont
  {K.}~\bibnamefont {{Inayoshi}}}, \bibinfo {author} {\bibfnamefont
  {Y.}~\bibnamefont {{Inoue}}}, \bibinfo {author} {\bibfnamefont
  {K.}~\bibnamefont {{Ioka}}}, \bibinfo {author} {\bibfnamefont
  {K.}~\bibnamefont {{Ito}}}, \bibinfo {author} {\bibfnamefont
  {Y.}~\bibnamefont {{Itoh}}}, \bibinfo {author} {\bibfnamefont
  {K.}~\bibnamefont {{Izumi}}}, \bibinfo {author} {\bibfnamefont
  {C.}~\bibnamefont {{Jeon}}}, \bibinfo {author} {\bibfnamefont {H.-B.}\
  \bibnamefont {{Jin}}}, \bibinfo {author} {\bibfnamefont {K.}~\bibnamefont
  {{Jung}}}, \bibinfo {author} {\bibfnamefont {P.}~\bibnamefont {{Jung}}},
  \bibinfo {author} {\bibfnamefont {K.}~\bibnamefont {{Kaihotsu}}}, \bibinfo
  {author} {\bibfnamefont {T.}~\bibnamefont {{Kajita}}}, \bibinfo {author}
  {\bibfnamefont {M.}~\bibnamefont {{Kakizaki}}}, \bibinfo {author}
  {\bibfnamefont {M.}~\bibnamefont {{Kamiizumi}}}, \bibinfo {author}
  {\bibfnamefont {N.}~\bibnamefont {{Kanda}}}, \bibinfo {author} {\bibfnamefont
  {G.}~\bibnamefont {{Kang}}}, \bibinfo {author} {\bibfnamefont
  {K.}~\bibnamefont {{Kashiyama}}}, \bibinfo {author} {\bibfnamefont
  {K.}~\bibnamefont {{Kawaguchi}}}, \bibinfo {author} {\bibfnamefont
  {N.}~\bibnamefont {{Kawai}}}, \bibinfo {author} {\bibfnamefont
  {T.}~\bibnamefont {{Kawasaki}}}, \bibinfo {author} {\bibfnamefont
  {C.}~\bibnamefont {{Kim}}}, \bibinfo {author} {\bibfnamefont
  {J.}~\bibnamefont {{Kim}}}, \bibinfo {author} {\bibfnamefont {J.~C.}\
  \bibnamefont {{Kim}}}, \bibinfo {author} {\bibfnamefont {W.~S.}\ \bibnamefont
  {{Kim}}}, \bibinfo {author} {\bibfnamefont {Y.-M.}\ \bibnamefont {{Kim}}},
  \bibinfo {author} {\bibfnamefont {N.}~\bibnamefont {{Kimura}}}, \bibinfo
  {author} {\bibfnamefont {N.}~\bibnamefont {{Kita}}}, \bibinfo {author}
  {\bibfnamefont {H.}~\bibnamefont {{Kitazawa}}}, \bibinfo {author}
  {\bibfnamefont {Y.}~\bibnamefont {{Kojima}}}, \bibinfo {author}
  {\bibfnamefont {K.}~\bibnamefont {{Kokeyama}}}, \bibinfo {author}
  {\bibfnamefont {K.}~\bibnamefont {{Komori}}}, \bibinfo {author}
  {\bibfnamefont {A.~K.~H.}\ \bibnamefont {{Kong}}}, \bibinfo {author}
  {\bibfnamefont {K.}~\bibnamefont {{Kotake}}}, \bibinfo {author}
  {\bibfnamefont {C.}~\bibnamefont {{Kozakai}}}, \bibinfo {author}
  {\bibfnamefont {R.}~\bibnamefont {{Kozu}}}, \bibinfo {author} {\bibfnamefont
  {R.}~\bibnamefont {{Kumar}}}, \bibinfo {author} {\bibfnamefont
  {J.}~\bibnamefont {{Kume}}}, \bibinfo {author} {\bibfnamefont
  {C.}~\bibnamefont {{Kuo}}}, \bibinfo {author} {\bibfnamefont {H.-S.}\
  \bibnamefont {{Kuo}}}, \bibinfo {author} {\bibfnamefont {Y.}~\bibnamefont
  {{Kuromiya}}}, \bibinfo {author} {\bibfnamefont {S.}~\bibnamefont
  {{Kuroyanagi}}}, \bibinfo {author} {\bibfnamefont {K.}~\bibnamefont
  {{Kusayanagi}}}, \bibinfo {author} {\bibfnamefont {K.}~\bibnamefont
  {{Kwak}}}, \bibinfo {author} {\bibfnamefont {H.~K.}\ \bibnamefont {{Lee}}},
  \bibinfo {author} {\bibfnamefont {H.~W.}\ \bibnamefont {{Lee}}}, \bibinfo
  {author} {\bibfnamefont {R.}~\bibnamefont {{Lee}}}, \bibinfo {author}
  {\bibfnamefont {M.}~\bibnamefont {{Leonardi}}}, \bibinfo {author}
  {\bibfnamefont {K.~L.}\ \bibnamefont {{Li}}}, \bibinfo {author}
  {\bibfnamefont {T.~G.~F.}\ \bibnamefont {{Li}}}, \bibinfo {author}
  {\bibfnamefont {C.-Y.}\ \bibnamefont {{Lin}}}, \bibinfo {author}
  {\bibfnamefont {F.-K.}\ \bibnamefont {{Lin}}}, \bibinfo {author}
  {\bibfnamefont {F.-L.}\ \bibnamefont {{Lin}}}, \bibinfo {author}
  {\bibfnamefont {H.~L.}\ \bibnamefont {{Lin}}}, \bibinfo {author}
  {\bibfnamefont {L.~C.-C.}\ \bibnamefont {{Lin}}}, \bibinfo {author}
  {\bibfnamefont {G.~C.}\ \bibnamefont {{Liu}}}, \bibinfo {author}
  {\bibfnamefont {L.-W.}\ \bibnamefont {{Luo}}}, \bibinfo {author}
  {\bibfnamefont {E.}~\bibnamefont {{Majorana}}}, \bibinfo {author}
  {\bibfnamefont {M.}~\bibnamefont {{Marchio}}}, \bibinfo {author}
  {\bibfnamefont {Y.}~\bibnamefont {{Michimura}}}, \bibinfo {author}
  {\bibfnamefont {N.}~\bibnamefont {{Mio}}}, \bibinfo {author} {\bibfnamefont
  {O.}~\bibnamefont {{Miyakawa}}}, \bibinfo {author} {\bibfnamefont
  {A.}~\bibnamefont {{Miyamoto}}}, \bibinfo {author} {\bibfnamefont
  {Y.}~\bibnamefont {{Miyazaki}}}, \bibinfo {author} {\bibfnamefont
  {K.}~\bibnamefont {{Miyo}}}, \bibinfo {author} {\bibfnamefont
  {S.}~\bibnamefont {{Miyoki}}}, \bibinfo {author} {\bibfnamefont
  {Y.}~\bibnamefont {{Mori}}}, \bibinfo {author} {\bibfnamefont
  {S.}~\bibnamefont {{Morisaki}}}, \bibinfo {author} {\bibfnamefont
  {Y.}~\bibnamefont {{Moriwaki}}}, \bibinfo {author} {\bibfnamefont
  {K.}~\bibnamefont {{Nagano}}}, \bibinfo {author} {\bibfnamefont
  {S.}~\bibnamefont {{Nagano}}}, \bibinfo {author} {\bibfnamefont
  {K.}~\bibnamefont {{Nakamura}}}, \bibinfo {author} {\bibfnamefont
  {H.}~\bibnamefont {{Nakano}}}, \bibinfo {author} {\bibfnamefont
  {M.}~\bibnamefont {{Nakano}}}, \bibinfo {author} {\bibfnamefont
  {R.}~\bibnamefont {{Nakashima}}}, \bibinfo {author} {\bibfnamefont
  {Y.}~\bibnamefont {{Nakayama}}}, \bibinfo {author} {\bibfnamefont
  {T.}~\bibnamefont {{Narikawa}}}, \bibinfo {author} {\bibfnamefont
  {L.}~\bibnamefont {{Naticchioni}}}, \bibinfo {author} {\bibfnamefont
  {R.}~\bibnamefont {{Negishi}}}, \bibinfo {author} {\bibfnamefont
  {L.}~\bibnamefont {{Nguyen Quynh}}}, \bibinfo {author} {\bibfnamefont
  {W.-T.}\ \bibnamefont {{Ni}}}, \bibinfo {author} {\bibfnamefont
  {A.}~\bibnamefont {{Nishizawa}}}, \bibinfo {author} {\bibfnamefont
  {S.}~\bibnamefont {{Nozaki}}}, \bibinfo {author} {\bibfnamefont
  {Y.}~\bibnamefont {{Obuchi}}}, \bibinfo {author} {\bibfnamefont
  {W.}~\bibnamefont {{Ogaki}}}, \bibinfo {author} {\bibfnamefont {J.~J.}\
  \bibnamefont {{Oh}}}, \bibinfo {author} {\bibfnamefont {K.}~\bibnamefont
  {{Oh}}}, \bibinfo {author} {\bibfnamefont {S.~H.}\ \bibnamefont {{Oh}}},
  \bibinfo {author} {\bibfnamefont {M.}~\bibnamefont {{Ohashi}}}, \bibinfo
  {author} {\bibfnamefont {N.}~\bibnamefont {{Ohishi}}}, \bibinfo {author}
  {\bibfnamefont {M.}~\bibnamefont {{Ohkawa}}}, \bibinfo {author}
  {\bibfnamefont {H.}~\bibnamefont {{Ohta}}}, \bibinfo {author} {\bibfnamefont
  {Y.}~\bibnamefont {{Okutani}}}, \bibinfo {author} {\bibfnamefont
  {K.}~\bibnamefont {{Okutomi}}}, \bibinfo {author} {\bibfnamefont
  {K.}~\bibnamefont {{Oohara}}}, \bibinfo {author} {\bibfnamefont {C.~P.}\
  \bibnamefont {{Ooi}}}, \bibinfo {author} {\bibfnamefont {S.}~\bibnamefont
  {{Oshino}}}, \bibinfo {author} {\bibfnamefont {S.}~\bibnamefont {{Otabe}}},
  \bibinfo {author} {\bibfnamefont {K.}~\bibnamefont {{Pan}}}, \bibinfo
  {author} {\bibfnamefont {H.}~\bibnamefont {{Pang}}}, \bibinfo {author}
  {\bibfnamefont {A.}~\bibnamefont {{Parisi}}}, \bibinfo {author}
  {\bibfnamefont {J.}~\bibnamefont {{Park}}}, \bibinfo {author} {\bibfnamefont
  {F.~E.}\ \bibnamefont {{Pe na Arellano}}}, \bibinfo {author} {\bibfnamefont
  {I.}~\bibnamefont {{Pinto}}}, \bibinfo {author} {\bibfnamefont
  {N.}~\bibnamefont {{Sago}}}, \bibinfo {author} {\bibfnamefont
  {S.}~\bibnamefont {{Saito}}}, \bibinfo {author} {\bibfnamefont
  {Y.}~\bibnamefont {{Saito}}}, \bibinfo {author} {\bibfnamefont
  {K.}~\bibnamefont {{Sakai}}}, \bibinfo {author} {\bibfnamefont
  {Y.}~\bibnamefont {{Sakai}}}, \bibinfo {author} {\bibfnamefont
  {Y.}~\bibnamefont {{Sakuno}}}, \bibinfo {author} {\bibfnamefont
  {S.}~\bibnamefont {{Sato}}}, \bibinfo {author} {\bibfnamefont
  {T.}~\bibnamefont {{Sato}}}, \bibinfo {author} {\bibfnamefont
  {T.}~\bibnamefont {{Sawada}}}, \bibinfo {author} {\bibfnamefont
  {T.}~\bibnamefont {{Sekiguchi}}}, \bibinfo {author} {\bibfnamefont
  {Y.}~\bibnamefont {{Sekiguchi}}}, \bibinfo {author} {\bibfnamefont
  {L.}~\bibnamefont {{Shao}}}, \bibinfo {author} {\bibfnamefont
  {S.}~\bibnamefont {{Shibagaki}}}, \bibinfo {author} {\bibfnamefont
  {R.}~\bibnamefont {{Shimizu}}}, \bibinfo {author} {\bibfnamefont
  {T.}~\bibnamefont {{Shimoda}}}, \bibinfo {author} {\bibfnamefont
  {K.}~\bibnamefont {{Shimode}}}, \bibinfo {author} {\bibfnamefont
  {H.}~\bibnamefont {{Shinkai}}}, \bibinfo {author} {\bibfnamefont
  {T.}~\bibnamefont {{Shishido}}}, \bibinfo {author} {\bibfnamefont
  {A.}~\bibnamefont {{Shoda}}}, \bibinfo {author} {\bibfnamefont
  {K.}~\bibnamefont {{Somiya}}}, \bibinfo {author} {\bibfnamefont {E.~J.}\
  \bibnamefont {{Son}}}, \bibinfo {author} {\bibfnamefont {H.}~\bibnamefont
  {{Sotani}}}, \bibinfo {author} {\bibfnamefont {R.}~\bibnamefont
  {{Sugimoto}}}, \bibinfo {author} {\bibfnamefont {J.}~\bibnamefont
  {{Suresh}}}, \bibinfo {author} {\bibfnamefont {T.}~\bibnamefont {{Suzuki}}},
  \bibinfo {author} {\bibfnamefont {T.}~\bibnamefont {{Suzuki}}}, \bibinfo
  {author} {\bibfnamefont {H.}~\bibnamefont {{Tagoshi}}}, \bibinfo {author}
  {\bibfnamefont {H.}~\bibnamefont {{Takahashi}}}, \bibinfo {author}
  {\bibfnamefont {R.}~\bibnamefont {{Takahashi}}}, \bibinfo {author}
  {\bibfnamefont {A.}~\bibnamefont {{Takamori}}}, \bibinfo {author}
  {\bibfnamefont {S.}~\bibnamefont {{Takano}}}, \bibinfo {author}
  {\bibfnamefont {H.}~\bibnamefont {{Takeda}}}, \bibinfo {author}
  {\bibfnamefont {M.}~\bibnamefont {{Takeda}}}, \bibinfo {author}
  {\bibfnamefont {H.}~\bibnamefont {{Tanaka}}}, \bibinfo {author}
  {\bibfnamefont {K.}~\bibnamefont {{Tanaka}}}, \bibinfo {author}
  {\bibfnamefont {K.}~\bibnamefont {{Tanaka}}}, \bibinfo {author}
  {\bibfnamefont {T.}~\bibnamefont {{Tanaka}}}, \bibinfo {author}
  {\bibfnamefont {T.}~\bibnamefont {{Tanaka}}}, \bibinfo {author}
  {\bibfnamefont {S.}~\bibnamefont {{Tanioka}}}, \bibinfo {author}
  {\bibfnamefont {E.~N.}\ \bibnamefont {{Tapia San Martin}}},\ and\ \bibinfo
  {author} {\bibfnamefont {S.}~\bibnamefont {{Telada}}},\ }\bibfield  {title}
  {\bibinfo {title} {{Overview of KAGRA: KAGRA science}},\ }\href
  {https://doi.org/10.1093/ptep/ptaa120} {\bibfield  {journal} {\bibinfo
  {journal} {Progress of Theoretical and Experimental Physics}\ }\textbf
  {\bibinfo {volume} {2021}},\ \bibinfo {eid} {05A103} (\bibinfo {year}
  {2021})},\ \Eprint {https://arxiv.org/abs/2008.02921} {arXiv:2008.02921
  [gr-qc]} \BibitemShut {NoStop}\bibitem [{\citenamefont {{Yelikar}}\ \emph {et~al.}(2024)\citenamefont
  {{Yelikar}}, \citenamefont {{O'Shaughnessy}}, \citenamefont {{Wysocki}},\
  and\ \citenamefont {{Wade}}}]{gwastro-ns-eos-pop_systematics-Yelikar2024}\BibitemOpen
  \bibfield  {author} {\bibinfo {author} {\bibfnamefont {A.~B.}\ \bibnamefont
  {{Yelikar}}}, \bibinfo {author} {\bibfnamefont {R.}~\bibnamefont
  {{O'Shaughnessy}}}, \bibinfo {author} {\bibfnamefont {D.}~\bibnamefont
  {{Wysocki}}},\ and\ \bibinfo {author} {\bibfnamefont {L.}~\bibnamefont
  {{Wade}}},\ }\bibfield  {title} {\bibinfo {title} {{Effects of waveform
  systematics on inferences of neutron star population properties and the
  nuclear equation of state}},\ }\href
  {https://doi.org/10.48550/arXiv.2410.14674} {\bibfield  {journal} {\bibinfo
  {journal} {Submitted to PRD; available as arxiv:2410.14674}\ ,\ \bibinfo
  {eid} {arXiv:2410.14674}} (\bibinfo {year} {2024})},\ \Eprint
  {https://arxiv.org/abs/2410.14674} {arXiv:2410.14674 [gr-qc]} \BibitemShut
  {NoStop}\bibitem [{\citenamefont {{Anik}}\ \emph {et~al.}(2026)\citenamefont {{Anik}},
  \citenamefont {{Steiner}},\ and\ \citenamefont
  {{O'Shaughnessy}}}]{gwastro-nsnuc-AnikSteiner-EOSPopMaxMass-2025}\BibitemOpen
  \bibfield  {author} {\bibinfo {author} {\bibfnamefont {M.~H.}\ \bibnamefont
  {{Anik}}}, \bibinfo {author} {\bibfnamefont {A.~W.}\ \bibnamefont
  {{Steiner}}},\ and\ \bibinfo {author} {\bibfnamefont {R.}~\bibnamefont
  {{O'Shaughnessy}}},\ }\bibfield  {title} {\bibinfo {title} {{Inference of
  Neutron Star Mass Distributions and the Dense Matter Equation of State from
  Multimessenger Observations}},\ }\href
  {https://doi.org/10.3847/1538-4357/ae75e4} {\bibfield  {journal} {\bibinfo
  {journal} {\apj}\ }\textbf {\bibinfo {volume} {1005}},\ \bibinfo {pages}
  {148} (\bibinfo {year} {2026})},\ \Eprint {https://arxiv.org/abs/2512.12130}
  {arXiv:2512.12130 [astro-ph.HE]} \BibitemShut {NoStop}\bibitem [{\citenamefont {Sorensen}\ \emph {et~al.}(2024)\citenamefont
  {Sorensen}, \citenamefont {Agarwal}, \citenamefont {Brown}, \citenamefont
  {Chajecki}, \citenamefont {Danielewicz}, \citenamefont {Drischler},
  \citenamefont {Gandolfi}, \citenamefont {Holt}, \citenamefont {Kaminski},
  \citenamefont {Ko} \emph {et~al.}}]{SorensenEtAl2024}\BibitemOpen
  \bibfield  {author} {\bibinfo {author} {\bibfnamefont {A.}~\bibnamefont
  {Sorensen}}, \bibinfo {author} {\bibfnamefont {K.}~\bibnamefont {Agarwal}},
  \bibinfo {author} {\bibfnamefont {K.~W.}\ \bibnamefont {Brown}}, \bibinfo
  {author} {\bibfnamefont {Z.}~\bibnamefont {Chajecki}}, \bibinfo {author}
  {\bibfnamefont {P.}~\bibnamefont {Danielewicz}}, \bibinfo {author}
  {\bibfnamefont {C.}~\bibnamefont {Drischler}}, \bibinfo {author}
  {\bibfnamefont {S.}~\bibnamefont {Gandolfi}}, \bibinfo {author}
  {\bibfnamefont {J.~W.}\ \bibnamefont {Holt}}, \bibinfo {author}
  {\bibfnamefont {M.}~\bibnamefont {Kaminski}}, \bibinfo {author}
  {\bibfnamefont {C.-M.}\ \bibnamefont {Ko}}, \emph {et~al.},\ }\bibfield
  {title} {\bibinfo {title} {Dense nuclear matter equation of state from
  heavy-ion collisions},\ }\href {https://doi.org/10.1016/j.ppnp.2023.104080}
  {\bibfield  {journal} {\bibinfo  {journal} {Progress in Particle and Nuclear
  Physics}\ }\textbf {\bibinfo {volume} {134}},\ \bibinfo {pages} {104080}
  (\bibinfo {year} {2024})},\ \Eprint {https://arxiv.org/abs/2301.13253}
  {arXiv:2301.13253} \BibitemShut {NoStop}\bibitem [{\citenamefont {Brown}\ \emph {et~al.}(2025)\citenamefont {Brown},
  \citenamefont {Gade}, \citenamefont {Stroberg}, \citenamefont {Escher},
  \citenamefont {Fossez}, \citenamefont {Giuliani}, \citenamefont {Hoffman},
  \citenamefont {Nazarewicz}, \citenamefont {Seng}, \citenamefont {Sorensen}
  \emph {et~al.}}]{BrownEtAl2025}\BibitemOpen
  \bibfield  {author} {\bibinfo {author} {\bibfnamefont {B.~A.}\ \bibnamefont
  {Brown}}, \bibinfo {author} {\bibfnamefont {A.}~\bibnamefont {Gade}},
  \bibinfo {author} {\bibfnamefont {S.~R.}\ \bibnamefont {Stroberg}}, \bibinfo
  {author} {\bibfnamefont {J.}~\bibnamefont {Escher}}, \bibinfo {author}
  {\bibfnamefont {K.}~\bibnamefont {Fossez}}, \bibinfo {author} {\bibfnamefont
  {P.}~\bibnamefont {Giuliani}}, \bibinfo {author} {\bibfnamefont {C.~R.}\
  \bibnamefont {Hoffman}}, \bibinfo {author} {\bibfnamefont {W.}~\bibnamefont
  {Nazarewicz}}, \bibinfo {author} {\bibfnamefont {C.-Y.}\ \bibnamefont
  {Seng}}, \bibinfo {author} {\bibfnamefont {A.}~\bibnamefont {Sorensen}},
  \emph {et~al.},\ }\bibfield  {title} {\bibinfo {title} {Motivations for early
  high-profile {FRIB} experiments},\ }\href
  {https://doi.org/10.1088/1361-6471/adb449} {\bibfield  {journal} {\bibinfo
  {journal} {Journal of Physics G: Nuclear and Particle Physics}\ }\textbf
  {\bibinfo {volume} {52}},\ \bibinfo {pages} {050501} (\bibinfo {year}
  {2025})},\ \Eprint {https://arxiv.org/abs/2410.06144} {arXiv:2410.06144}
  \BibitemShut {NoStop}\bibitem [{\citenamefont {Le~F{\`e}vre}(2026)}]{LeFevre2026}\BibitemOpen
  \bibfield  {author} {\bibinfo {author} {\bibfnamefont {A.}~\bibnamefont
  {Le~F{\`e}vre}},\ }\bibfield  {title} {\bibinfo {title} {Nuclear
  equation-of-state at high density and multi-messenger astronomy: contribution
  of heavy-ion collisions},\ }\href {https://doi.org/10.48550/arXiv.2606.29588}
  {\bibfield  {journal} {\bibinfo  {journal} {arXiv e-prints}\ ,\ \bibinfo
  {pages} {arXiv:2606.29588}} (\bibinfo {year} {2026})},\ \Eprint
  {https://arxiv.org/abs/2606.29588} {arXiv:2606.29588} \BibitemShut {NoStop}\bibitem [{\citenamefont {Annala}\ \emph {et~al.}(2018)\citenamefont {Annala},
  \citenamefont {Gorda}, \citenamefont {Kurkela},\ and\ \citenamefont
  {Vuorinen}}]{gwastro-ns-eos-170817-Annala2018}\BibitemOpen
  \bibfield  {author} {\bibinfo {author} {\bibfnamefont {E.}~\bibnamefont
  {Annala}}, \bibinfo {author} {\bibfnamefont {T.}~\bibnamefont {Gorda}},
  \bibinfo {author} {\bibfnamefont {A.}~\bibnamefont {Kurkela}},\ and\ \bibinfo
  {author} {\bibfnamefont {A.}~\bibnamefont {Vuorinen}},\ }\bibfield  {title}
  {\bibinfo {title} {Gravitational-wave constraints on the neutron-star-matter
  equation of state},\ }\href {https://doi.org/10.1103/PhysRevLett.120.172703}
  {\bibfield  {journal} {\bibinfo  {journal} {Phys. Rev. Lett.}\ }\textbf
  {\bibinfo {volume} {120}},\ \bibinfo {pages} {172703} (\bibinfo {year}
  {2018})}\BibitemShut {NoStop}\bibitem [{\citenamefont {Most}\ \emph {et~al.}(2018)\citenamefont {Most},
  \citenamefont {Weih}, \citenamefont {Rezzolla},\ and\ \citenamefont
  {Schaffner-Bielich}}]{gwastro-mergers-ns-MostNewConstraintsRadii2018}\BibitemOpen
  \bibfield  {author} {\bibinfo {author} {\bibfnamefont {E.~R.}\ \bibnamefont
  {Most}}, \bibinfo {author} {\bibfnamefont {L.~R.}\ \bibnamefont {Weih}},
  \bibinfo {author} {\bibfnamefont {L.}~\bibnamefont {Rezzolla}},\ and\
  \bibinfo {author} {\bibfnamefont {J.}~\bibnamefont {Schaffner-Bielich}},\
  }\bibfield  {title} {\bibinfo {title} {New constraints on radii and tidal
  deformabilities of neutron stars from {GW170817}},\ }\href
  {https://doi.org/10.1103/PhysRevLett.120.261103} {\bibfield  {journal}
  {\bibinfo  {journal} {Phys. Rev. Lett.}\ }\textbf {\bibinfo {volume} {120}},\
  \bibinfo {pages} {261103} (\bibinfo {year} {2018})},\ \Eprint
  {https://arxiv.org/abs/1803.00549} {arXiv:1803.00549 [gr-qc]} \BibitemShut
  {NoStop}\bibitem [{\citenamefont {Tews}\ \emph {et~al.}(2018)\citenamefont {Tews},
  \citenamefont {Margueron},\ and\ \citenamefont
  {Reddy}}]{gwastro-TewsCriticalExaminationConstraints2018}\BibitemOpen
  \bibfield  {author} {\bibinfo {author} {\bibfnamefont {I.}~\bibnamefont
  {Tews}}, \bibinfo {author} {\bibfnamefont {J.}~\bibnamefont {Margueron}},\
  and\ \bibinfo {author} {\bibfnamefont {S.}~\bibnamefont {Reddy}},\ }\bibfield
   {title} {\bibinfo {title} {Critical examination of constraints on the
  equation of state of dense matter obtained from {GW170817}},\ }\href
  {https://doi.org/10.1103/PhysRevC.98.045804} {\bibfield  {journal} {\bibinfo
  {journal} {Phys. Rev. C}\ }\textbf {\bibinfo {volume} {98}},\ \bibinfo
  {pages} {045804} (\bibinfo {year} {2018})},\ \Eprint
  {https://arxiv.org/abs/1804.02783} {arXiv:1804.02783 [nucl-th]} \BibitemShut
  {NoStop}\bibitem [{\citenamefont {Tews}\ \emph {et~al.}(2019)\citenamefont {Tews},
  \citenamefont {Margueron},\ and\ \citenamefont
  {Reddy}}]{gwastro-mergers-ns-TewsConfrontingGravitationalWave2019}\BibitemOpen
  \bibfield  {author} {\bibinfo {author} {\bibfnamefont {I.}~\bibnamefont
  {Tews}}, \bibinfo {author} {\bibfnamefont {J.}~\bibnamefont {Margueron}},\
  and\ \bibinfo {author} {\bibfnamefont {S.}~\bibnamefont {Reddy}},\ }\bibfield
   {title} {\bibinfo {title} {Confronting gravitational-wave observations with
  modern nuclear physics constraints},\ }\href
  {https://doi.org/10.1140/epja/i2019-12774-6} {\bibfield  {journal} {\bibinfo
  {journal} {Eur. Phys. J. A}\ }\textbf {\bibinfo {volume} {55}},\ \bibinfo
  {pages} {97} (\bibinfo {year} {2019})},\ \Eprint
  {https://arxiv.org/abs/1901.09874} {arXiv:1901.09874 [nucl-th]} \BibitemShut
  {NoStop}\bibitem [{\citenamefont {Mroczek}\ \emph {et~al.}(2023)\citenamefont
  {Mroczek}, \citenamefont {Miller}, \citenamefont {Noronha-Hostler},\ and\
  \citenamefont {Yunes}}]{Mroczek2023Transitions}\BibitemOpen
  \bibfield  {author} {\bibinfo {author} {\bibfnamefont {D.}~\bibnamefont
  {Mroczek}}, \bibinfo {author} {\bibfnamefont {M.~C.}\ \bibnamefont {Miller}},
  \bibinfo {author} {\bibfnamefont {J.}~\bibnamefont {Noronha-Hostler}},\ and\
  \bibinfo {author} {\bibfnamefont {N.}~\bibnamefont {Yunes}},\ }\href@noop {}
  {\bibinfo {title} {Searching for phase transitions in neutron stars with
  modified gaussian processes}} (\bibinfo {year} {2023}),\ \Eprint
  {https://arxiv.org/abs/2302.07978} {arXiv:2302.07978 [nucl-th]} \BibitemShut
  {NoStop}\bibitem [{\citenamefont {{Risti\'{c}}}\ \emph {et~al.}(2026)\citenamefont
  {{Risti\'{c}}}, \citenamefont {{O'Shaughnessy}}, \citenamefont {{Wagner}},
  \citenamefont {{Fontes}}, \citenamefont {{Fryer}}, \citenamefont
  {{Korobkin}}, \citenamefont {{Mumpower}},\ and\ \citenamefont
  {{Wollaeger}}}]{2026PhRvR...8b3117R}\BibitemOpen
  \bibfield  {author} {\bibinfo {author} {\bibfnamefont {M.}~\bibnamefont
  {{Risti\'{c}}}}, \bibinfo {author} {\bibfnamefont {R.}~\bibnamefont
  {{O'Shaughnessy}}}, \bibinfo {author} {\bibfnamefont {K.}~\bibnamefont
  {{Wagner}}}, \bibinfo {author} {\bibfnamefont {C.~J.}\ \bibnamefont
  {{Fontes}}}, \bibinfo {author} {\bibfnamefont {C.~L.}\ \bibnamefont
  {{Fryer}}}, \bibinfo {author} {\bibfnamefont {O.}~\bibnamefont {{Korobkin}}},
  \bibinfo {author} {\bibfnamefont {M.~R.}\ \bibnamefont {{Mumpower}}},\ and\
  \bibinfo {author} {\bibfnamefont {R.~T.}\ \bibnamefont {{Wollaeger}}},\
  }\bibfield  {title} {\bibinfo {title} {{Joint electromagnetic and
  gravitational wave inference of binary neutron star merger GW170817 using
  forward-modeling ejecta predictions}},\ }\href
  {https://doi.org/10.1103/gffk-xjtg} {\bibfield  {journal} {\bibinfo
  {journal} {Physical Review Research}\ }\textbf {\bibinfo {volume} {8}},\
  \bibinfo {eid} {023117} (\bibinfo {year} {2026})},\ \Eprint
  {https://arxiv.org/abs/2503.12320} {arXiv:2503.12320 [astro-ph.HE]}
  \BibitemShut {NoStop}\bibitem [{\citenamefont {{Abbott}}\ \emph {et~al.}(2021)\citenamefont
  {{Abbott}}, \citenamefont {{Abbott}}, \citenamefont {{Abraham}},
  \citenamefont {{Acernese}}, \citenamefont {{Ackley}}, \citenamefont
  {{Adams}}, \citenamefont {{Adhikari}}, \citenamefont {{Adya}}, \citenamefont
  {{Affeldt}}, \citenamefont {{Agathos}},\ and\ \citenamefont
  {et~al.}}]{ligo-O1O2-opendata}\BibitemOpen
  \bibfield  {author} {\bibinfo {author} {\bibfnamefont {R.}~\bibnamefont
  {{Abbott}}}, \bibinfo {author} {\bibfnamefont {T.~D.}\ \bibnamefont
  {{Abbott}}}, \bibinfo {author} {\bibfnamefont {S.}~\bibnamefont {{Abraham}}},
  \bibinfo {author} {\bibfnamefont {F.}~\bibnamefont {{Acernese}}}, \bibinfo
  {author} {\bibfnamefont {K.}~\bibnamefont {{Ackley}}}, \bibinfo {author}
  {\bibfnamefont {C.}~\bibnamefont {{Adams}}}, \bibinfo {author} {\bibfnamefont
  {R.~X.}\ \bibnamefont {{Adhikari}}}, \bibinfo {author} {\bibfnamefont
  {V.~B.}\ \bibnamefont {{Adya}}}, \bibinfo {author} {\bibfnamefont
  {C.}~\bibnamefont {{Affeldt}}}, \bibinfo {author} {\bibfnamefont
  {M.}~\bibnamefont {{Agathos}}},\ and\ \bibinfo {author} {\bibnamefont
  {et~al.}},\ }\bibfield  {title} {\bibinfo {title} {{Open data from the first
  and second observing runs of Advanced LIGO and Advanced Virgo}},\ }\href
  {https://doi.org/10.1016/j.softx.2021.100658} {\bibfield  {journal} {\bibinfo
   {journal} {SoftwareX}\ }\textbf {\bibinfo {volume} {13}},\ \bibinfo {eid}
  {100658} (\bibinfo {year} {2021})},\ \Eprint
  {https://arxiv.org/abs/1912.11716} {arXiv:1912.11716 [gr-qc]} \BibitemShut
  {NoStop}\bibitem [{\citenamefont {{Miller}}\ \emph {et~al.}(2019)\citenamefont
  {{Miller}} \emph {et~al.}}]{Miller2019}\BibitemOpen
  \bibfield  {author} {\bibinfo {author} {\bibfnamefont {M.~C.}\ \bibnamefont
  {{Miller}}} \emph {et~al.},\ }\bibfield  {title} {\bibinfo {title} {{PSR
  J0030+0451 Mass and Radius from NICER Data and Implications for the
  Properties of Neutron Star Matter}},\ }\href
  {https://doi.org/10.3847/2041-8213/ab50c5} {\bibfield  {journal} {\bibinfo
  {journal} {\apjl}\ }\textbf {\bibinfo {volume} {887}},\ \bibinfo {pages}
  {L24} (\bibinfo {year} {2019})},\ \Eprint {https://arxiv.org/abs/1912.05705}
  {1912.05705} \BibitemShut {NoStop}\bibitem [{\citenamefont {{Vinciguerra}}\ \emph {et~al.}(2024)\citenamefont
  {{Vinciguerra}} \emph {et~al.}}]{Vinciguerra2024}\BibitemOpen
  \bibfield  {author} {\bibinfo {author} {\bibfnamefont {S.}~\bibnamefont
  {{Vinciguerra}}} \emph {et~al.},\ }\bibfield  {title} {\bibinfo {title} {{An
  Updated Mass-Radius Analysis of the 2017-2018 NICER Data Set of PSR
  J0030+0451}},\ }\href {https://doi.org/10.3847/1538-4357/acfb83} {\bibfield
  {journal} {\bibinfo  {journal} {\apj}\ }\textbf {\bibinfo {volume} {961}},\
  \bibinfo {pages} {62} (\bibinfo {year} {2024})},\ \Eprint
  {https://arxiv.org/abs/2308.09469} {2308.09469} \BibitemShut {NoStop}\bibitem [{\citenamefont {{Miller}}\ \emph {et~al.}(2026)\citenamefont
  {{Miller}} \emph {et~al.}}]{MillerJ0437_2025}\BibitemOpen
  \bibfield  {author} {\bibinfo {author} {\bibfnamefont {M.~C.}\ \bibnamefont
  {{Miller}}} \emph {et~al.},\ }\bibfield  {title} {\bibinfo {title} {{The
  Radius of PSR J0437-4715 from NICER Data}},\ }\href@noop {} {\bibfield
  {journal} {\bibinfo  {journal} {\apjl}\ }\textbf {\bibinfo {volume} {1000}},\
  \bibinfo {pages} {L48} (\bibinfo {year} {2026})},\ \Eprint
  {https://arxiv.org/abs/2512.08790} {2512.08790} \BibitemShut {NoStop}\bibitem [{\citenamefont {{Salmi}}\ \emph {et~al.}(2024)\citenamefont {{Salmi}}
  \emph {et~al.}}]{Salmi2024J1231}\BibitemOpen
  \bibfield  {author} {\bibinfo {author} {\bibfnamefont {T.}~\bibnamefont
  {{Salmi}}} \emph {et~al.},\ }\bibfield  {title} {\bibinfo {title} {{A NICER
  View of PSR J1231-1411: A Complex Case}},\ }\href@noop {} {\bibfield
  {journal} {\bibinfo  {journal} {\apj}\ } (\bibinfo {year} {2024})},\ \Eprint
  {https://arxiv.org/abs/2409.14923} {2409.14923} \BibitemShut {NoStop}\bibitem [{\citenamefont {{Saffer}}\ \emph {et~al.}(2025)\citenamefont
  {{Saffer}}, \citenamefont {{Fonseca}}, \citenamefont {{Ransom}},
  \citenamefont {{Stairs}}, \citenamefont {{Lynch}}, \citenamefont {{Good}},
  \citenamefont {{Masui}}, \citenamefont {{McKee}}, \citenamefont {{Meyers}},
  \citenamefont {{Patil}},\ and\ \citenamefont {{Tan}}}]{Saffer2025}\BibitemOpen
  \bibfield  {author} {\bibinfo {author} {\bibfnamefont {A.}~\bibnamefont
  {{Saffer}}}, \bibinfo {author} {\bibfnamefont {E.}~\bibnamefont {{Fonseca}}},
  \bibinfo {author} {\bibfnamefont {S.}~\bibnamefont {{Ransom}}}, \bibinfo
  {author} {\bibfnamefont {I.}~\bibnamefont {{Stairs}}}, \bibinfo {author}
  {\bibfnamefont {R.}~\bibnamefont {{Lynch}}}, \bibinfo {author} {\bibfnamefont
  {D.}~\bibnamefont {{Good}}}, \bibinfo {author} {\bibfnamefont {K.~W.}\
  \bibnamefont {{Masui}}}, \bibinfo {author} {\bibfnamefont {J.~W.}\
  \bibnamefont {{McKee}}}, \bibinfo {author} {\bibfnamefont {B.~W.}\
  \bibnamefont {{Meyers}}}, \bibinfo {author} {\bibfnamefont {S.~S.}\
  \bibnamefont {{Patil}}},\ and\ \bibinfo {author} {\bibfnamefont {C.~M.}\
  \bibnamefont {{Tan}}},\ }\bibfield  {title} {\bibinfo {title} {{A Lower Mass
  Estimate for PSR J0348+0432 Based on CHIME/Pulsar Precision Timing}},\ }\href
  {https://doi.org/10.3847/2041-8213/adc25e} {\bibfield  {journal} {\bibinfo
  {journal} {\apjl}\ }\textbf {\bibinfo {volume} {983}},\ \bibinfo {pages}
  {L20} (\bibinfo {year} {2025})},\ \Eprint {https://arxiv.org/abs/2412.02850}
  {2412.02850} \BibitemShut {NoStop}\bibitem [{\citenamefont {Antoniadis}\ \emph {et~al.}(2013)\citenamefont
  {Antoniadis} \emph {et~al.}}]{Antoniadis2013}\BibitemOpen
  \bibfield  {author} {\bibinfo {author} {\bibfnamefont {J.}~\bibnamefont
  {Antoniadis}} \emph {et~al.},\ }\bibfield  {title} {\bibinfo {title} {A
  massive pulsar in a compact relativistic binary},\ }\href@noop {} {\bibfield
  {journal} {\bibinfo  {journal} {Science}\ }\textbf {\bibinfo {volume}
  {340}},\ \bibinfo {pages} {1233232} (\bibinfo {year} {2013})}\BibitemShut
  {NoStop}\bibitem [{\citenamefont {{Romani}}\ \emph {et~al.}(2026)\citenamefont
  {{Romani}}, \citenamefont {{Beleznay}}, \citenamefont {{Filippenko}},
  \citenamefont {{Brink}},\ and\ \citenamefont {{Zheng}}}]{J0952_2025}\BibitemOpen
  \bibfield  {author} {\bibinfo {author} {\bibfnamefont {R.~W.}\ \bibnamefont
  {{Romani}}}, \bibinfo {author} {\bibfnamefont {M.}~\bibnamefont
  {{Beleznay}}}, \bibinfo {author} {\bibfnamefont {A.~V.}\ \bibnamefont
  {{Filippenko}}}, \bibinfo {author} {\bibfnamefont {T.~G.}\ \bibnamefont
  {{Brink}}},\ and\ \bibinfo {author} {\bibfnamefont {W.}~\bibnamefont
  {{Zheng}}},\ }\bibfield  {title} {\bibinfo {title} {{PSR J0952-0607:
  Tightening a Record-high Neutron Star Mass}},\ }\href
  {https://doi.org/10.3847/1538-4357/ae28c5} {\bibfield  {journal} {\bibinfo
  {journal} {\apj}\ }\textbf {\bibinfo {volume} {996}},\ \bibinfo {pages} {101}
  (\bibinfo {year} {2026})},\ \Eprint {https://arxiv.org/abs/2512.05099}
  {2512.05099} \BibitemShut {NoStop}\bibitem [{\citenamefont {Tsang}\ \emph {et~al.}(2012)\citenamefont {Tsang},
  \citenamefont {Stone}, \citenamefont {Camera}, \citenamefont {Danielewicz},
  \citenamefont {Gandolfi}, \citenamefont {Hebeler}, \citenamefont {Horowitz},
  \citenamefont {Lee}, \citenamefont {Lynch}, \citenamefont {Kohley},
  \citenamefont {Lemmon}, \citenamefont {M{\"o}ller}, \citenamefont {Murakami},
  \citenamefont {Riordan}, \citenamefont {Roca-Maza}, \citenamefont
  {Sammarruca}, \citenamefont {Steiner}, \citenamefont {Vida{\~n}a},\ and\
  \citenamefont {Yennello}}]{Tsang2012}\BibitemOpen
  \bibfield  {author} {\bibinfo {author} {\bibfnamefont {M.~B.}\ \bibnamefont
  {Tsang}}, \bibinfo {author} {\bibfnamefont {J.~R.}\ \bibnamefont {Stone}},
  \bibinfo {author} {\bibfnamefont {F.}~\bibnamefont {Camera}}, \bibinfo
  {author} {\bibfnamefont {P.}~\bibnamefont {Danielewicz}}, \bibinfo {author}
  {\bibfnamefont {S.}~\bibnamefont {Gandolfi}}, \bibinfo {author}
  {\bibfnamefont {K.}~\bibnamefont {Hebeler}}, \bibinfo {author} {\bibfnamefont
  {C.~J.}\ \bibnamefont {Horowitz}}, \bibinfo {author} {\bibfnamefont
  {J.}~\bibnamefont {Lee}}, \bibinfo {author} {\bibfnamefont {W.~G.}\
  \bibnamefont {Lynch}}, \bibinfo {author} {\bibfnamefont {Z.}~\bibnamefont
  {Kohley}}, \bibinfo {author} {\bibfnamefont {R.}~\bibnamefont {Lemmon}},
  \bibinfo {author} {\bibfnamefont {P.}~\bibnamefont {M{\"o}ller}}, \bibinfo
  {author} {\bibfnamefont {T.}~\bibnamefont {Murakami}}, \bibinfo {author}
  {\bibfnamefont {S.}~\bibnamefont {Riordan}}, \bibinfo {author} {\bibfnamefont
  {X.}~\bibnamefont {Roca-Maza}}, \bibinfo {author} {\bibfnamefont
  {F.}~\bibnamefont {Sammarruca}}, \bibinfo {author} {\bibfnamefont {A.~W.}\
  \bibnamefont {Steiner}}, \bibinfo {author} {\bibfnamefont {I.}~\bibnamefont
  {Vida{\~n}a}},\ and\ \bibinfo {author} {\bibfnamefont {S.~J.}\ \bibnamefont
  {Yennello}},\ }\bibfield  {title} {\bibinfo {title} {Constraints on the
  symmetry energy and neutron skins from experiments and theory},\ }\href
  {https://doi.org/10.1103/PhysRevC.86.015803} {\bibfield  {journal} {\bibinfo
  {journal} {Phys. Rev. C}\ }\textbf {\bibinfo {volume} {86}},\ \bibinfo
  {pages} {015803} (\bibinfo {year} {2012})},\ \Eprint
  {https://arxiv.org/abs/1204.0466} {arXiv:1204.0466} \BibitemShut {NoStop}\bibitem [{\citenamefont {Alford}\ \emph {et~al.}(2013)\citenamefont {Alford},
  \citenamefont {Han},\ and\ \citenamefont {Prakash}}]{AlfordHanPrakash2013}\BibitemOpen
  \bibfield  {author} {\bibinfo {author} {\bibfnamefont {M.~G.}\ \bibnamefont
  {Alford}}, \bibinfo {author} {\bibfnamefont {S.}~\bibnamefont {Han}},\ and\
  \bibinfo {author} {\bibfnamefont {M.}~\bibnamefont {Prakash}},\ }\bibfield
  {title} {\bibinfo {title} {Generic conditions for stable hybrid stars},\
  }\href {https://doi.org/10.1103/PhysRevD.88.083013} {\bibfield  {journal}
  {\bibinfo  {journal} {Phys. Rev. D}\ }\textbf {\bibinfo {volume} {88}},\
  \bibinfo {pages} {083013} (\bibinfo {year} {2013})},\ \Eprint
  {https://arxiv.org/abs/1302.4732} {arXiv:1302.4732} \BibitemShut {NoStop}\bibitem [{\citenamefont
  {{O'Shaughnessy}}(2026)}]{whitepaper-transparent-agentic-science}\BibitemOpen
  \bibfield  {author} {\bibinfo {author} {\bibfnamefont {R.}~\bibnamefont
  {{O'Shaughnessy}}},\ }\href@noop {} {\bibinfo {title} {Transparent agentic
  science: Documentation, reliability, and scientific judgment}},\ \bibinfo
  {howpublished}
  {\url{https://github.com/oshaughnessy-junior/whitepaper-transparent-agentic-science}}
  (\bibinfo {year} {2026}),\ \bibinfo {note} {technical note and workflow
  skills}\BibitemShut {NoStop}\end{thebibliography}
\end{document}


\title{Supplemental Material: Adaptive proposals for nonparametric equation-of-state inference with RIFT}
\author{R.~O'Shaughnessy}\affiliation{Rochester Institute of Technology}
\author{Z.~Zhu}\affiliation{Rochester Institute of Technology}
\maketitle
This supplement develops the stellar-solver accuracy check
(Sec.~\ref{app:numerics}).  The sampling-adequacy check for an explicit
transition-EOS mass--radius likelihood is in
Appendix~{E} of the main text.  We then establish
what the compressed observables retain (Sec.~\ref{app:dimension}) and account
for proposal support, normalization, and approximation error
(Sec.~\ref{app:operating}).  References to sections and figures in the paper are identified explicitly.

\section{Numerical foundations and solver validation}\label{app:numerics}
\subsection{Accuracy and fixed-mass derivatives}
\begin{figure}[t]\centering
  \includegraphics[width=\columnwidth]{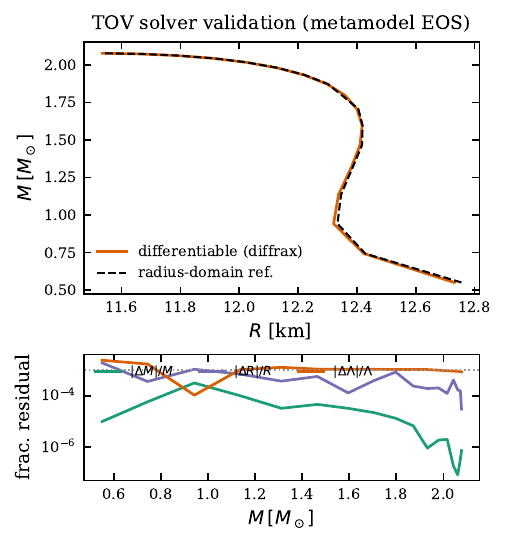}
  \caption{\label{fig:val_tov} TOV solver validation for a metamodel EOS. Top:
  $M(R)$ from the differentiable (\texttt{diffrax}) solver and the independent
  radius-domain reference. Bottom: fractional residuals in $M$, $R$, and
  $\Lambda$ along the stable branch; the dotted line marks the validation
  tolerance.\llmfig}
\end{figure}

Solver systematics are calibrated against three independent references, since EOS
inference at current and next-generation~\cite{WalkerEtAl2024,
BandopadhyayEtAl2024,WilliamsEtAl2026}
sensitivities demands they sit well below statistical errors.
(i)~An in-repository, deliberately distinct radius-domain integrator (SciPy RK45,
$(H,\beta)$ tidal system) is a regression target; the differentiable solver agrees
with it to $M<0.05\%$, $R<0.02\%$, and $\Lambda<0.1\%$.  Figure~\ref{fig:val_tov}
shows that comparison for a metamodel EOS: the mass--radius sequences overlie,
and the fractional residuals along the stable branch stay below the production
tolerance.  (ii)~Named tabulated EOS
are cross-checked against the LALSuite~\cite{lalsuite-soft} TOV routines used by
the downstream RIFT consumer.  (iii)~Against RePrimAnd~\cite{Kastaun2021reprimand},
the thermodynamically consistent surface continuation reduces the radius residual
from $\sim200$\,m to $\sim3$--$10$\,m.  Fixed-mass sensitivities
(Sec.~{II\,A\,1}) are checked against finite differences with the
central enthalpy re-rooted to the target mass for every perturbed EOS.
Agreement is within a few percent for tidal derivatives and below one percent
for radius and maximum-mass derivatives.

\subsection{Parametric metamodel for the differentiated forecast}\label{app:metamodel}
The forecast of Sec.~{II\,F} uses a parametric EOS.  The same
metamodel supplies the low-density part of the bridge ensemble, with its empirical
parameters resampled per draw (Sec.~{II\,A\,2}).  With
$x=(n-\nsat)/(3\nsat)$, the isoscalar energy per nucleon and the symmetry energy are
Taylor expansions about saturation~\cite{Margueron2018}:
\begin{align}
e_{\rm sat}(n)&=E_{\rm sat}+\tfrac12K_{\rm sat}x^2+\tfrac16Q_{\rm sat}x^3,\nonumber\\
S(n)&=E_{\rm sym}+L_{\rm sym}x+\tfrac12K_{\rm sym}x^2+\tfrac16Q_{\rm sym}x^3.
\label{eq:metamodel}
\end{align}
For proton fraction $x_p$ and $\delta=1-2x_p$, the energy density is
\begin{equation}
\epsilon=n\left[x_pm_p+(1-x_p)m_n+e_{\rm sat}(n)+S(n)\delta^2\right]
+\epsilon_e(x_pn),
\end{equation}
with $\epsilon_e$ a relativistic electron gas.  Beta equilibrium,
$\mu_e=m_n-m_p+4S\delta$, fixes $x_p(n)$.  Pressure follows from
$p=n\,\partial\epsilon/\partial n|_{x_p}-\epsilon$.  We differentiate through
$x_p$ with the implicit derivative at the converged root.

Above $n_m=\result{MatchNsat}\nsat$, with matching values $\epsilon_m$ and $p_m$, the map appends a
constant-sound-speed segment:
\begin{align}
p&=p_m+\cs^2(\epsilon-\epsilon_m),\nonumber\\
n&=n_m\left(\frac{\epsilon+p}{\epsilon_m+p_m}\right)^{1/(1+\cs^2)}.
\label{eq:csstail}
\end{align}
Thus $L_{\rm sym}$ and the other varied parameters act above $n_m$ only through
$(n_m,\epsilon_m,p_m)$.  This segment has no random structure, does not impose the
perturbative-QCD endpoint, and has no counterpart to the bridge correlation length.
We vary $(E_{\rm sym},L_{\rm sym},K_{\rm sym},K_{\rm sat})$ and fix $\nsat=\result{MetaNsat}$\,fm$^{-3}$,
$E_{\rm sat}=\result{MetaEsat}$\,MeV, $Q_{\rm sat}=\result{MetaQsat}$\,MeV, $Q_{\rm sym}=\result{MetaQsym}$, and $\cs^2=\result{AdCsSqTail}$.
Below $n=\result{MetaNmin}$\,fm$^{-3}$ the map continues to the surface with a
polytrope of index $\Gamma=\result{MetaGammaCrust}$, a polytropic crust as in the archive.  As a
sensitivity test, the forecast also uses the SLy crust below $n_t=\result{CrustNt}$\,fm$^{-3}$.
We evaluate the Jacobian by central differences of $\ln\Lambda$ at fixed mass, solving
for the central enthalpy at each perturbed EOS, with a step of $\result{AdFdStep}$ prior widths.
Steps from $0.01$ to $0.1$ prior widths agree.  With the SLy crust, forward-mode
derivatives through the implicit-function theorem give the same Jacobian.  With the
polytropic continuation they do not, so we use differences there.

\subsection{Nuclear-parameter conditioning checks}
\label{app:symmetry-checks}
The numerical checks supporting the large-catalogue interpretation in
Sec.~{III\,B} use the local conditioning of
Sec.~{II\,F}.  We report the kernel sensitivity,
leave-one-out calibration, cross-model comparison, and mass-coverage tests
here to separate validation details from the main physical result.

The catalogue measures $\Lambda(1.2)$ and $\Lambda(1.4)$ to
\result{SymLamTwelvePct}\% and \result{SymLamFourteenPct}\%, with correlation
\result{SymLamCorr}.  We condition the nuclear parameters on this measurement with
Eq.~\textup{(32)}, using $h=\result{SymKernelH}$
($n_{\rm eff}=\result{SymKernelEss}$).  The slope narrows from
$\sigma_{L_{\rm sym}}=\result{SymSlopePriorSd}$ to \result{SymSlopePostSd}\,MeV.
Kernel widths from \result{SymHMin} to \result{SymHMax} give
\result{SymSlopeHMin}--\result{SymSlopeHMax}\,MeV.  The curvature narrows from
$\sigma_{K_{\rm sym}}=\result{SymCurvPriorSd}$ to \result{SymCurvPostSd}\,MeV, with an
$L_{\rm sym}$--$K_{\rm sym}$ correlation of $\result{SymPostCorr}$, and $E_{\rm sym}$
and $K_{\rm sat}$ narrow by less than $2$\,MeV.  A perfect measurement of the same two
masses would give \result{SymSlopePerfectSd}\,MeV, the scatter of $L_{\rm sym}$
among draws with the same $\Lambda(1.2)$ and $\Lambda(1.4)$; Fig.~{5}a
shows the width approaching this floor.

The leave-one-out test covers the \result{SymLooN} archive draws near the measured
point.  Its rms error in $L_{\rm sym}$ is \result{SymLooRms}\,MeV, against a
predicted \result{SymLooPred}\,MeV, with a bias of \result{SymLooBias}\,MeV.  The
global linear-Gaussian conditioning of all draws claims
\result{SymGlobalSd}\,MeV for the same measurement.  Its actual rms error near this
point is \result{SymGlobalLooRms}\,MeV, with a bias of \result{SymGlobalLooBias}\,MeV.
The posterior mean, $L_{\rm sym}=\result{SymSlopePostMean}$\,MeV, lies
\result{SymSlopeTruthZ}$\sigma$ above the injected \result{SymSlopeTruth}\,MeV.  The
leave-one-out test shows no comparable mean bias.  This single injection
does not establish coverage across the EOS prior.

We cross-check this result with the parametric EOS model of
Sec.~{II\,F}, which has a fixed constant-sound-speed tail above
$n_L$.  For $\Lambda(1.4)$ alone, measured to about $\result{SymLamFourteenPct}\%$,
Eq.~\textup{(34)} with the polytropic crust gives
$\sigma_{L_{\rm sym}}=\result{SymSlopeAdOneMass}$\,MeV, and
Eq.~\textup{(32)} gives \result{SymSlopeOneMassSd}\,MeV.  For both masses
with the measured covariance, Eq.~\textup{(34)} gives
$\sigma_{L_{\rm sym}}=\result{SymFisherSlopeSd}$\,MeV,
$\sigma_{K_{\rm sym}}=\result{SymFisherCurvSd}$\,MeV and an
$L_{\rm sym}$--$K_{\rm sym}$ correlation of $\result{SymFisherCorr}$.  The samples give
\result{SymSlopePostSd}\,MeV, \result{SymCurvPostSd}\,MeV and $\result{SymPostCorr}$.
In both calculations two values of $\Lambda$ constrain one combination of slope and
curvature.  Figure~{5} overlays the forecast, which matches the
sample-based posterior in orientation and nearly in width.

We repeat the conditioning with different sets of masses, each measured to $\result{SymScanPct}\%$
about the injected EOS.  A single $\Lambda(1.4)$ gives
$\sigma_{L_{\rm sym}}=\result{SymSlopeSingleMass}$\,MeV.  Adding $\Lambda(1.2)$ gives
\result{SymSlopePairMass}\,MeV.  Adding instead a lower-density probe at
$1.0\,\msun$ gives \result{SymSlopeLowMass}\,MeV.  The gain comes from measuring how
$\Lambda$ changes with mass.  Measuring $\result{SymSpanMasses}\,\msun$ gives
\result{SymSlopeSpanMass}\,MeV, wider than two masses.  Its leave-one-out rms error,
\result{SymSpanLooRms}\,MeV, matches the predicted \result{SymSpanLooPred}\,MeV, so
the width is calibrated; we have not isolated why the further masses widen it with
this archive.  This archive-based test does not establish a general limit on the
information supplied by low-mass neutron stars.

\subsection{Stellar integration and phase matching}
Writing the auxiliary tidal variable as
$y\equiv rH'/H$ with $H$ the metric perturbation~\cite{Postnikov2010}, and
defining
\begin{align}
F(r) &= \frac{1 - 4\pi r^2(\epsilon - p)}{1 - 2m/r}, \label{eq:tidalF}\\
Q(r) &= \frac{4\pi}{1-2m/r}\!\left(5\epsilon + 9p + \frac{\epsilon + p}{\cs^2}
- \frac{6}{4\pi r^2}\right)
\nonumber \\ &
        - \left[\frac{2(m + 4\pi r^3 p)}{r^2(1-2m/r)}\right]^2,
        \label{eq:tidalQ}
\end{align}
the tidal variable obeys the Riccati equation
\begin{equation}
r\,\frac{dy}{dr} + y^2 + y\,F(r) + r^2 Q(r) = 0 .
\label{eq:tidaly}
\end{equation}
The term $(\epsilon + p)/\cs^2$ in Eq.~\eqref{eq:tidalQ} enters because the
fluid's response to the external tidal field is governed by
$d\epsilon/dp = 1/\cs^2$.  At the surface the Love number $k_2(y,C)$ follows
from $y$ and the compactness $C=M/R$ via the standard
expression~\cite{Hinderer2008,Postnikov2010,DamourNagar2009}.

\emph{Branches, transitions, and the surface.}  A smooth approximation to a strong first-order transition has small
$\cs^2$ and stiffens the integrand.  An exact jump instead occupies a single
enthalpy and requires the density-discontinuity condition
\begin{equation}
y_{\rm out} = y_{\rm in}
  - \frac{4\pi r_d^3\,(\epsilon_{\rm in}-\epsilon_{\rm out})}
         {m_d + 4\pi r_d^3 p_t},
\label{eq:yjunction}
\end{equation}
where $\epsilon_{\rm in}>\epsilon_{\rm out}$ for outward integration.
The junction is evaluated at the discontinuity radius $r_d$, enclosed mass $m_d$, and transition
pressure $p_t$.  A dedicated hybrid solver applies Eq.~\eqref{eq:yjunction} for a
single transition~\cite{TakatsyKovacs2020}; the default adaptive solver is
validated against RePrimAnd for smooth EOS.  At the stellar surface we optionally append a single-polytrope
continuation as a \emph{convention} for the bare metamodel, used only for
cross-code comparison; the solver hardcodes no low-density approximation, and realistic crust models extending to normal-matter densities can be
supplied directly.

The TOV integrand requires $(\epsilon,p,\cs^2)$ as functions of $h$.  We evaluate
them with a thermodynamically consistent cubic-Hermite interpolation of $p(h)$
whose node slopes are set to the exact first law $dp/dh = \epsilon + p$; the
energy density is then \emph{derived} as $\epsilon = dp/dh - p$ rather than
interpolated independently, so the table is thermodynamically consistent by
construction.  Relative to independent log-linear interpolation at equal table
resolution this improves $\Lambda$ accuracy by roughly $25\times$.  Tables
truncated near neutron drip underestimate $R$ by $\sim100$--$200$\,m purely
through the integration-stop convention; the optional single-polytrope surface
continuation derived from the cold first law (Sec.~{II\,A\,1}) closes this gap
for cross-code comparisons but represents a convention, not physics.

\section{Representation and observational information}
\label{app:dimension}
The information diagnostics use the smoothed bridge ensemble and its compressed
stellar sequences.  We first state the physical support and reconstruction
accuracy of that representation, then test sound-speed recovery and local
consistency in the directions measured by an experiment.

\subsection{Thermodynamic support and compression}
The unsmoothed bridge satisfies the Komoltsev--Kurkela consistency conditions.
Diffusive smoothing preserves causality but changes the pressure integral in
Eq.~\textup{(8)}, shifting the pQCD endpoint by a few percent
at production settings~\cite{OShaughnessyZhu2026prior}.  The ensemble retains
sound speeds close to the causal limit, while $\mu$ and $p$ remain monotone
and $d\epsilon/dn=\mu$ holds to numerical precision.

\result{SeqComponents} PCA components per channel reconstruct the full $\Lambda(m)$ curve with
$\result{PcaLamRmsErrPct}\%$ rms and $\result{PcaLamMaxErrPct}\%$ maximum relative error for
90\% of the bridge ensemble (Fig.~{6}).

\subsection{Sound-speed recovery and held-out prediction}
The analysis behind Fig.~{7} asks how much sound-speed
structure the observable coefficients retain.  For draw $i$, let $a_{ik}$ be
the principal-component amplitudes of $c_{s,i}^2(u)$, with $u$ the normalized
central-enthalpy coordinate of Eq.~\textup{(13)}.  We fit each
$a_{ik}$ linearly from the $\result{DimNumRegressors}$ observable coefficients: \result{SeqComponents} each for
$M$, $R$, and $\log\Lambda$.  If mode $k$ carries fraction $f_k$ of the
sound-speed variance and the fit explains fraction $R_k^2$ of its amplitude
variance, the unrecovered fraction is
\begin{equation}
 \mathcal U=\sum_k f_k(1-R_k^2).
\label{eq:unrecovered}
\end{equation}
The main-text value initially uses the same $N_{\rm EOS}=\result{GenZeroDraws}$
draws for fitting and evaluation.  Held-out tests distinguish this descriptive
recovery from prediction on new EOS.

Let $\mathbf X$ contain the $\result{DimNumRegressors}$ observable coefficients and a constant column.
Its least-squares projection is
\begin{equation}
 \mathbf P_X=\mathbf X(\mathbf X^T\mathbf X)^{-1}\mathbf X^T.
\label{eq:hatmatrix}
\end{equation}
The diagonal element $\ell_i=(\mathbf P_X)_{ii}$, or leverage, identifies
isolated draws whose removal requires extrapolation.  Such draws make
held-out predictions sensitive to the fold assignment.

We therefore remove the largest-leverage draw, recompute the projection, and
repeat until $\max_i\ell_i<\result{DimLeverageMax}$.  This geometry-only rule removes
\result{DimLevDropped} of \result{GenZeroDraws} draws and gives a stable
held-out unrecovered fraction of $\result{DimCvPruned}\%$.  The screen uses only coefficient geometry and is applied to the reconstruction
test.

Resampling and three independently generated, TOV-solved ensembles recover
similar results: about one fifth of the sound-speed variation remains
unrecovered in held-out predictions.  The inverse regression characterizes
sound-speed recovery within this family; the forward-map test below instead
compares local EOS changes with an explicit observational metric.

\subsection{Local EOS consistency of sampled stellar sequences}\label{supp:consistency}
This is an optional diagnostic, not a filter used in the reported analyses.
At a TOV-solved reference EOS, let
$\mathsf J_{ai}=\partial c_{{\rm obs},a}/\partial c_{{\rm eos},i}$.
The matrix contains derivatives of sequence coefficients with respect to EOS
coefficients.  A small physical change has the form
$\delta c_{{\rm obs},a}=\sum_i\mathsf J_{ai}\delta c_{{\rm eos},i}$.
To test a proposed sequence, choose the EOS perturbation that minimizes its
squared residual, with observational uncertainties defining the scale:
\begin{equation}
D_\perp^2=\min_{\delta c_{\rm eos}}\sum_{ab}W_{ab}\,r_a r_b,
\end{equation}
where
\begin{equation}
r_a=\delta c_{{\rm obs},a}-\sum_i\mathsf J_{ai}\delta c_{{\rm eos},i}.
\end{equation}
Equivalently, the weighted projection onto the reachable directions is
\begin{align}
\mathsf P_J &=\mathsf J(\mathsf J^{\!\top}\mathsf W\mathsf J)^+
\mathsf J^{\!\top}\mathsf W,\label{eq:PJ}\\
r_\perp &=(\mathsf I-\mathsf P_J)\delta c_{\rm obs},\qquad
D_\perp^2=r_\perp^{\!\top}\mathsf W r_\perp.\label{eq:subspaceconsistency}
\end{align}
Here $+$ is a pseudoinverse.  The residual is the part of the proposed change
that no nearby EOS perturbation in the retained basis can produce.

\subsubsection{Measurement scale and limits of the test}
The metric compares physical quantities at their measurement accuracies.
We evaluate the combinations used by the likelihood, such as
$\widetilde\Lambda(m_1,m_2)$ for a binary, the correlated $(M,R)$ constraints
for a pulsar, or $M_{\rm max}$.  If these quantities are $z_\alpha$ and their
measurement covariance is $C_{\alpha\beta}^{\rm meas}$, the corresponding
metric on the observable coefficients is
\begin{equation}
 W_{ab}=
 \frac{\partial z_\alpha}{\partial c_{{\rm obs},a}}
 \left(C^{\rm meas}+C^{\rm TOV}\right)^{+}_{\alpha\beta}
 \frac{\partial z_\beta}{\partial c_{{\rm obs},b}},
\label{eq:metricW}
\end{equation}
with repeated indices summed.  Here $C^{\rm TOV}$ is the covariance of the held-out residuals
$\delta z_\alpha-(\partial z_\alpha/\partial c_{{\rm obs},a})
J_{ai}\delta c_{{\rm eos},i}$, and accounts for local linearization and basis
error in the measured quantities themselves.  Equation~\eqref{eq:metricW}
therefore includes the experimental scales and correlations; the resulting
distance is unchanged by a change of units.  In a forecast, $C^{\rm meas}$ must instead be a stated
design covariance at representative masses; the result is then specific to
that experiment.

The threshold is calibrated using held-out TOV residuals and measurement
errors at a chosen false-rejection rate.  The test checks compatibility with
nearby EOS perturbations; large changes require an updated Jacobian or an
explicit TOV solve.

\section{Proposal support and inference accounting}\label{app:operating}
The numerical and coverage checks above concern the model and its sampling.
The remaining distinctions concern what distribution is estimated: fixed-prior
importance sampling, validity-conditioned proposals, and sequential Gaussian
projection need separate normalizations and error checks.

\subsection{Refinement schedule and Monte Carlo precision}
The synthetic headline run uses a redraw threshold $N_{\rm eff}^{\min}=\result{HeroRefitThreshold}$
and batches of $\result{HeroDrawsPerGen}$ draws.  Thresholds and batch sizes balance posterior
precision against evaluation cost; in observational analyses, the per-EOS RIFT
evidence dominates this cost.

For approximately Gaussian sampling, the fractional Monte Carlo uncertainty
of a standard deviation is $[2(N_{\rm eff}-1)]^{-1/2}$.  For a posterior
quantile $x_p$, the sampling error is approximately
$\sqrt{p(1-p)/N_{\rm eff}}/f(x_p)$, where $f$ is the posterior density.
These estimates guide checks of moment and credible-interval stability.

\subsection{Validity truncation and proposal normalization}
\label{app:truncation}
Rejection of invalid sequences changes each Gaussian proposal's normalization.
If batch $g$ has acceptance probability $P_g$, its density on the accepted set
is $q_g^A=q_g/P_g$.  The probabilities generally differ between batches and
must be included in Eq.~\textup{(24)}, as follows.

Let $A(c)$ be one when the reconstructed sequence is finite, has positive mass
and radius, and contains a nontrivial rising-mass branch, and zero otherwise.
The operational base prior is therefore the Gaussian conditioned on this valid
set,
\begin{equation}
  \pi_0^A(c)=\frac{\pi_0(c)A(c)}{P_0},
  \qquad P_0=\int dc\,\pi_0(c)A(c).
\label{eq:truncatedbase}
\end{equation}
The conditioned prior includes these structural restrictions.

Rejection sampling from proposal $q_g$ produces the density
$q_g^A(c)=q_g(c)A(c)/P_g$, where $P_g=\int dc\,q_g(c)A(c)$ is its acceptance
probability.  Consequently the proposal mixture that actually generated the
pooled sample is
\begin{equation}
 q_{\rm mix}^A(c)=\sum_g \alpha_g\frac{q_g(c)A(c)}{P_g},
 \qquad \alpha_g=\frac{N_g}{\sum_{g'}N_{g'}},
\label{eq:truncatedmix}
\end{equation}
and, on the accepted set, the fixed-target weight is equivalently
\begin{equation}
 w(c)\propto
 \mathcal{L}(c)\,
 \frac{\pi_0(c)}{\sum_g\alpha_g q_g(c)/P_g}.
\label{eq:truncatedweight}
\end{equation}
For a single proposal $P_g$ is a common factor and cancels from self-normalized
posterior moments.  It also cancels across a mixture when all $P_g$ are equal;
otherwise each generation's normalization must be retained as in
Eq.~\eqref{eq:truncatedweight}.  The target normalization $P_0$ likewise cancels
from posterior moments, although it would be required for an absolute evidence
under the conditioned prior.

A separate approximation enters when the initial draws are reused from the
TOV-solved calibration archive: we represent that archive's density by the
fitted coefficient Gaussian.  The importance correction is exact for draws
from the stated conditioned Gaussian densities, but it cannot remove this
calibration-archive approximation.  For GW170817 the approximation fails: the
multi-branch draws lie outside the core of the fitted Gaussian, and the
two-generation weights concentrate on a few draws.  The main-text GW170817
results therefore weight the archive directly, with selected evidence runs
(main-text Appendix~{A\,3}).

\begin{table}[t]
\caption{\label{tab:truncation}
Checks on validity truncation.  The last row bounds the combined error of the
sequential Gaussian projection, including truncation.}
\begin{ruledtabular}
\begin{tabular}{p{0.31\columnwidth}p{0.61\columnwidth}}
Check & Outcome \\
\hline
Accepted support & Finite positive $M$ and $R$ and a nontrivial rising branch;
no $\Lambda(C)$ veto. \\
GW170817 proposals &
$P_g=\result{CombineAcceptLo}$ and $\result{CombineAcceptHi}$. \\
Single-Gaussian combine &
$\Lambda(1.4)=\result{CombineTotalLam}\pm\result{CombineTotalLamSd}$,
$N_{\rm eff}=\result{CombineTotalEss}$; the archive approximation dominates. \\
Normalization ablation &
Omitting $P_g$ leaves the mean and width unchanged at the quoted precision. \\
Sequential projection &
Against the converged fixed-prior reference, the $\gamma=0$ run has a mean
that differs by $\result{HeroAdaptiveMeanDeltaPct}\%$ and a standard deviation
$\result{HeroAdaptiveWidthRatio}$ times larger. \\
\end{tabular}
\end{ruledtabular}
\end{table}

Validity rejection increases proposal cost by roughly $1/P_g$.  The
sequential catalogue calculation additionally replaces the accumulated
posterior by a Gaussian moment projection at each refit
(Sec.~{II\,D\,1}).  The fixed-prior comparisons in
Sec.~{III\,C} quantify the resulting error.

\subsection{Structural support and the maximum-mass prior}\label{supp:proposal-support}
The structural filters require finite, positive curves and a nontrivial
rising-mass branch.  For the broad fitted Gaussian,
$\result{BaseStructuralRejectPct}\%$ of attempted draws fail these checks;
the acceptance normalization is included as above.

\begin{figure}[!ht]\centering
  \includegraphics[width=\columnwidth]{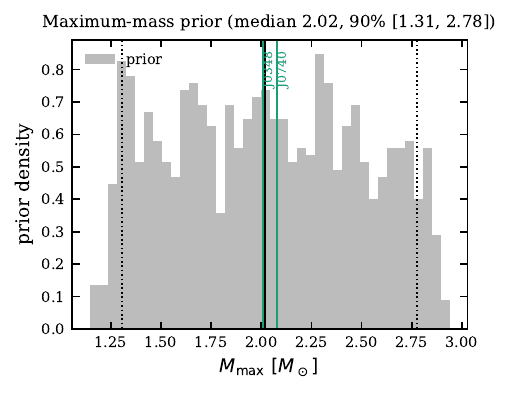}
  \caption{\label{fig:val_mmax} Maximum-mass prior of the (filtered) bridge
  ensemble: broad by construction (median $\result{MmaxPriorMedian}\,\msun$;
  $90\%$ $[\result{MmaxPriorLo},\result{MmaxPriorHi}]$). The
  radio maximum-mass factors, rather than a hard prior cut, downweight the soft tail
  (Sec.~{III}).\llmfig}
\end{figure}

The companion quantity is the prior on the maximum mass (Fig.~\ref{fig:val_mmax}).
It is set by the high-density bridge rather than the sub-saturation symmetry
sector and is deliberately broad, with median $\result{MmaxPriorMedian}\,\msun$
and $90\%$ range $[\result{MmaxPriorLo},\result{MmaxPriorHi}]\,\msun$.  We impose
no maximum-mass cut: radio timing factors downweight EOS that do not support the
precisely measured massive pulsars during inference (Sec.~{III}).

\subsection{Compactness--Love consistency}
\label{app:other-rejection}
Compactness--Love relations provide a diagnostic of the fitted cross-channel
covariance.  Their EOS-dependent residuals
\cite{YagiYunes2013,SuleimanRead2024,LegredEtAl2024Universal,WilliamsEtAl2026}
make a population-fitted relation unsuitable as a universal physical veto.
We retain tidal deformability as an independent channel and apply only the
structural filters above.